\documentclass[11pt,a4paper]{report}

\usepackage[bitstream-charter]{mathdesign} 

\usepackage{enumerate}
\usepackage{lipsum}

\usepackage[parfill]{parskip} 		
\usepackage{amsmath}				
\DeclareMathAlphabet{\pazocal}{OMS}{zplm}{m}{n}
\numberwithin{equation}{section}	

\usepackage[nottoc]{tocbibind}			
\usepackage{breakcites}					

\usepackage[yyyymmdd]{datetime}

\usepackage{fancyhdr}
\usepackage{graphicx}
\graphicspath{{img/}}

\usepackage{placeins} 			

\usepackage{caption}			
\usepackage{subcaption}

\usepackage{tikz}
\usepgfmodule{nonlineartransformations}
\usetikzlibrary{curvilinear,patterns,decorations.pathreplacing,spy,calc}
\usepackage{pgfplots}
\usepackage{pgfplotstable}
\usepgfplotslibrary{groupplots}
\pgfplotsset{/pgfplots/table/search path={dat}}

\definecolor{green}{RGB}{0,150,0}
    
\usepackage{hyperref}
\hypersetup{
    colorlinks=true,
    linkcolor=blue,
    filecolor=blue,      
    urlcolor=blue,
    citecolor=darkgray,
    pdftitle={Overleaf Example},
    pdfpagemode=FullScreen,
    }

\usepackage{listings}

\definecolor{codegreen}{rgb}{0,0.6,0}
\definecolor{codegray}{rgb}{0.5,0.5,0.5}
\definecolor{codepurple}{rgb}{0.6,0.2,0.6}
\definecolor{backcolour}{rgb}{0.96,0.96,0.96}

\lstdefinestyle{mystyle}{
    backgroundcolor=\color{backcolour},   
    commentstyle=\color{codegray},
    keywordstyle=\color{codepurple},
    numberstyle=\tiny\color{codegray},
    stringstyle=\color{codepurple},
    basicstyle=\ttfamily\footnotesize,
    breakatwhitespace=false,         
    breaklines=true,                 
    captionpos=b,                    
    keepspaces=true,                 
    numbers=left,                    
    numbersep=5pt,                  
    showspaces=false,                
    showstringspaces=false,
    showtabs=false,                  
    tabsize=2
}

\lstdefinelanguage{OpenCL}
{
  language     = C++,
  morekeywords = {float3}
}

\usepackage{bm}				

\renewcommand{\b}{\mathbf{b}}      
\newcommand{\f}{\mathbf{f}}      
\newcommand{\g}{\mathbf{g}}  
\newcommand{\n}{\mathbf{n}}     
\renewcommand{\r}{\mathbf{r}}   
\renewcommand{\t}{\mathbf{t}}       
     
\renewcommand{\u}{\mathbf{u}}    
\renewcommand{\v}{\mathbf{v}}   
\newcommand{\w}{\mathbf{w}}   
\newcommand{\x}{\mathbf{x}} 

\newcommand{\0}{\mathbf{0}}     
\newcommand{\A}{\mathbf{A}}  
\newcommand{\B}{\mathbf{B}}  
\newcommand{\C}{\mathbf{C}}    
\newcommand{\E}{\mathbf{E}}   
\newcommand{\F}{\mathbf{F}}   
\newcommand{\N}{\mathbf{N}}
\renewcommand{\P}{\mathbf{P}}   
\renewcommand{\S}{\mathbf{S}}   
   
\newcommand{\U}{\mathbf{U}}

\newcommand{\R}{\mathbb{R}}
\renewcommand{\N}{\mathbb{N}}
\newcommand{\I}{\mathbb{I}}          
     
\renewcommand{\hat}{\widehat}   

\renewcommand{\phi}{\varphi}   

\newcommand{\SIG}{\bm{\Sigma}}

\DeclareMathOperator{\tr}{tr}

\begin{document}

\title{Computational Tools for Cardiac Simulation\\GPU-Parallel Multiphysics}   	
\author{Toby Simpson}         
\date{\today}   


\begin{titlepage}
\begin{center}
\vfill
\Large\textbf{Computational Tools for Cardiac Simulation}\\[2mm]  
{GPU-Parallel Multiphysics}\\
\vspace{1.0cm}
\normalsize{Doctoral Dissertation submitted to the \\[2mm]
Faculty of Informatics of the Universit\`a della Svizzera Italiana}\\[2mm]
{in partial fulfillment of the requirements for the degree of}\\[2mm]
{Doctor of Philosophy \\[32mm]}
\normalsize{presented by}\\[2mm]
\Large{Toby Simpson}
\vfill
\normalsize{under the supervision of}\\[2mm]
\Large{Rolf Krause}
\vfill
\normalsize{January 2022}\\
\end{center}
\end{titlepage}

\newpage
\begin{abstract}

Cardiovascular disease affects millions of people worldwide and its social and economic cost clearly motivates scientific research. Computer simulation can lead to a better understanding of cardiac physiology, and for pathology presents opportunities for low-cost and low-risk design and testing of therapies, including surgical and pharmacological intervention as well as automated diagnosis and screening.

Currently, the simulation of a whole heart model, including the interaction of electrophysiology, solid mechanics and fluid dynamics is the subject of ongoing research in computational science. Typically, the  computation of a single heartbeat requires many processor hours on a supercomputer. The financial and ultimately environmental cost of such a computation prevents it from becoming a viable clinical or research solution. 

We re-formulate the standard mathematical models of continuum mechanics, such as the Bidomain Model, Finite Strain Theory and the Navier-Stokes Equations, specifically for parallel processing and show proof-of-concept of a computational approach that can generate a complete description of a human heartbeat on a single Graphics Processing Unit (GPU) within a few minutes. 

The approach is based on a Finite Volume Method (FVM) discretisation which is both matrix- and mesh-free, ideally suited to voxel-based medical imaging data.  The solution of nonlinear ordinary and partial differential equations proceeds via the method of lines and operator-splitting. The resulting algorithm is implemented in the OpenCL standard and can run on almost any platform. It does not perform any CPU processing and has no dependence on third-party software libraries. 

The implementation is simple and computationally cheap enough to be used as the kernel of more complex software. Used iteratively and in parallel across an array of GPUs, it would allow movement through a solution space of parameterised hearts in optimisation problems for parameter estimation, patient-specific fitting, or in the training of Artificial Neural Networks (ANN). The algorithm presents new opportunities in research and clinical practice and is readily extensible as a simulation tool for a broad range of multiphysics problems.
\end{abstract}


\newpage
\tableofcontents 


\part{Aim}

\chapter{Introduction}\label{chp:intro}

\section{Motivation}

Cardiovascular disease affects millions of people worldwide, reducing both quality and duration of life.  The high social and economic cost of the disease clearly motivate research into its treatment. 

Computer simulation can replace physical procedures, reducing cost and removing associated risks. When simulation is performed accurately and cheaply with respect to the key computational resources of energy and time, then it can increase efficiency across a wide range of scientific and industrial disciplines. 

In medical research and and clinical practice this advantage becomes even more apparent, since an \textit{in-silico} medical intervention can be repeated and refined iteratively in a way that is not possible in the physical world. Similarly, the insight gained from experiment can be shared and re-used more effectively through repositories of standardised results. As technology improves such resources may be used to inform the automation of diagnosis and clinical decision-making.

\section{Review}

Even in its simplest form the simulation of a human heartbeat represents a significant computational challenge.  We first try to define what constitutes a whole heart simulation by summarising some of the features that it might contain, most of which are common to the literature review which follows:
\begin{itemize}
\item Geometry that includes the torso, myocardium, four chambers of the heart, proximal blood vessels and the valves that connect them. It may be extended to include pericardium and coronary blood supply although this is rarely present. The geometry may be schematic or generated from segmented medical imaging data.
\item Microscopic Electrophysiology representing the cell membrane and ion channels that support a travelling wave of transmembrane potential through the myocardium, with anisotropic conductivity determined by tissue fibre directions. Models for this behaviour include the Bidomain and Monodomain reaction-diffusion equations \cite{Keener2009} and ionic membrane currents described by the comprehensive \cite{TenTusscher2006} or simplified Fitzhugh-Nagumo (FHN) \cite{FitzHugh1955} \cite{Nagumo1962} Ordinary Differential Equations (ODEs).
\item Macroscopic Electrophysiology includes behaviour of larger electrophysiological structures such as the Sino-Atrial (SA) node, the Purkinje network and the Bundles of Bachmann and His.
\item Excitation-Contraction (EC) coupling linking the action potential to actin and myosin filaments via intracellular Calcium (Ca$^{2+}$) dynamics that lead to contractile stress in the mechanically active myocardium. Models include  \cite{Nash2004} and \cite{Courtemanche1998}\cite{OHara2011}. Mechano-Electric feedback also leads to additional ionic currents dependent upon extension and velocity \cite{Trayanova2004}.
\item Hyperelastic Material models for the stress-strain relations of the myocardium include Mooney-Rivlin model \cite{Mooney1940} \cite{Rivlin1948}, Guccione \cite{Guccione1995} Ogden \cite{Ogden1972} \cite{Holzapfel2002}.  These are generally modified to include passive anisotropy and active contraction related to tissue fibre directions, as well as incompressibility, in accordance with experimental data.
\item Fluid Dynamics of the blood modelled via the incompressible Navier-Stokes equations, sometimes modified to be non-Newtonian in small capillaries. 
\item Fluid-Structure Interaction (FSI) allowing changes in pressure and velocity to drive blood flow, which in turn feeds back onto solid tissue of the heart, valve leaflets and blood vessels. The problem may be strongly coupled with two-way feedback or weakly coupled in only one direction, usually with a prescribed solid motion.  For the strongly-coupled dynamics there are two common approaches: firstly the Arbitrary Lagrangian Eulerian (ALE) \cite{Hughes1981} formulation, and secondly the Immersed Boundary (IB) method \cite{Kohl2001}.
\item Circulatory System has a significant effect on the behaviour of the heart with distal blood flow regulating pressure and velocity at the blood vessels connected to the heart.  The closed-loop systemic arterial tree and pulmonary system is usually modelled as a windkessel \cite{Westerhof1971}. Some researchers use a similar system of ODEs within the heart chambers as a 0D substitute for the Fluid dynamics and FSI problems outlined above.
\end{itemize}

The following literature review considers some of the many publications that indicate the progress made towards the objective of whole heart simulation:

\cite{Watanabe2004} Describes an idealised left ventricle and proximal aorta, with FHN electrophysiology, EC coupling, anisotropic Guccione constitutive relation, strongly coupled FSI via ALE form of the Navier-Stokes equations, and windkessel circulation.  The problem is discretized and solved via the Galerkin Finite Element Method (FEM) on a tetrahedral mesh (~100k DOF), to give a full description of fluid velocity, pressure and volume. 

\cite{Tang2007} Models left and right ventricles with respect to inflow and outflow after surgery. Using Navier-Stokes, ALE, FEM with a modified anisotropic Mooney-Rivlin model. Dynamic material parameters simulate systolic and diastolic tissue mechanics.

\cite{Lee2009} Provides a review of multi-physics and coupled models, including strongly and weakly coupled FSI and gives results for an idealised ventricle at low and high resolution. He notes the computational intractability of fine model and considers the tradeoff between accuracy and computational cost.

\cite{Formaggia2010} A book with very complete derivation of solid and fluid dynamics, and their coupling through the ALE formulation. The formulation and notation guides much of this work. Solution is via FEM and is subject to many simplifying assumptions. There is an anecdotal description of a stable algorithm for FSI time iteration. 

\cite{Nordsletten2011a} Studies an idealised ventricle with Finite elasticity, anisotropic fibre directions, parameters from ex-vivo studies, a model for calcium transfer, and a Lagrange multiplier form of ALE FSI coupling.  FEM discretisation is solved with nonlinear Newton-Raphson iteration, no timings are given.

\cite{Nordsletten2011b} is a key review paper for continuum mechanics, the ALE formulation, fibre directions giving the  Guccione material model and electromechanical transfer, illustrated over single ventricle, discretised and solved with ALE FEM.

\cite{Trayanova2011} presents a review focussed mainly on electrophysiology and some electro-mechanical feedback for arrhythmias. Applications are given with respect to clinical, research and therapeutic applications.

\cite{Khalafvand2011} Another review paper, considering Computational Fluid Dynamics (CFD) for valve leaflets.  The work considers strongly and weakly coupled approaches including prescribed geometry with one-way coupling, ALE and fictitious methods such as the IB approach. There is no consideration of computational performance.

\cite{Niederer2011} Provodes a key benchmark study comparing the results of 11 submitted codes over the same electrophysiology problems. It forms the basis for the validation of electrophysiology in this work (see Chapter~\ref{chp:exp}) and several others.

\cite{Niederer2011a} A Bidomain electrophysiology study over a whole ventricle made with view to clinical use.  The publication includes timings, introducing the metric of \textit{real-time computational lag} (computation time divided by simulation time).  The study uses a FEM tetrahedral mesh with width 0.25mm and 26 million Degrees of Freedom (DOF), a preconditioned Conjugate Gradient (CG) solver processed in parallel using the PETSc library \cite{Balay1998} with Message Passing Interface (MPI) over 16384 cores. The result is a value of 240, or 1 second in 5 minutes and the authors note that 50\% of the time is used for communication.

\cite{Neic2012} gives the first consideration of GPU acceleration. An electrophysiology problem equivalent to the previous paper is accelerated at the lowest level, unrolling parallel loops within the PETSc library and passing to them to CUDA Fortran kernels. The strong scaling study reports a speedup of of 10 times using 20 GPUs on the coarsest grid.  The authors note that this is achieved while \textit{`minimally perturbing the code base'}.

\cite{Sugiura2012} A prototype FEM ALE study with fine heart mesh with coarse torso mesh for ECG, Calcium and sarcomere dynamics via explicit ODE, automatic re-meshing for valves. Gives physiological values for ejection fractions, pressure and volume. RIKEN/Fujitsu k-computer with 88,128 CPUs, 6 hours per heartbeat.

\cite{Land2015} Another key benchmark study which aggregates and 11 submitted mechanics code for simple problems. It is the basis for validation of the mechanics calculation in this work (see Chapter~\ref{chp:exp}).

\cite{Zhang2016} A review of multi-scale simulation including fluid mechanics.  It specifies the different problems but does not demonstrate any solution.

\cite{Quarteroni2017} A large and comprehensive paper gives full details of models and assembly and time integration for electrophysiology, electro-mechanical coupling, solid dynamics and FSI. Studies reproduce benchmarks electrophysiology, and mechanics test problems. An idealised ventricle is shown but without fully-coupled FSI, using pre-determined structure and 0D fluid rather than a two-way solution. It focuses on convergence rather than performance concluding that a full model at all scales is \textit{`still far beyond reach'}. 

\cite{Niederer2018} A broad overview of research and clinical applications written for a wide audience with many useful references. 

\cite{Santiago2018} A fully-coupled heart model including fluid mechanics over a complete geometry including atria and aortae and sliding pericardium. There is no Purkinje network, valves or atrial contraction. The ALE formulation refers to the prior literature for electrophysiology, solid and fluid models. Discretisation via in-house FEM code, with a mixture of forward, backward Euler and Newmark schemes, Generalized Minimal Residual (GMRES) method Picard method for velocity-pressure, FSI stabilised with quasi-Newton iterations. MPI-parallel solver alternates between CFD and electrophysiology/solid mechanics with 800k DOFs. The authors show qualitative results and a scaling study up to 4800 cores with speedup of 0.85,  but do not publish wall clock timings, noting that this is a validation of the first instance of their model.

\cite{Kaboudian2019} A GPU Lattice Bolzmann electrophysiology simulation based on WebGL Javascript API, demonstrating 2D/3D spiral waves, and low resolution rabbit ventricle at 1/3 real-time, also examples of fluid dynamics, and crystal formation.  This is the first published attempt to re-formulate the problem specifically for GPU processing.

\cite{Viola2020} Studies an idealised left heart with atrium, ventricle, aorta and mitral valve. Navier-Stokes is treated with a Finite Difference Method (FDM) on a cartesian mesh with circulatory effects prescribed as boundary conditions. FSI is via interpolated IB onto valve leaflets modelled as 2D mass-spring membranes with Adams-Bashforth integration for strong coupling. There is a Fung mechanical model \cite{Fung1981} with fibre anisotropy,  full bidomain with FHN for atria and  TenTusscher-Panfilov electrophysiology for ventrcle. Bachman and His effects are prescribed as timed stimuli. The simulation is timed at 1500 CPU Hours per heartbeat (850ms of simulation).

\cite{Gerach2021} Model a four-chamber geometry with a 0D closed-loop circulation, with the effects of ablation scar, solving Bidomain with EC and anisotropic Guccione material. The FEM discretisation mixes of implicit and explicit schemes, with activation of atria and ventricles prescribed via external stimulus. They note that experimentally derived parameters from literature must be adjusted to give  physiological results. The mechanics mesh has 136k DOFs (Authors note that it is too coarse), but is finer for Electrophysiology 0.6mm. Results include PV loops for atria and ventricles, using a 2019 Apple iMac with 8 MPI processes requires 20-24 hours per heart beat.

\cite{Augustin2021} Simulate a whole canine heart with a 0D lumped ODE circulation over 25 heartbeats.  Electrophysiology uses Reaction Eikonal \cite{Neic2017} approach.  Mechanics via FEM with Newton-Raphson using GMRES on PETSc framework. Results include PV loops, activation maps and EGC. The mesh has 156k nodes at 1.25mm resolution and is processed on 256 cores requiring 30 minutes per heartbeat.

\cite{Viola2022} Updates prior work on an idealised heart with GPU acceleration.  The study includes a complete Fluid-Structure-Electrophysiology Interaction (FSEI) simulation using an in-house FEM library. An MPI-parallel in-house FEM library is processed with OpenACC-type compiler directives for CUDA Fortran acceleration. Weak and strong scaling studies are performed on 8 NVIDIA A100 cards showing 1-2 orders of magnitude speedup with respect to prior serial code. A single heartbeat requires 3-10 hours of processing depending on mesh resolution.

\cite{DelCorso2022} is a similar paper from the same group as \cite{Viola2022} which presents electrophysiology only, discretised with Finite Volume Model (FVM) and CUDA Fortran GPU acceleration of the GNRES solver. A Bidomain simulation at 0.5mm mesh width using the approach of \cite{Rush1978}, requires 8 hours per heartbeat, with an equivalent monodomain requiring 1.5 hours.

\cite{Verzicco2022} The most recent publication in the field is a comprehensive review paper giving the history of the field, with a focus on the inclusion of strongly-coupled FSI and its relevance for pathology.  It serves as an ideal and modern introduction to the subject giving a fair representation of the state-of-the-art.  It concludes with the following remark:

\textit{`The high computational cost of complex multiphysics heart models and the need of trained researchers to run them constitutes an insurmountable barrier to the use of digital twins in clinical practice. Furthermore, they rely on massive parallel supercomputers, have a time-to-solution of the order of days and produce terabytes of data which need extensive post-processing to extract relevant information; this is totally incompatible with clinical decisions that must be taken, at most, within hours and obtained by commodity computers.'}

As research has progressed, research groups have reached a broad consensus on the mathematical formulation of the models that represent the features of a whole heart simulation and on the computational approach taken to solve them.  The general method is summarised below:

\begin{itemize}
\item Model parameters are estimated from experiments carried out on small tissue samples, usually ex-vivo, from a variety of species.  
\item Medical imaging data is segmented and used to generate a tetrahedral mesh of the heart and surrounding tissue. 
\item The various differential operators related to the mathematical formulation are assembled into large in-memory linear systems via Finite Element discretisation.
\item Integration in time is accomplished via a mixture of explicit and implicit methods.  The hardest and most nonlinear part of the problem, fluid-structure interaction, is usually solved as a saddle-point sub-problem using Newton-type iterations.  
\item The solution of the many linear systems required for integration is carried out using well-established High Performance Computing (HPC) libraries implementing algorithms such as preconditioned CG or GMRES, amongst others. 
\item The computation of these algorithms requires the use of large supercomputing resources, dividing and balancing the problem over many (thousands) of processors. 
\end{itemize}

From the review above, a generous estimate of processing time for complete simulation of a single heartbeat may be about an hour. Another rough estimate of the financial cost of computation is 1 US Dollar per node hour. The HPC simulations described in the publications above are therefore consuming (approximately) tens-of-thousands of Dollars per second of cardiac simulation. 

Modern computer chip design has already hit the `Power Wall' in which power consumption and heat dissipation are the limiting factors for processing speed. Parallel computing is thus standard and the cost-of-communication between processors has become the bottleneck for performance. Attempts to accelerate per-core performance via the addition of GPUs has shown some speedup but ultimately does not overcome this limitation.

Fitting a heart simulation to an idealised reference subject would potentially require the simulation of thousands of heartbeats. To then reproduce simulations for single pathologies would require thousands more. To do this on a per-patient basis for diagnosis would incur thousands more and thousands again for treatment. The turnaround time between and admission and a decision about treatment could not be met. We therefore conclude that the economic and environmental cost of energy consumption associated with current approaches does not make them feasible for clinical use. We also do not expect to see improvements in processor design that will allow for their practical application in the foreseeable future.

\section{Objectives}

The motivation for this work is therefore to re-design tools for cardiac simulation that can be applied in clinical or research situations.  We therefore aim to provide \textit{proof-of-concept} of a computational approach that is sufficiently accurate to give insight at a clinical or academic level but which is not prohibitive in terms of computational cost.  

This work will deliver a piece of software that is designed to meet the following objectives:

\begin{itemize}
\item Provide a complete description of a heartbeat, representing Electrophysiology, Solid and Fluid dynamics and their interaction via FSI, as described above.
\item Run on a geometry that represents the whole heart, which can be assimilated directly from medical imaging data.
\item Allow for computation on a single small device, which is low-cost in terms of acquisition and energy consumption.
\item Make use of a novel dual grid discretisation that allows for matrix-free computation and makes optimal use of GPU memory access.
\item Run within seconds or minutes, allowing for repeated use over many heartbeats, or as the objective function for parameter estimation or patient model fitting.
\item Apply the well-established mathematical formulations as above, without requiring new or unproven mathematical approaches.
\item Provide results and biological indicators that are qualitatively similar to those of the studies outlined above.
\item Be simple enough in design to achieve the required performance, but extensible to allow the inclusion of more complex mathematical models and features.
\item Have no software dependency and thus give freedom of design to future developers and users.
\item Be simple enough to be configured and run by third parties such as clinicians and researchers, or incorporated into more complex software.  This document is designed to serve both as a thesis and software documentation.
\end{itemize}

The approach therefore starts from first principles, taking the simplest and most well-established models of solid and fluid dynamics and re-formulating them for parallel computation on a single small device, namely a Graphics Processing Unit (GPU). The formulation itself is based upon the Finite Volume Method combining the Lagrangian solid and Eulerian fluid formulations via the Arbitrary Lagrangian-Eulerian method (ALE). 

Geometry is encoded on a deforming structured grid via a Signed Distance Function (SDF), and as such the approach can be considered \textit{mesh-free}. The algorithm also uses a \textit{matrix-free} approach, with each GPU process assembling the various linear operators by row as required. Nonlinearity is handled via \textit{operator-splitting} in which the differential operators that drive the state of the system are solved separately and in turn. The removal large linear systems and operator splitting allow the use of a \textit{self-dual} grid which is ideal for both the FVM discretisation and the fast memory access of the GPU. 

Since the design objectives are guided by constraints and compromise, then it follows that it can not fully reproduce the accuracy or completeness of the existing studies in the literature. The results are general in their nature since this is the first prototype of the software. They demonstrate the validity of the algorithm rather than the model itself. Some features are not present:

\begin{itemize}
\item Electro-mechanical coupling is simplified, with contraction coupled directly to the action potential. The correct model for Calcium dynamics can be added in future in a straightforward way. 
\item There is no model for the circulation or proximal blood vessels. The effects can be modelled modelled directly by adding them to the geometry with similar results.
\item The current schematic heart does not include fully functioning valves.  Their action requires an extra model for structure-structure interaction which is not currently present.  This would is the subject of future work.
\end{itemize}

Given the objectives above, the work hopes in future to deliver software for practical use, rather than for purely academic investigation.

\section{Overview}

This document is laid out as follows:

Chapter~\ref{chp:prb} gives details of the mathematical models that will be employed, all of which are based upon the principles of continuum mechanics. The electrophysiology simulation includes the Mitchell-Schaffer ionic membrane model and the diffusive Bidomain and Monodomain models for the movement of charge through the myocardium and torso.  Solid mechanics, including the deformation of tissue and muscle contraction, is formulated via Finite Strain theory using various stress-strain relationships for hyperelastic materials. The fluid mechanics of blood flow is described by the incompressible Navier-Stokes equations. Fluid-stucture interaction is makes use of the Arbitrary Lagrangian-Eulerian formulation.
 
Chapter~\ref{chp:gpu} contains a brief introduction to GPU computing. Its aim is to show how architecture specialized for parallel processing can give advantages in computational efficiency. It will also to make clear the restrictions upon code design that arise as a result.  This in turn guides the design of the algorithm.

Chapter~\ref{chp:disc} shows how the geometry and continuous differential operators required by the mathematical models of Chapter~\ref{chp:prb} are expressed in the discrete setting of a computational mesh, leading to the assembly and solution of the linear systems resulting from these operators.

Chapter~\ref{chp:alg} describes how the continuum mechanics models of Chapter~\ref{chp:prb} and discrete operators of Chapter~\ref{chp:disc} are arranged for parallel computation, under consideration of the design constraints of Chapter~\ref{chp:gpu}.  It details the relationships between computational kernels and the mathematical formulation as well as memory use and therefore serves as documentation for the computer code.

Chapter~\ref{chp:exp} details the numerical experiments which are used to validate the quality of the discretisation of the operators in Chapter~\ref{chp:disc} and their combination and integration in Chapter~\ref{chp:alg}.  These include numerical convergence studies with respect to analytic solutions for the differential operators and the reproduction of benchmark studies for the simulation of electrophysiology and solid mechanics. 

Chapter~\ref{chp:sim} demonstrates the application of the software to cardiac simulation.  It includes qualitative studies of whole heart function as well as quantitative results such as a synthetic Electrocardiogram (ECG) and Pressure-Polume (PV) loop.

Chapter~\ref{chp:rev} provides a review of the work, considering its strengths and weaknesses and provides an outlook for future improvement and applications.

\chapter{Cardiac Multiphysics}\label{chp:prb}

During the normal heartbeat a wave of membrane depolarisation propagates at the Sino-Atrial (SA) node in the Right Atrium (RA) and travels through the myocardium in an orderly fashion, first to the Left Atrium (LA) and then to the Right and Left Ventricles (RV, LV).  The tissue has a property of  anisotropic conductivity that is dependent upon both the orientation of the muscle fibres and the presence of tissues of varying conductivity such as the Pukinje Fibre network and the Bundle of His.  

As this travelling wave of membrane depolarisation passes through the heart, changes in ion channels and receptor proteins within the membranes of the myocites mediate intracellular ionic changes that lead to a sequence of co-ordinated muscle contraction. 

The deformation resulting from movement of the contractile tissue results in forces that change the pressure in the chambers of the heart.  These pressure changes, together with viscous forces and advection cause the movement of blood through the heart and into surrounding vessels. The unidirectional flow and hence circulation of blood is maintained by the sequential contraction of the chambers and by valves within the heart and circulatory system that prevent the reversal of flow.

This chapter gives a mathematical description of the processes outlined above.  

\section{Electrophysiology}

Membrane depolarisation initiated at the SA node is propagated as a travelling wave through the myocardium via two mechanisms:

Firstly, active processes maintain chemical potential gradients across the membranes of myocardial cells. Changes in charge, and hence transmembrane potential, cause the opening and closing of voltage-dependent channels that allow ions (Na\textsuperscript{+}, K\textsuperscript{+}, Ca\textsuperscript{+}, Cl\textsuperscript{-}) to flow across the membrane, further depolarising it on a feedforward manner, until the ionic potential gradients have reached an equilibrium.  Thereafter there is a refractory period during which the membrane can not be excited until the potential gradient has been restored by the pumping of ions back across the cell membrane.

Secondly, the changes in potential caused by membrane depolarisation cause the diffusion of electrical charge between an excited cell and its neighbours.  This in turn stimulates ion channels as described above and causes ionic excitation to spread.  The refractory period ensures that a recently excited cell is unresponsive to further depolarisation and thus the unidirectional progress of the travelling wave is maintained.

The Partial Differential Equation (PDE) that models the spreading cycle of depolarisation and repolarisation is a \textit{reaction-diffusion} equation. We consider the reactive Mitchell-Schaeffer membrane model and the diffusive Bidomain and Monodomain equations. 

\subsection{Mitchell-Schaeffer Model}

The model of \cite {Mitchell2003} describes a \textit{uniformly polarized membrane patch} representing an action potential in a spatially clamped ventricular myocite. It considers a a pair of coupled time-dependent Ordinary Differential Equations (ODE)s with time $t \in \R^+$, for a dimensionless transmembrane voltage $v \in [0,1]$ and gating variable $w \in [0,1]$. The voltage $v$ is determined by the first ODE
\begin{equation}
\label{eqn:ms1}
\frac{dv}{dt} = J_{\text{in}}(v,w) + J_{\text{out}}(v)  + J_{\text{stim}}(t)
\end{equation}
where the three currents $J$ are defined as follows:  The inward current $J_{\text{in}}$ combines ionic movements (Na\textsuperscript{+}, Ca\textsuperscript{+}) that raise the membrane voltage $v$. The feedforward nature of the ion channels gives it a cubic dependence on the membrane voltage $v$ and a linear dependence on the gating variable $w$, scaled by a constant time parameter $\tau_{\text{in}}$:
\begin{equation}
J_{\text{in}}(v,w) = \dfrac{1}{\tau_{\text{in}}}v^2(1 - v)w.
\end{equation}
The outward current  $J_{\text{out}}$ combines those ions that lower membrane voltage (primarily K\textsuperscript{+}) and depends negatively on $v$ but is not gated, scaled by a time constant $\tau_{\text{out}}$ 
\begin{equation}
J_{\text{out}}(v)  = - \dfrac{1}{\tau_{\text{out}}} v.
\end{equation}
The stimulus current $J_{\text{stim}}$ can applied by the experimenter. 

The second ODE for the gating variable $w$ is as follows:
\begin{equation}
\label{eqn:ms2}
\frac{dw}{dt} =
\left\{\begin{array}{cc}
\dfrac{1-w}{\tau_{\text{open}}} & \text{, if } v < v_{\text{gate}}\\[10pt]
\dfrac{1-w}{\tau_{\text{close}}} & \text{, if } v \geq v_{\text{gate}}
\end{array}\right.
\end{equation}
where $\tau_{\text{open}}$ and $\tau_{\text{close}}$ are time constants and $v_{\text{gate}}$ is a threshold voltage for activation. The nonlinear and opposing nature of the equations for $v$ and $w$ place the system into the class of \textit{activator-inhibitor} ODEs.

The authors note the relationships between their model and those of \cite{Fenton1998} , \cite{Luo1994} and \cite{FitzHugh1955} \cite{Nagumo1962}, and point out two improvements: Firstly it contains four physically meaningful constants, corresponding to the four phases of the action potential: initiation, plateau, decay and recovery.  Secondly, that it avoids the non-physical voltage overshoot of its predecessors. It is also simpler than the more comprehensive \cite{TenTusscher2004}. The parameters used by the model have been experimentally fitted into a physical range \cite{Ngoma2017}. 

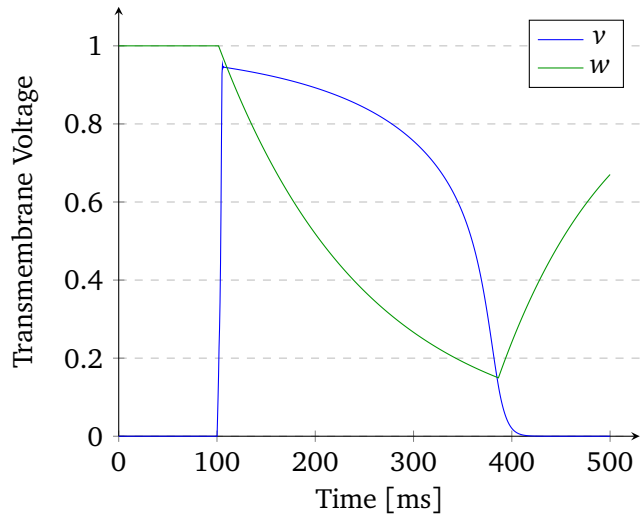
\begin{figure}[h]
\centering
\begin{tikzpicture}
\begin{axis}[
    xlabel={Time [ms]},
    ylabel={Transmembrane Voltage},
    xmin=0, xmax=530,
    ymin=0, ymax=1.1,
    xtick={0,100,200,300,400,500},
    ytick={0,0.2,0.4,0.6,0.8,1.0},
    legend pos=north east,
    ymajorgrids=true,
    grid style=dashed,
    axis x line = bottom,
    axis y line = left
]
\addplot [color=blue,mark=none] table[x index=0,y index=1] {ms1.csv};
\addplot [color=green,mark=none] table[x index=0,y index=2] {ms1.csv};
\legend{$v$,$w$}
\end{axis}
\end{tikzpicture}
\caption{Mitchell-Schaeffer action potential, Explicit Euler {$\Delta t = 0.5$ms}.}
\label{fig:ms1}
\end{figure}

An example of a numerical solution of the system is shown in Figure \ref{fig:ms1}.  It is worth noting that the equations are highly nonlinear or \textit{stiff}, given the steep rise in onset membrane voltage.  As such they illustrate a key point in the design of this work.  The nonlinearity makes it too costly to solve the system implicitly.  It is acceptable therefore to perform an explicit time integration and accept a maximum time step for stability.

\subsection{Bidomain Model}

Given a description of the action potential in time we next consider the movement of the traveling wave in space.  The Bidomain model considers a small region of tissue as a single point $\x \in \R^d$ where $d \in \N$ is the dimension, that contains both intra- and extra-cellular spaces, along with variables for the potential differences and current flows between them.  The propagation of electrical activity from one cell to another is represented as a diffusion of ionic charge.  This construction leads to a system of  PDEs which govern the movement of charge both through the \underline{i}ntracellular and \underline{e}xtracellular spaces of the heart and also into the surrounding tors\underline{o} (underlined characters correspond to variable subscripts). For this reason the Bidomain model was instrumental in the development of defibrillation. The standard reference for this material is the book of \cite{Keener2009}.

The model is derived as follows, we consider a point in space $\x \in \R^d$, an element of the following sets: \\ \\
\begin{tabular}{lll}
$\mathbb{H}$ & & heart tissue with boundary $\partial \mathbb{H}$ \\ 
$\mathbb{T}$ & & surrounding tissue of the body with boundary $\partial \mathbb{T}$
\end{tabular}\\ \\
For a point in $\mathbb{H}$: \\ \\
\begin{tabular}{lll}
$v$ 					& $\in \R$ 						& transmembrane voltage \\
$v_{i}, v_{e}$ 			& $\in \R$ 						& intracellular and extracellular voltage \\
$\SIG_{i}, \SIG_{e}$	& $\in \R^{d \times d}$			& intracellular and extracellular conductivity tensors\\
$J_{i}, J_{e}, J_{t}$	& $\in \R^d$ 		 			& intracellular, extracellular and transmembrane current density\\
$C_m$ 					& $\in \R$  					& membrane capacitance per unit area\\
$G_m$ 					& $\in \R$   					& membrane conductivity per unit area \\
$\chi_m$				& $\in \R$  					& membrane surface area to volume ratio\\
$I_\text{m}$ 			& $\in \R$  	 				& ionic current over membrane per unit area \\
\end{tabular} \\ \\
For a point in $\mathbb{T}$: \\ \\
\begin{tabular}{lll}
$J_{o}$ 				& $\in \R^d$ 					& tissue current density \\
$v_{o}$ 				& $\in \R$						& tissue voltage \\
$\SIG_{o}$				& $\in \R^{d \times d}$	 		& tissue conductivity tensor 
\end{tabular} \\

\subsubsection{Derivation}

First apply the current-voltage relationship over the intracellular and extracellular domains using Ohm's law:
\begin{eqnarray}
J_{i} &=& - \SIG_{i} \nabla v_{i} \\
J_{e} &=& - \SIG_{e} \nabla  v_{e}
\end{eqnarray}
Then require that there is no accumulation of charge in $\mathbb{H}$, setting the current densities equal and opposite in sign:
\begin{equation}
-\nabla \cdot J_i = \nabla \cdot J_e = \chi_m I_m 
\end{equation}
Substituting gives the first model equation, which states that all current exiting one domain must enter the other:
\begin{equation}
\nabla \cdot (\SIG_{i} \nabla v_i) + \nabla \cdot (\SIG_{e} \nabla v_e) = 0
\end{equation}
By convention, the transmembrane current $J_t$ and voltage $v$ are measured with respect to the intracellular domain:
\begin{eqnarray}
\label{eqn:bidomain2}
J_t &=& \nabla \cdot (\SIG_{i} \nabla v_i)  = - \nabla \cdot (\SIG_{e} \nabla v_e) \\
v &=& v_i - v_e
\end{eqnarray}
The model for transmembrane current depends on the dynamic behaviour of the membrane and is based on the cable equation: 
\begin{equation}
J_t  = \chi_m \left( C_m \frac{\partial v}{\partial t}  + I_\text{m}  \right)
\end{equation}
The ionic current $I_{\text{ion}}$ is where the Mitchell-Schaeffer model enters the Bidomain equations. Scaled by membrane conductivity $\chi_m$ it models the dynamic properties of the transmembrane voltage.
The expressions for $J_t$ are combined for the second model equation:
\begin{equation}
\nabla \cdot (\SIG_{i} \nabla v_i) = \chi_m \left( C_m \frac{\partial v}{\partial t}  + I_\text{m}  \right)
\end{equation}

\subsubsection{Boundary Conditions}

We next repeat a similar process with the body tissue domain $\mathbb{T}$, first applying Ohm's law for the current-voltage relationship:
\begin{equation}
J_{o} = - \SIG_{o} \nabla v_{o} 
\end{equation}
then preventing the accumulation of charge in $\mathbb{T}$:
\begin{equation}
\nabla \cdot (\SIG_{o} \nabla v_{o})  = 0
\end{equation}
Next, we electrically isolate the body tissue $\mathbb{T}$ by setting current flow to zero in the unit normal direction $\mathbf{n} \in \R^d$ normal to $\partial \mathbb{T}$:
\begin{equation}
(\SIG_{o} \nabla v_{o}) \cdot \mathbf{n}  = 0, \quad \x \in \partial \mathbb{T}
\end{equation}
Couple the extracellular voltage $v_e$ to the tissue voltage $v_{o}$ on the boundary of the heart $\partial \mathbb{H}$:
\begin{equation}
v_e = v_{o}, \quad \x \in \partial \mathbb{H}
\end{equation}
Couple the current flow from the extracellular to the tissue domain on $\partial \mathbb{H}$:
\begin{equation}
(\SIG_{e} \nabla v_e) \cdot \mathbf{n} = (\SIG_{o} \nabla v_{o}) \cdot \mathbf{n} , \quad \x \in \partial \mathbb{H}
\end{equation}
Isolate the intracellular domain from the body tissue:
\begin{equation}
(\SIG_{i} \nabla v_i) \cdot \mathbf{n} = 0 , \quad \x \in \partial \mathbb{H}
\end{equation}
Grouping these equations gives the fully determined Bidomain system:
\begin{equation}
\label{eqn:bidomain1}
\begin{array}{rcll}
\nabla \cdot (\SIG_{i} \nabla v_i) &=& \chi_m \left( C_m \frac{\partial v}{\partial t}  + I_\text{m}  \right) 		& \x \in \mathbb{H}  \\
\nabla \cdot (\SIG_{i} \nabla v_i) + \nabla \cdot (\SIG_{e} \nabla v_e)  &=&  0							& \x \in \mathbb{H}  \\
\nabla \cdot (\SIG_{o} \nabla v_{o})  							&=& 0  							& \x \in \mathbb{T}  \\
\mathbf{n}   \cdot (\SIG_{o} \nabla v_{o})						&=& 0							& \x \in \partial \mathbb{T} \\
\mathbf{n}   \cdot (\SIG_{e} \nabla v_e) - \mathbf{n}  \cdot (\SIG_{o} \nabla v_{o}) &=& 0					& \x \in \partial \mathbb{H} \\
(\SIG_{i} \nabla v_i) \cdot \mathbf{n} 							&=& 0 							& \x \in \partial \mathbb{H}
\end{array}
\end{equation}

\subsection{Monodomain Model}

The Monodomain model is a simplification of the Bidomain model with a single anisotropy field for both the intra and extracellular spaces. Continuing from the definition of the Bidomain model (\ref{eqn:bidomain1}) it makes the assumption that the intracellular $\SIG_{i}$ and extracellular $\SIG_{e}$ conductivity tensors are equal up to scaling.  The two anisotropy fields are thus combined via a constant $\lambda \in \R$ which gives the intracellular to extracellular conductivity ratio:
\begin{equation}
\SIG_{e}  = \lambda \SIG_{i} 
\end{equation}
From the derivation of the Bidomain model we take expressions (\ref{eqn:bidomain2}) for transmembrane current $J_t$ and transmembrane voltage $v$:
\begin{eqnarray}
J_t &=& \nabla \cdot (\SIG_{i} \nabla v_i)  = - \nabla \cdot (\SIG_{e} \nabla v_e) \\
v &=& v_i - v_e
\end{eqnarray}
Combining them via the conductivity ratio $\lambda$ and re-arranging gives the formulation of the Monodomain model:
\begin{equation}
\label{eqn:monodomain1}
\frac{\lambda}{1 + \lambda} \nabla \cdot (\SIG \nabla v) = \chi_m \left( C_m \frac{\partial v}{\partial t}  + I_\text{m}  \right).
\end{equation}
The equation (\ref{eqn:monodomain1}) calculates the change in membrane potential difference $v$ with respect to time $t$.  The reaction term on the right is  the transmembrane current $I_{\text{ion}}$ generated by the membrane potential difference.  The diffusion term on the left is the flow of current into neighbouring regions, where $\SIG$ is the anisotropic diffusivity tensor that relates potential gradients $\nabla v$ to current flow. There are also constants $\chi_m$, a surface area to volume ratio for the cell membrane, and $C_m$ the membrane capacitance.   

It remains to quantify the conductivity tensor $\SIG$. In practice our model defines a vector field $\f \in \R^d$ of unit length fibre directions as well as longitudinal  and transversal conductivities $\sigma_L, \sigma_T \in \R$ respectively, from which the conductivity tensor is derived via an outer product as follows,
\begin{equation}
\label{eqn:sigma1}
\bm{\SIG} = \sigma_T \I + (\sigma_L - \sigma_T) \f \otimes \f.
\end{equation}
The same field of fibre directions also determines the stresses arising from muscular contraction. 

\section{Continuum Mechanics}

\subsection{Deformation}\label{sec:deform1}

The movement and interaction of both the solid and fluid materials of the heart are described within the formalism of continuum mechanics, effectively a set of scalar, vector and tensor fields defined over time in physical space. This brief introduction follows \cite{Gonzalez2008}, \cite{Reddy2013} and \cite{Formaggia2010}.
\begin{figure}[h]
\centering
\makeatletter
\def\wavetransformation
{
\pgfmathsincos@{\pgf@sys@tonumber\pgf@x} 
\pgfmathsetmacro{\myx}{\pgf@x + \pgfmathresultx cm}
\pgfmathsetmacro{\myy}{\pgf@y + \pgfmathresulty cm}
\pgf@x= \myx pt%
\pgf@y= \myy pt%
}
\makeatother
\begin{tikzpicture}[grid/.pic={\draw (0,0) grid [step=0.5] (4,2);}]
\pic (a) {grid};
{
    \pgftransformnonlinear{\wavetransformation}
    \pic[rotate=45,shift={(5 cm,-6 cm)}] (b) {grid};
}
\node[anchor=east] (A) at (0,2) {$\hat{\Omega}$};
\node[anchor=east] (B) at (5.6,1) {$\Omega(t)$};
\draw[->,blue,thick,out=30,in=150] (2.1,1.1) to node [midway,above,black] {$\phi(\hat{\x},t)$} (7.8,0.6) ;
\draw  [fill=black]  (2,1)  circle (0.05cm); 
\draw  [fill=black]  (8,0.55)  circle (0.05cm); 
\node (C) at (1.8, 1.25) {$\hat{\x}$};
\node (D) at (8,0.3) {$\x$};
\end{tikzpicture}
\caption{The reference $\hat{\Omega}$ and deformed $\Omega(t)$ configurations, with a material point $\hat{\x}$ and its current image $\x$ under the motion $\phi(\hat{\x},t)$.}
\label{fig:deform1}
\end{figure}
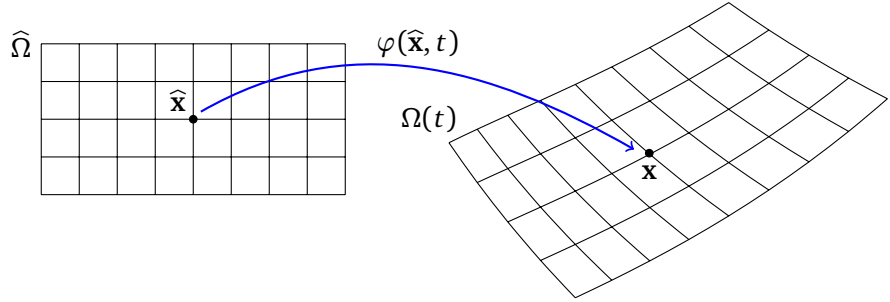

Figure \ref{fig:deform1} shows a body in its reference configuration $\hat{\Omega} \subset \R^d$, described by material coordinates $\hat{\x} \in \R^d$  which undergoes a smooth admissible deformation over time $t \in \R$, called a motion $\varphi:  \R^d \times \R \mapsto \R^d$, into a current configuration $\Omega(t) = \Omega_t \subset \R^d$ described by spatial coordinates $\x \in \R^d$.

This allows the definition of the following fields:
\begin{equation}
\begin{array}{ll}
\text{Motion}  						&  \x  	= \phi(\hat{\x},t)						\\[8pt]
\text{Velocity}  						& \v 	=  \dfrac{\partial \phi}{\partial t}(\hat{\x},t) = \dfrac{\partial \x}{\partial t}	\\[11pt]
\text{Deformation gradient}  \quad		& \F	=  \dfrac{\partial\phi}{\partial\hat{\x}} (\hat{\x},t) = \dfrac{\partial \x}{\partial \hat{\x}}	\\[11pt]
\text{Jacobian determinant} 			& J		= \det \F(\hat{\x},t). 
\end{array}
\label{eqn:mot1}
\end{equation}
The Lagrangian description gives a field with respect to the material coordinates and the Eulerian description is made with respect to the spatial coordinates. 
The motion gives the relationship between the two descriptions, for instance for the density $\rho:\R^d \times \R \mapsto \R$,
\begin{equation}
\rho(\x,t) = \hat{\rho}\left(\phi^{-1}(\x,t),t\right).
\end{equation}
Key to the formulation are the deformation gradient $\F$ and its determinant $J$. Due to its regularity we can express the motion as a Taylor series for some vector $d\hat{\x}$:
\begin{equation}
\phi(\hat{\x} + d\hat{\x},t) = \hat{\x} + \F(\hat{\x},t)d\hat{\x} + \mathcal{O}(d\hat{\x}^2).
\end{equation}
The change in length of a vector under the motion is thus approximated by:
\begin{equation}
\label{eqn:def_len1}
\|d\hat{\x}\|_2 \approx \sqrt{d\hat{\x}^\top \F(\hat{\x},t)^\top \F(\hat{\x},t) d\hat{\x}}.
\end{equation}
For a subvolume $\hat{\Omega} \subset \R^d$ and its image $\Omega_t$ the determinant $J = \det(\F(\hat{\x},t))$ gives a relationship for volume between the reference and deformed configurations.
\begin{equation}
\label{eqn:def_vlm1}
|\Omega| = \int_\Omega d\x = \int_{\hat{\Omega}} J d\hat{\x}
\end{equation}
 Nanson's formula relates surface area in the reference and deformed configurations, where $\hat{\n}$ and $\n$ are the two unit normals respectively:
\begin{equation}
\label{eqn:def_area1}
|\partial \Omega| = \int_{\partial \Omega} \n\ dA = \int_{\partial \hat{\Omega}} J \F^{-\top} \hat{\n} \ d\hat{A}.
\end{equation}
It combines the two previous results, effectively dividing volume by length for surface area. Transposition arises since normals are orthogonal to the surface and integration thus involves a dot product.

\subsection{Conservation Laws}

A complete description  of a continuum body is based upon a set of conservation laws for mass, momentum and energy, as well as laws for intertia and thermodynamics. In order to continue we give expressions for the conserved quantities of mass $m \in \R$ and linear momentum $\mathbf{l} \in \R^d$ for a body $\Omega \subset\R^d$. 
\begin{eqnarray}
\label{eqn:cons1}
m 		&=& \int_\Omega \rho(\x,t) \ dV \\
\label{eqn:lin_mom1}
\mathbf{l} 	&=& \int_\Omega \rho(\x,t) \v(\x,t) \ dV .
\end{eqnarray}

\subsection{Cauchy's Theorem}

A body in a continuum can be subject to two types of force: body forces ${\b(\x) \in \R^d}$ such as gravity, act per unit volume, whereas traction forces $\t(\n,\x) \in \R^d$ act per unit area across surfaces that may form the exterior surface of the body itself or pass through its interior.  Cauchy's Theorem shows the existence of a second order tensor field and the dependence of the traction force upon it:
\begin{equation}
\t(\n,\x) = \bm{\sigma}(\x) \n(\x),
\end{equation}
where $\bm{\sigma}(\x)$, as defined in the Eulerian description, is known as the Cauchy Stress tensor. The resultant force $\r \in \R^d$ acting on a body $\Omega \subset \R^d$ is thus a combination of surface and volume integrals:
\begin{equation}
\label{eqn:fres1}
\r(\x,t) = \int_{\partial \Omega} \bm{\sigma}(\x,t) \n(\x,t) \ dA+ \int_\Omega \rho(\x,t) \b(\x,t) \ dV,
\end{equation}
where $\rho(\x,t)$ is the density at a given spatial point. It can be shown that the symmetry of the Cauchy stress tensor $\bm{\sigma}$ ensures conservation of angular momentum.

\subsection{Newton's Second Law}

Newton's second law \cite{Newton1687} expresses a change in linear momentum (\ref{eqn:lin_mom1}) in terms of resultant force (\ref{eqn:fres2}):
\begin{equation}
\label{eqn:nsl1}
\frac{d}{dt} \int_\Omega \rho(\x,t) \v(\x,t) \ dV  
=
\int_{\partial \Omega} \bm{\sigma}(\x,t) \n(\x,t) \ dA + \int_\Omega \rho(\x,t) \b(\x,t) \ dV,
\end{equation}
giving the basis for the description of both solid and fluid dynamics.

\section{Solid Dynamics}\label{sec:solid1}

In a solid body forces arise from deformation. The material or constitutive model relates a measure of deformation or \textit{strain} to a potential function for \textit{strain energy density}. The derivative of strain energy with respect to deformation gives a tensor field for \textit{stress}. Conservation of linear momentum and the \textit{stress-strain} relation give the Elastodynamics equation, from which the time evolution of a deforming solid can be determined.

\subsection{Constitutive Models}

The expression for the change in length under deformation (\ref{eqn:def_len1}) leads to the Right Cauchy-Green strain tensor $\C$. It contains information about stretching but removes rotation. The Green-Lagrange strain tensor $\E$ measures how much $\C$ varies from the identity tensor $\I$. 
\begin{eqnarray} 
\C &=& \F^\top \F \\
\E &=& \frac{1}{2}(\F^\top \F - \I)
\end{eqnarray}
A hyperelastic material is one in which a strain energy density function $W$ can be defined in terms of the three measures of strain $\F, \C$ and $\E$. This family of models include those of Saint Venant-Kirchhoff \cite{Holzapfel2002}, Mooney-Rivlin \cite{Mooney1940}, \cite{Rivlin1948}, Ogden \cite{Ogden1972}.

We take as an example the St. Venant Kirchoff material model, which defines strain energy as quadratic in relation to the the Green-Lagrange strain:
\begin{equation}
W(\hat{\x},t) = \frac{\lambda}{2}\tr(\E)^2 + \mu \tr(\E^2).
\end{equation}
Lam\'e constants $\lambda$ and $\mu$, quantify the material's resistance to compression and shear respectively and are determined experimentally.

Taking derivatives with respect to strain gives expressions for the First and Second Piola-Kirchhoff stress tensors, and applying the inverse Piola Transform gives the Cauchy stress.
\begin{equation}
\label{eqn:stress1}
\begin{array}{llllll}
\text{Second Piola-Kirchhoff stress} 	& \S 			&=& \frac{\partial W}{\partial \E} &=& \lambda \tr(\E) \I + 2 \mu \E 	\\[5pt] 
\text{First Piola-Kirchhoff stress} 	& \P 			&=& \frac{\partial W}{\partial \F} &=& \F\S 						\\[5pt] 
\text{Cauchy stress} 				& \bm{\sigma} 	&&  					&=& J^{-1} \P\F^{\top} 				\\[5pt] 
\end{array}
\end{equation}
The material models are extended to allow for muscle contraction coupled to the monodomain equation (\ref{eqn:monodomain1}) by the addition of contractile stress to the material model corresponding to fibre directions $\hat{\f}$ in the reference configuration.  Other models will be introduced as needed, such as \cite{Guccione1995} and \cite{Tang2007} which allow for passive isotropy due to the fibre orientation of the myocardium.  

\subsection{Elastodynamics Equation}

Until now these results have been given in their Eulerian form, but we now combine them in their Lagrangian form. 

Taking the reference configuration at $t = 0$ we have $\x = \phi(\hat{\x},0) = \hat{\x}$, thus (\ref{eqn:cons1}) and (\ref{eqn:def_vlm1}) imply that mass is conserved over time:
\begin{eqnarray}
\int_{\Omega} \rho(\x,t) \ dV &=&  \int_{\hat{\Omega}} J(\hat{\x},t) \rho(\phi(\hat{\x},t),t)  \ d\hat{V} \\
\label{eqn:cons2}
&=&  \int_{\hat{\Omega}} \rho(\hat{\x},0)  \ d\hat{V}.
\end{eqnarray}
We convert the Eulerian description of resultant force (\ref{eqn:fres1}) into its Lagrangian form using integral relations (\ref{eqn:def_area1}) and (\ref{eqn:def_vlm1}) and substitute mass conservaiton (\ref{eqn:cons2}):
\begin{equation}
\label{eqn:fres2}
\r(\x,t) = \int_{\partial \hat{\Omega}} \underbrace{J(\hat{\x},t) \bm{\sigma}(\phi(\hat{\x},t),t) \F(\hat{\x},t)^{-\top}}_{\P(\hat{\x},t)} \hat{\n}(\hat{\x}) \ d\hat{A} 
+  \int_{\hat{\Omega}} \rho(\hat{\x},0)  \b(\phi(\hat{\x},t),t) d\hat{V}
\end{equation}
The tensor field $\P(\hat{\x},t)$ is the First Piola-Kirchhoff stress tensor which encodes the \textit{stress-strain} relation.

Substituting into Newton's Second Law (\ref{eqn:nsl1}) relates the change in linear momentum (\ref{eqn:lin_mom1}) to body and traction forces.
\begin{equation}
\label{eqn:elas1}
\frac{d}{dt} \int_\Omega \rho(\x,t) \v(\x,t) \ dV  
=
\int_{\partial \hat{\Omega}} \P(\hat{\x},t) \hat{\n}(\hat{\x}) \ d\hat{A}  
+
\int_{\hat{\Omega}}  \rho(\hat{\x},0)  \b(\phi(\hat{\x},t),t) d\hat{V}
\end{equation}
This is the Elastodynamics equation, which we will integrate in time for the solution of the solid body dynamics problem.

\section{Fluid Dynamics}

We model blood as an incompressible Newtonian fluid and apply conservation of mass and linear momentum as before. The fluid flow is naturally expressed in its Eulerian form and calculation of the time derivative leads to a \textit{transport} term. Fluids do not support shear forces, and thus Cauchy stress depends upon \textit{strain rate} and \textit{pressure}.  The combination of transport, stress and body forces result in the Navier-Stokes equations for incompressible flow. 

This section is based upon the previous texts in continuum mechanics as well as  \cite{Versteeg1995} and \cite{Patankar1980}.

\subsection{Continuity Equation}

The Eulerian description of a fluid flow is spatial velocity field $\v(\x,t)$. Change in density is equivalent to the accumulation of mass by velocity. Conservation of mass can thus be expressed in differential form as follows:
\begin{equation}
\frac{\partial \rho }{\partial t} + \nabla \cdot (\rho  \v) = 0.
\end{equation}
For an incompressible fluid, density $\rho$ is constant and the expression reduces to the continuity equation:
\begin{equation}
\label{eqn:cont1}
\nabla \cdot \v = 0.
\end{equation}

\subsection{Advection}\label{sub:adv1}

Consider a scalar function $q : \R^d \times \R \mapsto \R$, of a spatial point $q(\x,t)$ in motion such that $\x = \phi(\hat{\x},t)$. The total time derivative of $q$ has a contribution from changes in both space and time. Applying the Chain Rule recovers the inner product of velocity $\v$ and spatial gradient $\nabla q$:
\begin{eqnarray}
\frac{Dq}{Dt} 
&=& \frac{\partial q}{\partial t} + \frac{\partial q}{\partial \phi} \cdot \frac{\partial \phi}{\partial t} \\
&=& \frac{\partial q}{\partial t} + \v \cdot \nabla q.
\end{eqnarray}
The value of $q$ is thus transported by velocity. For the fluid $q$ is replaced by $\v$, and the velocity field transports itself by advection:
\begin{equation}
\label{eqn:adv1}
\frac{D\v}{Dt} = \frac{\partial \v}{\partial t} + (\v \cdot \nabla) \v.
\end{equation}
It is worth noting that if the velocity $\v$ of the fluid is known, then its evolution can be described without knowledge of its motion $\phi$.  

\subsection{Cauchy Stress}

In an incompressible fluid, forces arise from changes in velocity and pressure. The Cauchy stress tensor is thus a function of spherical pressure field $p(\x,t)$ and the \textit{strain rate} tensor for viscous stress:
\begin{equation}
\label{eqn:cauchy1}
\bm{\sigma}(p,\v) = -p \I + \mu \left[\nabla \v + \nabla \v^\top \right]
\end{equation}
where $\mu$ is the dynamic viscosity, considered constant in the Newtonian case and $\I$ is an identity matrix.

\subsection{Navier-Stokes Equations}

The expression for Newton's Second Law (\ref{eqn:nsl1}) is written in its integral form over an arbitrary set $\Omega$. Since the balance of momentum holds for any $\Omega$ and density $\rho$ is constant, the law can be expressed in its differential form as follows:
\begin{equation}
\rho \frac{D}{Dt}  \v(\x,t)  
=
\nabla \cdot \bm{\sigma}(\x,t) + \rho \b(\x,t).
\end{equation}
We substitute for advection (\ref{eqn:adv1}) on the left and fluid Cauchy stress (\ref{eqn:cauchy1}) on the right
\begin{equation}
\rho \left[ \frac{\partial \v}{\partial t} + (\v \cdot \nabla) \v \right]
=
\nabla \cdot \left(-p \I + \mu \left[\nabla \v + \nabla \v^\top \right] \right)
+ \rho \b,
\end{equation}
then make use of the following calculus identities, 
\begin{eqnarray}
\nabla \cdot p \I &=& (\nabla p) \I + p (\nabla \cdot \I) = \nabla p, \\
\nabla \cdot (\nabla \v) &=& \Delta \v, \\
\nabla \cdot  (\nabla \v^\top) &=& \nabla (\nabla \cdot \v) = 0,
\end{eqnarray}
where, in the last case,  $\nabla \cdot \v = 0$ by continuity (\ref{eqn:cont1}) . The resulting momentum equation, along with continuity are the Navier-Stokes equations for incompressible flow:
\begin{eqnarray}
\nabla \cdot \v &=& 0 \\
\label{eqn:ns1}
\rho \left[ \frac{\partial \v}{\partial t} + (\v \cdot \nabla) \v \right]
&=&
- \nabla p + \mu \Delta \v + \rho \b.
\end{eqnarray}
The integration of these equations in time will give the solution of the fluid dynamics problem.

\section{Fluid-Structure Interaction}

Given the continuum description of solid and fluid dynamics it now remains to define the dynamics at their interface.  The combined description of the Lagrangian properties of the solid and the Eulerian properties of the fluid lead to the Arbitrary Lagrangian-Eulerian framework (ALE). This material follows \cite{Formaggia2010} and \cite{Nordsletten2011}.

\subsection{ALE Formulation}\label{sub:ale1}

\tikzset{pics/grid/.style={code={
\fill[draw=gray, pattern=north east lines,pattern color=lightgray] (-2,-2) rectangle (+2,-1);
\fill[draw=gray, pattern=north east lines,pattern color=lightgray] (-2,+1) rectangle (+2,+2);
\draw [gray]  (-2,-2) grid [step=0.5] (2,2);
\draw[draw=blue,thick] (-2,-1) to (2,-1);
\draw[draw=blue,thick] (-2,+1) to (2,+1);
}}}

\makeatletter
\def\mytransformation{%
    \pgfmathsetmacro{\myX}{\pgf@x + 4*cos(3*\pgf@y)}
    \pgfmathsetmacro{\myY}{\pgf@y - 5*sin(3*\pgf@x)}
    \setlength{\pgf@x}{\myX pt} 
    \setlength{\pgf@y}{\myY pt}
}
\makeatother

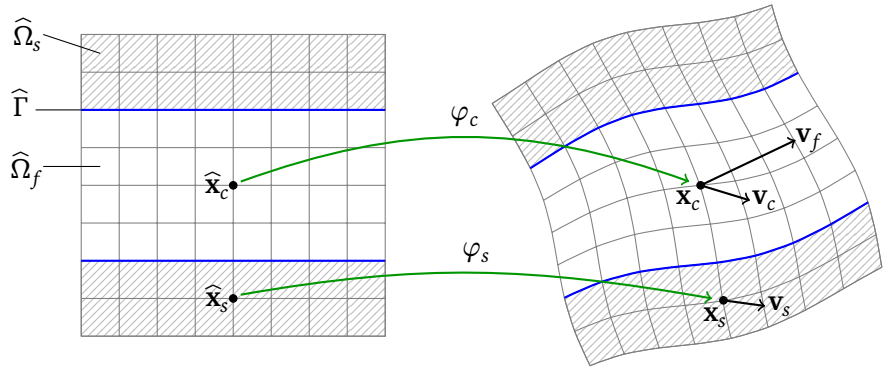
\begin{figure}[h]
\centering
\begin{tikzpicture}
\pic[scale=1,shift={(-6,0)}] (a) {grid};
\begin{scope}
	\pgftransformnonlinear{\mytransformation}
	\pic[scale=1,rotate=20] (b) {grid};
\end{scope}
\draw  [fill=black]  (-6,0)  circle (0.05cm); 
\draw  [fill=black]  (0.15,0)  circle (0.05cm); 
\node (C) at (-6.2,0) {$\hat{\x}_c$};
\node (D) at (-0.0,-0.2) {$\x_c$};
\draw[->,green,thick,out=20,in=160] (-5.9,0.05) to node [midway,above,black] {$\phi_c$} (-0.0,0.05) ;
\draw  [fill=black]  (-6,-1.5)  circle (0.05cm); 
\draw  [fill=black]  (0.45,-1.525)  circle (0.05cm); 
\node (C) at (-6.2,-1.5) {$\hat{\x}_s$};
\node (D) at (0.35,-1.75) {$\x_s$};
\draw[->,green,thick,out=10,in=170] (-5.9,-1.45) to node [midway,above,black] {$\phi_s$} (0.3,-1.5) ;
\node at (-8.7,2.0) {$\hat{\Omega}_s$};
\node at (-8.8,1.1) {$\hat{\Gamma}$};
\node at (-8.7,0.2) {$\hat{\Omega}_f$};
\draw[-] (-8.5,2.0) to (-7.75,1.75) ;
\draw[-] (-8.6,1.0) to (-8,1) ;
\draw[-] (-8.5,0.25) to (-7.75,0.25) ;
\draw[->,thick] (0.45,-1.525) to (1.0,-1.6) ;
\node at (1.2,-1.6) {$\v_s$};
\draw[->,thick] (0.15,0.0) to (0.8,-0.2) ;
\node at (1.0,-0.2) {$\v_c$};
\draw[->,thick] (0.15,0.0) to (1.4,0.6) ;
\node at (1.6,0.6) {$\v_f$};
\end{tikzpicture}
\caption{Solid, Fluid and Computational domains, velocities and motions in the ALE formulation.}
\label{fig:ale1}
\end{figure}

Figure \ref{fig:ale1} shows \underline{s}olid $\hat{\Omega}_s$  and \underline{f}luid $\hat{\Omega}_f$ domains in the reference configuration with their images $\Omega_s(t)$ and $\Omega_f(t)$ under deformation. Their boundaries $\partial\hat{\Omega}_s$ and $\partial\hat{\Omega}_f$ allow the definition of boundary conditions in the normal way. The fluid-structure interface or surface $\Gamma(t)$ is defined by their common boundary:
\begin{equation}
\Gamma(t) = \Omega_s(t) \cap \Omega_f(t)
\end{equation}

The current position $\x_s$ and velocity $\v_s$ of a point on the interior of the solid domain $\hat{\x}_s$ can be described via the elastodynamics equation (\ref{eqn:elas1}) via the solid motion $\phi_s(\hat{\x}_s,t)$.

As noted in Section \ref{sub:adv1}, for a point on the interior of the fluid domain, flow velocity $\v_f$ is an Eulerian field described by the Navier-Stokes equations (\ref{eqn:ns1}) without the need for a reference position or motion.

\subsection{Computational Domain}

In order to couple the conservation of mass and momentum on the solid and fluid domains the ALE formulation requires an extra definition. The computational domain $\Omega_c \subset \R^d$ represents both the solid and the void through which fluid flows. 
\begin{equation}
\Omega_c = \Omega_s \cup \Omega_f
\end{equation}
It thus occupies the same space as $\Omega$ but carries Lagrangian information about the current position $\x_c$ and velocity $\v_c$ of any point in the reference configuration $\hat{\x}$ via the computational motion $\phi_c(\hat{\x}_c,t)$. 

In the solid domain the solid and computational motions are equal but on the fluid domain the computational motion $\phi_c$ is independent of the fluid.
\begin{equation}
\phi_c = \phi_s, \ \forall \x \in \Omega_s
\end{equation}
The purpose of $\Omega_c$ is to encode the coupling conditions  between the solid and fluid domains on the surface $\Gamma$.  The computational motion does not have any physical meaning in itself and does not influence the solution. It is updated by any means that ensures the admissibility of the deformation.  

Since it has velocity on the fluid domain, this velocity $\v_c$ must be subtracted from fluid velocity $\v_f$ so that only relative velocity is used in the calculation of advection (\ref{eqn:adv1}), hence:
\begin{equation}
\label{eqn:adv2}
\rho \left[
\frac{\partial \v_f}{\partial t} 
+
 \left[(\v_f - \v_c) \cdot \nabla \right] \v_f 
\right]
= - \nabla p + \mu \Delta \v_f + \rho \b
\end{equation}
It is now possible to apply the coupling conditions to the solid, fluid and computational domains.

\subsection{Coupling Conditions}

Firstly, in the reference configuration the images of solid and computational deformation coincide for all points in the domain $\hat{\Omega} = \hat{\Omega}_s \cup \hat{\Omega}_f$.
\begin{equation}
\phi_s(\hat{\x},0) = \phi_c(\hat{\x},0), \ \forall \hat{\x} \in \hat{\Omega}
\end{equation}
Secondly, the solid and computational motion of each point on the surface $\hat{\Gamma}$ coincides at all times. This gives continuity of position.
\begin{equation}
\phi_s(\hat{\x},t) = \phi_c(\hat{\x},t), \ \forall \hat{\x} \in \hat{\Gamma}
\end{equation}
Thirdly, the velocity of each point on the surface $\Gamma$ coincides at all times. This applies a \textit{no-slip} boundary condition to the fluid-structure interface.
\begin{equation}
\label{eqn:couple2}
\v_s(\x,t) =\v_f(\x,t) =\v_c(\x,t) , \ \forall \x \in \Gamma
\end{equation}
Finally, traction forces are set equal and opposite on the surface $\Gamma$.  This gives continuity of stress:
\begin{equation}
\label{eqn:couple1}
\bm{\sigma}_s \n_s + \bm{\sigma}_f \n_f = \0, \ \forall \x \in \Gamma 
\end{equation}
In this way a point on the surface is subject to forces arising from the motion of both solid and fluid. 

\subsection{Fluid-Structure Summary}

With the inclusion of \underline{D}irichlet and \underline{N}eumann boundary conditions $\f,\g$ on the \underline{s}olid and \underline{f}luid domains, ignoring body forces, the coupled FSI problem can be written as follows \cite{Formaggia2010}:

Solid:
\begin{alignat}{5}
\rho_{s,0} \dfrac{d  \v_s}{dt}   - \nabla \cdot \bm{\sigma}_s 	&&\ =\ &\0,\quad&& \x&\ \in\ &\Omega_s\\
\v_s 											&&\ =\ &\f_s,\quad&& \x&\ \in\ &\partial\Omega_{s,D} \\[6pt]
\bm{\sigma}_s \n_s 								&&\ =\ &\g_s,\quad&& \x&\ \in\ &\partial\Omega_{s,N} 
\end{alignat}
Fluid:
\begin{alignat}{5}
\nabla \cdot \v_f 																	&&\ =\ &\0,\quad&&\x&\ \in\ &\Omega_f \\
\rho_f \left[\frac{\partial \v_f}{\partial t} + [(\v_f - \v_c) \cdot \nabla ] \v_f \right]  + \nabla p - \mu \Delta \v_f	&&\ =\ &\0,\quad&&\x&\ \in\ &\Omega_f \\
\v_f 																				&&\ =\ &\f_f,\quad&&\x&\ \in\ &\partial\Omega_{f,D} \\[6pt]
\bm{\sigma}_f \n_f 																	&&\ =\ &\g_f,\quad&&\x&\ \in\ &\partial\Omega_{f,N} 
\end{alignat}
FSI:
\begin{alignat}{5}
\x_s = \x_f																			&&\ =\ &\x_c,\quad&&\x&\ \in\ &\Gamma \\
\v_s = \v_f																			&&\ =\ &\v_c,\quad&&\x&\ \in\ &\Gamma \\
\bm{\sigma_f} \n + \bm{\sigma_s} \n													&&\ =\ &\0,\quad&&\x&\ \in\ &\Gamma 
\end{alignat}

We have thus defined a set of coupled problems in electrophysiology, solid and fluid dynamics.  In order to solve them we next consider the computational tools at our disposal.

\part{Method}

\chapter{GPU Computation}\label{chp:gpu}

\section{Introduction}

This chapter will briefly introduce GPU architecture and show how parallel processing can increase computing performance. The cost of this improved performance is a set of constraints upon code design. If these constraints are not followed carefully then the benefits of parallel computing may be lost. 

It is common for practitioners to \textit{parallelize} or \textit{accelerate} existing code by \textit{porting} it to GPU, and then to show \textit{speedup} with respect to their CPU implementation. There also exist frameworks and libraries which can be applied to standard problems, but without considering parallel processing in the design of a mathematical algorithm there is a limit to what these approaches can achieve.

The aim of this thesis is to re-consider existing mathematical and computational approaches and to show \textit{proof-of-concept} for a multi-physics solver with parallel computing at the centre of its design. As such it is important to make clear the objectives and compromises that must be made, since they will guide the subsequent work.

This chapter is based largely upon the OpenCL language specification \cite{Khronos2022}, programming \cite{Nvidia2012}, and performance guides \cite{Nvidia2011}.  There are many guide books including \cite{Scarpino2011}.

\section{GPU Architecture}

Originally, the Graphics Processing Unit (GPU) was a specialised piece of computer hardware designed to manipulate and alter memory for output to the screen buffer for display. The specialism arose because frequently required tasks, such as interpolation, were often parallel in their nature, usually with respect to pixels, and required low latency, specifically the screen buffer should be refreshed at the frame rate of the display.

Once present, the use of the units was extended to include other computationally intensive and parallel tasks, such as linear operations on polygonal meshes and texture mapping in graphics rendering, as well as video and sound processing.  Success in these tasks, drew interest for scientific applications and languages were developed that permitted General Purpose (GPGPU) computing.  

One such language is OpenCL \cite{Khronos2022}, which is a based upon set of agreed standards for functionality agreed by the GPU vendors and consumers, comprising the Khronos Working Group, and implemented separately by vendors. The group is also responsible for other standards including OpenGL shader language, widely used in computer graphics. The adoption of these standards was beneficial to the growth of GPU computing since it allowed elasticity of supply and demand.  Since then however GPGPU has become a large and very significant market which now includes multimedia processing transforms, the training and forward evaluation of neural networks and cryptocurrency mining.  Vendors have extended functionality into proprietary languages such as NVIDIA's CUDA, Apple's Metal Shader Language and Microsoft's DirectCompute Extensions, claiming improved programming flexibility and performance, largely for business purposes. Increasing competition has reduced the vendors' enthusiasm for open standards such as OpenCL, but it is still pre-installed and supported on almost all new and existing machines.
 
At the same time that GPU languages have evolved, so have the units themselves.  Starting as specialised circuits they became larger off-chip devices with their own DRAM memory.  Now that their benefits have been demonstrated both in terms of speed and power consumption, there has been some focus on improving the integration of CPU and GPU memory, which is a significant bottleneck, and is discussed later in this chapter. 

OpenCL 3.0 includes instructions that allow simultaneous co-processing of memory by both CPU and GPU.   

We start by introducing the main aspects of GPU computation then consider how this determines code design.

\subsection{Hardware Multi-threading}

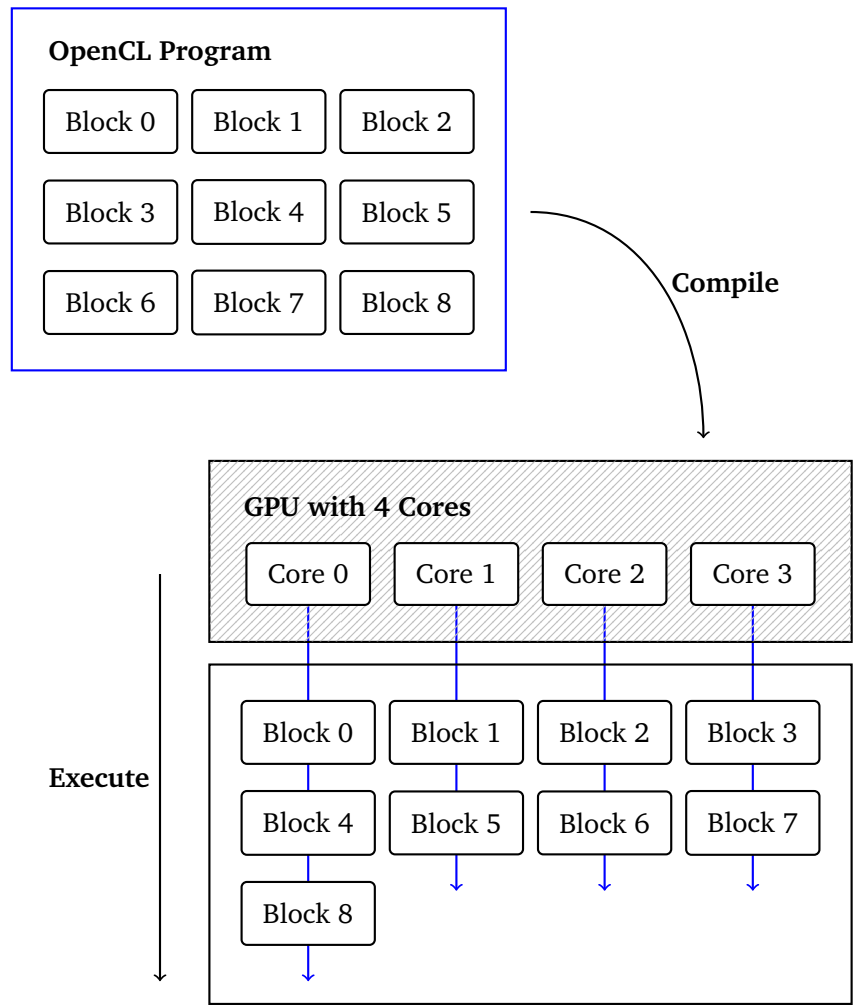
\begin{figure}[h]
\centering
\begin{tikzpicture}[xscale=1.3,yscale=1.2]
\def\block(#1,#2,#3,#4){\node[draw=black,thick,rounded corners=2,inner sep=8,fill=white] (#1) at (#2,#3) [anchor=center] {#4};}

\begin{scope}[shift={(0,8)}]
\draw[thick,blue](-1,-0.75) rectangle (4,3.25);
\node at (0.5,2.75) {\bf{OpenCL Program}};
\block(A, 0.0, 0, Block 6)
\block(B, 1.5, 0, Block 7)
\block(C, 3.0, 0, Block 8)
\block(E, 0.0, 1, Block 3)
\block(F, 1.5, 1, Block 4)
\block(G, 3.0, 1, Block 5)
\block(E, 0.0, 2, Block 0)
\block(F, 1.5, 2, Block 1)
\block(G, 3.0, 2, Block 2)
\end{scope}

\begin{scope}[shift={(2,1.25)}]
\draw[->,thick,blue] (0.0,4) to (0.0,-0.75);
\draw[->,thick,blue] (1.5,4) to (1.5,+0.25); 
\draw[->,thick,blue] (3.0,4) to (3.0,+0.25); 
\draw[->,thick,blue] (4.5,4) to (4.5,+0.25); 
\end{scope}

\begin{scope}[shift={(2,5)}]
\draw[thick,black](-1,-0.75) rectangle (5.5,1.25);
\draw[pattern=north east lines,pattern color=lightgray](-1,-0.75) rectangle (5.5,1.25);
\node at (0.5,0.75) {\bf{GPU with 4 Cores}};
\block(A, 0.0, 0, Core 0)
\block(B, 1.5, 0, Core 1)
\block(C, 3.0, 0, Core 2)
\block(E, 4.5, 0, Core 3)
\end{scope}

\begin{scope}[shift={(2,1.25)}]
\draw[thick,black](-1,-1) rectangle (5.5,2.75);
\block(A, 0.0, 2, Block 0)
\block(B, 1.5, 2, Block 1)
\block(C, 3.0, 2, Block 2)
\block(E, 4.5, 2, Block 3)
\block(A, 0.0, 1, Block 4)
\block(B, 1.5, 1, Block 5)
\block(C, 3.0, 1, Block 6)
\block(E, 4.5, 1, Block 7)
\block(A, 0.0, 0, Block 8)
\end{scope}

\draw[->,black,thick,out=0,in=90] (4.25,9) to node [midway,above,black,anchor=west] {\bf{Compile}} (6,6.5) ;
\draw[->,black,thick] (0.5,5) to node [midway,above,black,anchor=east] {\bf{Execute}} (0.5,0.5) ;

\end{tikzpicture}
\caption{Thread blocks in a program are processed independently and in parallel by a multicore GPU.}
\label{fig:gpu1}
\end{figure}

Both CPU and GPU execution pipelines make use of threads. A four quad-core CPU can process 16 threads in parallel (or 32 with hyperthreading). There may be a delay while the results of a floating point operations (FLOPs) or memory access becomes available in register memory.  For FLOPs this may be as much as 24 cycles while for off-chip memory access 400-800 cycles. If this latency causes the pipeline to stall, then the environment can switch to another to keep the processor busy. The pipeline is thus filled via \textit{multi-threading}.  The cost of switching threads on a CPU may be high, since each is fully independent and carries with it a large amount context and register information stored both on and off-chip.

Figure \ref{fig:gpu1} gives a schematic representation of an OpenCL program.  The program consists of a large number of threads, grouped into thread blocks (or work groups).  Each block contains as many as 1024 threads and is allocated to a compute unit (or core) at compile time.  At execution the thread blocks are processed sequentially but independently by the compute units.  It is worth noting that the order in which the blocks are processed may not have a close physical relation to their layout in the program, as shown in the figure.

The execution context, counters and register information are preserved on-chip for all threads for the entire life of the block. Within the block there are \textit{thread groups} with a minimum size of 32 threads. If any thread group is stalled the processor can switch to another and continue execution at zero cost.  As a result the register latency for floating point operations and memory access can, in theory, be  hidden. 

This architecture is known as Single Instruction Multiple Thread (SIMT).  In addition the GPU may have vector processing units, which can perform concurrent floating point operations (usually four) in the same number of cycles as a single FLOP, thus implementing a Single Instruction Multiple Data (SIMD) architecture. GPU languages have vector and matrix data types for this purpose.

A good GPU may have 100 compute units and can thus support the concurrent processing of 100,000 threads, with billions scheduled at any one time.  Thus the GPU gives an opportunity for a huge increase in computing performance. For this improvement to be achieved, the action of the program must be \textit{parallel}, that is the design of the program should allow the threads to operate on a large number of data elements at the same time. 

\subsection{Memory}\label{sub:memory1}

\subsubsection{Host/Device Memory Transfer}

In general, CPU memory is not directly accessible to the GPU. The original model for computation involved populating a memory buffer on the CPU \textit{host}, copying it via the Peripheral Component Interconnect (PCI) bus to GPU \textit{device} for processing and rendering, then filling the screen buffer for the display. The one-way nature of this process allowed for read-only and write-only buffers which removes the need for cache management.

With the advent of GPGPU, users generally need to copy results back to the CPU to store results as files, requiring another pass through the PCI bus. If procedures require intermediate processing or memory access then there must be further transfers and cache management.  

The bandwidth of the PCI bus between CPU and GPU memory is around 8GBps, compared to 128GBps between GPU memory and processor, or 100GBps between CPU memory and processor. The cost of memory transfer is therefore a key bottleneck for GPU computation.

\subsubsection{Device Memory Spaces}

Table \ref{tbl:gpu1} shows features of GPU device memory spaces. They are ordered roughly by preference, with Register and Shared memory being ideal but limited in size and scope.  Constant memory is cached and therefore has zero latency on second read and can be considered as good as a register.  Thereafter Texture memory, which is discussed later, benefits from caching and optimised memory memory access. 
\begin{table}[h]
\begin{center}
\begin{tabular}{llllll}
\bf{Memory} & \bf{On-chip} & \bf{Cached} 	& \bf{Access} 	& \bf{Scope} 	& \bf{Lifetime} 	\\
\hline
Register 	& yes		& - 			& read/write 	& thread 		& thread 			\\
Shared 		& yes 		& - 			& read/write 	& block 		& block				\\
Constant 	& no 		& yes 			& read			& all			& all 				\\
Texture 	& no 		& yes 			& read/write* 	& all 			& all 				\\
Local 		& no		& yes/no*		& read/write 	& thread 		& thread 			\\
Global	 	& no 		& no 			& read/write 	& all  			& all 			
\end{tabular}
\end{center}
\caption{Principle features of GPU memory spaces, reproduced from \cite{Nvidia2011}.  *Texture memory write access and Local memory caching are version and vendor-dependent.}
\label{tbl:gpu1}
\end{table}

The vast majority of memory is Global, where large working buffers reside and latency is roughly 100 times that of registers. During execution the programmer can actively manage the movement of data from Global to Shared memory and as such it is a user-controlled cache.  Some devices will cache data into Shared memory directly. When Global memory is accessed, however, it is not possible to avoid the cost of one read or write per kernel execution.

When memory requirements exceed the number of available registers, the processor will first make use of on-chip Shared then off-chip Global memory, with associated performance reduction called \textit{register spill}. If a kernel requires more resources than are available then it will fail to launch.

\subsubsection{Texture Memory}

Texture mapping is the process in computer graphics rendering by which 2D bitmap images are applied to the surfaces of 3D polygonal meshes. It is a fundamental part of the graphics pipeline which makes multiple reads of many such bitmaps and various techniques such as interpolation, reflection and repetition, all of which are ideal for GPU processing.  As such there is a special portion of cached memory available for these operations which has several hardware optimisations for fast access. 

\begin{figure}[h]
     \centering
     \begin{subfigure}{0.4\textwidth}
		\centering
		\includegraphics[width=0.8\textwidth]{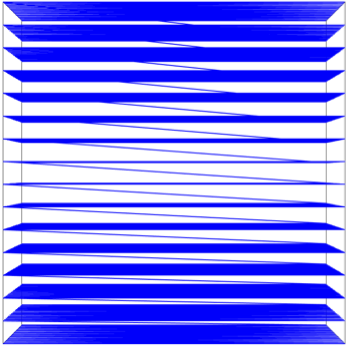}
		\includegraphics[width=0.8\textwidth]{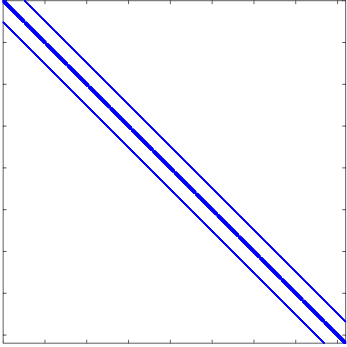}
         \caption{Canonical}
     \end{subfigure}
     \begin{subfigure}{0.4\textwidth}
		\centering
		\includegraphics[width=0.8\textwidth]{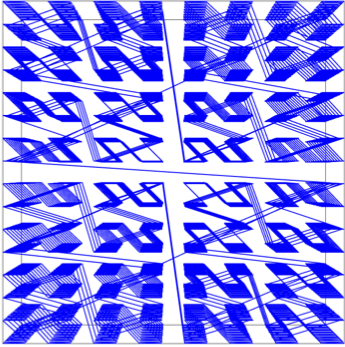}
		\includegraphics[width=0.8\textwidth]{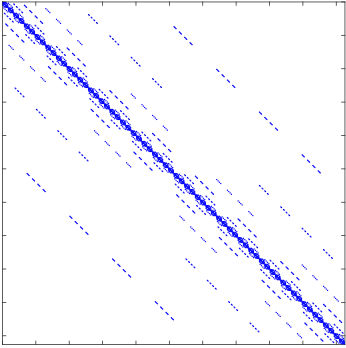}
         \caption{Z-Ordered}
     \end{subfigure}
	\caption{Canonical and Z-Ordered memory indexing in 3D, and corresponding sparsity patterns for 7-point stencil.}
    \label{tbl:zorder1}
\end{figure}

Space-filling curves map high dimensional spaces into one-dimension. Texture memory uses this technique to re-index memory addresses.  When assembling a stencil operation, \cite{Morton1966} showed that the technique is equivalent to re-ordering a sparse matrix such that bandwidth is optimally reduced. An example for a 3D cube with a 7-point stencil operation is shown in Figure \ref{tbl:zorder1}.

The canonical scheme in 2D generates a scalar memory index from zero-based row and column indices by taking the row index multiplied by the row size plus the column index. Z-ordering takes the row and column indices as binary numbers and interleaves their digits in alternating fashion. The method extends naturally into higher dimensions resulting in the distinctive z-shaped curves that give the method its name.

In addition to this, texture memory has some other attractive features. Firstly, Row and column indices can be normalised into the range $[0,1]^d$ and called using floating point coordinate values.  There is specialised hardware that calculates memory addresses for the surrounding values and performs bilinear or trilinear interpolation. This is equivalent to storing a scalar, vector or even tensor field in a bi- or tri-linear Lagrangian basis. The GPU can be set to handle out-of-bounds either returning zero, fixed, repeated or reflected values as necessary.

Secondly, Bitmap data in RGBA format is stored as a vector of four floating point values which is ideal for storing vector fields. The values can be stored as half, single, or double precision values or unsigned integers. When queried the values can be returned as higher precision floating point or integer types.

These two features allow data fields stored at different spatial resolutions and precision to be combined seamlessly into the same calculation.  This is important for multigrid methods, since a problem can be completely reconstructed at any scale.

\section{Design Objectives}

As discussed in the previous sections, the performance benefits of the SIMT and SIMD architectures, as well as optimised memory access of the GPU offer a potentially huge increase in performance over CPU processing. In order to benefit from these advantages, the computer program must have the following characteristics:
\begin{itemize}
\item The action of the program should be \textit{parallel}, that is that action of the threads can be performed on a large number of data elements at the same time. 
\item Traffic over the PCI bus between the CPU host and the GPU device should be minimised.
\item There should be \textit{coherence} in data access, that is there should be spatial and temporal locality in both read and write memory operations which allows the device to \textit{coalesce} memory operations.
\end{itemize}
The following sections describe some strategies by which these objectives can be achieved.  

\subsection{Parallel Execution}

Maximising parallel execution starts with structuring the algorithm in a way that exposes as much data parallelism as possible.

The floating point dot product $\R^d \times \R^d \mapsto \R$ is a fundamental operation in linear algebra. It comprises a product which can be computed in parallel and a sum which is inherently serial.  For the product both CPU and GPU can use independent threads.  As discussed in previous sections, there would be a cost in PCI memory transfer from host to device but for large vectors this would be outweighed by the increased thread count of the GPU (10k threads) compared to the CPU (32 threads). 

For the sum, the CPU compiler would use \textit{loop unrolling} to accelerate the operation, allocating memory for subtotals which could be computed in parallel. The GPU would use a similar \textit{parallel reduction} operation in which vectors of subtotals would hold results of parallel sums. In both cases, speedup would thus be a tradeoff between parallel computation and memory use, as well as the cost of memory transfer for the GPU.  For operations more complex than the dot product, algorithms and strategies grow in complexity, and a large quantity of literature and software exists that attempts to solve these problems.

The alternative is to remove the dot product from the algorithm entirely, and this is the approach taken here. At all stages the operations used by the solvers are data parallel.  The threads behave like stencil operations reading from their near neighbours, and writing to a single location in memory. There is no CPU calculation in the algorithm at all, and no data transfer between device or host during calculation except for the writing of output files.

\subsection{Control Flow}

Branching keywords (\textit{if, for, do, while, switch}) cause threads to follow different execution paths.  A CPU handles code branching in a straightforward way, executing only paths that are required. The SIMT architecture of the GPU relies on all multiprocessors carrying out the same instructions at any given time.  As a result code divergence is serialised by the compiler. Where conditional logic is used within a kernel, both branches are calculated and one value is discarded.  It is therefore important to design the algorithm such that this code branching contains a minimal amonut of processing.

For example, an \text{if} control condition will be allocated a per thread control predicate, which stores the result of the logical evaluation.  The code block within the if statement will execute for all threads, but any resulting operations such as register writes will only take place where the control predicate is true. This is known as \textit{bit-masking}.

When there is an \text{if...else} condition, the compiler will execute both code branches for all threads, applying bit-masking to the results, effectively serialising the divergence. This serialisation removes the advantage of parallel processing and greatly reduces performance. Therefore, the algorithm and code design in this work removes all branching statements.

\subsection{Memory Throughput}

For both CPU and GPU memory access represents the key limit to computational performance.

The first and most obvious optimisation is the use of single precision floating point variables. GPUs were not developed for high numerical precision and while some offer extensions for double precision calculations, most have single precision floating point registers and arithmetic units. As such single precision calculations halve instruction and memory throughput.

The range of values and accuracy of single precision floats is sufficient for the calculation physiological information that our model requires.  In some cases (such as fibre fields) half precision data is sufficient.

\subsubsection{PCI Bus}

As mentioned earlier, the relatively low bandwidth of the PCI bus (8MBps) compared to GPU the Global memory (128MBps) space is a key performance bottleneck. There are several strategies which minimise transfer over the PCI bus:

Memory should be allocated and persisted on the GPU wherever possible. Routines that allocate device memory allow flags that control read and write access by both host and device allowing the compiler to optimise cache use.  Memory transfers should \textit{coalesced} into a few bulk operations rather than many smaller ones. 

Data transfer can be avoided by re-calculation of values rather than storage.  This is especially true with reference to CPU and GPU where a kernel that is slower on GPU than CPU may be preferable as it avoids transfer.  It is also relevant to device-only storage where calculation may be faster than retreival.

Memory transfer can be achieved in a straightforward way by the allocation of resources on both host and device followed by a memory copy from one to the other.  Another method that is available is the use of \textit{pinned memory}.  Implemented in different ways by different vendors it is effectively a region of cache that is in the address space of both device and host and not subject to paging.  As such there are routines that will return a pointer to GPU memory that can be read or written by the CPU at a higher bandwidth than a traditional copy operation. 

In addition, pinned memory can be read in a synchronous or asynchronous fashion. Once a pointer is allocated, the application can either perform a \textit{blocking transfer} in which the GPU processing is paused until IO is complete, or a \textit{non-blocking transfer} in which the GPU continues its execution.  In the second case the latency of memory transfer is hidden but its contents may change during the operation.  This must be considered as part of the program design.

As discussed above, the continuing integration of CPU and GPU memory has led to \textit{Shared Virtual Memory} in which memory is accessible to both host and device and can be operated upon in a synchronous and asynchronous way, with out the need for explicit transfers.  For example, Apple's recent M1 processor has 64GB of shared memory for exactly this purpose and demonstrates that manufacturers are accepting data parallelism as a fundamental part of processor design.

As a result the design considerations outlined above, the code implementation in this work does not create any memory or perform any calculations on the on the CPU.  All allocations and calculations are carried out directly on the GPU. The only memory transfer over PCI bus is the reading of buffers to generate output files. This is performed asynchronously via pinned memory and is thus optimal with respect to memory transfer.

\subsubsection{Device Memory}

There are several considerations which relate to the efficient use of the GPU device memory spaces:

The use of shared memory should be maximised.  An example of blocked matrix multiplication is usually given, in which the block size allows multiple threads to share local memory access.  In this case data is loaded into shared memory manually. 

In general however, it can be assumed that the device will make one cached read and/or write per kernel execution.  Thus it is important that the kernel does not require multiple access to global memory.  Atomic writes are serialised, that is multiple writes to the same memory address will be carried out separately but the order of their execution is not defined, this results in a \textit{race condition} for both read and write. It is important that code either avoids race conditions or is robust to them.  For instance, when using Jacobi iteration to solve a linear system, a race condition would lead in some instances to a Gauss-Seidel iteration instead.  In this case the race condition is actually beneficial to the speed of convergence and the solvers in this work will take advantage of this behaviour.

Originally it was a requirement that a memory location was either read-only or write-only per kernel access, thus removing the need for cache coherence.  Modern languages and processors allow for read/write operations within a kernel but it is still beneficial to observe this design constraint.  Memory can be flagged as read or write only per kernel for increased performance.

The features of texture memory indicate the way in which \textit{coherent} memory access patterns allow for \textit{coalesced} memory access.  The SIMT architecture blocks threads on to multiprocessor cores which use and cache Shared memory.  It is important that all threads in a block are accessing memory from the same or similar cache lines. For this reason random or strided memory access will destroy the benefits of device caching. The next chapter describes the structured grid on which the problem is discretised and this design decision is made with coherent memory access in mind.

Register spill should be avoided by separating large complex kernel in to smaller units.  If a kernel requires more memory than the available registers then it will be allocated from Global resources, which introduces hundreds of clock cycles of latency. There is a low cost in launching a kernel (estimated around $300\mu s$) and so decomposition may be a better option.

As already mentioned, the kernels in this work will make multiple local reads per kernel and only write to a single unique location per memory buffer.  This coherent access pattern takes full advantage of spatial locality and cached global memory access. 

\subsection{Instruction Usage}

The final design considerations take place at the register and processor level with the aim of maximising processor throughput:

As mentioned before, the floating point operations of a GPU are naturally single precision and the macroscopic nature of the work allows for their use.

As with any good code, slow functions such as division and numerical root finding should be avoided. Similarly most GPUs do not have dedicated integer processing units and integer division and modulo operations should be avoided.

The OpenCL language has compiler directives which coalesce floating multiply-add operations which can achieve significant speedup with a small cost in accuracy.  There are also options for relaxed and fast mathematics calculations.  These are implemented with geometric operations on meshes with only visual quality in mind and thus should be avoided

The following Chapters \ref{chp:disc} and \ref{chp:alg} describe respectively the discrete formulation and GPU parallel solution of the continuous problem outlined in Chapter \ref{chp:prb}.

\chapter{Discretization and Solver}\label{chp:disc}

In general, the various continuous spatial differential operators specified in Chapter \ref{chp:prb} are discretized onto a computational mesh, and integrated in time via the method of lines. The use of the dual mesh combined with the Finite Volume Method for GPU assembly and processing is original to this work.

This Chapter has three sections, the first of which describes the discrete operators that will be required, first on the regular structured grid, and then in the deformed configuration. The next section explains how a signed distance function is used to encode geometry relating solid and fluid domains into the mesh. Finally, we show how a linear Laplacian solver is assembled with respect to both the geometry and deformation.

\section{Operators}

We consider the problem as described in Chapter \ref{chp:prb} on a domain $\Omega \subset \R^d$ as a set of scalar $\R^d \rightarrow \R$, vector $\R^d \rightarrow \R^d$ and second-order tensor $\R^d \rightarrow \R^{d \times d}$ fields on a regular structured grid. 

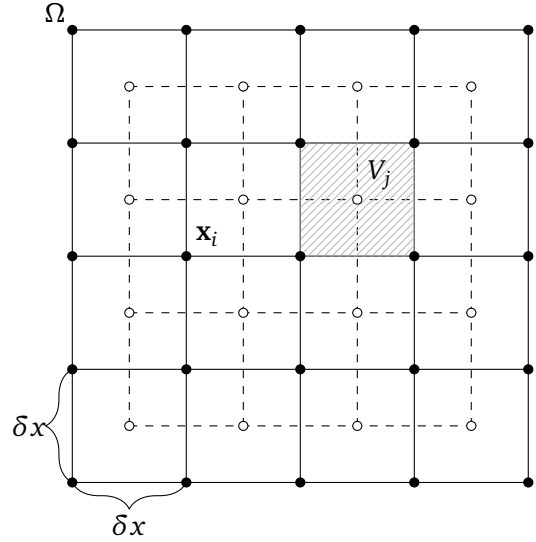
\begin{figure}[h]
\centering
\begin{tikzpicture}[scale=1.5]
\draw[step=1] (0,0) grid (4,4);		
\draw[step=1,dashed,shift={(0.5,0.5)}] (0,0) grid (3,3);	
\fill[draw=gray, pattern=north east lines,pattern color=lightgray] (2,2) rectangle (3,3);						
\foreach \x in {0,...,4} 
    \foreach \y in {0,...,4} 
      { 
        \draw  [fill=black]  (\x,\y) circle (0.04); 
      } 
\foreach \x in {0,...,3} 
    \foreach \y in {0,...,3} 
      { 
        \draw  [fill=white,shift={(0.5,0.5)}]  (\x,\y) circle (0.04); 
      } 
\draw [decorate,decoration={brace,amplitude=10}]  (0,0) -- (0,1) node [black,midway,xshift=-16] {$\delta x$};
\draw [decorate,decoration={brace,amplitude=10,mirror}]  (0,0) -- (1,0) node [black,midway,yshift=-16] {$\delta x$};
\node at (-0.15,4.15) {$\Omega$};
\node[anchor=south west] at (2.5,2.5) {$V_j$};
\node[anchor=south west] at (1,2) {$\x_i$};
\end{tikzpicture}
\caption{A regular structured grid over the domain $\Omega \subset \R^2$ with mesh width $\delta x$, vertices $\x_i$ and face dual at the centres of enclosed volumes $V_j$.}
\label{fig:mesh1}
\end{figure}

Some information is stored at mesh vertices $\x_i \subset \R^d$ while other values relate to the enclosed volumes $V_i$ and is stored at their centres.  As such both the mesh and its dual are used in the solution of the problem and this is illustrated in Figure \ref{fig:mesh1}.

The fields will be stored for optimised memory access and processed in parallel in a per-vertex or per-volume manner. This is in line with the design considerations for GPU processing and memory access outlined in Chapter \ref{chp:gpu}.

The mesh will deform according to the motions $\phi_s, \phi_c$ of the solid and computational subdomains $\Omega_s, \Omega_c$ as described in section \ref{sub:ale1}. Throughout this Chapter the calculations will be illustrated in $\R^2$ for simplicity, although everything described extends naturally into $\R^3$ and by the orthogonality of the coordinate system.

All of the operations in the following section are SIMT parallel in elements or vertices and SIMD parallel for vector and tensor fields. The first part of Chapter \ref{chp:exp} will demonstrate their convergence properties. 


\subsection{Reference Volume}

\begin{figure}[h]
\centering
\begin{tikzpicture}[scale=3]
\draw[step=1] (0,0) grid (1,1);							
\foreach \x in {0,...,1} 
    \foreach \y in {0,...,1} 
      { 
        \draw  [fill=black]  (\x,\y) circle (0.02); 
      } 
\draw[blue,dashed]  (0.0,0.4) to  (1.0,0.4);
\draw[blue,dashed]  (0.4,0.0) to  (0.4,1.0);
\draw  [fill=white]  (0.4,0.4) circle (0.02); 
\node[anchor= south west] at (0.4,0.4) {$\u(\hat{\x})$};
\node at (0.0,-0.1) {$\u_{00}$};
\node at (1.0,-0.1) {$\u_{10}$};
\node at (0.0,1.1) {$\u_{01}$};
\node at (1.0,1.1) {$\u_{11}$};
\end{tikzpicture}
\caption{The reference volume $\hat{V} \in \R^2$, showing function evaluation of $\u$ at at the vertices and a notional point $\hat{\x}$.}
\label{fig:ref1}
\end{figure}
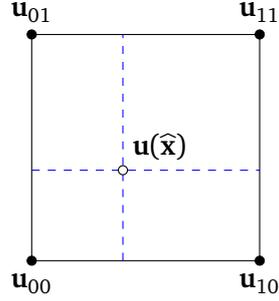

We consider a reference volume $\hat{V} \subset \R^2$ as a unit square, with local coordinate system for a point $\hat{\x} = [x_1,x_2]^\top \in [0,1]^2$. The continuously differentiable vector field $\u : \R^2 \rightarrow \R^2$ is defined over $\hat{V}$. Discrete differential operators will be derived for $\u$ but are also needed for scalar $u : \R^2 \rightarrow \R$ and tensor $\U: \R^2 \rightarrow \R^{2 \times 2}$ fields. Where necessary definitions will be expanded accordingly. These definitions are equivalent to the reference element in FEM and are summarised in Figure \ref{fig:ref1}.  

\subsubsection{Interpolation}

The value of $\u(\hat{\x})$ can be approximated by bilinear interpolation. We use an abbreviated notation for the values at the vertices $\u(0,1) = \u_{01}$ etc, as shown in Figure \ref{fig:ref1}:
\begin{equation}
\label{eqn:refintp1}
\u(\hat{\x}) \approx	(1-x_2)\left[(1-x_1) \u_{00} + x_1 \u_{10}\right] + x_2\left[(1-x_1) \u_{01} + x_1 \u_{11}\right]
\end{equation}
This is equivalent to applying a set of linear Lagrange basis functions to the vertices of $\hat{V}$ and weighting $\u$ by $\hat{\x}$. At the centre of the volume, where $\hat{\x} = [0.5,0.5]^\top$ the formula reduces as follows:
\begin{equation}
\label{eqn:refintp2}
\u(\hat{\x})  \approx	\frac{1}{4}\left(\u_{00} + \u_{10} + \u_{01} + \u_{11}\right).
\end{equation}
The interpolation operation is the same for the scalar $u$ and tensor $\U$ fields.
 
\subsubsection{Gradient}

Taking the gradient of the Lagrange basis functions yields a difference formula for the discrete gradient of $\u$, equivalent to interpolating differences along the edges of the volume.  Vertex values and difference weights are assembled into matrix form giving an expression for the Jacobian matrix:
\begin{equation}
\label{eqn:refgrad1}
\frac{\partial \u}{\partial \hat{\x}} 
\approx
\left[\begin{array}{cccc}\u_{00} & \u_{10} & \u_{01} & \u_{11}\end{array}\right]
\left[\begin{array}{rr}
- 1 + \hat{x}_2	& -1 + \hat{x}_1 \\
+ 1 - \hat{x}_2	& -  \hat{x}_1 \\
- 	\hat{x}_2		& + 1 - \hat{x}_1 \\
+ 	\hat{x}_2		& +  \hat{x}_1
\end{array}\right] \in \R^{2 \times 2}
\end{equation}
It is worth noting that the gradients are piecewise constant in their respective dimensions. At the centre of $\hat{V}$ the formula reduces as follows:
\begin{equation}
\label{eqn:refgrad2}
\frac{\partial \u}{\partial \hat{\x}}
\approx \frac{1}{2}
\left[\begin{array}{cccc}\u_{00} & \u_{10} & \u_{01} & \u_{11}\end{array}\right]
\left[\begin{array}{rr}
-1	& -1 \\
+1	& -1 \\
-1	& +1 \\
+1	& +1 
\end{array}\right]
\end{equation}
For a scalar field $u$ the expression gives a row vector approximating the gradient transposed $\nabla u(\hat{\x})^\top$. For the tensor field $\U$, the operator yields a second order tensor in $\R^{2 \times 2 \times 2}$:
\begin{equation}
\label{eqn:refgrad3}
\frac{\partial \U}{\partial \hat{\x}}
\approx
\left[\begin{array}{cc}
\dfrac{\partial \U}{\partial \hat{x}_1} & \dfrac{\partial \U}{\partial \hat{x}_2} 
\end{array}\right].
\end{equation}

\subsubsection{Divergence}

The divergence of a vector or tensor field is defined as the limit of outward flux per unit volume. 
\begin{equation}
\label{eqn:refdiv1}
\nabla \cdot \u = \lim_{\hat{V} \rightarrow 0}\frac{1}{|\hat{V}|} \oint_{\partial \hat{V}} \u \cdot \hat{\n} \ dA \\
\end{equation}
Using the FVM approach, the discrete integral is the sum over $j$ faces of the face midpoint values $\u_j$, interpolated via (\ref{eqn:refintp1}), dotted with the exterior normals $\hat{\n}_j$. 
%
The values at the face centres are the averages of the values at their incident vertices and thus the inner products represent average flux across each face. Since side lengths (face areas in $\R^3$) and volume $\hat{V}$ have unit value in the reference configuration the divergence can be stated simply as:
\begin{equation}
\label{eqn:refdiv2}
\nabla \cdot \u \approx \sum_j^4 \u_j \cdot \hat{\n}_j.
\end{equation}

For a scalar field $u$, this operation is not defined, although in FVM the discrete application of Green's theorem to a scalar field gives the Green-Gauss gradient. On the reference volume this is equal to both the standard FDM and FEM gradient approximations.  

For a tensor field $\U$ in the reference volume, the inner product is replaced by a matrix-vector product and the expression for discrete divergence is as follows:
\begin{equation}
\label{eqn:refdiv3}
\nabla \cdot \u \approx \sum_j^4 \U_j \cdot \hat{\n}_j.
\end{equation}

\subsubsection{Laplacian}

\begin{figure}[h]
\centering
\begin{tikzpicture}[scale=1.5]
\draw[step=1] (0,0) grid (2,2);		
\draw[step=1,dashed,shift={(0.5,0.5)}] (0,0) grid (1,1);	
\foreach \x in {0,...,2} 
    \foreach \y in {0,...,2} 
      { 
        \draw  [fill=black]  (\x,\y) circle (0.04); 
      } 
\foreach \x in {0,...,1} 
    \foreach \y in {0,...,1} 
      { 
        \draw  [fill=white,shift={(0.5,0.5)}]  (\x,\y) circle (0.04); 
      } 
\node[anchor=south west] at (1.0,1.0) {$\hat{\x}$};
\node[anchor=east] at (-1,1.0) {$\underbrace{\nabla \u_j}_{\text{Step 1}}$};
\draw[blue] (-1,1.0) -- (0.5,0.5);
\draw[blue] (-1,1.0) -- (1.5,0.5);
\draw[blue] (-1,1.0) -- (0.5,1.5);
\draw[blue] (-1,1.0) -- (1.5,1.5);
\node[anchor=west] at (2.5,0.5) {$\underbrace{\sum_{j=1}^{4}\nabla \u_j \cdot \hat{\n}_j = \Delta \u(\hat{\x})}_{\text{Step 2}}$};
\draw[color=blue] (1,1) -- (2.5,0.75);
\end{tikzpicture}
\caption{The Discrete Laplacian operator uses nine values of $\u$ at the vertices (black dots) to calculate four values of the gradient field $\nabla \u_j$ at the volume centres (white dots). It then calculates the divergence of the gradients at the centre of the dual, which is the Laplacian at vertex $\hat{\x}$ itself.}
\label{fig:ref2}
\end{figure}
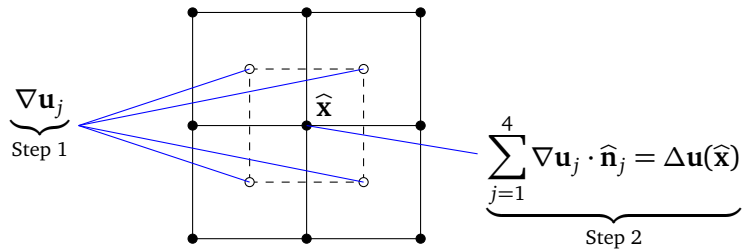

Due to the piecewise constant nature of the discrete gradient on the reference volume, its divergence is zero by definition.  For the Laplacian it is necessary to consider both the primal and dual in two steps, meshes as shown in Figure \ref{fig:ref2}. We make use of the fact that a regular structured grid is a \textit{self-dual} mesh.

The field $\u$ is stored at each of the primal mesh vertices (black dots).  In the first step,  (\ref{eqn:refgrad2}) is used to calculate a discrete gradient at the centre of each of the four volumes (white dots) adjacent to the central vertex $\hat{\x}$.

The volume centres are the vertices of the dual mesh.  In the second step, the divergence of the gradients is calculated via (\ref{eqn:refdiv3}).  Since the centre of the dual volume is a vertex of the primal mesh, this gives the Laplacian at the central vertex $\hat{\x}$.

It is important to note that an identical procedure can be used to find the Laplacian for a field on the dual mesh, stored at the centre of the volumes, via gradients the vertices of the primal mesh. The algorithm will make use of both the primal and dual Laplacian operators.

Again, in the reference configuration these definitions are equivalent to both those of FEM and FDM.

\subsection{Deformed Volume}

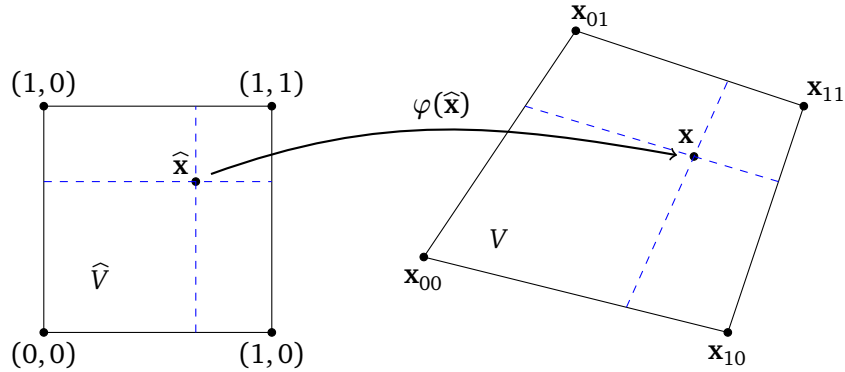
\begin{figure}
\centering
\begin{tikzpicture}
\draw (0,0) rectangle (3,3);
\draw (5,1) -- (9,0) -- (10,3) -- (7,4) -- cycle ;
\draw  [fill=black]  (0,0) circle (0.05cm); 
\draw  [fill=black]  (0,3) circle (0.05cm); 
\draw  [fill=black]  (3,0) circle (0.05cm); 
\draw  [fill=black]  (3,3) circle (0.05cm); 

\draw  [fill=black]  (5,1) circle (0.05cm); 
\draw  [fill=black]  (9,0) circle (0.05cm); 
\draw  [fill=black]  (10,3) circle (0.05cm); 
\draw  [fill=black]  (7,4) circle (0.05cm); 

\draw  [fill=black]  (2,2)  circle (0.05cm); 
\draw  [fill=black]  (8.555,2.333)  circle (0.05cm); 
\draw[->,thick,out=20,in=170] (2.2,2.1) to node [midway,above] {$\phi(\hat{\x})$} (8.333,2.35) ;
\draw[blue,dashed] (2,0) -- (2,3);
\draw[blue,dashed] (0,2) -- (3,2);
\node at (0,-0.3) {$(0,0)$};
\node at (3,-0.3) {$(1,0)$};
\node at (0,3.3) {$(1,0)$};
\node at (3,3.3) {$(1,1)$};
\node at (5,0.7) {$\x_{00}$};
\node at (9,-0.3) {$\x_{10}$};
\node at (7.2,4.2) {$\x_{01}$};
\node at (10.3,3.2) {$\x_{11}$};
\draw[blue,dashed]  (7.666,0.333) to  (9.000,3.333);
\draw[blue,dashed] (6.333,3.000) to  (9.666,2.000);
\node at (1.8,2.25) {$\hat{\x}$};
\node at (8.45,2.6) {$\x$};
\node at (0.75,0.75) {$\hat{V}$};
\node at (6,1.25) {$V$};
\end{tikzpicture}
\caption{The reference and deformed volume showing vertices, an interior point $\hat{\x}$ and its image $\x$.}
\label{fig:def1}
\end{figure}

In the deformed configuration, a Lagrangian description of position and velocity is provided by the solid $\phi_s$ and computational $\phi_c$ motions, as defined in section \ref{sub:ale1}. We will derive expressions for discrete operators at the centre of a volume under an admissible deformation $\phi : \R^2 \rightarrow \R^2$. Figure \ref{fig:def1} shows the reference volume $\hat{V}$ and its image $V$ under deformation.

\subsubsection{Interpolation}

The interpolation operator at the centre of the deformed volume is the same as for the undeformed volume (\ref{eqn:refintp1}).

\subsubsection{Deformation Gradient}

We now consider the deformed image $\x = \phi(\hat{\x})$ of a point $\hat{\x}$ in the reference volume $\hat{V}$ and apply the discrete gradient operator (\ref{eqn:refgrad1}) to the deformation itself.  The resulting discrete approximation to the deformation gradient $\F \in \R^{2 \times 2}$ and its determinant $J$ are key to the continuum mechanics definitions of (\ref{eqn:mot1}). At the centre of the volume (\ref{eqn:refgrad2}) gives the following expression:  
\begin{equation}
\label{eqn:defgrad1}
\F = \frac{\partial \x}{\partial \hat{\x}} 
= \frac{\partial x_i}{\partial \hat{x}_j} 
\approx 
\frac{1}{2}
\left[\begin{array}{cccc}
 \x_{00} & \x_{10} & \x_{01} & \x_{11}
\end{array}\right]
\left[\begin{array}{cc}
-1 & -1 \\
+1 & -1 \\
-1 & +1 \\
+1 & +1
\end{array}\right]
\end{equation}
It will frequently be necessary to calculate the inverse of the deformation gradient in various, such as $J \F^{-\top}$ in Nanson's relation (\ref{eqn:def_area1}).

In practice we work in $\R^3$ and calculate a transposed inverse and determinant using the vector cross product.  For a matrix $\mathbf{A} \in \R^{3 \times 3}$ with columns $\mathbf{A} = [\mathbf{a}_1 \  \mathbf{a}_2  \ \mathbf{a}_3 ]$ we have:
\begin{eqnarray}
\label{eqn:cross1}
\mathbf{A}^{-\top} &=& \frac{1}{\det \mathbf{A}} 
\left[\begin{array}{ccc}
\mathbf{a}_2  \times \mathbf{a}_3 & \mathbf{a}_3  \times \mathbf{a}_1  & \mathbf{a}_1  \times \mathbf{a}_2  \\
\end{array}\right] \\
\det \mathbf{A} &=& \mathbf{a}_1 \cdot (\mathbf{a}_2  \times \mathbf{a}_3)
\end{eqnarray}
The cost is three cross products, for which OpenCL has a built-in function.

\subsubsection{Gradient}

For the gradient operator in the deformed configuration, we consider the function $\u$ under the coordinate transformation $\phi$:
\begin{equation}
\u(\x) = \u(\phi(\hat{\x})).
\end{equation}
Applying the chain rule and recognising the deformation gradient $\F$:
\begin{equation}
\frac{\partial \u}{\partial \hat{\x}}
\ = \ 
\frac{\partial \u}{\partial \x} \frac{\partial \x}{\partial \hat{\x}}
\ \approx\ 
\frac{\partial \u}{\partial \x} \F.
\end{equation}
The gradient in the deformed configuration can thus be calculated via the reference gradient (\ref{eqn:refgrad2}) and the deformation gradient (\ref{eqn:defgrad1}), inverted as per (\ref{eqn:cross1}). 
\begin{equation}
\label{eqn:defgrad2}
\frac{\partial \u}{\partial \x} 
\ =\ 
\frac{\partial \u}{\partial \hat{\x}}
\frac{\partial \hat{\x}}{\partial \x}
\ \approx \ 
\frac{\partial \u}{\partial \hat{\x}} \F^{-1}
\end{equation}

\subsubsection{Divergence}

Starting from the discrete integral for divergence in the reference volume (\ref{eqn:refdiv2}), the expression for deformed divergence takes into account the effect of the deformation on the volume (\ref{eqn:def_vlm1}) and the face normals (\ref{eqn:def_area1}).
\begin{equation}
\label{eqn:defdiv1}
\nabla \cdot \u \approx \frac{1}{J} \sum_j^4 \u_j \cdot J_j \F_j^{-\top} \hat{\n}_j 
\end{equation}
where $J$ is the determinant at the centre of the volume. The relevant deformation gradients can be calculated from vertex positions and since the multiplication $J_j\F_j^{-\top} \hat{\n}_j$ simply selects columns, it is not necessary to perform full matrix inversion. The relevant column can be calculated directly via (\ref{eqn:cross1}) at the cost of one cross product.

For a tensor field $\U$ the process is similar, applying (\ref{eqn:refdiv3}) in the deformed configuration gives:
\begin{equation}
\label{eqn:defdiv2}
\nabla \cdot \U \approx \frac{1}{J} \sum_j^4 J_j \U_j \F^{-\top} \hat{\n}_j,
\end{equation}

\subsubsection{Laplacian}\label{sub:lap1}

Calculation of the Laplacian follows the same steps as for the reference volume, making use of the self-dual grid as shown in Figure \ref{fig:deflap1}. First, the vertex coordinates and values (black dots) are used to calculate the coordinates (\ref{eqn:refintp1}) and gradient (\ref{eqn:defgrad2}) at the centres of the volumes (white dots) adjacent to the vertex $\x$.  The values at the volume centres are then used to calculate the deformed divergence of the resulting tensor field at the vertex $\x$ itself via (\ref{eqn:defdiv2}).

\tikzset{pics/grid/.style={code={
\draw [gray]  (-1,-1) grid [step=1] (1,1);
\draw[step=1,dashed,shift={(-0.5,-0.5)}] (0,0) grid (1,1);
\foreach \x in {-1,...,1} 
    \foreach \y in {-1,...,1} 
      { 
        \draw  [fill=black]  (\x,\y) circle (0.04); 
      } 
\foreach \x in {0,...,1} 
    \foreach \y in {0,...,1} 
      { 
        \draw  [fill=white,shift={(-0.5,-0.5)}]  (\x,\y) circle (0.04); 
      } 
}}}

\makeatletter
\def\mytransformation{
    \pgfmathsetmacro{\myX}{\pgf@x + 4*cos(4*\pgf@y)}
    \pgfmathsetmacro{\myY}{\pgf@y - 5*sin(4*\pgf@x)}
    \setlength{\pgf@x}{\myX pt} 
    \setlength{\pgf@y}{\myY pt}
}
\makeatother

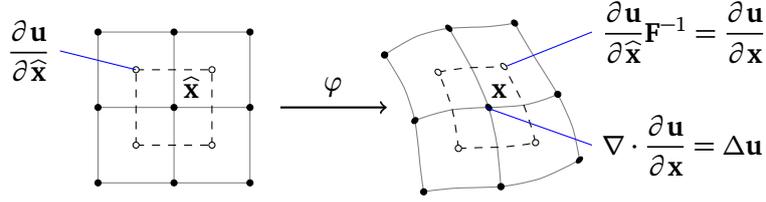
\begin{figure}
\centering
\begin{tikzpicture}
\pic[scale=1,shift={(-4,0)}] (a) {grid};
\begin{scope}
	\pgftransformnonlinear{\mytransformation}
	\pic[scale=1,rotate=20] (b) {grid};
\end{scope}
\draw[->,thick] (-2.6,0) to node [midway,above] {$\phi$} (-1.2,0) ;
\node[anchor=south west] at (0.05,0) {$\x$};
\node[anchor=south west] at (-4,0) {$\hat{\x}$};
\node[anchor=east] at (-5.5,0.75) {$\dfrac{\partial \u}{\partial \hat{\x}}$};
\draw[blue] (-5.5,0.75) -- (-4.5,0.5);
\node[anchor=west] at (1.5,1) {$\dfrac{\partial \u}{\partial \hat{\x}} \F^{-1} = \dfrac{\partial \u}{\partial \x}$};
\draw[blue] (0.38,0.55) -- (1.5,1);
\node[anchor=west] at (1.5,-0.5) {$\nabla \cdot\dfrac{\partial \u}{\partial \x} = \Delta \u$};
\draw[blue] (0.1,0) -- (1.5,-0.5);
\end{tikzpicture}
\caption{Calculation of the deformed Laplacian operator. First vertex values and coordinates (black dots) are used to calculate deformed gradients at the volume centres (white dots), then these are used to calculate the deformed divergence at vertex $\x$.}
\label{fig:deflap1}
\end{figure}

Under deformation however, the centre of the dual volume may not coincide with the vertex $\x$.  There are various approaches to this problem: Firstly, to calculate the position of the vertex in the reference volume (by solving a quadratic equation) and calculate the gradient and deformation at this point via (\ref{eqn:refgrad1}). Secondly, the position of the vertex can be moved to centre of the dual volume.  This is effectively a form of damping, which is in line with the continuum assumption.  Thirdly, the deformed Laplacian can be applied without adjustment, which is in line with the assumption that the deformation is constant over the volume. In practice, all three approaches are reasonable.


\section{Linear Solver}\label{sec:solv1}

The algorithm described in the following Chapter requires the solution of various linear systems which arise from the discretisation of differential operators.  We thus give a brief description of the matrix-free assembly and solution of a deformed vector Laplacian operator as defined in Section \ref{sub:lap1}. 

This solver serves as a prototype for all of the linear solvers used in the algorithm, and is applied both to the primal mesh and its dual. That is the Laplacian may be solved both with respect to vertex properties, such as velocity in the application of viscosity, or with repect to volumes as in the case of pressure or electrical charge.

\subsection{Jacobi Method}

The linear solver is based on the standard Jacobi method, which is parallel in its execution.  As discussed in Chapter \ref{chp:gpu}, race conditions in GPU memory will cause the solver to degenerate into a Gauss-Seidel smoother, which has faster convergence, but when this occurs depends upon the scheduler of the GPU and can not be controlled. Many discussions of iterative solvers include  \cite{Golub1996}, \cite{Briggs2000}, and \cite{Trottenberg2000}. 

The Jacobi method solves the linear system $\mathbf{A} \u = \b$ where $\mathbf{A} \in \R^{n \times n}$ and $\u,\b \in \R^{n}$.  The method considers the lower triangular, diagonal and upper triangular parts of $\mathbf{A}$: 
\begin{equation}
\mathbf{A} = \mathbf{L} + \mathbf{D} + \mathbf{U}.
\end{equation}
Given some $k$-th approximation $\u^{k}$ the following update step
\begin{equation}
\label{eqn:jacobi1}
\u^{k+1} = \mathbf{D}^{-1}\left[\b - (\mathbf{L} + \mathbf{U})\u^{k} \right],
\end{equation}
converges to the solution under the following condition on the spectral radius $\rho$ of the iteration matrix:
\begin{equation}
\rho (\mathbf{D}^{-1}\left[\mathbf{L}+\mathbf{U}\right]) < 1.
\end{equation}
It is sometimes preferable to apply a damped Jacobi update according to the parameter $\alpha \in \R$, usually chosen such that $\alpha \in [0,1]$ generates a convex combination of the current and previous iterates:
\begin{equation}
\label{eqn:dampjac1}
\u^{k+1} = (1-\alpha)\u^{k} + (\alpha) \mathbf{D}^{-1}\left[\b - (\mathbf{L} + \mathbf{U})\u^{k} \right].
\end{equation}

\subsubsection{Implicit Euler}\label{sec:ie1}

When the Jacobi iteration is employed with respect to the Implicit Euler time integration method there is an outer iteration over time $\u^{t} \rightarrow \u^{t+1}$, and in inner iterate for the solution of the system $\u^{k} \rightarrow \u^{k+1}$.  We derive the Jacobi step as follows, with operator $\mathbf{A}$ and generic constant $\alpha = \nu  \frac{\delta t}{\delta x^{2}}$:
\begin{eqnarray}
\u^{t+1} &=& \u^{t} + \alpha \mathbf{A} \u^{t+1} \\
\left(\I - \alpha \mathbf{A}\right)\u^{t+1} &=& \u^{t} \\
\left[\I - \alpha \mathbf{D} - \alpha (\mathbf{L} + \mathbf{U}) \right]\u^{t+1} &=& \u^{t}
\end{eqnarray}
Separating the diagonal gives an expression for the update to $\u^{k}$, the iterative the solution of $\u^{t+1}$ which is used frequently in the algorithm:
\begin{equation}
\label{eqn:iejac1}
\u^{k+1} = \left[\I - \alpha \mathbf{D}\right]^{-1}  \left[ \u^{t} +  \alpha (\mathbf{L} + \mathbf{U}) \u^{k}\right].
\end{equation}
It is worth noting that the operator includes the identity matrix $\I$ and the value $\alpha$ decreases with decreasing time step length.  As a result, the operator is diagonally dominant and thus well-conditioned for small time steps and the Jacobi iteration converges relatively quickly.

\subsection{Matrix-Free Assembly}\label{sec:ass1}

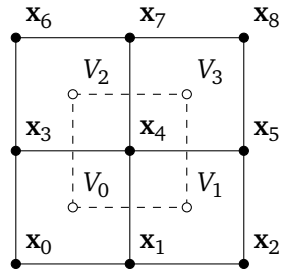
\begin{figure}[h]
\centering
\begin{tikzpicture}[scale=1.5]
\draw[step=1] (0,0) grid (2,2);		
\draw[step=1,dashed,shift={(0.5,0.5)}] (0,0) grid (1,1);	
\foreach \x in {0,...,2} 
    \foreach \y in {0,...,2} 
      { 
        \draw  [fill=black]  (\x,\y) circle (0.04); 
      } 
\foreach \x in {0,...,1} 
    \foreach \y in {0,...,1} 
      { 
        \draw  [fill=white,shift={(0.5,0.5)}]  (\x,\y) circle (0.04); 
      } 
\node[anchor=south west] at (0.5,0.5) {$V_0$};
\node[anchor=south west] at (1.5,0.5) {$V_1$};
\node[anchor=south west] at (0.5,1.5) {$V_2$};
\node[anchor=south west] at (1.5,1.5) {$V_3$};
\node[anchor=south west] at (0,0) {$\x_0$};
\node[anchor=south west] at (1,0) {$\x_1$};
\node[anchor=south west] at (2,0) {$\x_2$};
\node[anchor=south west] at (0,1) {$\x_3$};
\node[anchor=south west] at (1,1) {$\x_4$};
\node[anchor=south west] at (2,1) {$\x_5$};
\node[anchor=south west] at (0,2) {$\x_6$};
\node[anchor=south west] at (1,2) {$\x_7$};
\node[anchor=south west] at (2,2) {$\x_8$};
\end{tikzpicture}
\caption{Node numbering for assembly of the Laplacian operator.}
\label{fig:solv1}
\end{figure}
The iterative solver is applied in a matrix-free form that is parallel in matrix rows.  Thus, each GPU thread updates a single element of the solution vector, in accordance with good practice for memory access. It must therefore generate the coefficients corresponding to a single row of the Laplacian matrix. We begin by considering the assembly of the Laplacian operator \textit{on paper} and show that its implementation reduces to a computationally efficient calculation of stencil weights which can be expressed in closed form. 

As per the Figure \ref{fig:deflap1} the discrete formulation of the Laplacian operator in $\Omega \subset \R^2$ makes use of nine vertices with a point $\x$ at their centre. The implementation in $\R^3$ follows the same basic construction using 27 mesh points.

For a vector function $\u : \R^2 \rightarrow \R^2$ as defined earlier, the vertices and volume centres are numbered according to Figure \ref{fig:solv1}.  The the coefficients corresponding to the row of a matrix can be synthesised by concatenating the 9 function evaluations $\{\u_0 \ldots \u_8\}$ at vertices $\{\x_0 \ldots \x_8\}$ into a matrix $\mathbf{U} \in \R^{2 \times 9}$.  The elements of $\mathbf{U}$ are the only values that change from one Jacobi iteration to the next:
\begin{equation}
\mathbf{U} 
= \left[\begin{array}{ccc}\u_0 & \ldots & \u_8\end{array}\right] 
= \left[\begin{array}{ccc}
u_{11} & \cdots & u_{19} \\
u_{21} & \cdots & u_{29}
\end{array}\right].
\end{equation}
The undeformed gradient operator as per (\ref{eqn:refgrad2}) is applied as a difference matrix $\mathbf{D}_1 \in \R^{9 \times 8}$ to the vertex values, resulting in a matrix of four blocks corresponding to the four reference Jacobians at the centre of each volume $\{V_0,\ldots,V_3\}$.
\begin{equation}
\begin{split}
\left[\begin{array}{ccc}\u_0 & \cdots & \u_8\end{array}\right] 
\underbrace{
\frac{1}{2}
\small
\left[\begin{array}{rrrrrrrr}
-1&-1 &  0 &  0 &  0 &  0 &  0 &  0 \\
+1 & -1 & -1 & -1 &  0 &  0 &  0 &  0 \\
 0 &  0 & +1 & -1 &  0 &  0 &  0 &  0 \\
-1 & +1 &  0 &  0 & -1 & -1 &  0 &  0 \\
+1 & +1 & -1 & +1 & +1 & -1 & -1 & -1 \\
 0 &  0 & +1 & +1 &  0 &  0 & +1 & -1 \\
 0 &  0 &  0 &  0 & -1 & +1 &  0 &  0 \\
 0 &  0 &  0 &  0 & +1 & +1 & -1 & +1 \\
 0 &  0 &  0 &  0 &  0 &  0 & +1 & +1 
\end{array}\right]
}_{\mathbf{D}_{1}}\\[10pt]
\normalsize
=
\left[\begin{array}{cccc}
\dfrac{\partial \u}{\partial \hat{\x}}\Big|_{V_0} & 
\dfrac{\partial \u}{\partial \hat{\x}}\Big|_{V_1} & 
\dfrac{\partial \u}{\partial \hat{\x}}\Big|_{V_2} & 
\dfrac{\partial \u}{\partial \hat{\x}}\Big|_{V_3} 
\end{array}\right] 
\end{split}
\end{equation}
Next, the four inverse deformation gradients at each volume centre $\mathbf{F}_{V}^{-1}$, calculated via (\ref{eqn:defgrad1}) are applied in block form $\mathbf{F}_1 \in \R^{8 \times 8}$ for the deformed Jacobians at the volume centres:
\begin{equation}
\left[\begin{array}{cccc}
\dfrac{\partial \u}{\partial \hat{\x}}\Big|_{V_0} & 
\cdots &
\dfrac{\partial \u}{\partial \hat{\x}}\Big|_{V_3} 
\end{array}\right] 
\underbrace{
\left[\begin{array}{cccc}
\F_{V_0}^{-1} & 0 & 0 & 0 \\
0 & \F_{V_1}^{-1}  & 0 & 0 \\
0 & 0 & \F_{V_2}^{-1}  & 0 \\
0 & 0 & 0 & \F_{V_3}^{-1} 
\end{array}\right]
}_{\mathbf{F}_{1}}
=
\left[\begin{array}{cccc}
\dfrac{\partial \u}{\partial \x}\Big|_{V_0} & 
\ldots & 
\dfrac{\partial \u}{\partial \x}\Big|_{V_3} 
\end{array}\right] 
\end{equation}
An interpolation operator $\mathbf{D}_2 \in \R^{8 \times 8}$ as per (\ref{eqn:refintp1}) is then applied which gives four gradients interpolated at the face midpoints $\x_j = \{\x_0 \ldots \x_4\}$. Here, $\I \in \R^{2 \times 2}$ is the identity matrix.
\begin{equation}
\left[\begin{array}{cccc}
\dfrac{\partial \u}{\partial \x}\Big|_{V_0} & 
\ldots & 
\dfrac{\partial \u}{\partial \x}\Big|_{V_3} 
\end{array}\right] 
\underbrace{
\frac{1}{2} 
\left[\begin{array}{cccc}
\I & 0 & \I & 0 \\
0 & \I & \I & 0 \\
\I & 0 & 0 & \I \\
0 & \I & 0 & \I 
\end{array}\right]
}_{\mathbf{D}_{2}}
=
\left[\begin{array}{cccc}
\dfrac{\partial \u}{\partial \x}\Big|_{\x_0} & 
\ldots & 
\dfrac{\partial \u}{\partial \x}\Big|_{\x_4} 
\end{array}\right] 
\end{equation}

Finally, the products of the gradients and the deformed normals are summed to give the divergence of the gradient.  This is achieved by arranging the deformed face normals into a column vector, given by the product of the four deformation gradients at the face midpoints $\mathbf{F}_2 \in \R^{8 \times 8}$ divided by the volume determinant $J$, and the reference outer normal vectors $\mathbf{N}_1 \in \R^{8 \times 1}$ :
\begin{equation}
\left[\begin{array}{cccc}
\dfrac{\partial \u}{\partial \x}\Big|_{\x_0} & 
\ldots & 
\dfrac{\partial \u}{\partial \x}\Big|_{\x_4} 
\end{array}\right] 
\underbrace{
\frac{1}{J}
\left[\begin{array}{cccc}
J_0 \F_{0}^{-\top} & 0 & 0 & 0 \\
0 & J_1 \F_{1}^{-\top}  & 0 & 0 \\
0 & 0 & J_2 \F_{2}^{-\top}  & 0 \\
0 & 0 & 0 & J_3 \F_{3}^{-\top} 
\end{array}\right]
}_{\mathbf{F}_{2}}
\underbrace{
\left[\begin{array}{c}
\hat{\n}_0 \\ \hat{\n}_1 \\ \hat{\n}_2 \\ \hat{\n}_3 
\end{array}\right]
}_{\mathbf{N}_{1}}
\approx 
\Delta \u(\hat{\x})
\end{equation}
In summary, the forward evaluation of the discrete deformed Laplacian is a composition of matrix products:
\begin{equation}
\label{eqn:ass1}
\Delta \u  \approx
\U
\mathbf{D}_1
\mathbf{F}_1
\mathbf{D}_2
\mathbf{F}_2
\mathbf{N}_1.
\end{equation} 
We recall that the only values that change per Jacobi iteration are the elements of $\U$, while $\mathbf{D}_1$ and $\mathbf{D}_2$ are static difference and interpolation coefficients. The elements of $\mathbf{F}_1$ and $\mathbf{N}_1$ depend upon deformation gradients which only change per time step. Taking the product of all these static matrices gives a vector $\mathbf{c} \in \R^{9 \times 1}$ of coefficients:
\begin{equation}
\label{eqn:ass2}
\underbrace{\mathbf{D}_1}_{9 \times 8}
\underbrace{\mathbf{F}_1}_{8 \times 8}
\underbrace{\mathbf{D}_2}_{8 \times 8}
\underbrace{\mathbf{F}_2}_{8 \times 8}
\underbrace{\mathbf{N}_1}_{8 \times 1}
=
\underbrace{\mathbf{c}}_{9 \times 1}
\end{equation}
These are the entries of the row of the discrete Laplacian matrix, and are applied as weights to the columns of $\U$.

The assembly and multiplication of these matrices can be largely avoided if system is evaluated by composition from right to left, rather than left to right.
\begin{equation}
\label{eqn:ass3}
\underbrace{\U}_{2 \times 9}
\underbrace{\mathbf{D}_1
    \underbrace{\mathbf{F}_1
        \underbrace{\mathbf{D}_2
            \underbrace{
            	\mathbf{F}_2 \mathbf{N}_1
            }_{8 \times 1}
        }_{8 \times 1}
    }_{8 \times 1}
}_{9 \times 1}
=
\underbrace{\Delta \u}_{2 \times 1}
\end{equation}
As a result it is reduced to a small number of signed sums of difference quotients and entries of deformation matrices, which are easily hard-coded into computational kernels. This is both computationally intensive and low in memory bandwidth, ideal for GPU computation.

\FloatBarrier

\section{Geometry}\label{sec:geom}

\subsection{Signed Distance Function}

The geometry in the model is encoded via Signed Distance Functions (SDF).  This includes not only the solid and fluid domains, but regions of differing fibre direction and electrical conductivity, amongst others. Although the SDF is used here, the voxels that are used in the solution algorithm can be obtained directly from segmented medical imaging data.

\begin{figure}[h]
\centering
\includegraphics[width=0.45\textwidth,trim={2cm 8cm 2cm 8cm},clip]{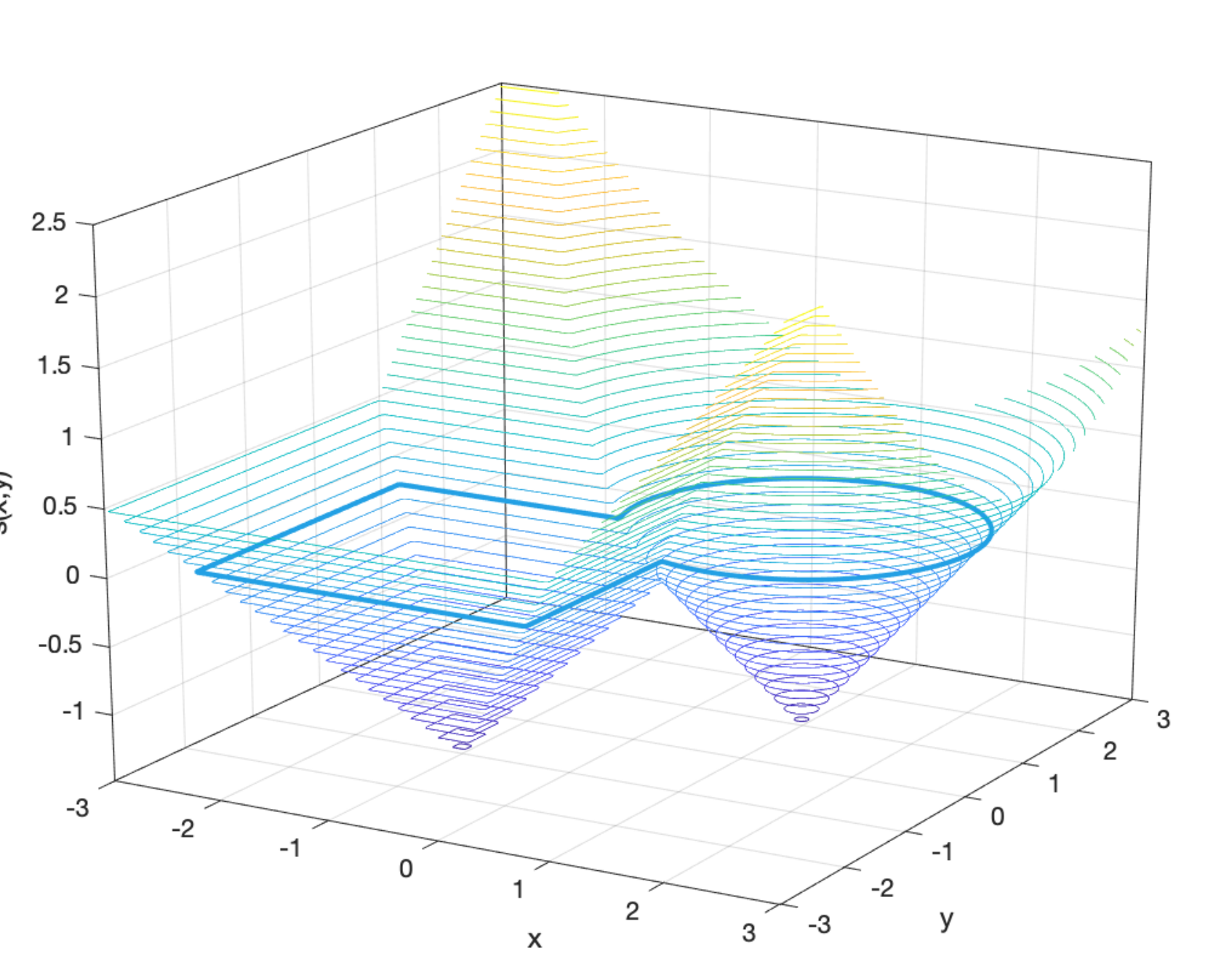} 
\includegraphics[width=0.45\textwidth,trim={2cm 8cm 2cm 8cm},clip]{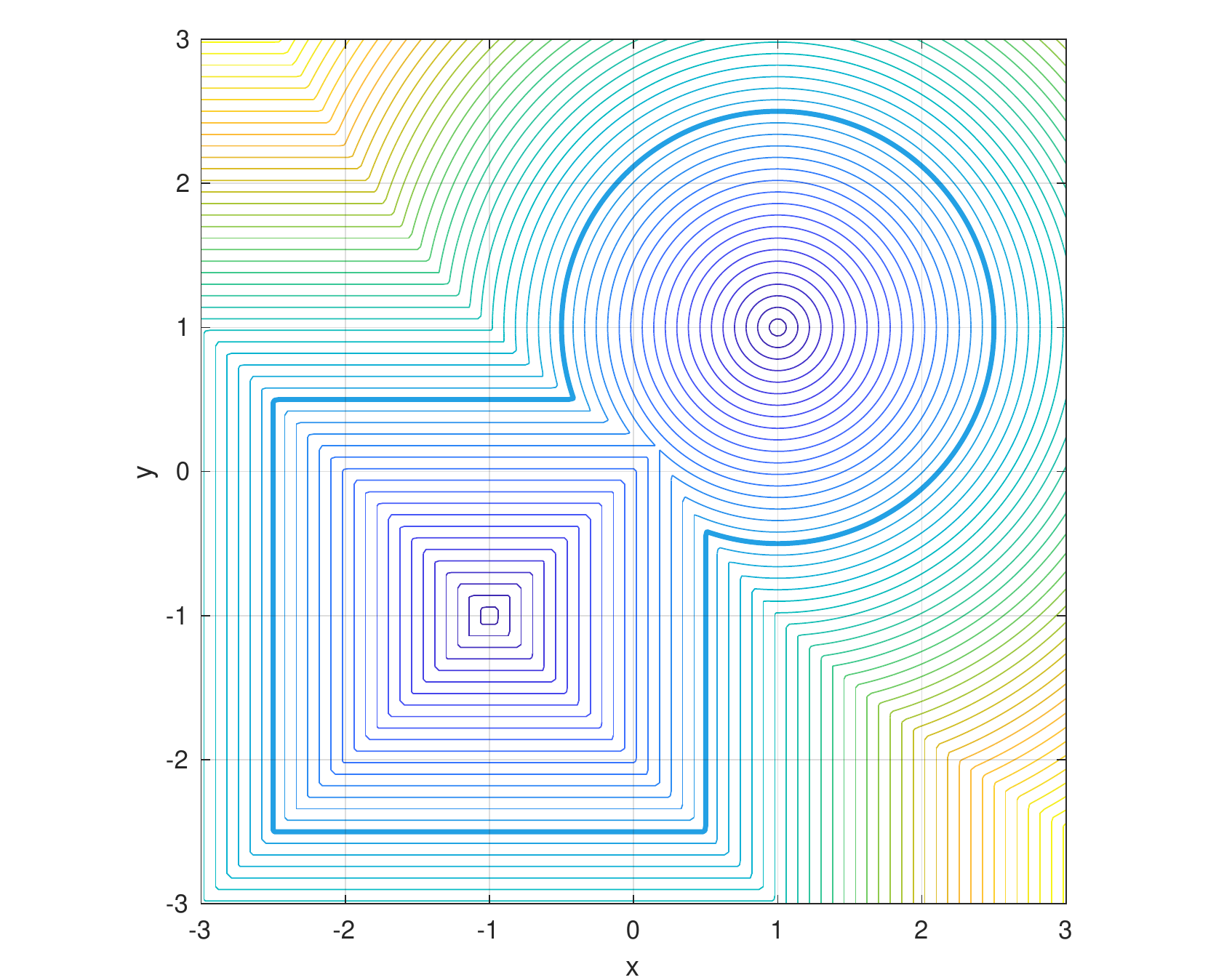} 
\caption{2D and 3D contour plots of the same Signed Distance Function $s(x,y)$. The function takes negative values on the interior of an object and positive values on its exterior. The surface (blue line) is defined as the set of points for which the value of the function is zero.}
\label{fig:sdf1}
\end{figure}

The SDF is a scalar function $s : \R^d \rightarrow \R$ over the domain that evaluates to a negative value on the interior of some subdomain and a positive value on its exterior.  The boundary or surface is the set of points for which the function evaluates to zero. Figure \ref{fig:sdf1} gives an example of a signed distance function for an intersecting circle and square, showing the positive and negative values of the function, as well as the boundary.

The use of SDFs originated in computer graphics since they have the important property that the gradient of the function on the boundary of the object is the exterior normal to the surface $\nabla s(\x) |_{s(\x)=0} = \n$. This is extremely useful when rendering scenes because it allows for the calculation of the paths of reflected light rays, during the application of lighting and surface effects, but is also ideal for the calculation of surface flux.

Mathematically the function derives from a norm. Consider a point $\mathbf{c} \in \R^d$. The Euclidean distance from $\mathbf{c}$ to any point $\x \in \R^d$ is given by the 2-norm $\|\x - \mathbf{c}\|_2$. If a scalar radius $r \in \R$ is subtracted from this distance the resulting function $s(\x,\mathbf{c},r) = \|\x - \mathbf{c}\|_2 - r$, defines a $d$-sphere with centre $\mathbf{c}$ and radius $r$. 
It is possible to define other functions based upon norms that define different shapes, for instance the 1-norm $\|\x - \mathbf{c}\|_1$ defines a $d$-cube. 

SDFs have several properties which are valuable in computation: 

They can be evaluated at any point and thus any resolution.  This allows the GPU to combine information stored at different resolutions in a straightforward manner, as outlined in section \ref{sub:memory1}. It is also ideal for multigrid solvers, since the entire problem can be fully reconstructed automatically on coarse grids.

They can be easily derived from medical imaging data, either by thresholding or by standard algorithms applied to voxel data files which are the output of MRI or CT scanners.

SDFs degrade to circles at long distances.  This allows the approximation of shapes which is useful for clipping (the exclusion of distant objects) or for collision detection and multi-body problems.

\subsection{Primitives}

Signed distance fields can be derived for various shapes and combined algebraically, allowing for the construction of complex shapes from simple objects. Much of this section is derived from the website \url{https://iquilezles.org}, although the functions have been adapted for simplicity.

\begin{figure}[h]
\centering
\includegraphics[width=\textwidth,trim={1cm 1cm 1cm 1cm},clip]{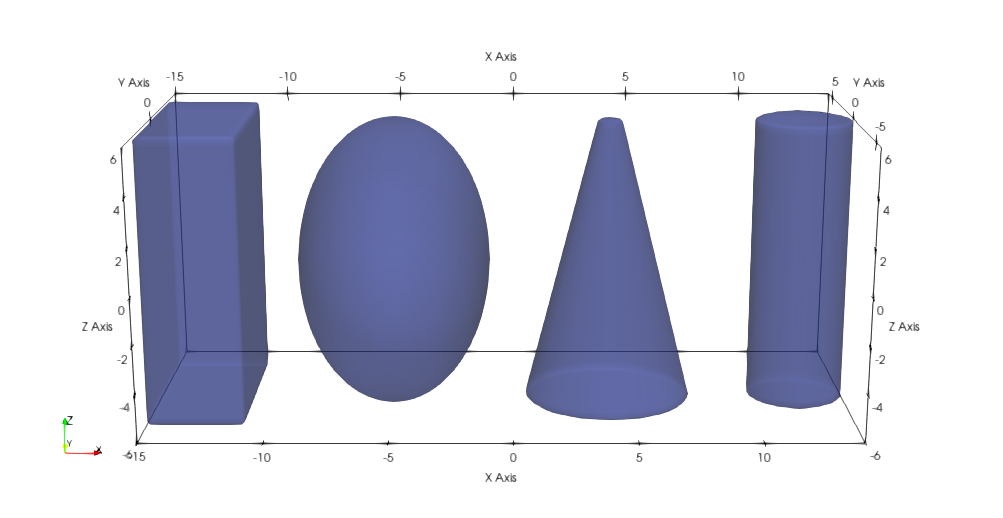} 
\caption{Cuboid, Ellipsoid, Cone and Cylinder primitives.}
\label{fig:prim1}
\end{figure}

Some examples of simple shapes, or primitives are shown in Figure \ref{fig:prim1} with code in Listing \ref{lst:prim1}. In each case the vector argument \texttt{p} is the coordinate of the point to be tested, and the function returns a scalar value (positive, negative or zero) dependent upon the location of the point with respect to the shape. The functions are mathematically simple and SIMT/SIMD parallel.

\begin{lstlisting}[language=OpenCL,label={lst:prim1}]
//cuboid, p=point, r=radius (per axis)
float sdf_cub(float3 p, float3 r)
{
    float3 q = fabs(p/r) - 1.0f;
    
    return max(q.x,max(q.y,q.z));
}

//ellipsoid, p=point, r=radius (per axis)
float sdf_ell(float3 p, float3 r)
{
    return length(p/r) - 1.0f;
}

//cylinder, p=point, r=radius, h=height
float sdf_cyl(float3 p, float r, float h)
{
    return max(length(p.xy) - r, fabs(p.z) - h);
}

//cone, p=point, r=radius, h=height, a=angle
float sdf_con(float3 p, float r, float h, float a)
{
    return max(length(p.xy) - a*p.z - r, fabs(p.z) - h);
}
\end{lstlisting}

Any affine or nonlinear transformation can be applied to the point \texttt{p} and thus the shape can be translated and deformed as required.  It is also possible to apply modulo functions to the coordinate to allow tiling.

The combination of primitive shapes is achieved by the implementation of simple set operations. We consider the two shapes $\pazocal{S}_1,\pazocal{S}_2$ and their respective SDFs $s_1, s_2 \in \R$ and implement the following set operations as follows:
\begin{equation}
\begin{array}{lccl}
\text{Union:}  			& \pazocal{S}_1 \cup 		\pazocal{S}_2 & \equiv & \min(s_1,s_2) \\
\text{Intersection:}  	& \pazocal{S}_1 \cap 		\pazocal{S}_2 & \equiv & \max(s_1,s_2) \\
\text{Complement:}  	& \pazocal{S}_1 \backslash  \pazocal{S}_2 & \equiv & \max(s_1,-s_2) 
\end{array}
\end{equation}

\begin{figure}[h]
     \centering
     \begin{subfigure}{0.3\textwidth}
     	\centering
		\includegraphics[width=\textwidth,trim={8cm 1cm 8cm 2cm},clip]{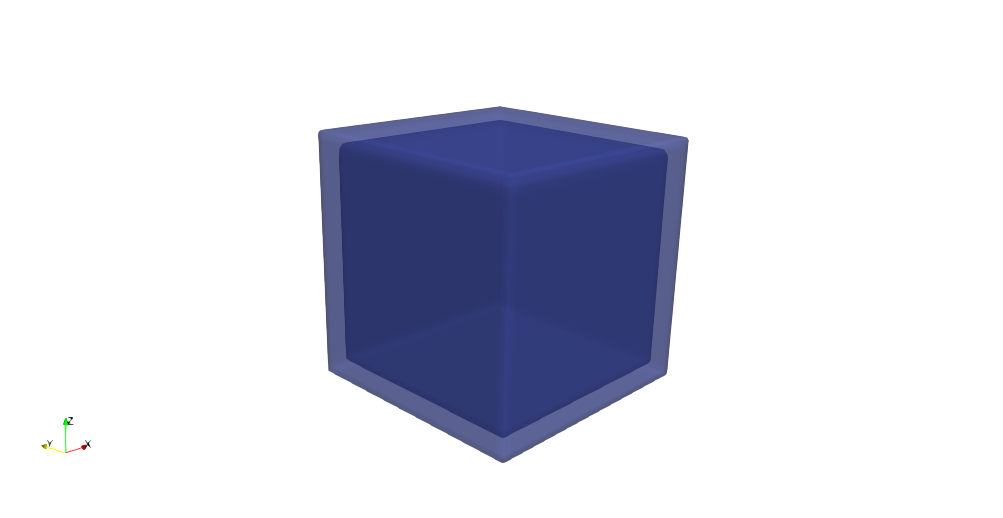}
         \caption{$\pazocal{S}_1$}
     \end{subfigure}
     \begin{subfigure}{0.3\textwidth}
     	\centering
		\includegraphics[width=\textwidth,trim={8cm 1cm 8cm 2cm},clip]{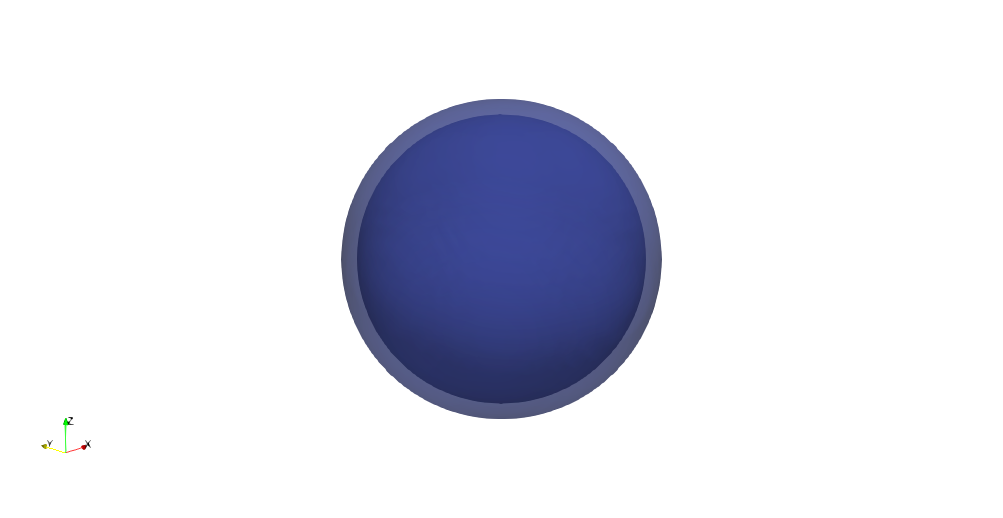}
         \caption{$\pazocal{S}_2$}
     \end{subfigure}
     \begin{subfigure}{0.3\textwidth}
		\centering
		\includegraphics[width=\textwidth,trim={8cm 1cm 8cm 2cm},clip]{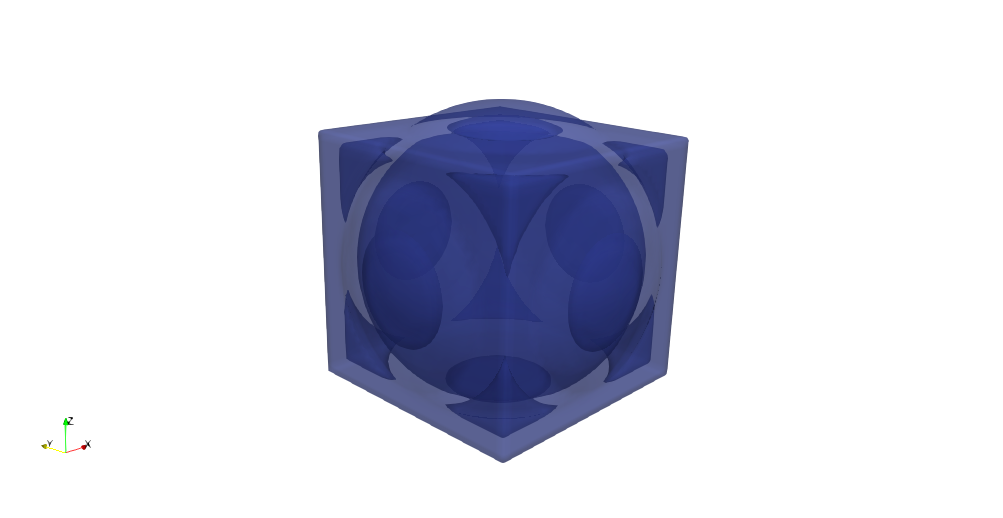}
         \caption{$\pazocal{S}_1 \cup \pazocal{S}_2$}
     \end{subfigure}
	\begin{subfigure}{0.3\textwidth}
     	\centering
		\includegraphics[width=\textwidth,trim={8cm 1cm 8cm 2cm},clip]{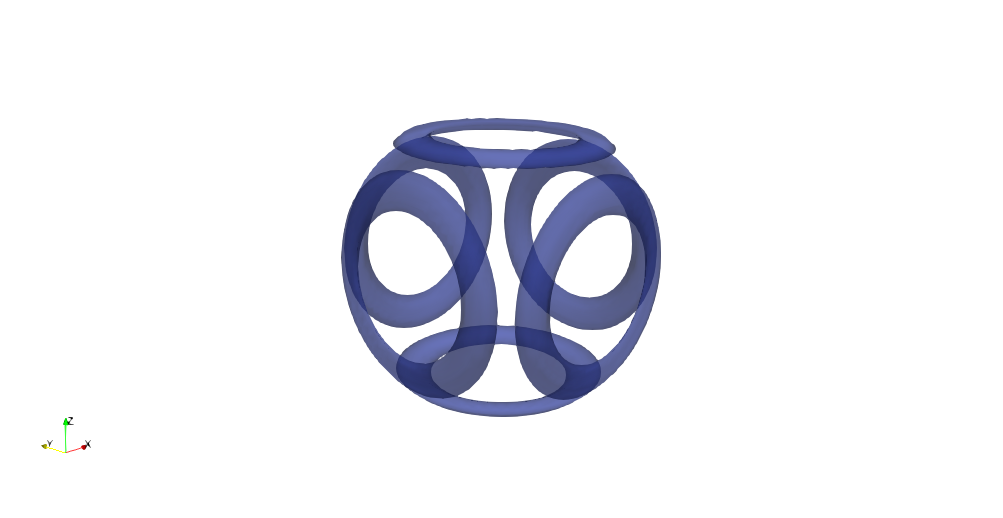}
         \caption{$\pazocal{S}_1 \cap \pazocal{S}_2$}
     \end{subfigure}
     \begin{subfigure}{0.3\textwidth}
     	\centering
		\includegraphics[width=\textwidth,trim={8cm 1cm 8cm 2cm},clip]{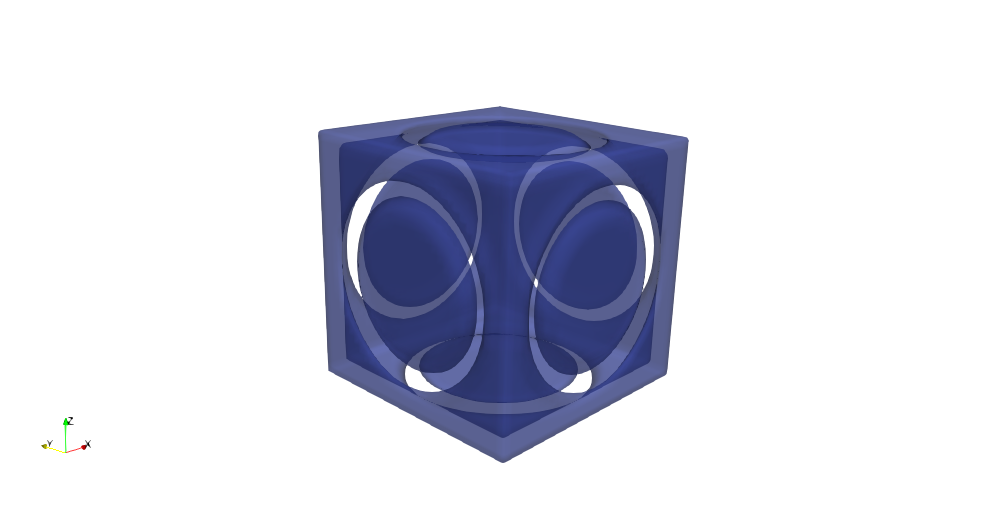}
         \caption{$\pazocal{S}_1 \backslash \pazocal{S}_2$}
     \end{subfigure}
     \begin{subfigure}{0.3\textwidth}
		\centering
		\includegraphics[width=\textwidth,trim={8cm 1cm 8cm 2cm},clip]{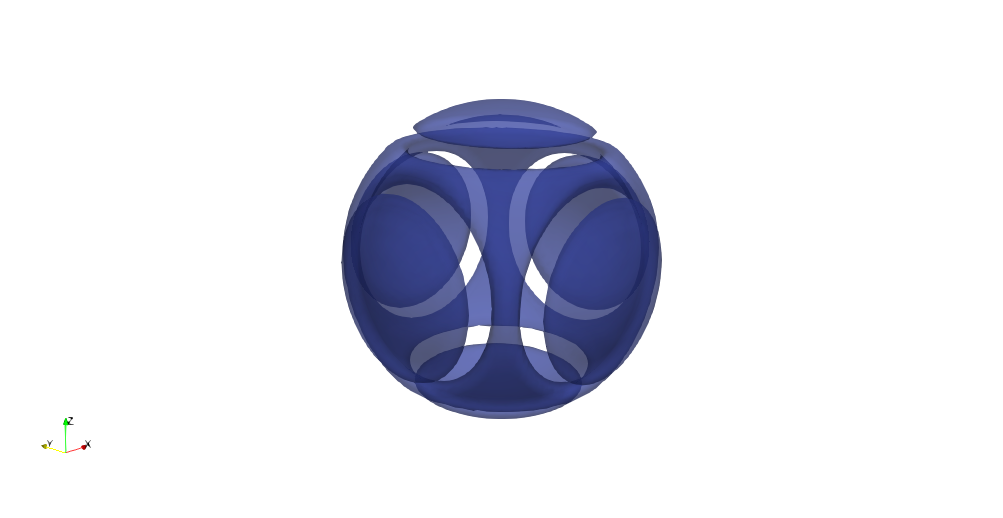}
         \caption{$\pazocal{S}_2 \backslash \pazocal{S}_1$}
     \end{subfigure}
	\caption{Set operations on Signed Distance Functions for a hollow cube $\pazocal{S}_1$ and sphere $\pazocal{S}_2$.}
    \label{fig:set1}
\end{figure}

It is also possible to define hollow shapes thresholding the absolute value of the SDF. Figure \ref{fig:set1} shows set operations acting on a hollow cube $\pazocal{S}_1$ and sphere $\pazocal{S}_2$. The code required to generate complex shapes and subregions is extremely simple.

\FloatBarrier

\begin{figure}[h]
     \centering
     \begin{subfigure}{0.32\textwidth}
     	\centering
		\includegraphics[width=\textwidth,trim={6cm 0cm 6cm 0cm},clip]{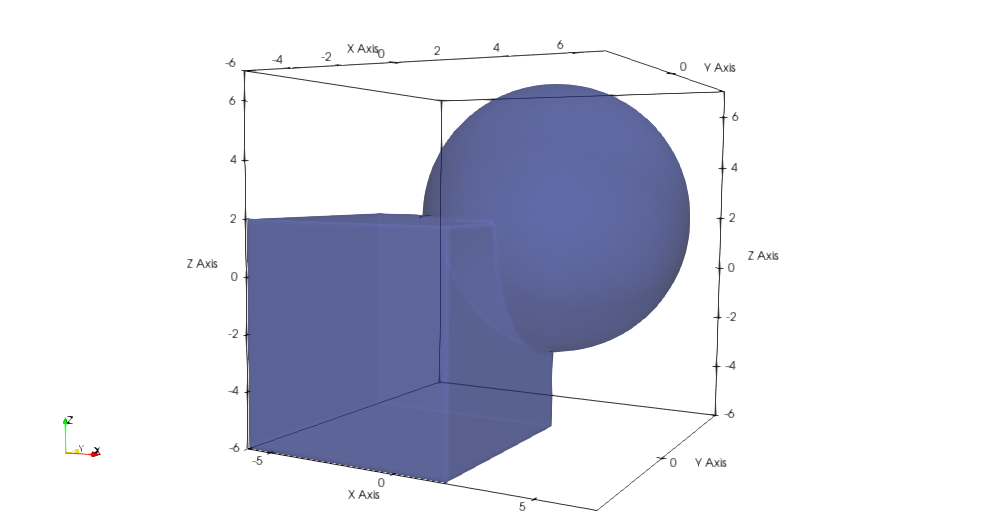}
         \caption{$k=0.0$}
     \end{subfigure}
     \begin{subfigure}{0.32\textwidth}
     	\centering
		\includegraphics[width=\textwidth,trim={6cm 0cm 6cm 0cm},clip]{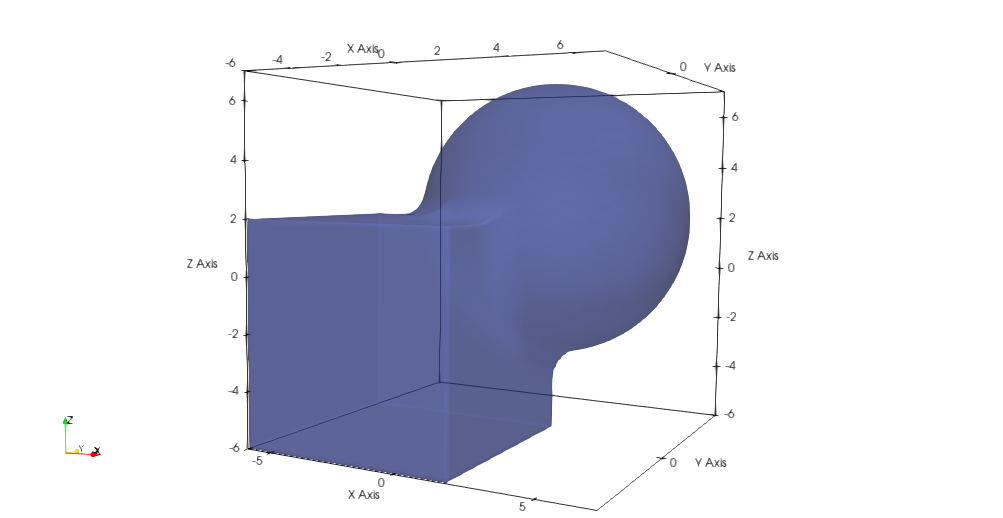}
         \caption{$k=0.25$}
     \end{subfigure}
     \begin{subfigure}{0.32\textwidth}
		\centering
		\includegraphics[width=\textwidth,trim={6cm 0cm 6cm 0cm},clip]{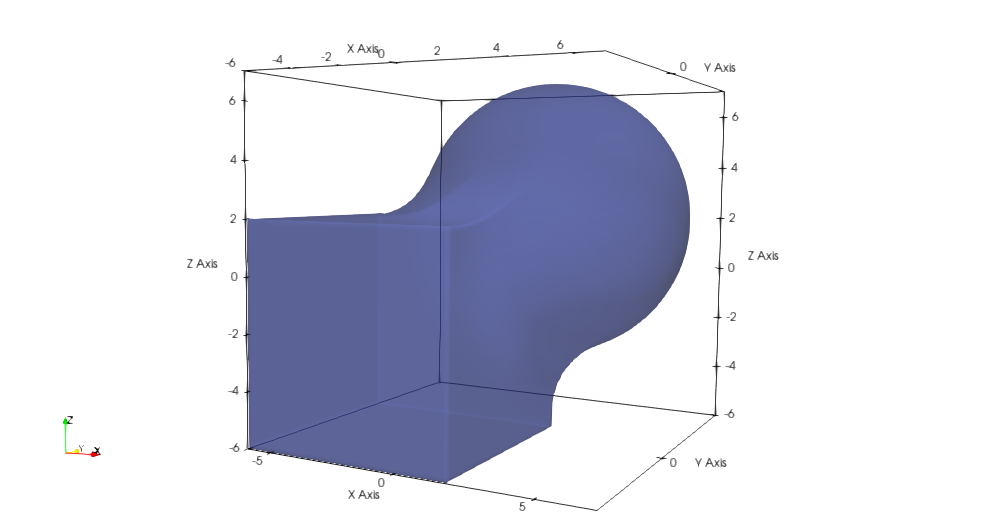}
         \caption{$k=0.5$}
     \end{subfigure}
	\caption{Smoothed minimum function for increasing values of parameter $k$.}
    \label{fig:smin1}
\end{figure}

A smoothed minimum function can be used to blend together two SDFs leading to more natural surfaces that better represent organic objects.  An example is shown in Figure \ref{fig:smin1}, with a varying smoothness parameter $k$.

\FloatBarrier

\subsection{Encoding}\label{sec:encoding1}

Having generated or captured a geometry it is now possible to encode it onto the mesh as detailed at the beginning of this chapter. Since the computational model makes use of both vertices and volumes, it important that the geometry has integrity with respect to these objects. As such a mesh volume $V$ can only belong to either the solid $\Omega_s$ or fluid $\Omega_f$ domains. A vertex, however, has no volume and can belong to one or the other, or both if it is on the fluid-structure interface. 

\begin{figure}[h]
\centering
\begin{subfigure}{0.49\textwidth}
    \centering
    \begin{tikzpicture}[scale=1.2]
    \draw[step=1] (0,0) grid (4,4);		
    \fill[draw=gray, pattern=north east lines,pattern color=lightgray] (0,0) to (0,4) to (1.2,4) to[out=-75,in=95] (2.8,0) to (0,0);					
    \begin{scope}[shift={(0.5,0.5)}]
    \draw  [fill=black] (0,0) circle (0.05); 
    \draw  [fill=black] (1,0) circle (0.05); 
    \draw  [fill=black] (2,0) circle (0.05); 
    \draw  [fill=white] (3,0) circle (0.05); 
    \draw  [fill=black] (0,1) circle (0.05); 
    \draw  [fill=black] (1,1) circle (0.05); 
    \draw  [fill=white] (2,1) circle (0.05); 
    \draw  [fill=white] (3,1) circle (0.05); 
    \draw  [fill=black] (0,2) circle (0.05); 
    \draw  [fill=black] (1,2) circle (0.05); 
    \draw  [fill=white] (2,2) circle (0.05); 
    \draw  [fill=white] (3,2) circle (0.05); 
    \draw  [fill=black] (0,3) circle (0.05); 
    \draw  [fill=white] (1,3) circle (0.05); 
    \draw  [fill=white] (2,3) circle (0.05); 
    \draw  [fill=white] (3,3) circle (0.05); 
    \end{scope}
    \draw[blue,thick] (1.2,4) to[out=-75,in=95] (2.8,0);
    \node[anchor=north west] at (0,4) {$\Omega_s$};
    \node[anchor=north east] at (4,4) {$\Omega_f$};
    \end{tikzpicture}
    \caption{Volume Pass}
\end{subfigure}
\begin{subfigure}{0.49\textwidth}
    \centering
    \begin{tikzpicture}[scale=1.2]
    \draw[step=1] (0,0) grid (4,4);		
    \fill[draw=gray, pattern=north east lines,pattern color=lightgray] (0,0) to (0,4) to (1,4) to (1,3) to (2,3) to (2,1) to (3,1) to (3,0) to (0,0);
    \draw[blue,thick] (1.2,4) to[out=-75,in=95] (2.8,0);		
    \draw  [fill=black] (0,0) circle (0.05); 
    \draw  [fill=black] (1,0) circle (0.05); 
    \draw  [fill=black] (2,0) circle (0.05); 
    \draw  [fill=green] (3,0) circle (0.05); 
    \draw  [fill=white] (4,0) circle (0.05); 
    \draw  [fill=black] (0,1) circle (0.05); 
    \draw  [fill=black] (1,1) circle (0.05); 
    \draw  [fill=green] (2,1) circle (0.05); 
    \draw  [fill=green] (3,1) circle (0.05); 
    \draw  [fill=white] (4,1) circle (0.05); 
    \draw  [fill=black] (0,2) circle (0.05); 
    \draw  [fill=black] (1,2) circle (0.05); 
    \draw  [fill=green] (2,2) circle (0.05); 
    \draw  [fill=white] (3,2) circle (0.05); 
    \draw  [fill=white] (4,2) circle (0.05); 
    \draw  [fill=black] (0,3) circle (0.05); 
    \draw  [fill=green] (1,3) circle (0.05); 
    \draw  [fill=green] (2,3) circle (0.05); 
    \draw  [fill=white] (3,3) circle (0.05); 
    \draw  [fill=white] (4,3) circle (0.05); 
    \draw  [fill=black] (0,4) circle (0.05); 
    \draw  [fill=green] (1,4) circle (0.05); 
    \draw  [fill=white] (2,4) circle (0.05); 
    \draw  [fill=white] (3,4) circle (0.05); 
    \draw  [fill=white] (4,4) circle (0.05); 

    \node[anchor=north west] at (0,4) {$\Omega_s$};
    \node[anchor=north east] at (4,4) {$\Omega_f$};
    \end{tikzpicture}
    \caption{Vertex Pass}
\end{subfigure}
\caption{The SDF is encoded into the mesh in two stages. In the first pass the SDF is evaluated at the volume centres. In the second, vertices count incident volumes and are assigned as solid (black), fluid (white) or surface (green).}
\label{fig:surf1}
\end{figure}
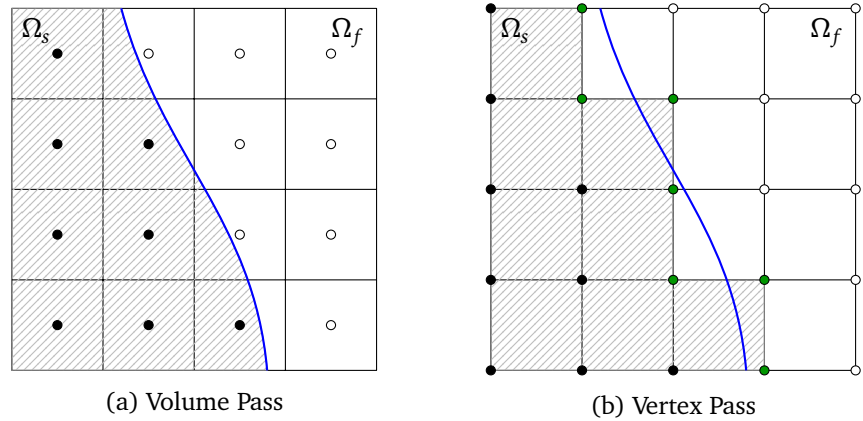

The algorithm that applies the SDF to the mesh thus makes two passes, in parallel over the mesh as follows:

\begin{enumerate}
\item A kernel runs per volume, evaluating the SDF at the centre point of each. As such the volumes $V$ are assigned to either the solid or fluid subdomains.
\item The next kernel runs per vertex $\x$ and iterates through all subvolumes $V$ incident to it. If all the adjacent subvolumes are in the solid subdomain the vertex is assigned as solid, similarly if all are fluid then the vertex is assigned as fluid.  If the vertex is incident upon both solid and fluid subvolumes it is assigned as surface.
\end{enumerate}

Figure \ref{fig:surf1} illustrates the relationship between the SDF and the mesh. 

A discussed in Chapter \ref{chp:gpu}, GPU memory access is SIMD parallel and it is efficient to store a 3-dimensional vertex coordinate $\x \in R^3$ in a \texttt{float4} vector variable\footnote{In OpenCL there is a \texttt{float3} variable, but it is an alias to \texttt{float4}.}. The fourth float in the vector is populated with the value of the SDF and it is thus available to test during any kernel execution that calls the vertex coordinates into register without cost. 

The geometry kernels also populate any other spatially-dependent data fields, such as the fibre directions, conductivity and varying tissue properties needed by the calculation. 

\begin{figure}[h]
\centering
\includegraphics[width=0.49\textwidth,trim={18cm 2cm 18cm 6cm},clip]{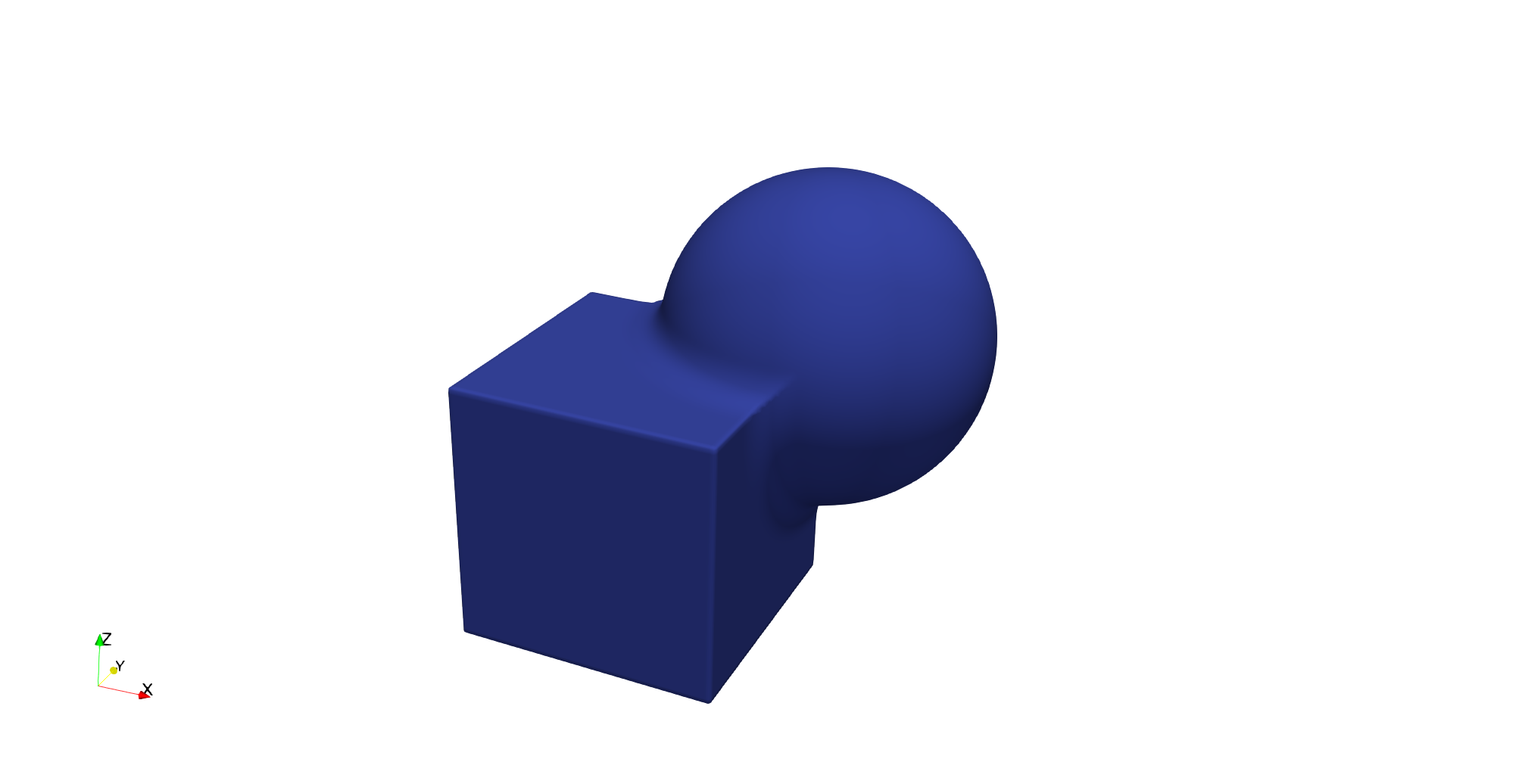}
\includegraphics[width=0.49\textwidth,trim={18cm 2cm 18cm 6cm},clip]{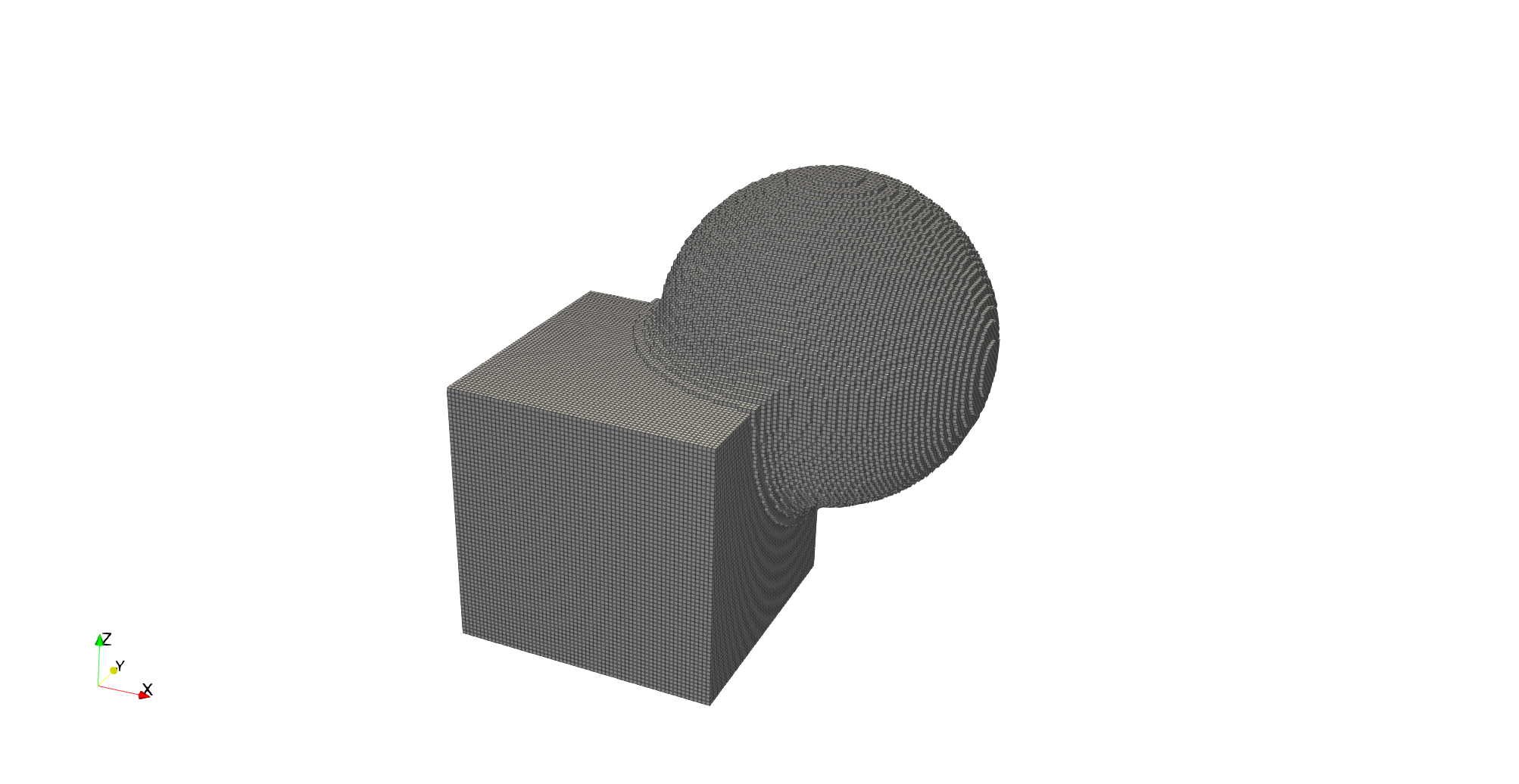}
\caption{A Signed Distance Function and its encoding into the mesh.}
\label{fig:surf2}
\end{figure}

Having assigned volumes and vertices, the geometry is encoded as voxel data in a regular structured grid.  This is illustrated in 3D in Figure \ref{fig:surf2}. When the input data is  voxel data from a medical scan, this reflects the incoming data exactly.  It would also be possible at this stage to fit the mesh precisely to the SDF.  This would be achieved by deforming the mesh, such that surface vertices were moved to the root of the SDF.  This would generate a smooth surface.  Then the application of a Laplacian to the non-surface vertex coordinates would regularise the volumes.  This has been successfully tested but is not yet implemented, since subsequent calculations would require extra storage and reference to the pre-deformed mesh.

\chapter{Algorithm Details}\label{chp:alg}

\section{Overview}

As described in Chapter \ref{chp:disc}, the discrete problem is processed in space on a regular structured grid with mesh width $\delta x \in \R$ and iterated over time $t \in \R^+$.  The memory for the calculation is a set of three-dimensional buffers of the OpenCL data type \texttt{float4}, which are vectors of four single-precision floating point values. Some of the buffers correspond to values at the vertices while others contain information relating to volumes.

The C program instantiates OpenCL objects, including the memory buffers on the GPU and their contents at time $t = 0$ are initialised according to input data, which includes the geometry and initial conditions. A main loop then iterates over a time step $\delta t \in \R$ and applies a sequence of computational \textit{kernels} which operate in parallel on the buffers, updating values relevant to the different steps of the algorithm. Kernels such as iterative solvers are applied in sub-loops to the main loop. 

At prescribed intervals, the contents of the buffers are copied from the CPU to GPU and written into output files. These files are post-processed and visualised as frames of video or plots of time series data. 

\section{Electophysiology}

The Monodomain equation (\ref{eqn:monodomain1}) is a PDE consisting of both reactive and diffusive terms, which are handled separately via \textit{operator-splitting}. The ODE corresponding to the Mitchell-Scheaffer Model (\ref{eqn:ms1}) for transmembrane potential is first updated via a step of Explicit Euler time integration, and this value is then used to update the diffusive part of the Monodomain equation via a step of Implicit Euler time integration. The regime is detailed below.

\subsection{Reaction}

First, a single kernel integrates the Mitchell-Scheaffer ODE with a step of Explicit Euler time integration, acting on the coupled scalar variables for transmembrane voltage $v$ in (\ref{eqn:ms1}) and the gating variable $w$ in (\ref{eqn:ms2}), which are stored as average values per volume:
\begin{eqnarray}
v_i^{t+1} &=& v_i^{t} + \delta t \frac{dv_i}{dt} \\
w_i^{t+1} &=& v_i^{t} + \delta t \frac{dw_i}{dt} ,
\end{eqnarray}
where in general, the discrete evaluation $v_i$ of a continuous variable $v$ is given a suffix $i \in \N$ which enumerates a vertex or, in this case Volume. The memory buffer contains a flag which indicates whether a volume is electrically active or not, in which case its value is preserved. 

\subsection{Diffusion}

Next a kernel applies an Implicit Euler step to the left hand side of (\ref{eqn:monodomain1}) corresponding to the diffusion of charge through the intracellular and extracellular domains. The resulting linear system is solved with a Jacobi iteration as described in Section \ref{sec:ie1}, again acting on the average value of $v_i$ per volume. 
\begin{equation}
v_i^{t+1} = v_i^{t} + \delta t \mathbf{A} v_i^{t+1}
\end{equation}
The solver must take into account the conductivity tensor $\bm{\SIG}$ calculated from the fibre directions $\f$ as in (\ref{eqn:sigma1}). Since the conductivity tensor acts on the eight discrete gradients $\nabla v$ surrounding $v_i$, it is derived from interpolated values of $\f$ at the corresponding points and appears in the assembly described in Section \ref{sec:ass1} as an additional block diagonal matrix $\mathbf{S}_1$ of eight conductivity tensors:
\begin{equation}
\nabla \cdot (\SIG \nabla v_i)  \approx
\v_i
\mathbf{D}_1
\mathbf{S}_1
\mathbf{F}_1
\mathbf{D}_2
\mathbf{F}_2
\mathbf{N}_1,
\end{equation}
where $\v_i$ is a row vector of the 27 values surrounding $v_i$ that are used by the stencil. The buffer of fibre directions uses all four elements of the \texttt{float4} variable such that the unit direction vector $\f \in \R^3$ and the longditudinal $\sigma_L$ and transverse $\sigma_T$ conductivities can be stored uniquely per volume $V_i$ of the domain. 

The model assumes that the fibres $\f$, and thus the conductivity field $\Sigma$, deform with the mesh.  The deformation gradients $\mathbf{F}_1,\mathbf{F}_2$ therefore need only to take into account the scaling and are thus the product of mesh width $\delta x$ and an identity matrix $\I \in \R^{3 \times 3}$. Since they are applied twice during assembly they can be removed and appear as the familiar scalar term $\frac{1}{\delta x^2}$. For the generation of the ECG, the elliptic part of the problem in the second line of (\ref{eqn:bidomain1}) is solved on the heart and torso via Jacobi iteration.

\section{Solid Dynamics}

\subsection{Stress}\label{sec:solid1}

In Section \ref{sec:solid1} it was shown that the principles of conservation of mass and momentum, along with a constitutive model for a solid continuum lead to the Elastodynamics equation (\ref{eqn:elas1}). The resultant force acting on an enclosed volume $V$, is thus the sum of traction and body forces, given by surface and volume integrals respectively.

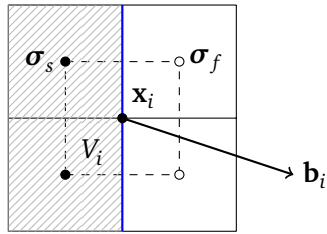
\begin{figure}[h]
\centering
\begin{tikzpicture}[scale=1.5]
\draw[step=1] (0,0) grid (2,2);		
\draw[step=1,dashed,shift={(0.5,0.5)}] (0,0) grid (1,1);	
\fill[draw=gray, pattern=north east lines,pattern color=lightgray] (0,0) rectangle (1,2);			
\draw[blue,thick] (1,0) to (1,2);			
\draw  [fill=black]  (1,1) circle (0.04); 	
\node[anchor=east] at (0.5,1.5) {$\bm{\sigma}_s$};
\node[anchor=west] at (1.5,1.5) {$\bm{\sigma}_f$};
\draw  [fill=black]  (0.5,0.5) circle (0.04); 
\draw  [fill=white]  (1.5,0.5) circle (0.04); 
\draw  [fill=black]  (0.5,1.5) circle (0.04); 
\draw  [fill=white]  (1.5,1.5) circle (0.04); 
\node[anchor=south west] at (1,1) {$\x_i$};
\node[anchor=south west] at (0.55,0.5) {$V_i$};
\draw[->,thick] (1,1) to (2.5,0.5);
\node[anchor=west] at (2.5,0.5) {$\mathbf{b}_i$};
\end{tikzpicture}
\caption{Spatial integration of the Elastodynamics equation at surface point $\x_i$ lying on the fluid structure interface $\Gamma$ (blue line), showing the volume  of integration $V_i$ (enclosed by dotted lines), and examples of the domain dependent stress tensors $\bm{\sigma}_s$ and $\bm{\sigma}_f$ (black and white dots respectively) and body forces $\mathbf{b}_i$.}
\label{fig:stress1}
\end{figure}

In the discrete case, these integrals are calculated in parallel per vertex as shown in Figure \ref{fig:stress1}.

Body forces $\mathbf{b}_i$ may be prescribed as averages over $V_i$ and thus sum directly into the integral. Traction forces are given by the negative divergence of Cauchy stress. The stress tensors surrounding a vertex $\x_i$ may be derived in one of two different ways:  
\begin{enumerate}
\item If the adjacent volume is part of the solid domain $\Omega_s$ then the tensor $\bm{\sigma}_s$ derives from the deformation gradient $\F$ in (\ref{eqn:defgrad1}) and solid material model such as Saint Venant-Kirchhoff (\ref{eqn:stress1}).
\item If the volume is in the fluid domain $\Omega_f$ then Cauchy stress $\bm{\sigma}_f$ is given by the pressure $p$ and strain-rate tensor as in the Navier-Stokes fluid (\ref{eqn:cauchy1}).
\end{enumerate}
In practice the kernels loop the volumes adjacent to the vertex and read the geometry information from the coordinate buffer as outlined in Section \ref{sec:encoding1}. The numerical calculation of the stress tensors in the deformed configuration is performed as per (\ref{eqn:defdiv2}). As a result, the coupling conditions for continuity of stress in the ALE formulation (\ref{eqn:couple1}) are satisfied.

\subsection{Muscle Contraction}

If tissue is mechanically active the Cauchy Stress tensor must include an additional component to represent muscle contraction. This is calculated by taking the unit vector for fibre direction $\f$ and multiplying it by a constant to represent the strength of contraction and the dimensionless value of transmembrane potential as given by (\label{eqn:monodomain1}). The components of the resulting vector are added to the diagonal of the Second Piola-Kirchhoff stress in the reference configuration (\ref{eqn:stress1}).  In this way the fibres and resulting contraction forces deform with the tissue as expected. 

This is the simplest possible implementation of Excitation-Contraction coupling and can be refined in future work.

\subsection{Leapfrog Integration}

Once the forces in the Elastodynamics equation are resolved, conservation of momentum allows the change in velocity to be calculated by consideration of mass, the integral of density over volume $V_i$. Integration in time is then achieved with the Leapfrog method. This explicit method has been derived separately across many disciplines and is, most importantly, symplectic in phase plane, that is it conserves angular momentum.  It is therefore suitable for the wave-like solutions of the Elastodynamics equation.

For position, velocity and acceleration $\x, \v = \dot{\x}, \mathbf{a} = \ddot{\x} \in \R^d$, in a classical dynamic system of differential equations, Leapfrog integration at time $t \in \R$ is as follows:
\begin{eqnarray}
\label{eqn:leap1}
\v_{t+\frac{1}{2}} 	&=& \v_{t-\frac{1}{2}} 	+  \delta t \mathbf{a}_t 		\\
\label{eqn:leap2}
\x_{t+1}			&=& \x_{t} 				+ \delta t \v_{t+\frac{1}{2}}  
\end{eqnarray}
It is worth noting that the position and velocity are given at different times, which gives the method its name. Also that the system is started from rest, which does not require any spcific management of the velocity half-step. The first step (\ref{eqn:leap1}) only is applied at this stage.

\subsection{Damping}

Solid damping can be applied as the diffusion of momentum via an Implicit Euler iteration on momentum as per Section \ref{sec:ie1}.  Damping and fluid viscosity can be applied to the whole domain separately or via the same kernel, taking care to use the correct coefficients for dynamic viscosity, damping and density.  The fluid viscosity can also be applied in the mechanics kernel as the strain rate in the Cauchy stress tensor $\bm{\sigma}_f$ of (\ref{eqn:cauchy1}), and natural damping is achieved via the interpolation of vertex position, as described in Section \ref{sec:solid1}.

\section{Fluid Dynamics}

\subsection{Helmholz Projection}\label{sec:helm1}

The projection method that is at the core of most modern fluid dynamics solvers was pioneered by Chorin \cite{Chorin2012} and is the basis for the well-known SIMPLE algorithm \cite{Patankar1980} and its subsequent variations. This section outlines the technique, following a useful introductions by \cite{Harris2004} and \cite{Griebel1998}, and those following show how it must be adapted to incorporate the deforming fluid-structure interface.

Helmholz-Hodge decomposition theorem states that a vector field $\w$ on domain $\Omega$, can be uniquely decomposed into the sum of two orthogonal fields:
\begin{equation}
\label{eqn:hh1}
\w = \u + \nabla p,
\end{equation}
where $\nabla p$ is the gradient of a scalar field, and $\u$ is parallel to $\partial \Omega$ and has zero divergence, that is: 
\begin{eqnarray}
\label{eqn:hh2}
\nabla \cdot \u &=& 0,\ \text{on}\ \Omega \\
\label{eqn:hh3}
\u\cdot \n &=& 0,\  \text{on}\ \partial \Omega. 
\end{eqnarray}
The unknown fields $\u$ and $p$ can be obtained from $\w$ by taking the divergence of (\ref{eqn:hh1}) and applying condition (\ref{eqn:hh2}) to give a Poisson problem for pressure:
\begin{eqnarray}
\label{eqn:hh4}
\nabla \cdot \w &=& \underbrace{\nabla \cdot \u}_{\nabla \cdot \u=0} + \nabla \cdot \nabla p \\
\Rightarrow \Delta p &=& \nabla \cdot \w
\end{eqnarray}
which can be solved up to the addition of a constant with the zero-Neumann boundary condition (\ref{eqn:hh3}). Taking the gradient of the solution $p$ and substituting into (\ref{eqn:hh1}) completes the solution:
\begin{equation}
\label{eqn:hh5}
\u = \w - \nabla p.
\end{equation}
The continuity equation (\ref{eqn:cont1}) requires that fluid velocity must be divergence-free for conservation of mass. The decomposition method is applied to the discrete time formulation of the Navier-Stokes equations (\ref{eqn:ns1}) as follows.  

We start by considering a discrete time step $\delta t$, where $\nu = \frac{\mu}{\rho}$ is the kinematic viscosity and body forces $\b$ are ignored:
\begin{equation}
\v^{t+1} = \v^{t} + \delta t \left[  (\v \cdot \nabla) \v  + \nu \Delta \v  - \frac{1}{\rho} \nabla p \right].
\end{equation}
The discrete advection and viscosity operators are applied in an explicit fashion to give an intermediate velocity field $\w$ with non-zero divergence:
\begin{equation}
\w = \v^{t} + \delta t \left[  (\v^{t} \cdot \nabla) \v^{t}  + \nu \Delta \v^{t} \right].
\end{equation}
The velocity at the next time step, which must be divergence-free by continuity, is expressed as a sum representing the Helmholz decomposition of $\w$, equivalent to (\ref{eqn:hh1}):
\begin{equation}
\w = \v^{t+1} +  \frac{\delta t}{\rho} \nabla p^{t+1}
\end{equation}
Taking the divergence as above and using $\nabla \cdot \v^{t+1} = 0$ as in (\ref{eqn:hh4}) yields a Poisson equation for pressure, with divergence of intermediate velocity as data:
\begin{equation}
\Delta p^{t+1} = \frac{\rho}{\delta t}  \nabla \cdot \w.
\end{equation}
Finally, the intermediate field $\w$ can be corrected by the subtraction of the gradient of the calculated pressure field $p^{t+1}$ as in (\ref{eqn:hh5}):
\begin{equation}
\label{eqn:poisson1}
\v^{t+1} = \w - \frac{\delta t}{\rho} \nabla p^{t+1}.
\end{equation}
The explicit integration of advection and viscosity with the implicit solution of the Poisson pressure equation constitutes another \textit{operator-splitting} approach, and the update of the intermediate field into its divergence-free form is a \textit{predictor-corrector} method. 

\subsection{Advection}

The discrete advection operator must also consider the computational mesh of the ALE formulation as described in (\ref{eqn:adv2}). It follows the conservation form of the advection operator and applies the deformed divergence as in (\ref{eqn:defdiv2}). 
\begin{equation}
\frac{\partial \v_f}{\partial t} + \nabla \cdot (\v_f \otimes (\v_f - \v_c)) = 0,
\end{equation}
where $\otimes$ indicates an outer product. This is a sum of matrix vector products with outer normals.  
\begin{equation}
\v_f^{t+1}  = \v_f^{t} - \frac{1}{J} \sum_j^4 J_j (\v_f^{t} \otimes (\v_f^{t} - \v_c^{t})) \F^{-\top} \hat{\n}_j,
\end{equation}
At the same time as the computation we test the sign of the product $\v_f^t \cdot \n_j > 0 $ and remove from the sum those instances where it is true. This results in an upwind scheme for the advection operator.

\subsection{Viscosity}

Similarly to damping, viscosity can be applied as via an implicit Euler step \ref{sec:ie1} for momentum on the fluid and surface domains. In situations where the Reynolds number is relatively low, viscosity can be applied explicitly in the fluid strain rate tensor as part of the fluid Cauchy stress $\bm{\sigma}_f$ as given in (\ref{eqn:cauchy1}). The discrete deformed gradients of fluid velocity are calculated as per (\ref{eqn:defgrad2}) and weighted by the constant of dynamic viscosity. The stress tensor is applied conditionally as outlined in Section \ref{sec:solid1}. 

\subsection{Pressure}\label{sec:press1}

The Helmholz projection method of Section \ref{sec:helm1}, requires only the gradient of pressure to correct fluid velocity. The definition of fluid Cauchy stress $\bm{\sigma}_f$ as given in (\ref{eqn:cauchy1}) requires an absolute value for pressure.  The Dirichlet boundary conditions necessary for the correct solution of the Poisson pressure equation. as thus contained in ghost cells at the perimiter of the domain. They are set at the beginning of the time iteration and can represent actual values or simply a ground state against which relative pressure is measured.

The application of the Helmholz projection method proceeds in four stages:
\begin{enumerate}
\item Given an initial velocity $\v_f$, the advection and viscosity operators are applied as described above, to gain an intermediate velocity field.
\item The discrete divergence of velocity is calculated as per (\ref{eqn:defdiv1}), taking values at the mesh vertices and returning an average divergence at the volume centres. 
\item The divergence is given as data to the right hand side of a Jacobi iteration (\ref{eqn:jacobi1}) for the solution of the poisson pressure (\ref{eqn:poisson1}) on the fluid domain. 
\item The resulting value of $p$ is used in the fluid Cauchy stress tensor (\ref{eqn:jacobi1}) and applied to solid or fluid as necessary.  In this way the continuity of stress between the solid and fluid (\ref{eqn:couple1}) is enforced as well as the fluid pressure correction (\ref{eqn:hh5}).
\end{enumerate}
In the derivation of the Helmholz projection method, the zero-Neumann boundary condition on $\u \cdot \n=0$ on $\partial \Omega$ in \ref{eqn:hh3} is applied to the fluid velocity in the Poisson pressure equation.  In the ALE formulation the velocity may not be zero, since the \textit{no-slip} condition (\ref{eqn:couple2}) allows for the coupled movement of solid and fluid on $\partial \Omega$. As a result a non-zero Neumann boundary condition is applied to the Poisson pressure equation as per \cite{Chorin2012}:
\begin{equation}
\frac{\partial p}{\partial n} = \w \cdot \n,\  \text{on}\ \partial \Omega.  
\end{equation}
In practice the pressure field from the previous time step is used as the initial guess for the Jacobi iteration. The resulting pressure field is also under-relaxed aiding stability on the fluid-structure interface.

\section{Computational Mesh}

The computational mesh does not play any physical role in the computation but it must deform with the solid structures in order to maintain the admissibility if the motion as mentioned in Section \ref{sec:deform1}.  Effectively the determinant $J$ of the deformation gradient $\F$ should be greater than zero on $\Omega_c$.

The motion of the computational mesh is resolved by applying Jacobi iteration (\ref{eqn:jacobi1}) of a Laplace equation to the computational velocity $\v_c$, using the solid velocity at the surface and ghost cells as Dirichlet boundary conditions. This is in line with the coupling condition for velocity on the fluid-structure interface (\ref{eqn:couple2}).

At this point in the algorithm the second step of the Leapfrog iteration is applied (\ref{eqn:leap2}), updating the position of the solid and computational mesh, and the main loop returns for its next iteration.

\part{Results}

\chapter{Numerical Experiments}\label{chp:exp}

\section{Convergence Studies}

\subsection{Overview}

We start by validating the discretization as introduced in Chapter \ref{chp:disc}.  Convergence studies in the $L_2$ error norm with respect to mesh width $\delta x$ are carried out for the forward action of the various discrete differential operators, as well as the iterative solution of a discretized Laplace operator.  All of the operations are found to be quadratically convergent as expected.  

The single precision arithmetic introduces a lower bound to the mesh width, below which results become numerically unstable. This is because the definition of the divergence operator (\ref{eqn:defdiv1}) includes a division by the measure of the control volume, $J \approx \frac{1}{\delta x^3}$, which quickly tends to machine zero in single precision arithmetic.

\begin{figure}[h!]
     \centering
     \begin{subfigure}{0.45\textwidth}
		\centering
		\includegraphics[width=1.0\textwidth,trim={6cm 1cm 6cm 2cm},clip]{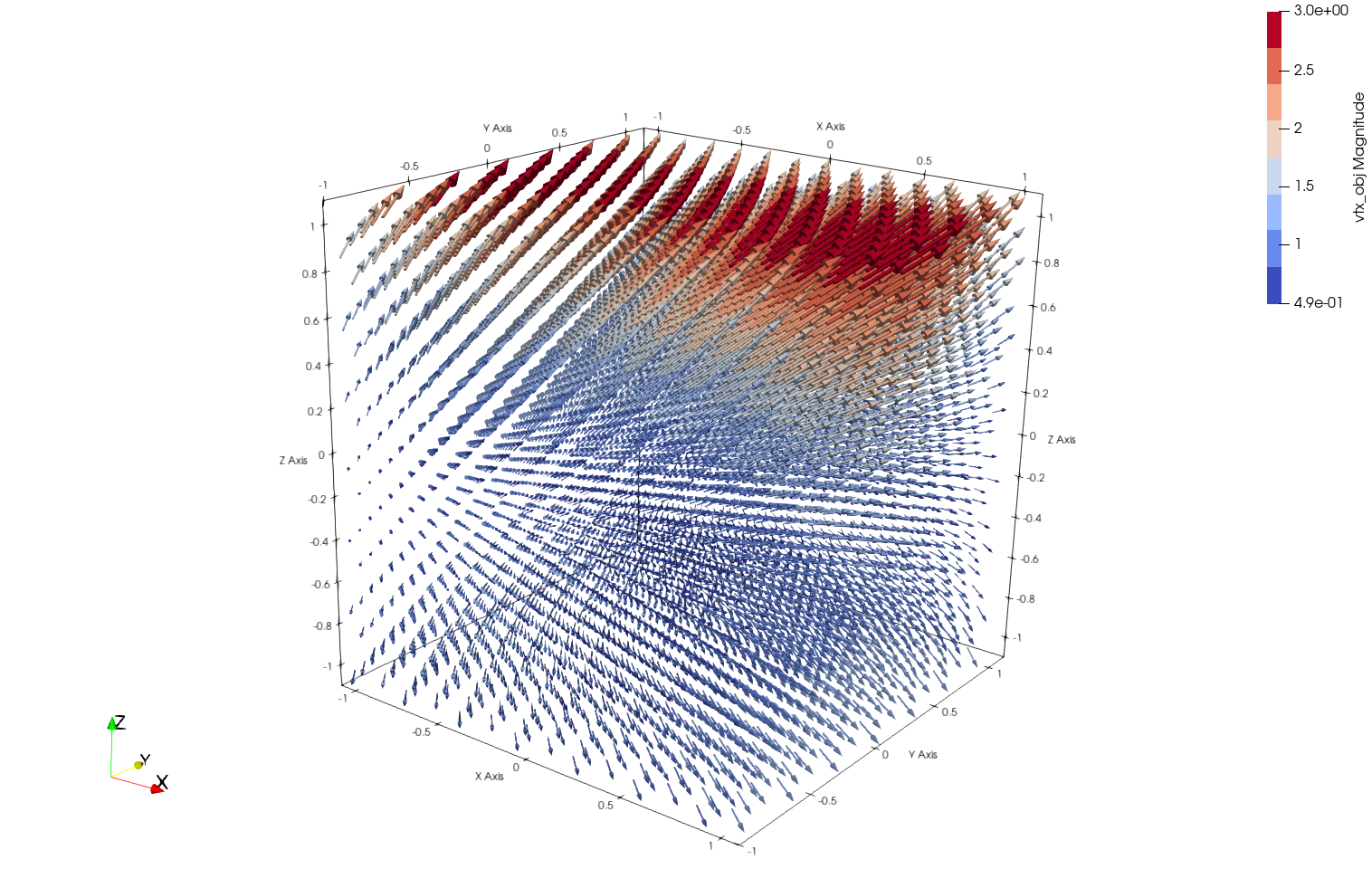}
         \caption{Reference}
     \end{subfigure}
     \begin{subfigure}{0.45\textwidth}
		\centering
		\includegraphics[width=1.0\textwidth,trim={6cm 1cm 6cm 2cm},clip]{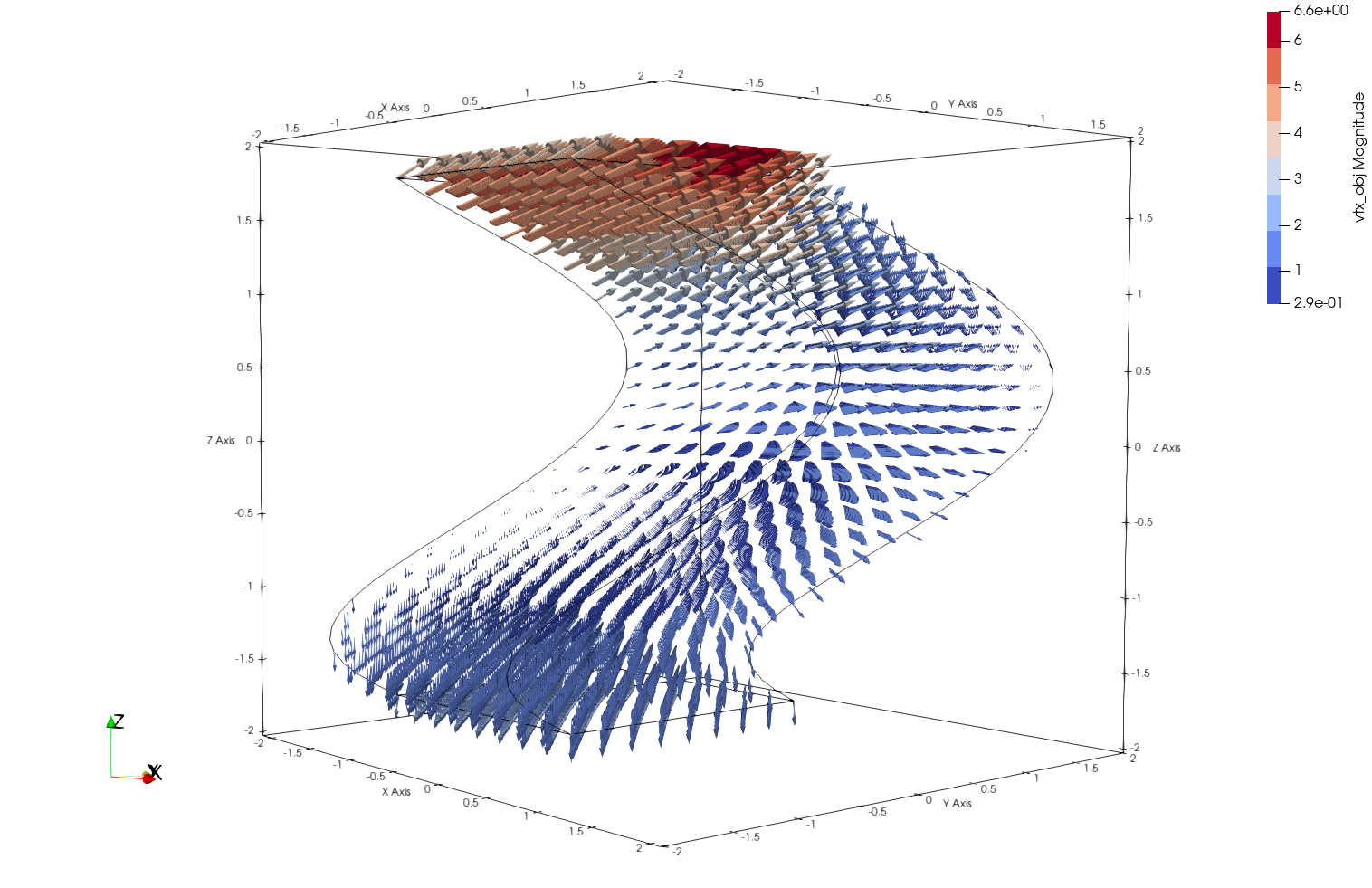}
         \caption{Deformed}
     \end{subfigure}
	\caption{Objective function $\u(\x)$ for the convergence study in reference and deformed configurations, $n = 32$.}
    \label{fig:conv1}
\end{figure}

\begin{figure}[h!]
     \centering
     \begin{subfigure}{0.4\textwidth}
		\centering
		\includegraphics[width=0.8\textwidth,trim={9cm 1cm 9cm 2cm},clip]{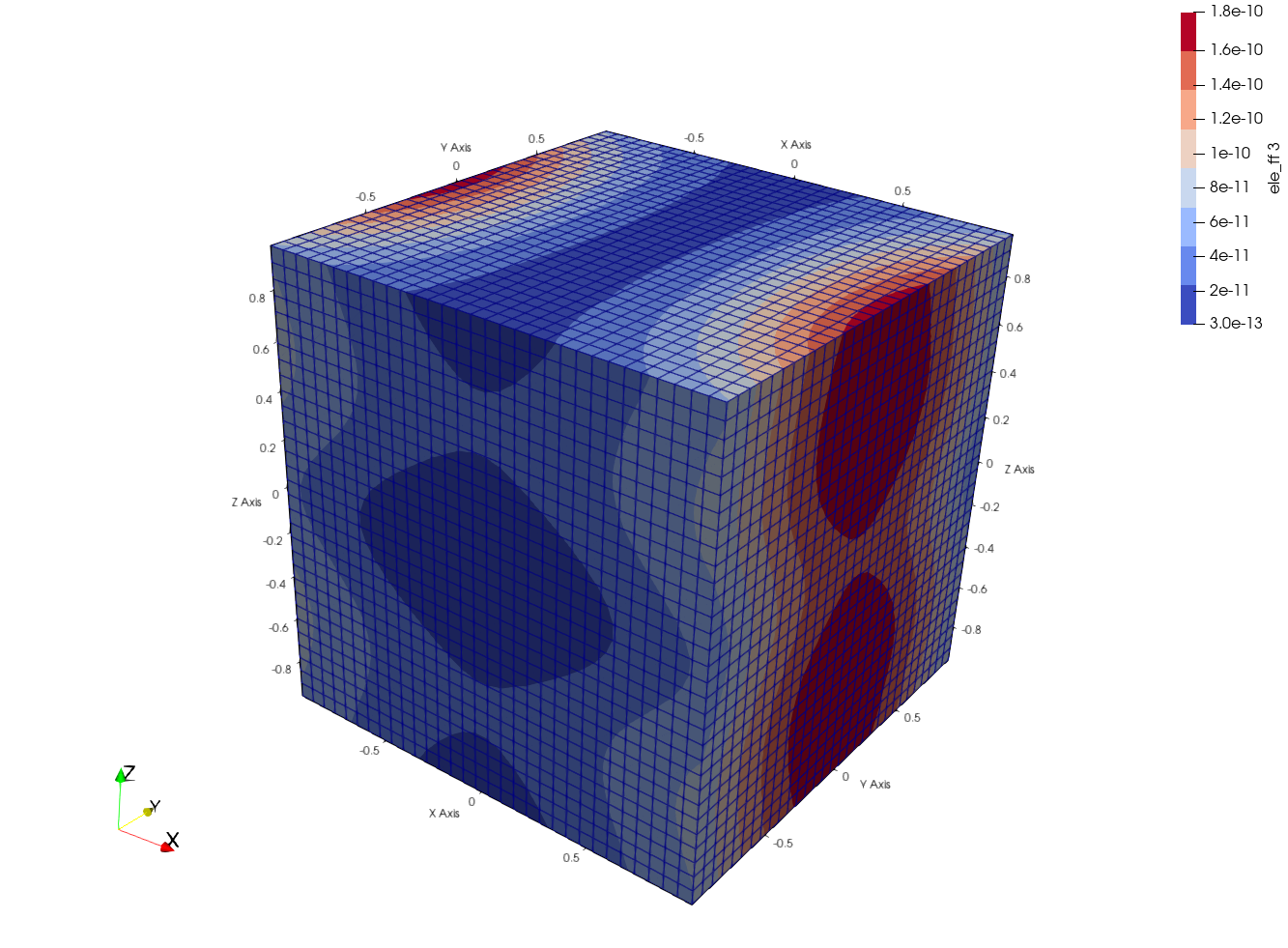}
         \caption{Reference}
     \end{subfigure}
     \begin{subfigure}{0.4\textwidth}
		\centering
		\includegraphics[width=0.8\textwidth,trim={9cm 1cm 9cm 2cm},clip]{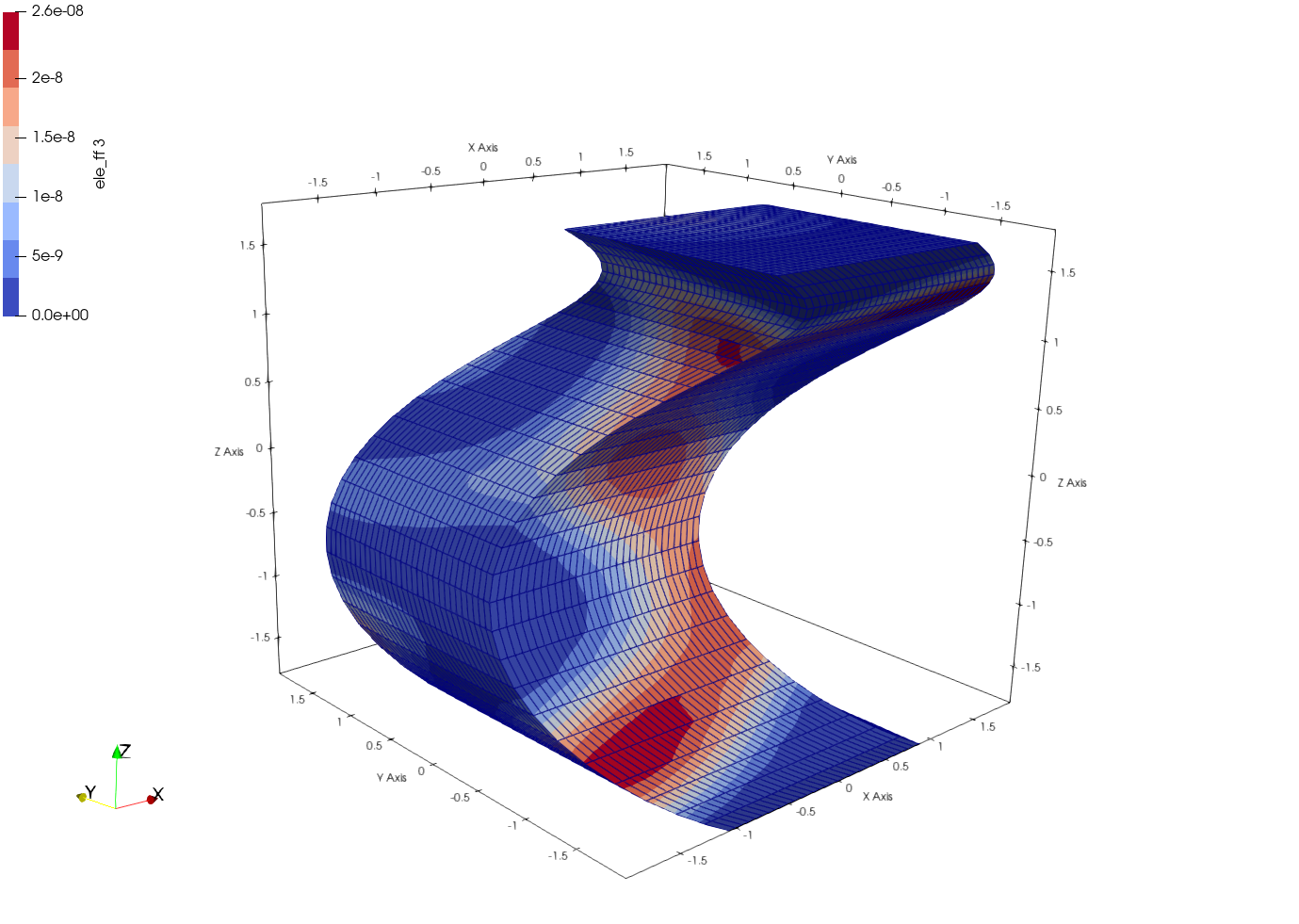}
         \caption{Deformed}
     \end{subfigure}
	\caption{Sample $L_2$ error norm for the convergence study in reference and deformed configurations, $n = 32$.}
    \label{fig:conv2}
\end{figure}

We consider the vector-valued objective function $\u : \R^3 \rightarrow \R^3$, where $\x = [x,y,z]^\top$ as follows:
\begin{equation}
\u(\x) =  
\left[
\begin{array}{c} 
\sin(x)\cos(y)\\ 
\cos(y)e^z\\ 
\mathrm{atan}\left(z\right) \end{array}
\right]
\end{equation}
To reference domain $\hat{\Omega} = [-1,+1]^3 \subset \R^3$, the following deformation $\phi : \R^3 \rightarrow \R^3$ was applied:
\begin{equation}
\x = \phi(\hat{\x}) = 
\left[
\begin{array}{c} 
x + \cos(\pi z)\\ 
y + \sin(\pi z)\\ 
z + \sin(z)
\end{array}
\right]
\end{equation}
The objective function is shown in reference $\u(\hat{\x})$ and deformed $\u(\phi(\hat{\x}))$configurations in Figure \ref{fig:conv1}.  
For each of the operators the error $\mathbf{e} \in \R^d$ is  the difference between analytic and numeric solutions $\u_\text{num},\u_\text{ana} \in \R^d$, which may be scalar or vector.  The $L_2$ norm is calculated in the ususal way as the sum of the integral of squared errors over each of the control volumes $V_i$ in the the domain, where $J$ is the determinant of the deformation gradient as in (\ref{eqn:mot1}):
\begin{eqnarray}
\mathbf{e} &=& \u_\text{num} - \u_\text{ana}\\
\label{eqn:l2norm1}
\|\mathbf{e}\|_{L_2} &=& \left( \sum_{i=1}^{V_{\text{tot}}} J_i (\mathbf{e}^{\top} \mathbf{e}) \right)^{\frac{1}{2}}
\end{eqnarray}
It is important to note that since $J_i$ is only an approximation of the actual volume measure $|V_i|$, this introduces an error in to the norm itself. Despite this error it is still possible to draw the most important conclusions from the convergence studies below. Examples of the error norm in the reference and deformed configurations are shown in Figure \ref{fig:conv2}. 
\FloatBarrier

\subsection{Discrete Operators}

\subsubsection{Interpolation}

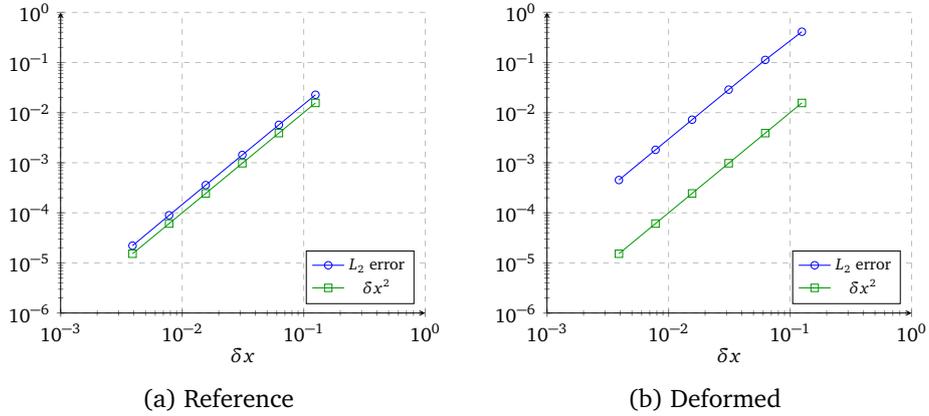
\begin{figure}[h!]
\begin{subfigure}{0.45\textwidth}
\begin{tikzpicture}[scale = 0.7]
\begin{axis}[
	xmode=log,
	ymode=log,
    xlabel={$\delta x$},
    legend pos=south east,
    legend style={font=\footnotesize},
    xmajorgrids=true,
    ymajorgrids=true,
    grid style=dashed,
    axis x line = bottom,
    axis y line = left,
	xmin = 1e-3,
	xmax = 1e-0,
	ymin = 1e-6,
	ymax = 1e-0]
\addplot [color=blue ,mark=o] table[x expr=\thisrowno{1}^-1, y index=5] {cnv/cnv_itp_ref.csv};
\addplot [color=green,mark=square] table[x expr=\thisrowno{1}^-1, y expr=\thisrowno{1}^-2] {cnv/cnv_itp_ref.csv};
\legend{$L_2$ error,$\delta x^2$}
\end{axis}
\end{tikzpicture}
\caption{Reference}
\end{subfigure}
\hspace{0.5cm}
\begin{subfigure}{0.45\textwidth}
\begin{tikzpicture}[scale = 0.7]
\begin{axis}[
	xmode=log,
	ymode=log,
    xlabel={$\delta x$},
    legend pos=south east,
    legend style={font=\footnotesize},
    xmajorgrids=true,
    ymajorgrids=true,
    grid style=dashed,
    axis x line = bottom,
    axis y line = left,
	xmin = 1e-3,
	xmax = 1e-0,
	ymin = 1e-6,
	ymax = 1e-0]
\addplot [color=blue ,mark=o] table[x expr=\thisrowno{1}^-1, y index=5] {cnv/cnv_itp_def.csv};
\addplot [color=green,mark=square] table[x expr=\thisrowno{1}^-1, y expr=\thisrowno{1}^-2] {cnv/cnv_itp_def.csv};
\legend{$L_2$ error,$\delta x^2$}
\end{axis}
\end{tikzpicture}
\caption{Deformed}
\end{subfigure}
\caption{$L_2$ error vs. mesh width $\delta x$ for the discrete interpolation operator.}
\label{fig:conv3}
\end{figure}

The discrete interpolation operator as given in (\ref{eqn:refintp1}) shows quadratic convergence with respect to mesh width.  The calculation is not susceptible to any numerical instability, indeed this task represents much of the work for which the GPU is designed. Figure \ref{fig:conv3} shows the results in reference and deformed configurations.

\FloatBarrier

\subsubsection{Gradient}

\begin{figure}[h!]
\begin{subfigure}{0.45\textwidth}
\begin{tikzpicture}[scale = 0.7]
\begin{axis}[
	xmode=log,
	ymode=log,
    xlabel={$\delta x$},
    legend pos=south east,
    legend style={font=\footnotesize},
    xmajorgrids=true,
    ymajorgrids=true,
    grid style=dashed,
    axis x line = bottom,
    axis y line = left,
	xmin = 1e-3,
	xmax = 1e-0,
	ymin = 1e-6,
	ymax = 1e-0]
\addplot [color=blue ,mark=o] table[x expr=\thisrowno{1}^-1, y index=5] {cnv/cnv_grd_ref.csv};
\addplot [color=green,mark=square] table[x expr=\thisrowno{1}^-1, y expr=\thisrowno{1}^-2] {cnv/cnv_grd_ref.csv};
\legend{$L_2$ error,$\delta x^2$}
\end{axis}
\end{tikzpicture}
\caption{Reference}
\end{subfigure}
\hspace{0.5cm}
\begin{subfigure}{0.45\textwidth}
\begin{tikzpicture}[scale = 0.7]
\begin{axis}[
	xmode=log,
	ymode=log,
    xlabel={$\delta x$},
    legend pos=south east,
    legend style={font=\footnotesize},
    xmajorgrids=true,
    ymajorgrids=true,
    grid style=dashed,
    axis x line = bottom,
    axis y line = left,
	xmin = 1e-3,
	xmax = 1e-0,
	ymin = 1e-6,
	ymax = 1e-0]
\addplot [color=blue ,mark=o] table[x expr=\thisrowno{1}^-1, y index=5] {cnv/cnv_grd_def.csv};
\addplot [color=green,mark=square] table[x expr=\thisrowno{1}^-1, y expr=\thisrowno{1}^-2] {cnv/cnv_grd_def.csv};
\legend{$L_2$ error,$\delta x^2$}
\end{axis}
\end{tikzpicture}
\caption{Deformed}
\end{subfigure}
\caption{$L_2$ error vs. mesh width $\delta x$ for the discrete gradient operator.}
\label{fig:conv4}
\end{figure}
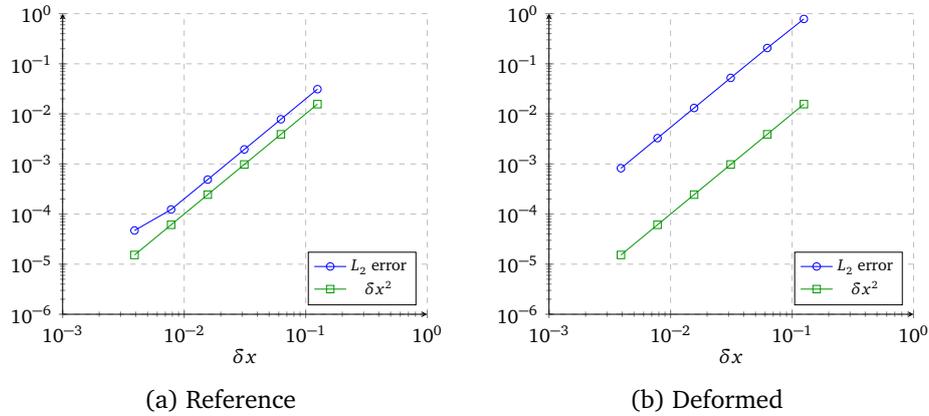

The gradient operator as given in (\ref{eqn:defgrad2}) again shows quadratic convergence with respect to mesh width.  The results of the convergence study are shown in Figure \ref{fig:conv4}. 

In this case, since the result $\frac{\partial \u}{\partial \x} \in \R^{3 \times 3}$ is a matrix, we use the Frobenius norm $e = \tr(\A^\top \A)$ to calculate a scalar error, which is then integrated as per (\ref{eqn:l2norm1}) for the $L_2$ norm.

In the reference configuration we see the beginning of some numerical instability. The calculation (\ref{eqn:defgrad2}) involves multiplication by the inverse of the deformation gradient $\F^{-1}$, and hence division by its determinant $J$.  As the mesh width $\delta x$ becomes small, the determinant tends to machine zero. The resulting division is close to zero and as a result we see an increase in the error as mesh width becomes small. In the deformed configuration there is sufficient error in the representation of the deformation itself to mask this effect.

In general the convergence of the discrete operator is quadratic, and as long as the mesh width is not too small the calculation is valid and numerically stable.

\FloatBarrier

\subsubsection{Divergence}

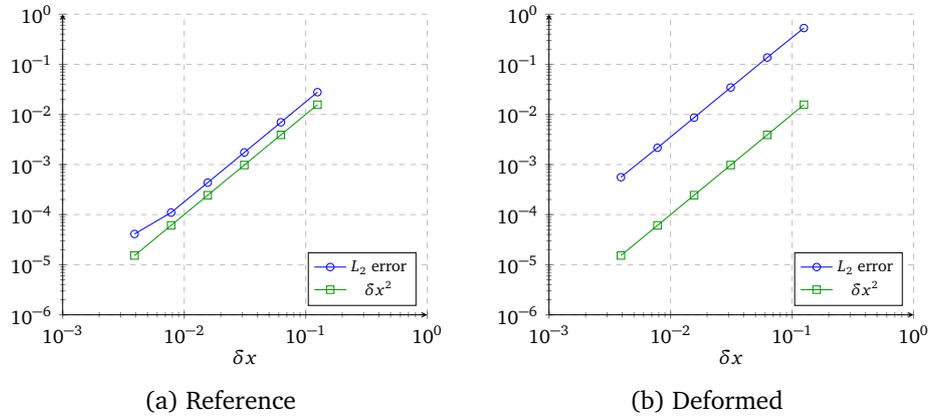
\begin{figure}[h!]
\begin{subfigure}{0.45\textwidth}
\begin{tikzpicture}[scale = 0.7]
\begin{axis}[
	xmode=log,
	ymode=log,
    xlabel={$\delta x$},
    legend pos=south east,
    legend style={font=\footnotesize},
    xmajorgrids=true,
    ymajorgrids=true,
    grid style=dashed,
    axis x line = bottom,
    axis y line = left,
	xmin = 1e-3,
	xmax = 1e-0,
	ymin = 1e-6,
	ymax = 1e-0]
\addplot [color=blue ,mark=o] table[x expr=\thisrowno{1}^-1, y index=5] {cnv/cnv_div_ref.csv};
\addplot [color=green,mark=square] table[x expr=\thisrowno{1}^-1, y expr=\thisrowno{1}^-2] {cnv/cnv_div_ref.csv};
\legend{$L_2$ error,$\delta x^2$}
\end{axis}
\end{tikzpicture}
\caption{Reference}
\end{subfigure}
\hspace{0.5cm}
\begin{subfigure}{0.45\textwidth}
\begin{tikzpicture}[scale = 0.7]
\begin{axis}[
	xmode=log,
	ymode=log,
    xlabel={$\delta x$},
    legend pos=south east,
    legend style={font=\footnotesize},
    xmajorgrids=true,
    ymajorgrids=true,
    grid style=dashed,
    axis x line = bottom,
    axis y line = left,
	xmin = 1e-3,
	xmax = 1e-0,
	ymin = 1e-6,
	ymax = 1e-0]
\addplot [color=blue ,mark=o] table[x expr=\thisrowno{1}^-1, y index=5] {cnv/cnv_div_def.csv};
\addplot [color=green,mark=square] table[x expr=\thisrowno{1}^-1, y expr=\thisrowno{1}^-2] {cnv/cnv_div_def.csv};
\legend{$L_2$ error,$\delta x^2$}
\end{axis}
\end{tikzpicture}
\caption{Deformed}
\end{subfigure}
\caption{$L_2$ error vs. mesh width $\delta x$ for the discrete divergence operator.}
\label{fig:conv5}
\end{figure}

The discrete divergence operator is given in (\ref{eqn:defdiv2}) and the results of the convergence study are shown in Figure \ref{fig:conv5}.  Again, the approximation shows quadratic converge with mesh width and for similar reasons (the multiplication by $\F^{-\top}$) it is susceptible to numerical instability for small $\delta x$.

\FloatBarrier

\subsubsection{Laplacian}

\begin{figure}[h!]
\begin{subfigure}{0.45\textwidth}
\begin{tikzpicture}[scale = 0.7]
\begin{axis}[
	xmode=log,
	ymode=log,
    xlabel={$\delta x$},
    legend pos=south east,
    legend style={font=\footnotesize},
    xmajorgrids=true,
    ymajorgrids=true,
    grid style=dashed,
    axis x line = bottom,
    axis y line = left,
	xmin = 1e-3,
	xmax = 1e-0,
	ymin = 1e-6,
	ymax = 1e+1]
\addplot [color=blue ,mark=o] table[x expr=\thisrowno{1}^-1, y index=5] {cnv/cnv_lap_ref.csv};
\addplot [color=green,mark=square] table[x expr=\thisrowno{1}^-1, y expr=\thisrowno{1}^-2] {cnv/cnv_lap_ref.csv};
\legend{$L_2$ error,$\delta x^2$}
\end{axis}
\end{tikzpicture}
\caption{Reference}
\end{subfigure}
\hspace{0.5cm}
\begin{subfigure}{0.45\textwidth}
\begin{tikzpicture}[scale = 0.7]
\begin{axis}[
	xmode=log,
	ymode=log,
    xlabel={$\delta x$},
    legend pos=south east,
    legend style={font=\footnotesize},
    xmajorgrids=true,
    ymajorgrids=true,
    grid style=dashed,
    axis x line = bottom,
    axis y line = left,
	xmin = 1e-3,
	xmax = 1e-0,
	ymin = 1e-6,
	ymax = 1e+1]
\addplot [color=blue ,mark=o] table[x expr=\thisrowno{1}^-1, y index=5] {cnv/cnv_lap_def.csv};
\addplot [color=green,mark=square] table[x expr=\thisrowno{1}^-1, y expr=\thisrowno{1}^-2] {cnv/cnv_lap_def.csv};
\legend{$L_2$ error,$\delta x^2$}
\end{axis}
\end{tikzpicture}
\caption{Deformed}
\end{subfigure}
\caption{$L_2$ error norm vs. mesh width $\delta x$ for the discrete Laplacian operator.}
\label{fig:conv6}
\end{figure}
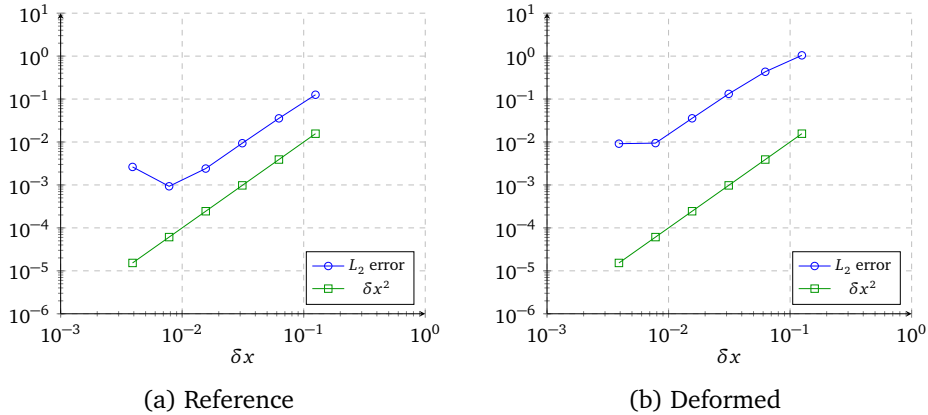

The derivation of the discrete Laplacian operator is given in Section \ref{sub:lap1} and it shows the expected quadratic convergence with respect to mesh width. The results of the convergence study are shown in Figure \ref{fig:conv6}.

Again the approximation is subject to numerical instability for small mesh widths, but now it is more pronounced. This is because the Laplacian requires the application of both gradient and divergence operators, and thus division by $J^2 \approx \delta x^6$. As a result it is important to design actual simulations with units that result in a mesh width within stable bounds.  For this reason, in the simulations that follow, normal SI units are converted such that values stay within reasonable range, distances are in millimetres, rather than metres, for instance. This is an important insight from the study.

\FloatBarrier

\subsection{Jacobi Solver}

The details of the assembly and iterative solution of the discrete Laplace operator are given in Section \ref{sec:solv1}. In this case we give the analytic values as data on the right hand side and solve for the objective function.

\begin{figure}[h!]
\begin{subfigure}{0.45\textwidth}
\begin{tikzpicture}[scale = 0.7]
\begin{axis}[
	xmode=normal,
	ymode=log,
    xlabel={Iterations},
    ylabel={$L_2$ error},
    legend pos=north east,
    legend style={font=\footnotesize},
    legend cell align={left},
    xmajorgrids=true,
    ymajorgrids=true,
    grid style=dashed,
    axis x line = bottom,
    axis y line = left,
	xmin = 0,
	xmax = 30000,
	ymin = 1e-4,
	ymax = 1e+1
]
\addplot [color=cyan ] table[x index=0, y index=5] {jac/cnv_jac_ref016.csv};
\addplot [color=green] table[x index=0, y index=5] {jac/cnv_jac_ref032.csv};
\addplot [color=blue ] table[x index=0, y index=5] {jac/cnv_jac_ref064.csv};
\addplot [color=red  ] table[x index=0, y index=5] {jac/cnv_jac_ref128.csv};
\legend{$n=16$,$n=32$,$n=64$,$n=128$}
\end{axis}
\end{tikzpicture}
\caption{Reference}
\end{subfigure}
\hspace{0.5cm}
\begin{subfigure}{0.45\textwidth}
\begin{tikzpicture}[scale = 0.7]
\begin{axis}[
	xmode=normal,
	ymode=log,
    xlabel={Iterations},
    ylabel={$L_2$ error},
    legend pos=north east,
    legend style={font=\footnotesize},
    legend cell align={left},
    xmajorgrids=true,
    ymajorgrids=true,
    grid style=dashed,
    axis x line = bottom,
    axis y line = left,
	xmin = 0,
	xmax = 40000,
	ymin = 1e-4,
	ymax = 1e+1
]
\addplot [color=cyan ] table[x index=0, y index=5] {jac/cnv_jac_def016.csv};
\addplot [color=green] table[x index=0, y index=5] {jac/cnv_jac_def032.csv};
\addplot [color=blue ] table[x index=0, y index=5] {jac/cnv_jac_def064.csv};
\addplot [color=red  ] table[x index=0, y index=5] {jac/cnv_jac_def128.csv};
\legend{$n=16$,$n=32$,$n=64$,$n=128$}
\end{axis}
\end{tikzpicture}
\caption{Deformed}
\end{subfigure}
\caption{$L_2$ error norm vs. iterations for the Laplacian Jacobi solver.}
\label{fig:conv7}
\end{figure}
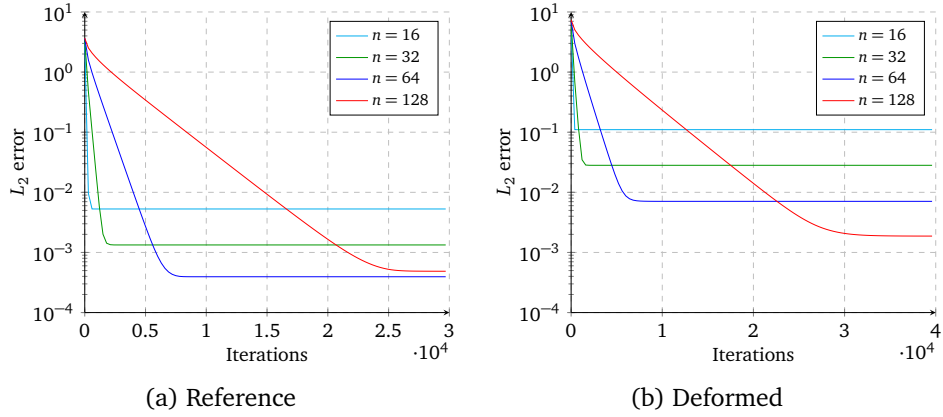

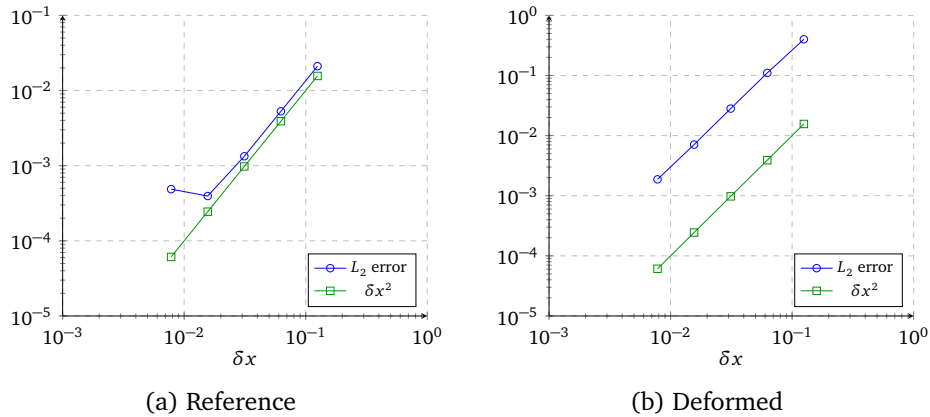
\begin{figure}[h!]
\begin{subfigure}{0.45\textwidth}
\begin{tikzpicture}[scale = 0.7]
\begin{axis}[
	xmode=log,
	ymode=log,
    xlabel={$\delta x$},
    legend pos=south east,
    legend style={font=\footnotesize},
    xmajorgrids=true,
    ymajorgrids=true,
    grid style=dashed,
    axis x line = bottom,
    axis y line = left,
	xmin = 1e-3,
	xmax = 1e-0,
	ymin = 1e-5,
	ymax = 1e-1]
\addplot [color=blue ,mark=o] table[x expr=\thisrowno{1}^-1, y index=5] {jac/cnv_jac_ref_all.csv};
\addplot [color=green,mark=square] table[x expr=\thisrowno{1}^-1, y expr=\thisrowno{1}^-2] {jac/cnv_jac_ref_all.csv};
\legend{$L_2$ error,$\delta x^2$}
\end{axis}
\end{tikzpicture}\caption{Reference}
\end{subfigure}
\hspace{0.5cm}
\begin{subfigure}{0.45\textwidth}
\begin{tikzpicture}[scale = 0.7]
\begin{axis}[
	xmode=log,
	ymode=log,
    xlabel={$\delta x$},
    legend pos=south east,
    legend style={font=\footnotesize},
    xmajorgrids=true,
    ymajorgrids=true,
    grid style=dashed,
    axis x line = bottom,
    axis y line = left,
	xmin = 1e-3,
	xmax = 1e-0,
	ymin = 1e-5,
	ymax = 1e-0]
\addplot [color=blue ,mark=o] table[x expr=\thisrowno{1}^-1, y index=5] {jac/cnv_jac_def_all.csv};
\addplot [color=green,mark=square] table[x expr=\thisrowno{1}^-1, y expr=\thisrowno{1}^-2] {jac/cnv_jac_def_all.csv};
\legend{$L_2$ error,$\delta x^2$}
\end{axis}
\end{tikzpicture}\caption{Deformed}
\end{subfigure}
\caption{$L_2$ error norm vs. mesh width $\delta x$ for the Laplacian Jacobi solver.}
\label{fig:conv8}
\end{figure}

Figure \ref{fig:conv7} shows the convergence of the Jacobi solver with respect to iterations.  This well-known result shows how the number of iterations required to reach convergence increases with. the number of degrees of freedom. The accuracy of the converged result with respect to mesh width is shown in Figure \ref{fig:conv8}.  As before the convergence is quadratic but there is numerical instability for small $\delta x$ due to the division by $J^2$ as for the forward operator. The effect is masked by the error introduced by the deformation but is still present, and the Jacobi solver must also be damped as per \ref{eqn:dampjac1} which requires more iterations.

Again it is important to recognise the lower bound on mesh width, which can be mitigated by careful choice of units.

The Jacobi iterative solver is used four times during the algorithm described in Chapter \ref{chp:alg}. In the case of Electrophysiology, Damping and Viscosity it is used to solve for a step of implicit Euler time integration as per \ref{eqn:iejac1} and, as discussed in Section \ref{sec:ie1}, it is well conditioned converges in relatively few iterations for small time steps. When used for the Poisson Pressure equation as discussed in Section \ref{sec:press1}, it is important to use the pressure solution from the last time step as the initial guess in order to minimise the number of iterations.

\FloatBarrier

\section{Benchmark Studies}

\subsection{Electophysiology}

The implementation of the Electrophysiology simulation was validated against the benchmark study of \cite{Niederer2011}.  The study defines a problem in anisotropic electrophysiology and compares the results of 11 different codes for its solution, each submitted by different research groups.  

The problem considers a cuboid of 20mm $\times$ 7mm $\times$ 3mm of myocardial tissue with an active electrophysiology in accordance with the monodomain equation (\ref{eqn:monodomain1}), repeated here:
\begin{equation}
\frac{\lambda}{1 + \lambda} \nabla \cdot (\SIG \nabla v) = \chi_m \left( C_m \frac{\partial v}{\partial t}  + I_\text{m}  \right).
\end{equation}
Instead of the Mitchell Schaeffer model (\ref{eqn:ms1}) for the transmembrane potential $I_m$, the study uses the more detailed model of  \cite{TenTusscher2004}, which consists of 11 coupled ODEs and describes all major ion channels as well as intracellular calcium dynamics. 

\begin{figure}[h]
\centering
\includegraphics[width=0.49\textwidth,trim={0cm 0cm 10cm 6cm},clip]{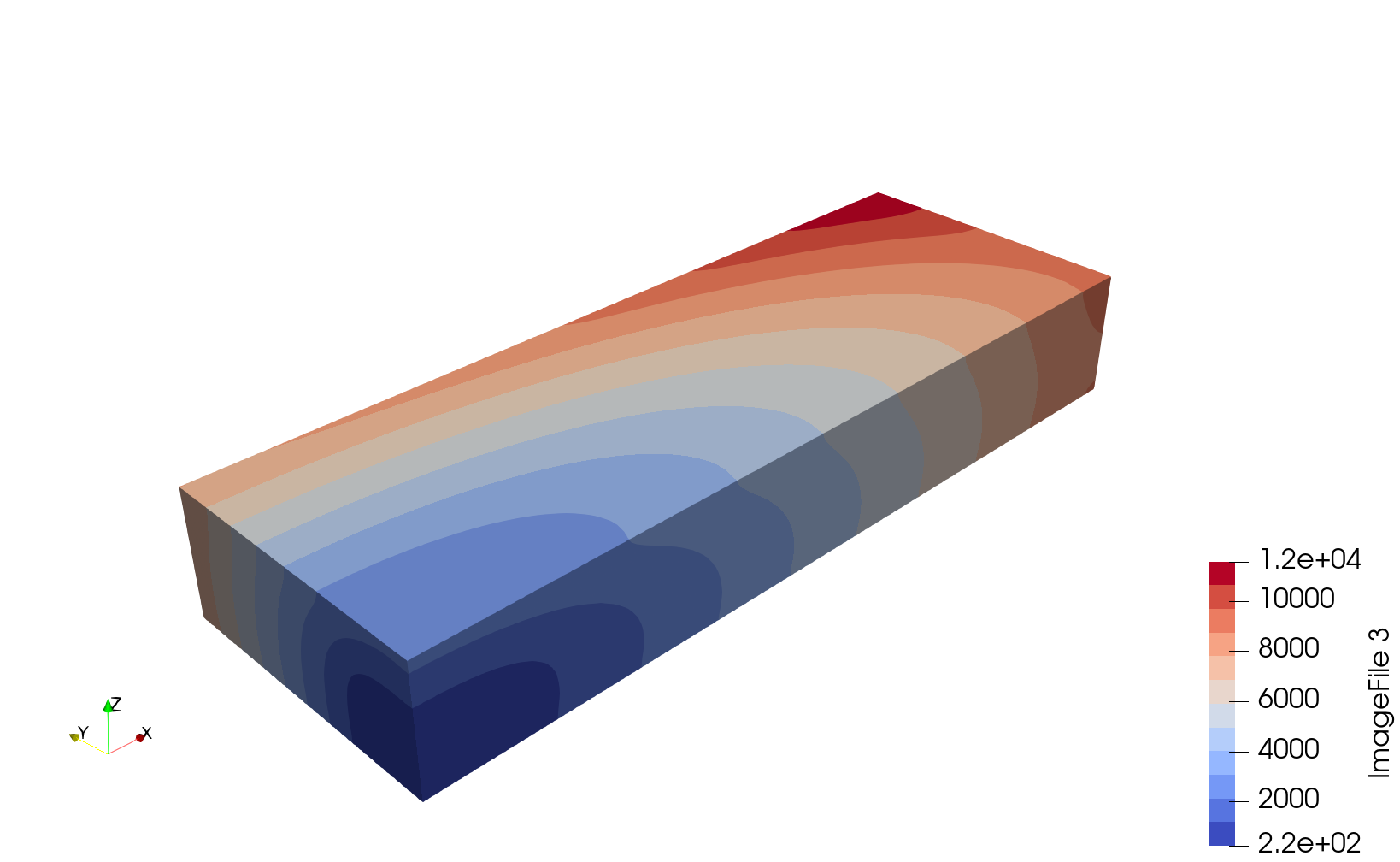}
\includegraphics[width=0.49\textwidth,trim={0cm 0cm 10cm 6cm},clip]{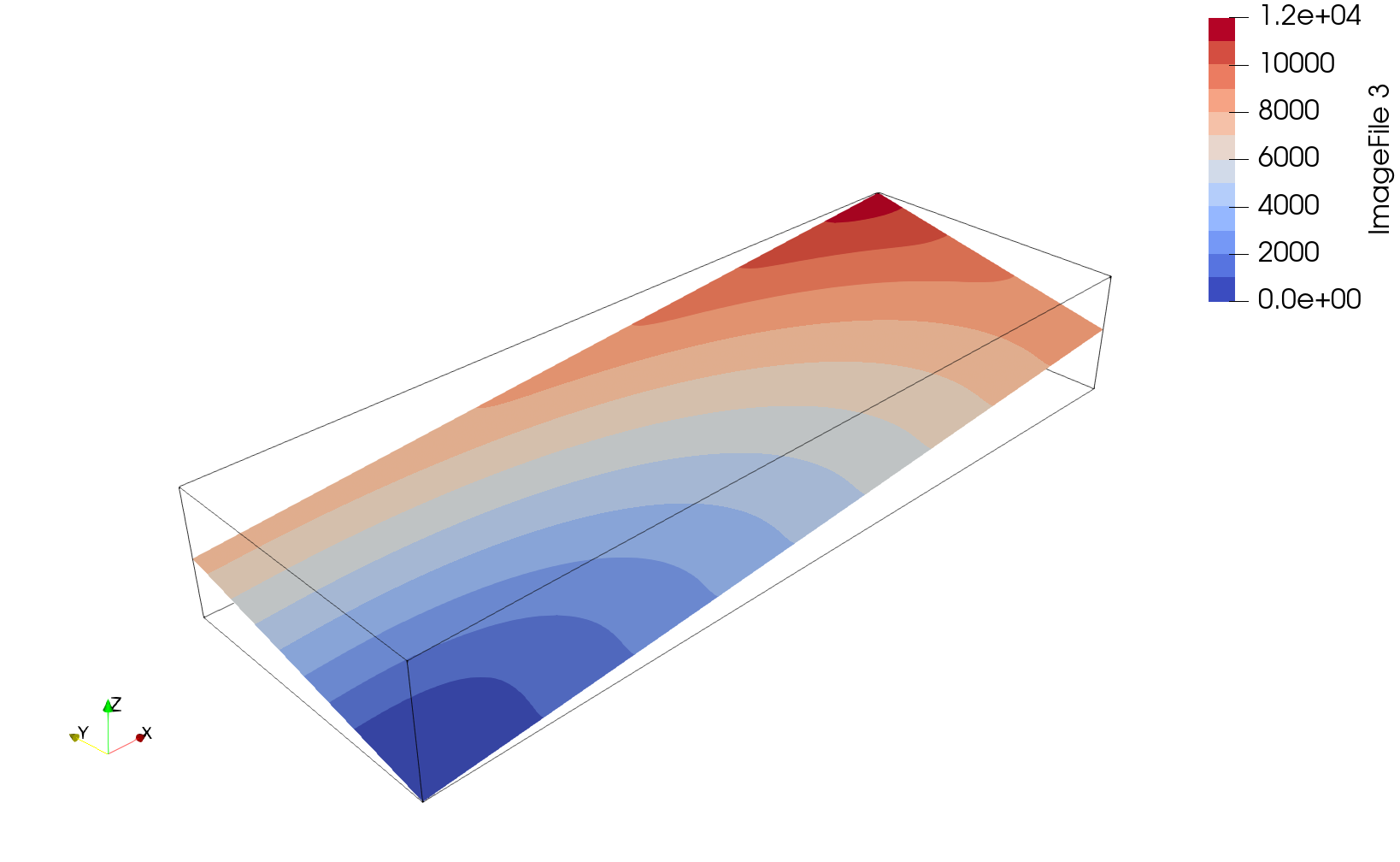}
\caption{Sample results for the eletrophysiology benchmark experiment, showing activation times in $ms$ through the cuboid. The stimulus is applied in the lower left corner and the latest activation is in the top right corner.}
\label{fig:niederer1}
\end{figure}

The cuboid is anisotropic, having greater conductivity along its long axis, and receives an initial stimulus in a cubic region of 1.5mm $\times$ 1.5mm $\times$ 1.5mm at its corner.  With all parameters of the model fully specified, the experiment records the activation time in milliseconds throughout the cuboid generated by each of the submitted codes.  An example result is shown in Figure \ref{fig:niederer1}. 

\begin{figure}[h]
\centering
\includegraphics[width=\textwidth]{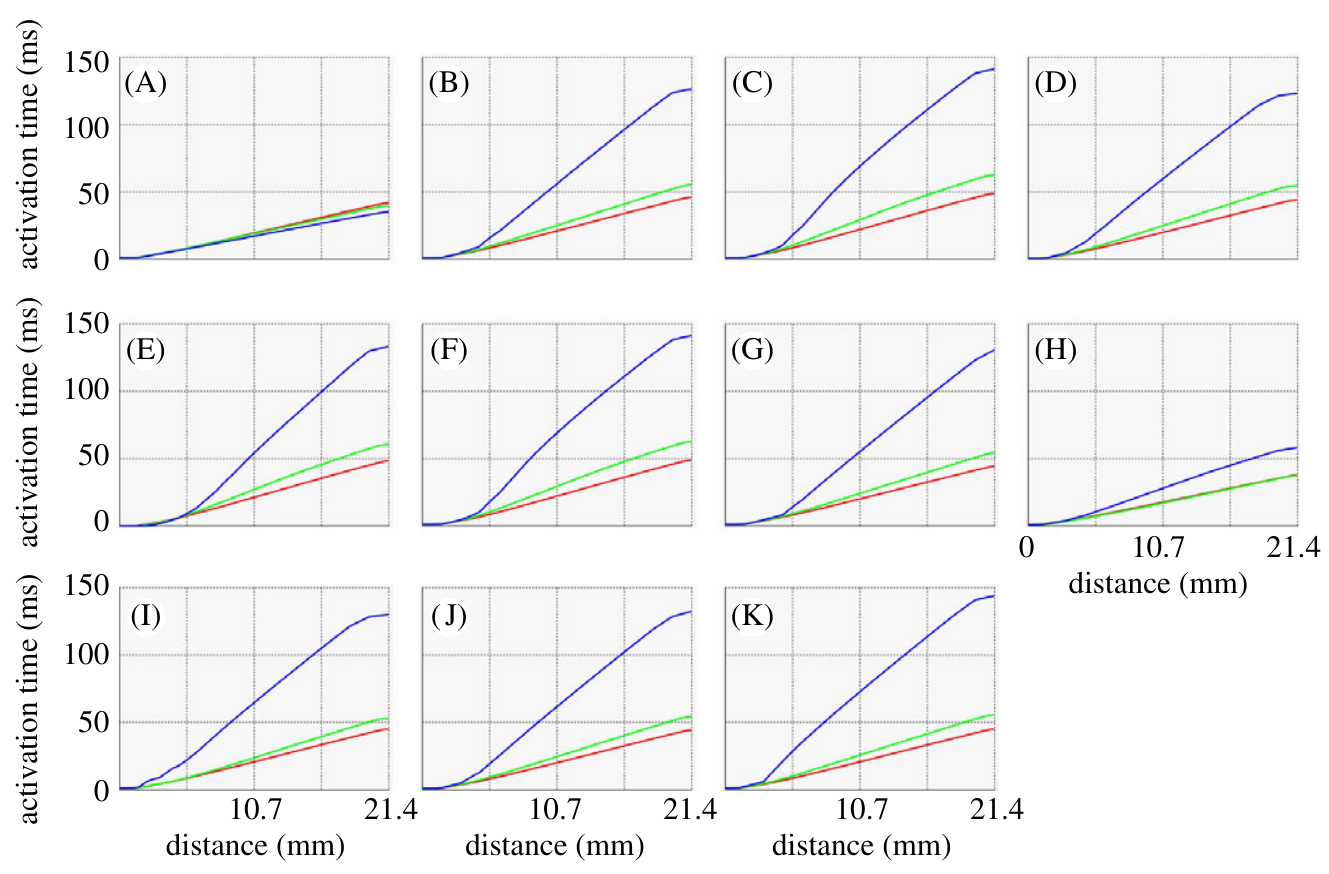}
\caption{Reproduction from \cite{Niederer2011}, showing activation time (ms) vs. distance (mm) along the diagonal of the cuboid, with $\delta t$ = 0.005ms and $\delta x =$ 0.1mm (red), $\delta x =$ 0.2mm (green) and $\delta x =$ 0.5mm (blue).}
\label{fig:niederer2}
\end{figure}

The results were compared for different spatial $\delta x \in$ \{0.5mm, 0.2mm, 0.1mm\} and temporal $\delta t \in$ \{0.05ms, 0.01ms, 0.005ms\} resolutions by plotting activation time (ms) vs. distance (mm) along the long diagonal of the cuboid from the stimulus to the opposite corner. Figure \ref{fig:niederer2} reproduces the results from the paper for the 11 submitted codes.

The discussion in the paper notes that the results are broadly similar, but subject to a variation of around 10\% even at high-resolution convergence. Some of the results also exhibit mesh-dependent variation. 

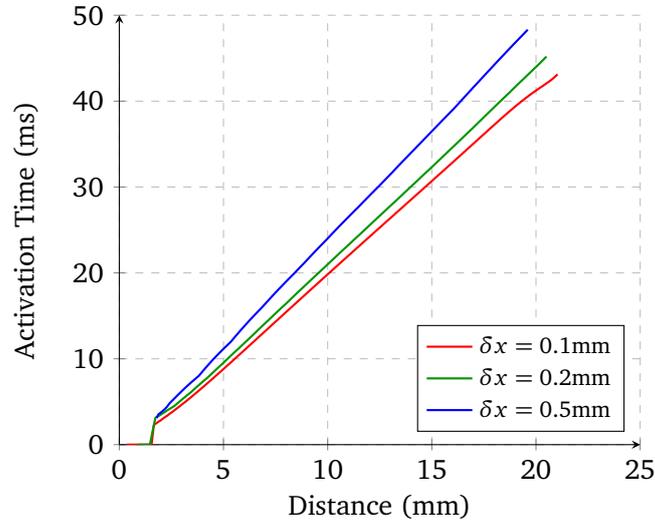
\begin{figure}[h]
\centering
\begin{tikzpicture}[scale = 1.0]
\begin{axis}[
    xlabel={Distance (mm)},
    ylabel={Activation Time (ms)},
    legend pos=south east,
    legend style={font=\footnotesize},
    xmajorgrids=true,
    ymajorgrids=true,
    grid style=dashed,
    axis x line = bottom,
    axis y line = left,
    xmin=0,
    xmax=25,
    ymin=0,
    ymax=50]
\addplot [color=red  ,thick] table[x index=13, y index=6,col sep=comma] {niederer/act_01_0005.csv};
\addplot [color=green,thick] table[x index=21, y index=14,col sep=comma] {niederer/act_02_0005.csv};
\addplot [color=blue ,thick] table[x index=13, y index=6,col sep=comma] {niederer/act_05_0005.csv};
\legend{$\delta x$ = 0.1mm,$\delta x$ = 0.2mm,$\delta x$ = 0.5mm}
\end{axis}
\end{tikzpicture}
\caption{Results for the experiment in \cite{Niederer2011}, using our model, showing activation time (ms) vs. distance (mm) along the diagonal of the cuboid, with $\delta t$ = 0.005ms and $\delta x$ as shown.}
\label{fig:niederer3}
\end{figure}

The equivalent plot using the algorithm as defined in this work is shown in Figure \ref{fig:niederer3}.  It is worth noting that the y-axis range is zero to 50ms, and so the plot most closely resembles subplot (A) in Figure  \ref{fig:niederer2}. As such the result of our experiment is broadly in line with the result at the finest resolution for the benchmarking results and shows little mesh-dependent variation. The paper gives the high-resolution convergence result at the top corner as between 42.5-43ms, and our result is within that range for the finer resolutions. This is despite the fact that our implementation uses the simpler model of \cite{Mitchell2003}.

\begin{figure}[h!]
\begin{subfigure}{0.45\textwidth}
\begin{tikzpicture}[scale = 0.7]
\begin{axis}[
    xlabel={Distance (mm)},
    ylabel={Activation Time (ms)},
    legend pos=south east,
    legend style={font=\footnotesize},
    xmajorgrids=true,
    ymajorgrids=true,
    grid style=dashed,
    axis x line = bottom,
    axis y line = left,
    xmin=0,
    xmax=25,
    ymin=0,
    ymax=60]
\addplot [color=red  ] table[x index=21, y index=14,col sep=comma] {niederer/act_02_0005.csv};
\addplot [color=green] table[x index=21, y index=14,col sep=comma] {niederer/act_02_0010.csv};
\addplot [color=blue ] table[x index=21, y index=14,col sep=comma] {niederer/act_02_0050.csv};
\legend{$\delta t$ = 0.005ms,$\delta t$ = 0.01ms,$\delta t$ = 0.05ms}
\end{axis}
\end{tikzpicture}
\caption{$\delta x$ = 0.2mm}
\end{subfigure}
\hspace{0.5cm}
\begin{subfigure}{0.45\textwidth}
\begin{tikzpicture}[scale = 0.7]
\begin{axis}[
    xlabel={Distance (mm)},
    ylabel={Activation Time (ms)},
    legend pos=south east,
    legend style={font=\footnotesize},
    xmajorgrids=true,
    ymajorgrids=true,
    grid style=dashed,
    axis x line = bottom,
    axis y line = left,
    xmin=0,
    xmax=25,
    ymin=0,
    ymax=60]
\addplot [color=red  ] table[x index=4, y index=7, col sep=comma] {niederer/act_05_0010.csv};
\addplot [color=green] table[x index=4, y index=7, col sep=comma] {niederer/act_05_0050.csv};
\addplot [color=blue] table[x index=4, y index=7, col sep=comma] {niederer/act_05_0100.csv};
\legend{$\delta t$ = 0.01ms,$\delta t$ = 0.05ms,$\delta t$ = 0.1ms}
\end{axis}
\end{tikzpicture}
\caption{$\delta x$ = 0.5mm}
\end{subfigure}
\caption{A comparison or activation time vs. distance for coarser meshes.  The overall accuracy is reduced with increasing $\delta x$ but is less sensitive to $\delta t$ and thus requires far fewer iterations.}
\label{fig:niederer4}
\end{figure}
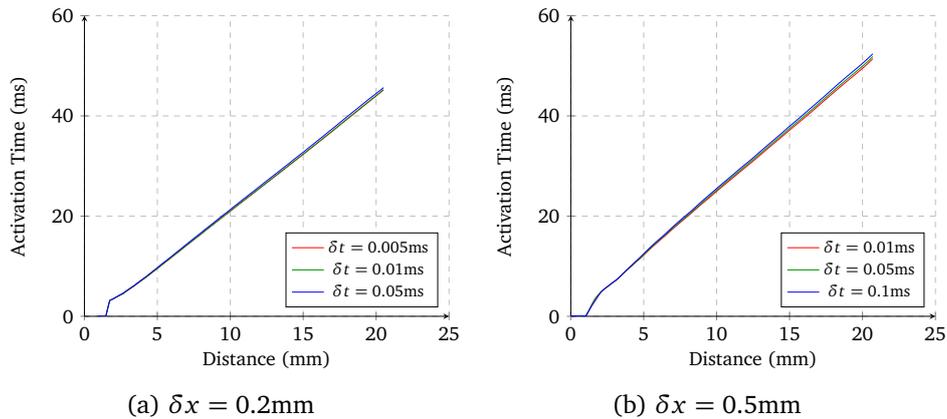

Furthermore, we consider the results of the simulation at coarser resolutions in both space and time. The results are shown in Figure \ref{fig:niederer4}(a) where $\delta x$ = 0.2mm and \ref{fig:niederer4}(b) where $\delta x$ = 0.5mm.  In both cases there is a reduction in overall accuracy for the activation time with respect to mesh width.  This can be seen by a higher activation time overall, but the solution is not sensitive to time resolution.  This is very useful since it allows for a longer timestep and fewer iterations.

In conclusion, the CFL condition determines a maximum time step $\delta t$. Since the size, and therefore resolution $\delta x$, of the domain is limited by the memory capacity of the processor, careful consideration can allow for an efficient compromise between spatial and temporal resolution and time to solution.

\subsection{Solid Mechanics}

For the verification of the solid mechanics simulation we reproduce some results from the benchmark study of \cite{Land2015}. The paper sets benchmark problems for the simulation of deformation which occurs in a solid body with varying spatial characteristics under a fluid pressure, either as a passive response or with active stress corresponding to muscle contraction. 

The problems were solved by code submitted by 11 different research groups and arrived at consensus solutions using differing approaches to formulation, discretisation and processing. We will reproduce the two experiments corresponding to the response of an idealised ventricle under passive expansion and active contraction.

\subsubsection{Geometry}

\begin{figure}[h!]
\centering
\includegraphics[width=0.6\textwidth,trim={0cm 5cm 0cm 2cm},clip]{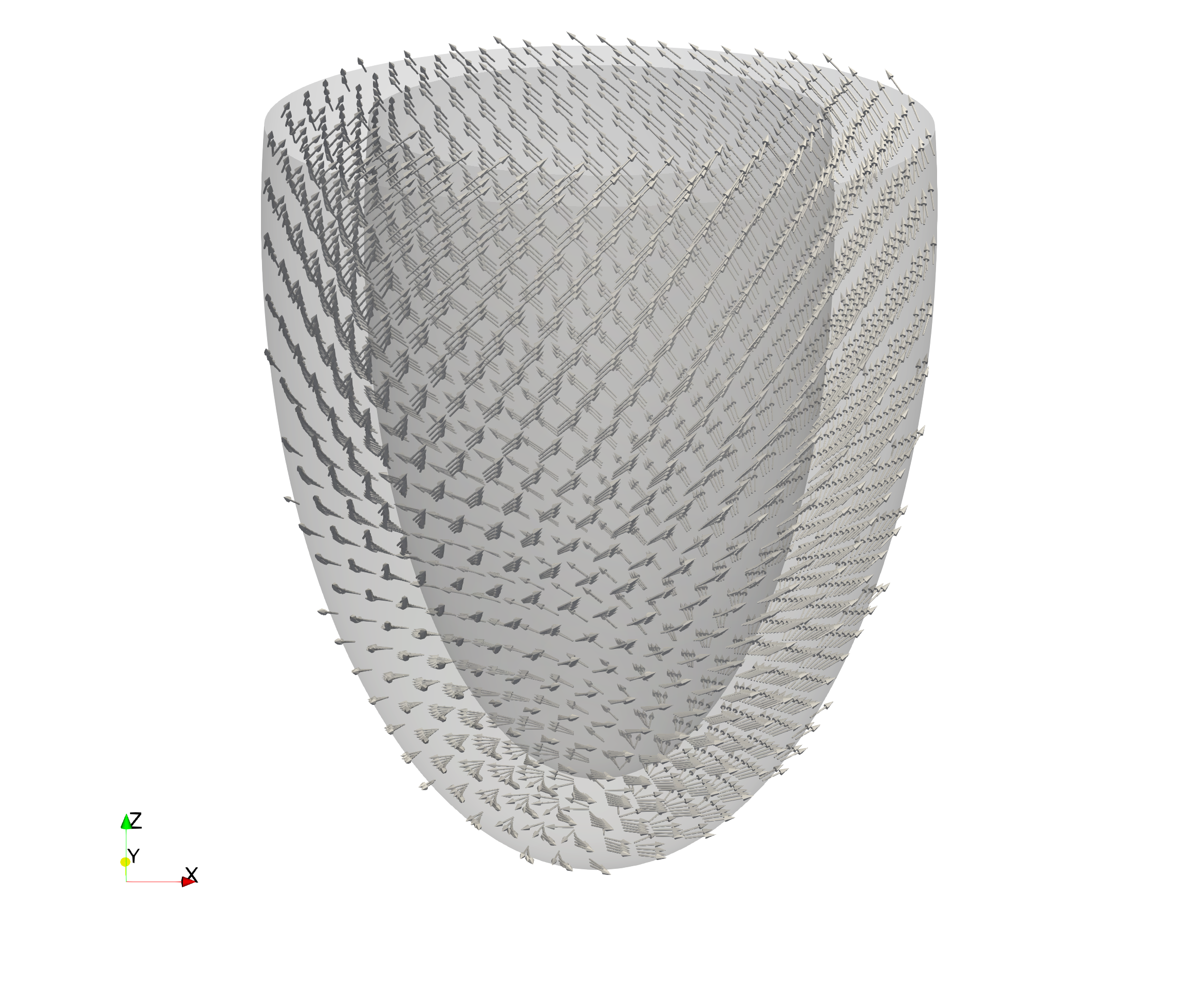}
\caption{Model geometry of \cite{Land2015}, representing an idealised ventricle in its reference configuration showing fibre directions.}
\label{fig:land6}
\end{figure}

The two experiments both make use of the same geometry, given in its reference configuration and shown in Figure \ref{fig:land6}.  An idealised ventricle is modelled as a truncated ellipsoid, fixed on its upper plane. The solid part of the geometry contain a vector field corresponding to fibre directions which vary spatially through the tissue in a spiral fashion.  The dimensions and properties of the ventricle are given as algebraic expressions which can be easily reproduced in our model via signed distance functions and stored vector fields.

A static spherical pressure field was set on the interior of the ventricle, corresponding to the fluid pressure of the blood.  On the exterior of the ventricle no boundary conditions were applied, except at the top plane, which was fixed in space.

\subsubsection{Material Model}\label{sec:moooney}

For the material model, the study makes use of the transversely isotropic constitutive law of \cite{Guccione1995}.  The strain energy function includes the parameterisation of an anisotropic material response determined by the fibre field within the tissue. This model has been widely used both in the determination of model parameters and in the simulation of tissue behaviour, see \cite{Tang2007},\cite{Hassaballah2013} amongst others.

We reproduce the results using a different material formulation, the Hyperelastic solid model of \cite{Mooney1940} and \cite{Rivlin1948} for the stress strain relationship, which is passively isotropic and incompressible, and add an anisotropic stress determined by the vector field of fibre directions. We briefly summarise the model as it will be used again in cardiac simulation.

The strain energy density function of a Mooney-Rivlin material is given as:
\begin{equation}
W = c_1(\overline{I}_1 - 3) + c_2(\overline{I}_2 - 3),
\end{equation}
where $c_1$ and $c_2$ are empirically determined material constants and $I_1$ and $I_2$ are the first and second invariants of $\overline{\mathbf{B}}$ the unimodular part of the left Cauchy-Green deformation tensor $\B$:
\begin{eqnarray}
\mathbf{B}  &=& \F \F^\top \\
\overline{\B} &=& \det(\B)^{-\frac{1}{3}} \B 
\end{eqnarray}
The unimodular part of $\B$ has a determinant $\det(\overline{\B}) = 1$. The material model is thus formulated directly in the deformed configuration and separates deformation from change in volume. The invariants are defined as follows and can be calculated quickly from the invariants $I_1, I_2$ of $\B$.
\begin{eqnarray}
\overline{I}_1 &=& J^{-\frac{2}{3}} I_1 \\
\overline{I}_2 &=& J^{-\frac{4}{3}} I_2 ,
\end{eqnarray}
where the standard tensor invariants are defined as follows:
\begin{eqnarray}
I_1 &=& \tr(\B) \\
I_2 &=& \frac{1}{2} \left[\tr(\B)^2 - \tr(\B^2) \right],
\end{eqnarray}
and $J = \det(\F)$ is the determinant of the deformation gradient as usual.

For an incompressible Mooney-Rivlin material $J = 1$ and thus $\overline{\B} = \B$. The Cauchy stress $\bm{\sigma} $ can therefore be expressed as:
\begin{equation}
\bm{\sigma} = - p^* \I + 2 ( c_1 \B - 2 c_2 \B^{-1} )
\end{equation}
Where $p^*$ represents a corrective pressure given by:
\begin{equation}
p^* = \frac{2}{3} (c_1 I_1 - c_2 I_2)
\end{equation}
As a result the highly nonlinear behaviour of an incompressible solid can be modelled using a stable, explicit iteration.  This the reason for the choice of the Mooney-Rivlin material and a key part of the design of the cardiac simulation.

For the anisotropic contraction a synthetic deformation is introduced, equivalent to an extension in the longditudinal fibre direction $\f \otimes \f$, as per (\ref{eqn:sigma1}).  The extension is added to the existing deformation gradient $\F$ and flows into the calculation of stress, resulting in a contractile response that is equivalent to muscle contraction, but without affecting the incompressible response of the material model. Again this is an important part of the design, in maintaining the integrity and stability of the mechanical simulation.

\subsubsection{Passive Expansion}

\begin{figure}[h!]
\begin{subfigure}{0.5\textwidth}
\includegraphics[width=\textwidth,trim={15cm 10cm 15cm 10cm},clip]{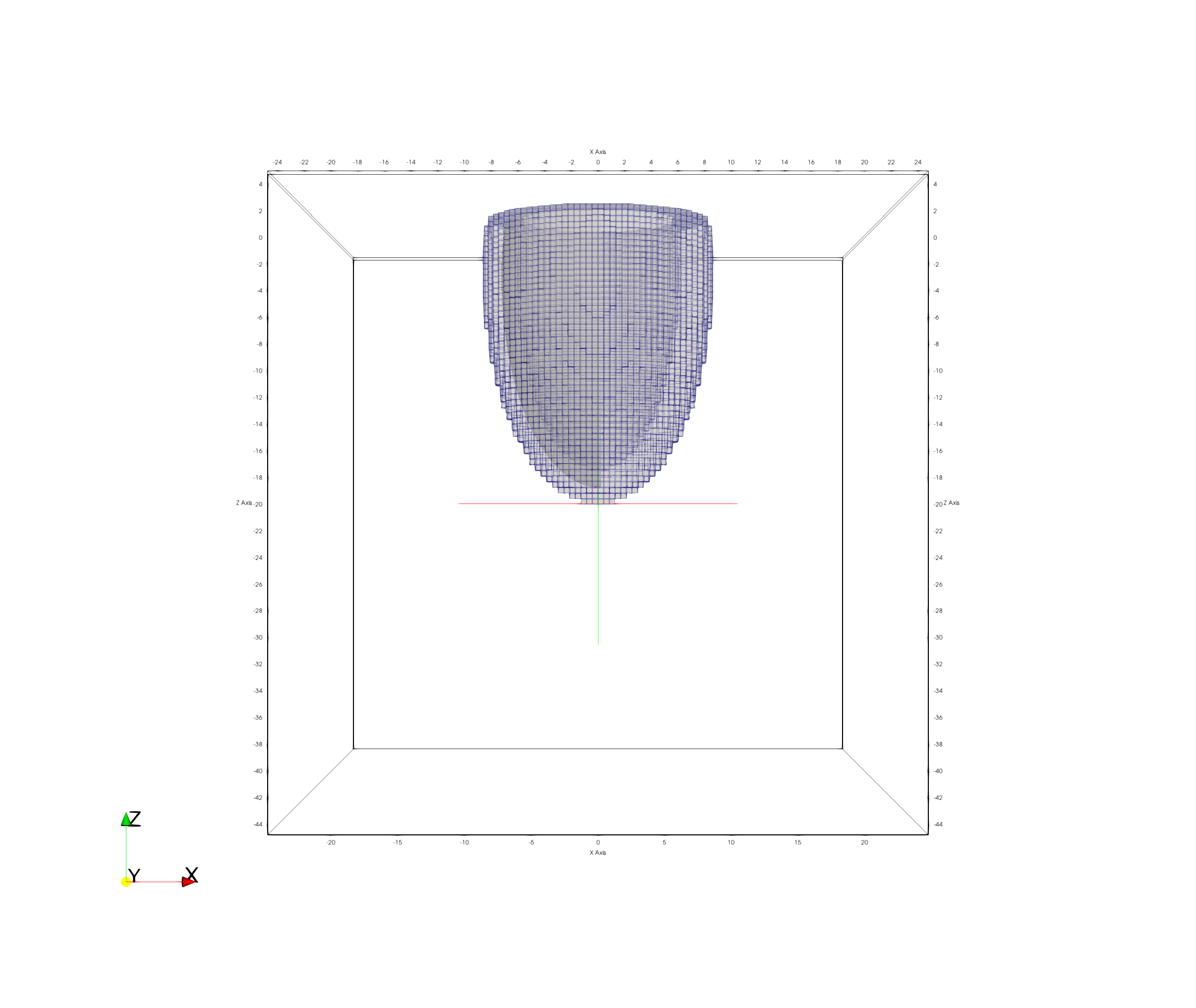}
\caption{Reference}
\end{subfigure}
\hspace{0.1cm}
\begin{subfigure}{0.5\textwidth}
\includegraphics[width=\textwidth,trim={15cm 10cm 15cm 10cm},clip]{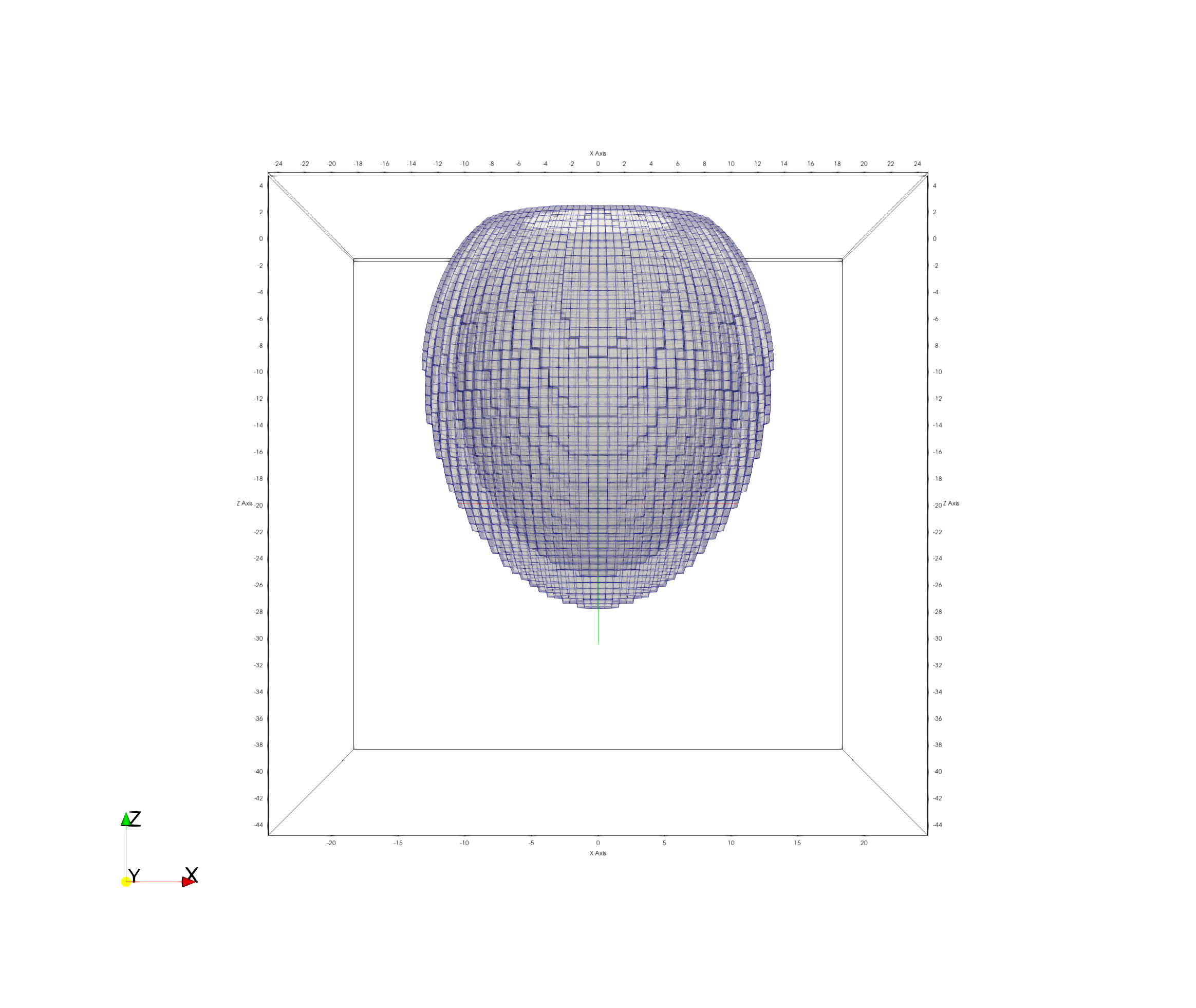}
\caption{Passive expansion}
\end{subfigure}
\caption{The reference configuration (a) and passive expansion (b) of an idealised ventricle.}
\label{fig:land4}
\end{figure}

In the first experiment a pressure of 10kPa was applied to interior surface of the solid in its reference configuration and the ventricle is allowed to inflate. The contributors to the paper solve a static problem but we run the dynamic solution until it reaches equilibrium, as shown in Figure \ref{fig:land4}.

The simulation was run on a three-dimensional regular structured grid of $100^3$ cells. The mesh width $\delta x$ = 0.5mm, and time step $\delta t$ = 0.005ms.  The computation output a file completely describing the state of the model every 500 iterations, producing 100 files in total.  The simulation time was thus $100 \times 500 \times 0.005$ = 250ms, which was sufficient for the solid to reach a stationary equilibrium. The GPU processor used was an NVIDIA A100 PCI with 40GB memory and the processing time was around 160 seconds.

\begin{figure}[h!]
\begin{subfigure}{0.5\textwidth}
\begin{tikzpicture}[scale = 0.9]
\begin{axis}[
    xlabel={$x$ (mm)},
    ylabel={$z$ (mm)},
    legend pos=north east,
    xlabel style={font=\footnotesize},
    ylabel style={font=\footnotesize},
    xmajorgrids=true,
    ymajorgrids=true,
    grid style=dashed,
    axis x line = bottom,
    axis y line = left,
    xmin=-15,
    xmax=0,
    ymin=-28,
    ymax=+5,
    width=6.8cm,
    height=13.7cm]
\addplot [color=gray , thick, dashed] table[x index=1, y index=3, col sep=comma] {land/land0_100.csv};
\addplot [color=blue , thick] table[x index=1, y index=3, col sep=comma] {land/land2_100.csv};
\legend{Reference,Deformed}
\end{axis}
\end{tikzpicture}
\caption{GPU Finite Volume}
\end{subfigure}
\hspace{0.5cm}
\begin{subfigure}{0.5\textwidth}
\includegraphics[width=0.9\textwidth]{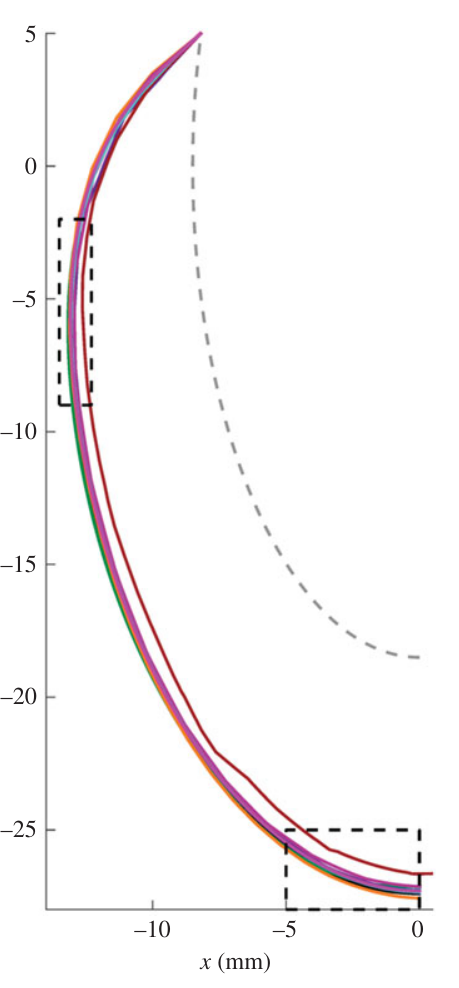}
\caption{Land et al. 2015}
\end{subfigure}
\caption{The reference configuration and passive deformation of an idealised ventricle with (a) generated by this work and (b) an extract from \cite{Land2015}, showing comparable results.}
\label{fig:land2}
\end{figure}
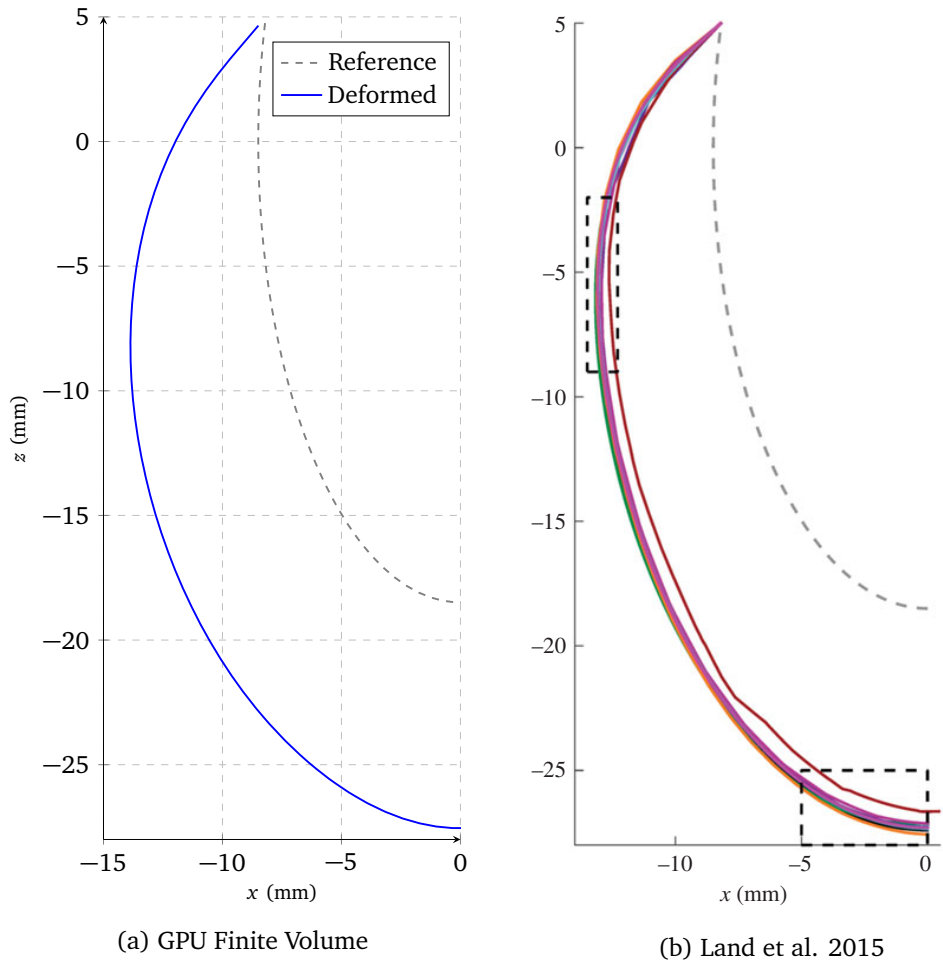

The paper reports the final position of the midline of the solid.  The experiment was repeated and the Mooney-Rivlin parameters updated until the results were in agreement with the consensus solution, as shown in Figure \ref{fig:land2}.  The values of $c_1$ and $c_2$ were thus both set to 0.5. These values were retained for subsequent experiments.

\FloatBarrier

\subsubsection{Active Contraction}

\begin{figure}[h!]
\begin{subfigure}{0.5\textwidth}
\includegraphics[width=\textwidth,trim={15cm 10cm 15cm 10cm},clip]{land/land0.png}
\caption{Reference}
\end{subfigure}
\hspace{0.1cm}
\begin{subfigure}{0.5\textwidth}
\includegraphics[width=\textwidth,trim={15cm 10cm 15cm 10cm},clip]{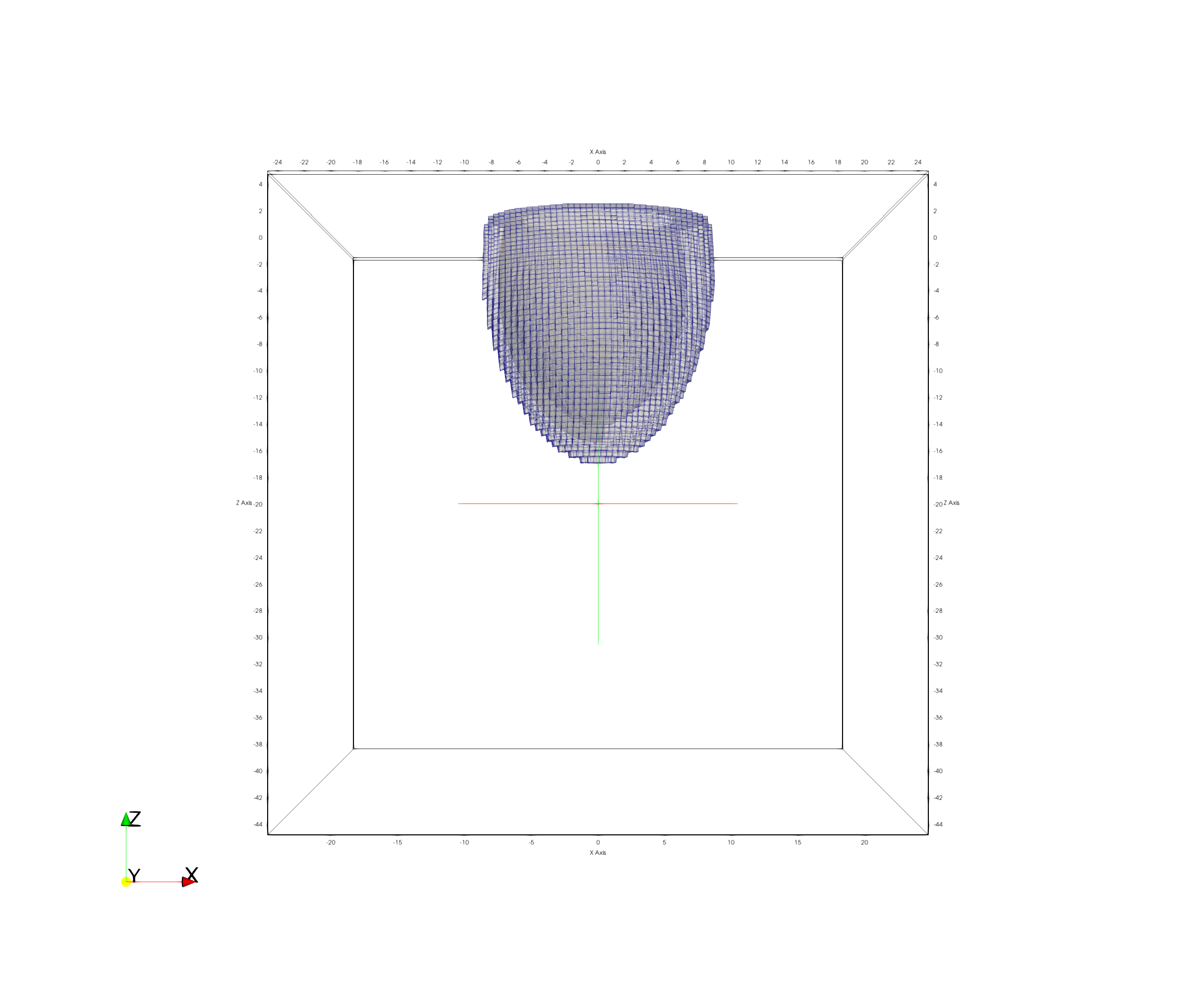}
\caption{Active contraction}
\end{subfigure}
\caption{The reference configuration (a) and active deformation (b) of an idealised ventricle under muscle contraction.}
\label{fig:land5}
\end{figure}

In the second experiment a pressure of 15KPa was applied to interior surface of the solid in its reference configuration as well as a contractile Piola Kirkhoff stress of 60kPa orientated along the fibre directions in the solid. Again, the contributors solve a static problem and we run a dynamic solution to equilibrium, as shown in Figure \ref{fig:land5}.

The simulation was run on the same three-dimensional regular structured grid of $100^3$ cells with mesh width $\delta$x = 0.5mm, but the larger forces required a shorter time step $\delta t$ = 0.001ms.  The computation output a file every 1000 iterations, producing 100 files in total for a simulation time of $100 \times 1000 \times 0.01 = 100$ms, again reaching a static equilibrium. The same NVIDIA A100 processor completed the calculation in 160 seconds.  This illustrates an important point, that despite the model making twice as many iterations (50,000 in the first case and 100,000 in the second) there is no difference in the computation time.  This is because the time to solution is completely dominated by traffic across the PCI bus and IO on the CPU. This is in line with the observations in Chapter \ref{chp:gpu}.

\begin{figure}[t!]
\begin{subfigure}{0.5\textwidth}
\begin{tikzpicture}[scale = 0.9]
\begin{axis}[
    xlabel={$x$ (mm)},
    ylabel={$z$ (mm)},
    legend pos=north east,
    xlabel style={font=\footnotesize},
    ylabel style={font=\footnotesize},
    xmajorgrids=true,
    ymajorgrids=true,
    grid style=dashed,
    axis x line = bottom,
    axis y line = left,
    ytick distance = 5,
    xmin=-10,
    xmax=0,
    ymin=-19,
    ymax=+5,
    width=6.5cm,
    height=14.6cm]
\addplot [color=gray , thick, dashed] table[x index=1, y index=3, col sep=comma] {land/land0_100.csv};
\addplot [color=blue , thick] table[x index=1, y index=3, col sep=comma] {land/land3_100.csv};
\legend{Reference,Deformed}
\end{axis}
\end{tikzpicture}
\caption{GPU Finite Volume}
\end{subfigure}
\hspace{0.5cm}
\begin{subfigure}{0.5\textwidth}
\includegraphics[width=0.9\textwidth]{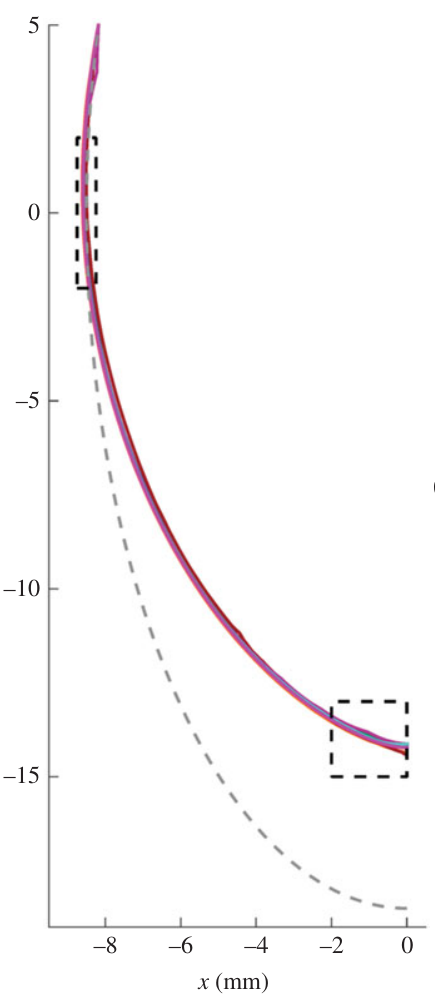}
\caption{Land et al. 2015}
\end{subfigure}
\caption{The reference configuration and active deformation of an idealised ventricle under muscle contraction, with (a) generated from this work and (b) an extract from \cite{Land2015}, showing comparable results, $\delta x$ = 0.5mm.}
\label{fig:land3}
\end{figure}
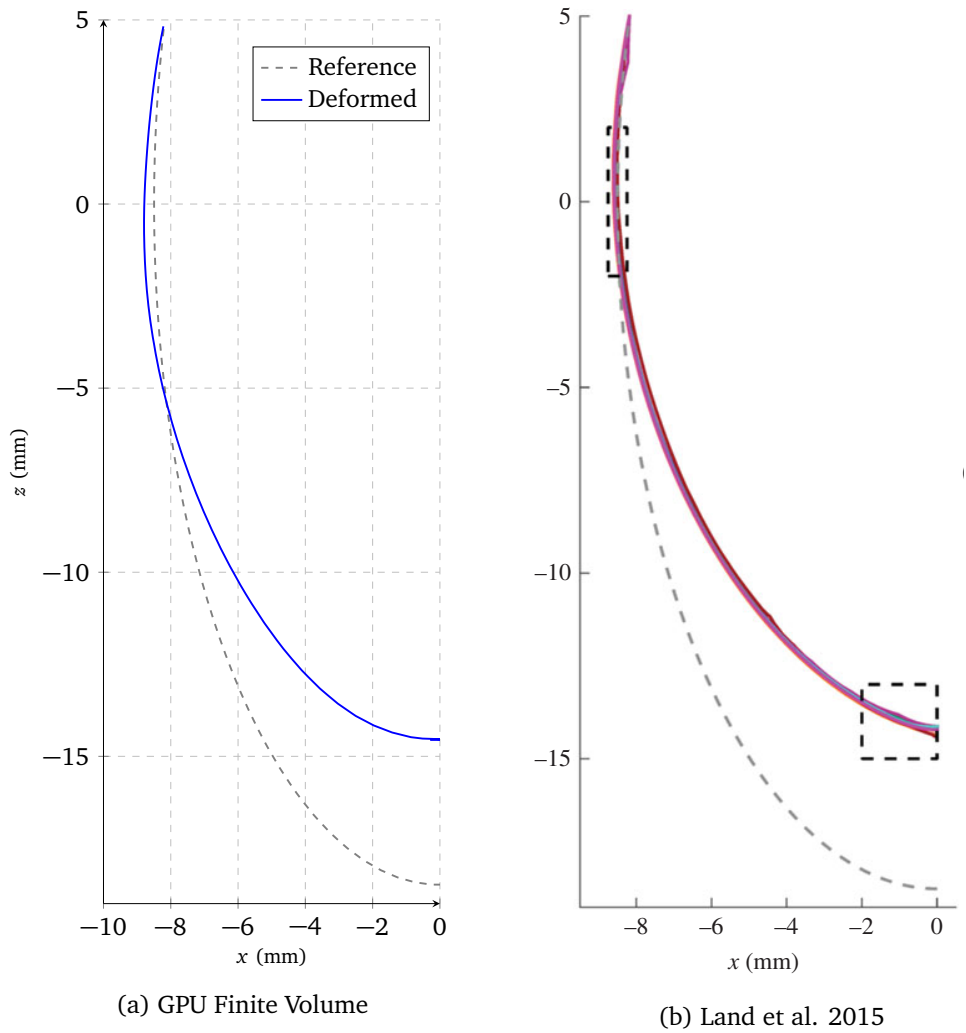

The experiment was repeated until the synthetic expansion parameter for muscle contraction was identified as 1.0, and it is retained for future experiments. The results for the final position of the midpoint are shown in Figure \ref{fig:land3}.

\vfill

\chapter{Cardiac Simulation}\label{chp:sim}

\section{Overview}

In this section we combine all of the elements of the work into a single whole heart simulation.  The objective is to provide a complete set of descriptive data, including position $\x$, solid $\v_s$ and fluid velocity $\v_f$, membrane depolarisation $u$ and fluid pressure $p$ over the course of a single heartbeat. The simulation time is thus one second ($t = 1000$ms) with 100 data files, each corresponding to a frame of video output at 10ms intervals. The results are generated at at spatial resolutions of $\delta x = 1.0$mm and $\delta t \in [0.05, 0.001]$ms respectively. This allows the algorithm to make many iterations between each use of the PCI Bus, hiding latency as discussed in Chapter \ref{chp:gpu}.

The resulting data can be sampled over subsets of the domain and post-processed to generate synthetic analogues to physiological recordings.  The transmembrane voltage is sampled at three positions and used to simulate the output of a three-lead Electrocardiogram (ECG). Position data and pressure within a ventricle will be used to generate a synthetic Pressure-Volume (PV) loop.

It is important to note that the code is a first prototype and this simulation serves as a proof-of-concept for the algorithm itself, rather than presenting the experiment and its results as a state-of-the-art simulation. Validation and the addition of richer models is left as future work.

The following sections will outline the construction of the cardiac geometry using the signed distance functions of Section \ref{sec:geom}, and the adaptation of the solvers and algorithm of Chapters \ref{chp:disc} and \ref{chp:alg} respectively. Finally, we display and discuss some of results.

\section{Geometry}

\subsection{Myocardium}

\begin{figure}[h!]
\centering
\includegraphics[width=0.45\textwidth,trim={15cm 5cm 15cm 5cm},clip]{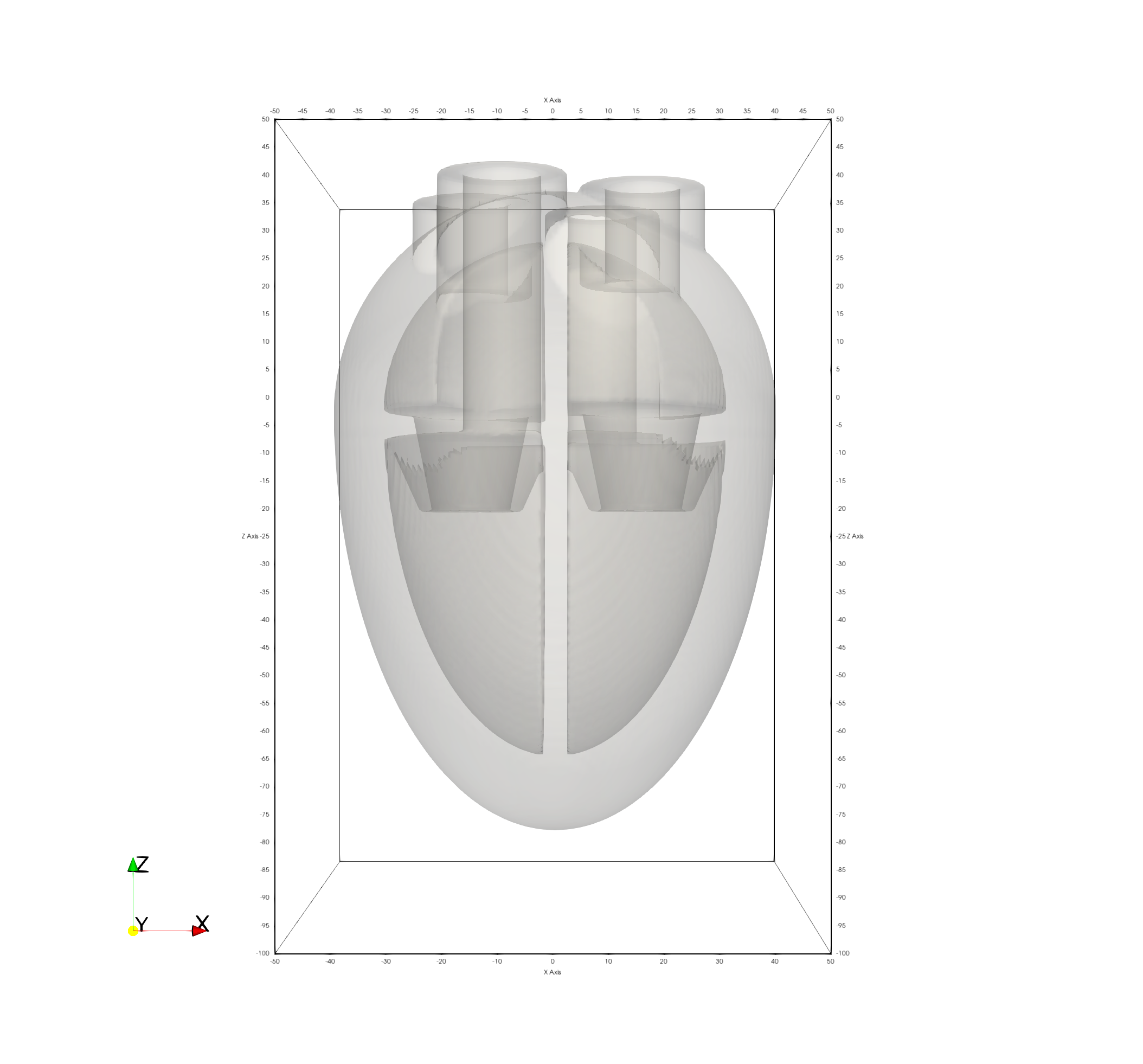}
\includegraphics[width=0.45\textwidth,trim={15cm 5cm 15cm 5cm},clip]{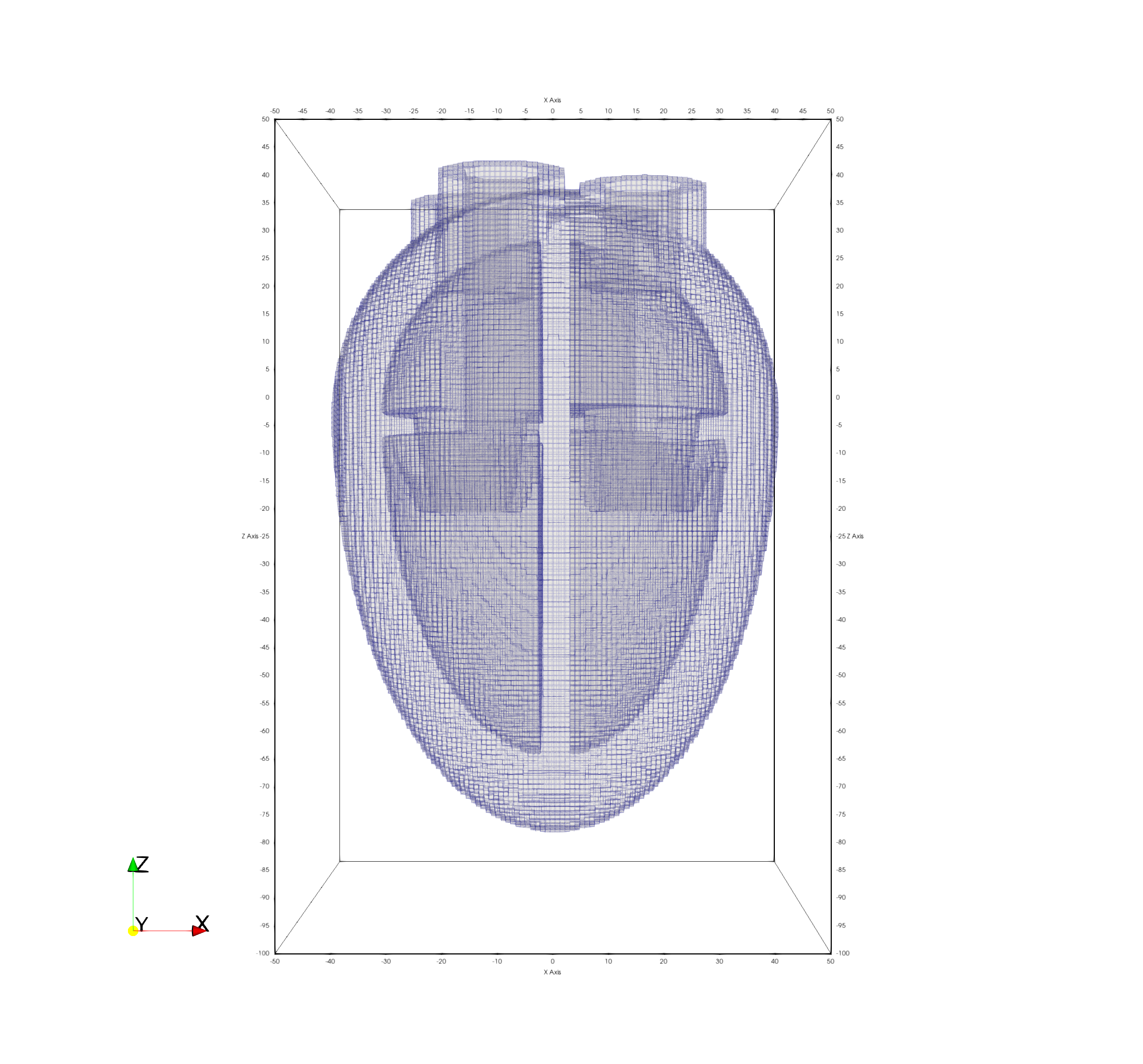}
\caption{The schematic whole heart in signed distance function and voxel representation.}
\label{fig:hgeom0}
\end{figure}

\begin{figure}[h!]
     \centering
     \begin{subfigure}{0.45\textwidth}
		\centering
		\includegraphics[width=\textwidth,trim={15cm 5cm 15cm 5cm},clip]{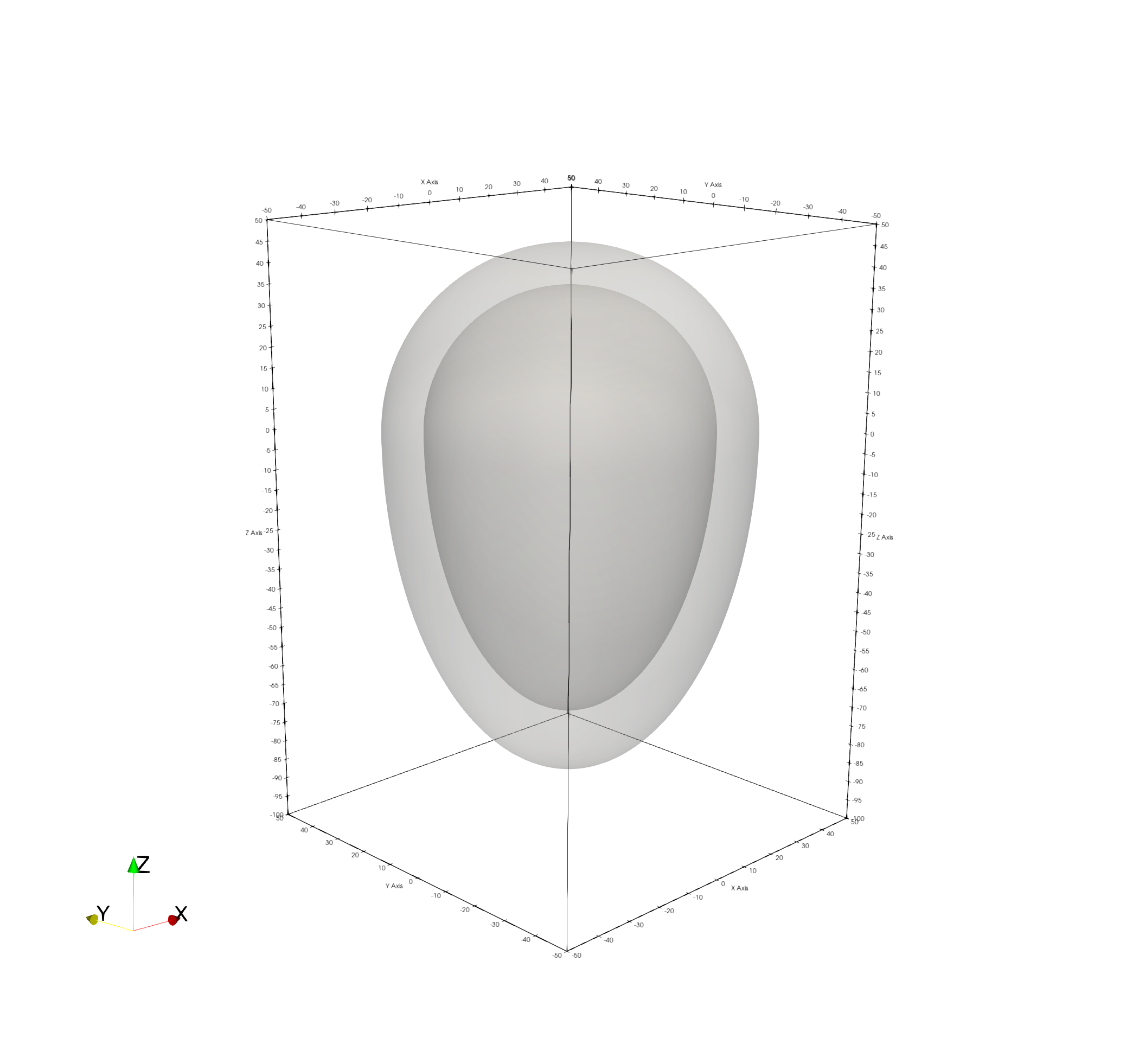}
         \caption{Myocardium}
         \label{fig:hgeom1}
     \end{subfigure}
     \begin{subfigure}{0.45\textwidth}
		\centering
		\includegraphics[width=\textwidth,trim={15cm 5cm 15cm 5cm},clip]{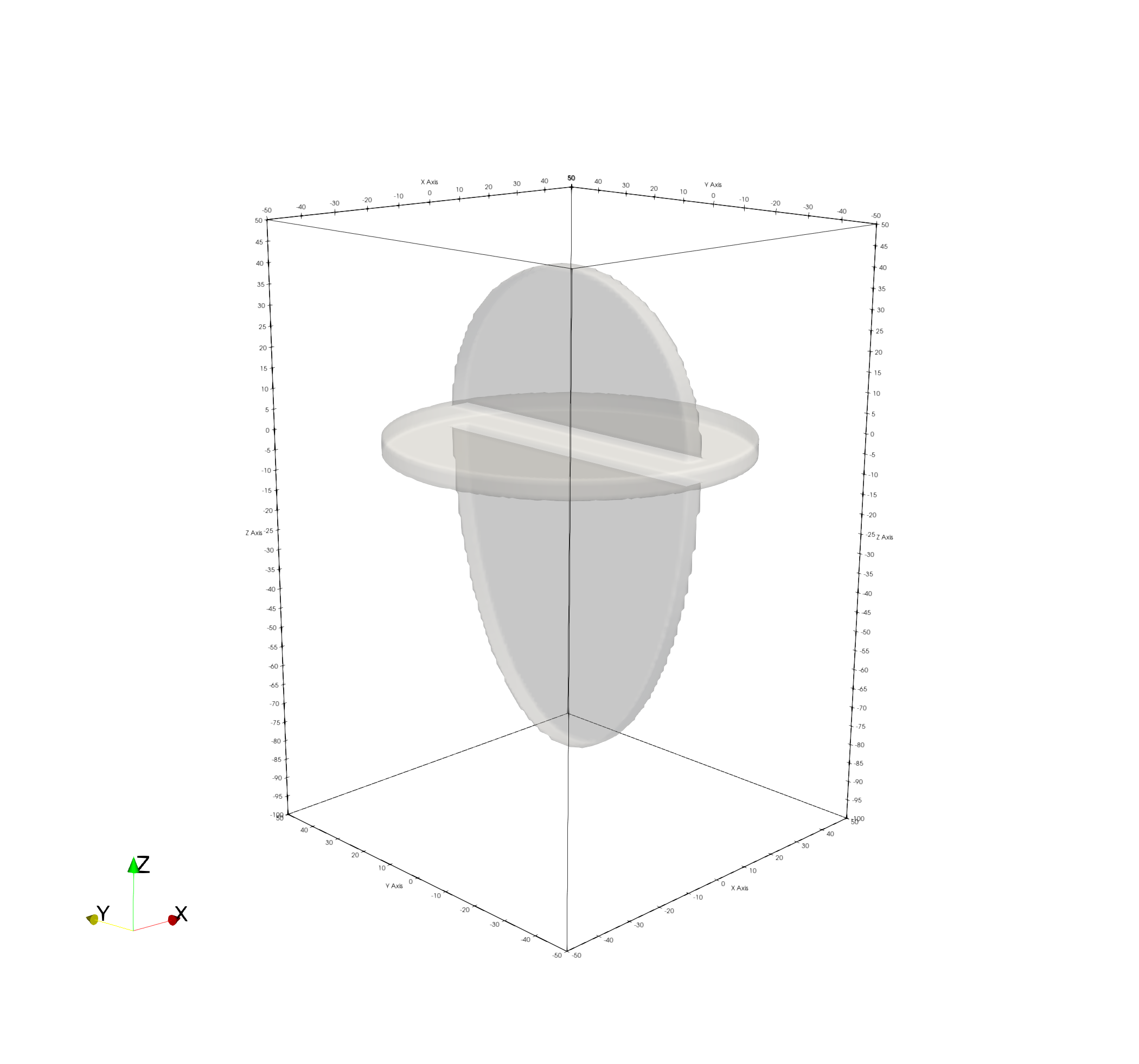}
         \caption{Septum}
         \label{fig:hgeom2}
     \end{subfigure}
     \begin{subfigure}{0.45\textwidth}
		\centering
		\includegraphics[width=\textwidth,trim={15cm 5cm 15cm 5cm},clip]{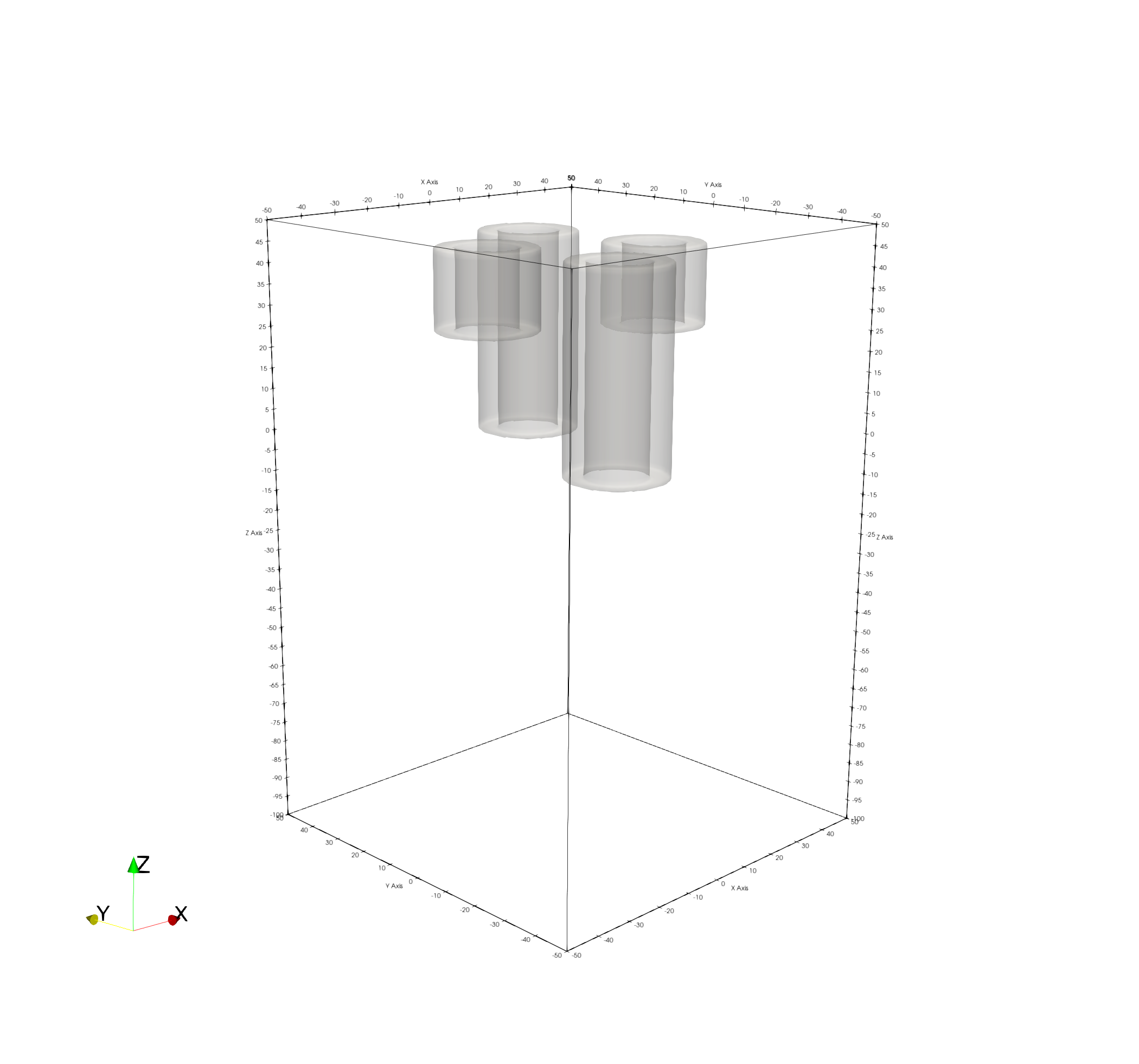}
         \caption{Blood vessels}
         \label{fig:hgeom3}
     \end{subfigure}
     \begin{subfigure}{0.45\textwidth}
		\centering
		\includegraphics[width=\textwidth,trim={15cm 5cm 15cm 5cm},clip]{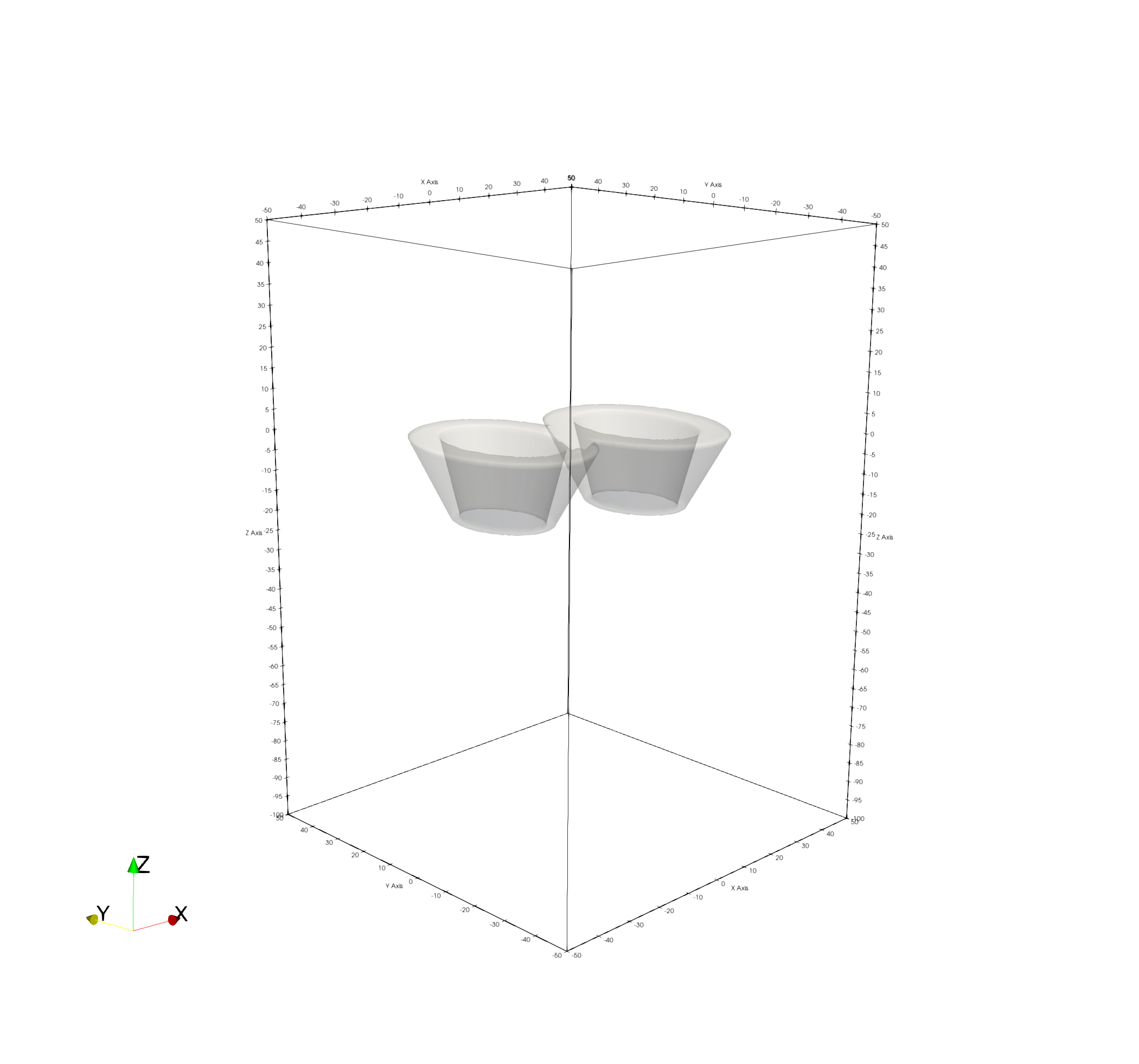}
         \caption{Valves}
         \label{fig:hgeom4}
     \end{subfigure}
	\caption{Geometric elements comprising the whole heart geometry}
\end{figure}

The geometry of a schematic heart is shown in Figure \ref{fig:hgeom0}. It is constructed around a central origin and is comprised of several elements defined by their own SDFs, which are combined according to the set algebra of Section \ref{sec:geom}. The myocardium is a combination of four SDFs, two half-spheres above the origin and two ellipsoids below it, as shown in Figure \ref{fig:hgeom1}. The intra-ventricular septum is and a horizontal septum are defined as elliptic discs as in Figure \ref{fig:hgeom2}.  Hollow vertical cylinders representing the veins and aorta entering the atria and leaving the ventricles respectively are shown in Figure \ref{fig:hgeom3}. Finally the mitral and tricuspid valves are represented as hollow elliptic cones which penetrate the horizontal septum as shown in Figure \ref{fig:hgeom4}.  The geometry does not include any other blood vessels or valves at the entrance of the atria or ventricles, and as such their action is significantly absent from the results of the simulation.

\subsection{Fibre Directions}

\begin{figure}[h!]
\centering
	\includegraphics[width=\textwidth,trim={5cm 5cm 5cm 5cm},clip]{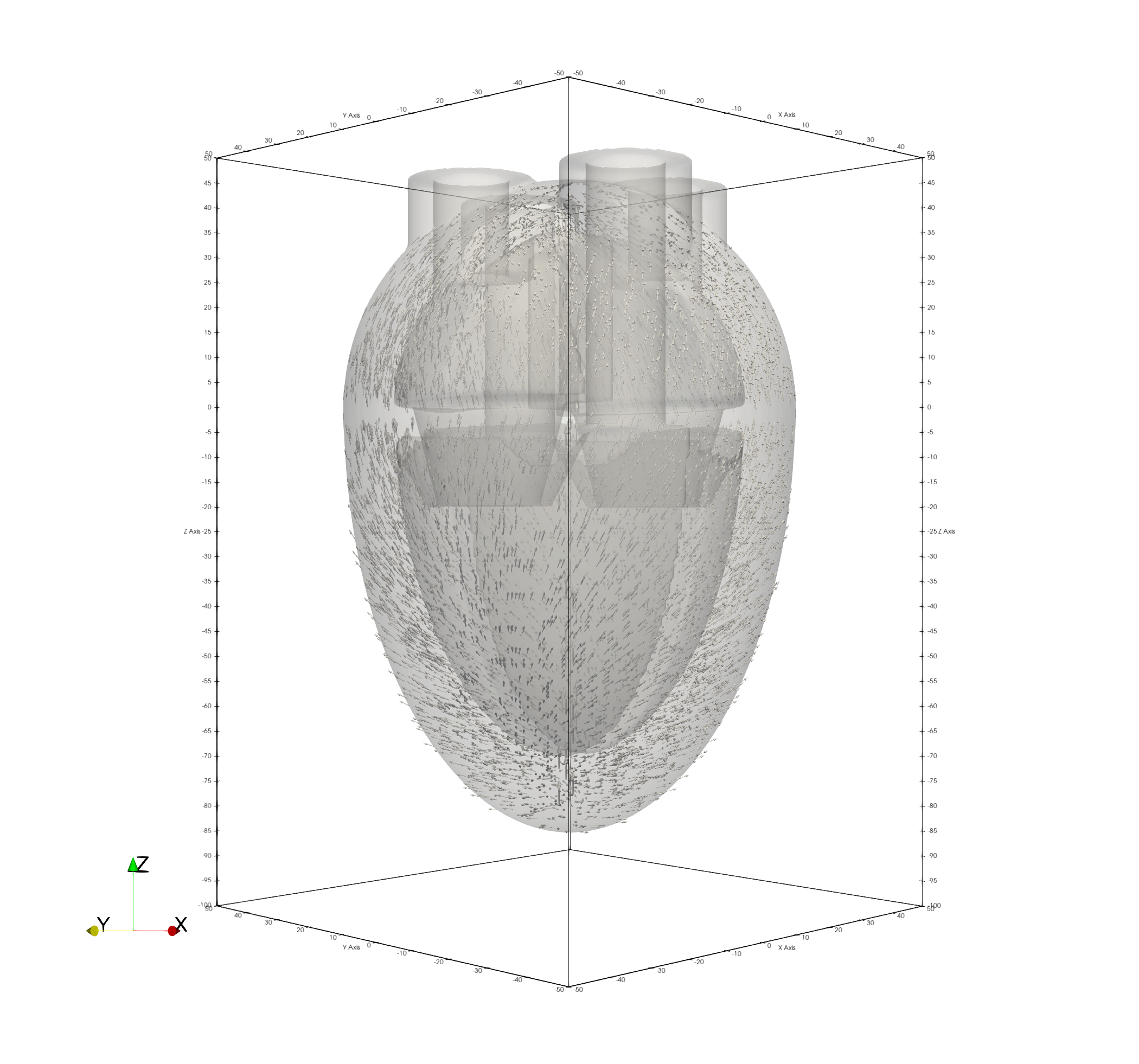}
\caption{Cardiac geometry showing the vector field of fibre directions.}
\label{fig:fibre01}
\end{figure}

Fibres are present in the the tissue of the myocardium (Figure \ref{fig:hgeom1}) which determine both the conductivity of the myocardial tissue as well as the axis along which it contracts.  The fibre field is stored a set of vectors, shown in Figure \ref{fig:fibre01}.  The conductivity is determined via (\ref{eqn:sigma1}), where $\f \in \R^3$ is the stored fibre direction. The values of londitudinal $\sigma_L$ and transverse $\sigma_T$ conductivity are given as follows:
\begin{eqnarray*}
\sigma_L &=& 0.20 \text{ mS mm}^{-1}\\
\sigma_T &=& 0.05 \text{ mS mm}^{-1}
\end{eqnarray*}
Where S indicates Siemens, the SI unit of conductivity, given as 1S $= \mu$A mV$^{-1}$ mm$^{-1}$.  Where no fibres are present the conductivity in all directions is set equal to $\sigma_T$. The form of the fibre direction is constructed as a unit vector lying initially in the horizontal (xy) plane, initially orthogonal to the position vector of the point in the myocardium at which the fibre is set.  The vector is then rotated by 0.4$\pi$ in the axis of its position vector to give the generally accepted helical form for the fibres \cite{Land2015} \cite{Potse2006} \cite{Tang2007}, among others.  It is possible to set the angle of fibre rotation with a transmural parameter, but this did not affect the results for either electrophysiology or mechanics and it is not used at this stage.

In addition to the fibres of the myocardium there is a fibre running vertically through the centre of the heart which represents the higher conductivity of the Purkinje network. The fibre has twice the longditudinal conductivity of the regular fibres.  It works in cunjuction with the horizontal septum (Figure \ref{fig:hgeom2}) which has its conductivity set to zero.  Since the fast fibre is the only point at which electrical stimulation can pass from the atria to ventricle the movement of depolarisation through the centre of the heat reproduces the action of the atrio-ventricular node, separating the contraction of the upper and lower sections of the heart into two distinct events. 

\section{Algorithm}

\subsection{Electrophysiology}

The heart receives a stimulus of 20mV at a position near its apex located in the left atrium, equivalent to the SA node.  The Mitchell Schaeffer and Monodomain models are used to model calculate the electrophysiology, as defined earlier in this work and implemented in the reproduction of the benchmark paper of \cite{Niederer2011} in Chapter \ref{chp:exp}. The diffusive behaviour of the Monodomain model includes a Laplacian operator which can be iterated in time either with an Implicit Euler integration, using a Jacobi or Richardson iteration.  The time step $\delta t = 0.001$ms is sufficiently short to allow for an Explicit Euler integration  at a spatial resolution $\delta x = 1.0$mm.  In this case the implicit and explicit results are equivalent. If an ECG is required the elliptic problem on the second line of (\ref{eqn:bidomain1}) is solved via a Jacobi or Richardson iteration.

\subsection{Solid Mechanics}

For the solid mechanics the incompressible Mooney Rivlin model is used with the same parameters as in the reproduction of the benchmark study of  \cite{Land2015} in Chapter \ref{chp:exp}.  As described in Section \ref{sec:moooney}, a synthetic extension of the tissue is coupled to the transmembrane voltage.  This induces a contractile stress in the tissue which does not interfere with the incompressible dynamics of the material model.  As a result the contraction is stable under explicit iteration.

\subsection{Fluid Dynamics}

\begin{figure}[h!]
\centering
\begin{tikzpicture}[scale=1.25]
\fill[pattern=north east lines, pattern color=lightgray] (-4,-2) -- (+2,-2) -- (+2,-1) -- (-0,-1)  -- (-0,+0)-- (-1,+0) -- (-1,+2) -- (-4,+2);  
\draw[step=1,help lines] (-4,-2) grid (+4,+2);				
\draw[draw=black,line width=0.2mm] (-2.5,-0.5) rectangle (-1.5,+0.5);
\draw[draw=black,line width=0.2mm] (-0.5,-0.5) rectangle (+0.5,+0.5);
\draw[draw=black,line width=0.2mm] (+1.5,-0.5) rectangle (+2.5,+0.5);
\draw[blue,thick] (-1,2) to (-1,0) to (0,0) to (0,-1) to (2,-1) to (2,-2);	
\foreach \x in {-4,...,+4} 
    \foreach \y in {-2,...,+2} 
      { 
        \draw  [fill=white]  (\x,\y) circle (0.05cm); 
      } 
\node at (-3.5,-1.5) {$\Omega_s$};
\node at (+3.5,+1.5) {$\Omega_f$};
\node[anchor=center] at (-2.25,-0.25) {$\sigma_s$};
\node[anchor=center] at (-2.25,+0.25) {$\sigma_s$};
\node[anchor=center] at (-1.75,-0.25) {$\sigma_s$};
\node[anchor=center] at (-1.75,+0.25) {$\sigma_s$};
\node[anchor=center] at (+2.25,-0.25) {$\sigma_f$};
\node[anchor=center] at (+2.25,+0.25) {$\sigma_f$};
\node[anchor=center] at (+1.75,-0.25) {$\sigma_f$};
\node[anchor=center] at (+1.75,+0.25) {$\sigma_f$};
\node[anchor=center] at (+0.25,-0.25) {$\sigma_f$};
\node[anchor=center] at (+0.25,+0.25) {$\sigma_f$};
\node[anchor=center] at (-0.25,-0.25) {$\sigma_s$};
\node[anchor=center] at (-0.25,+0.25) {$\sigma_f$};
\end{tikzpicture}
\caption{The contribution of solid and fluid Cauchy stress tensors to the velocity of solid, fluid and surface vertices.}
\label{fig:fsi1}
\end{figure}
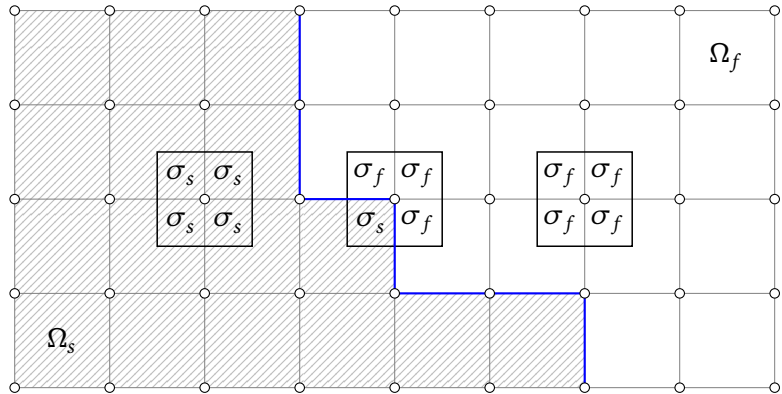

The behaviour of the fluid is as described in \ref{chp:alg}.  For the solution of the Poisson pressure equation resulting from the Helmholz projection of Section \ref{sec:helm1}, either a Jacobi or Richardson iteration is used. The current pressure field is used as the initial guess for the iterative solve and a small number of iterations are performed such that the solution is incomplete.  This is equivalent to the semi-implicit solution given in \cite{Formaggia2010} but takes advantage of the Finite Volume discretisation.  Since the solution is incomplete, the corrected fluid velocity is not entirely divergence-free and the gradient of the pressure field is under-estimated. This in turn enters the fluid structure interaction as Cauchy stress as shown in Figure \ref{fig:fsi1}.

Due to the conservative nature of FVM this error is retained in the solution and remains at the next solution of the Poisson equation.  As such the pressure correction is applied over more than one time step. The resulting error in the solution is equivalent to a fluid that is both compressible and damped.  This is not entirely physically incorrect and contributes to the stability of the fluid-structure interaction, which otherwise is the main source of error in the model.

\section{Results}

\subsection{Electrophysiology}

\begin{figure}[h!]
\centering
\includegraphics[width=0.24\textwidth,trim={15cm 4cm 15cm 4cm},clip]{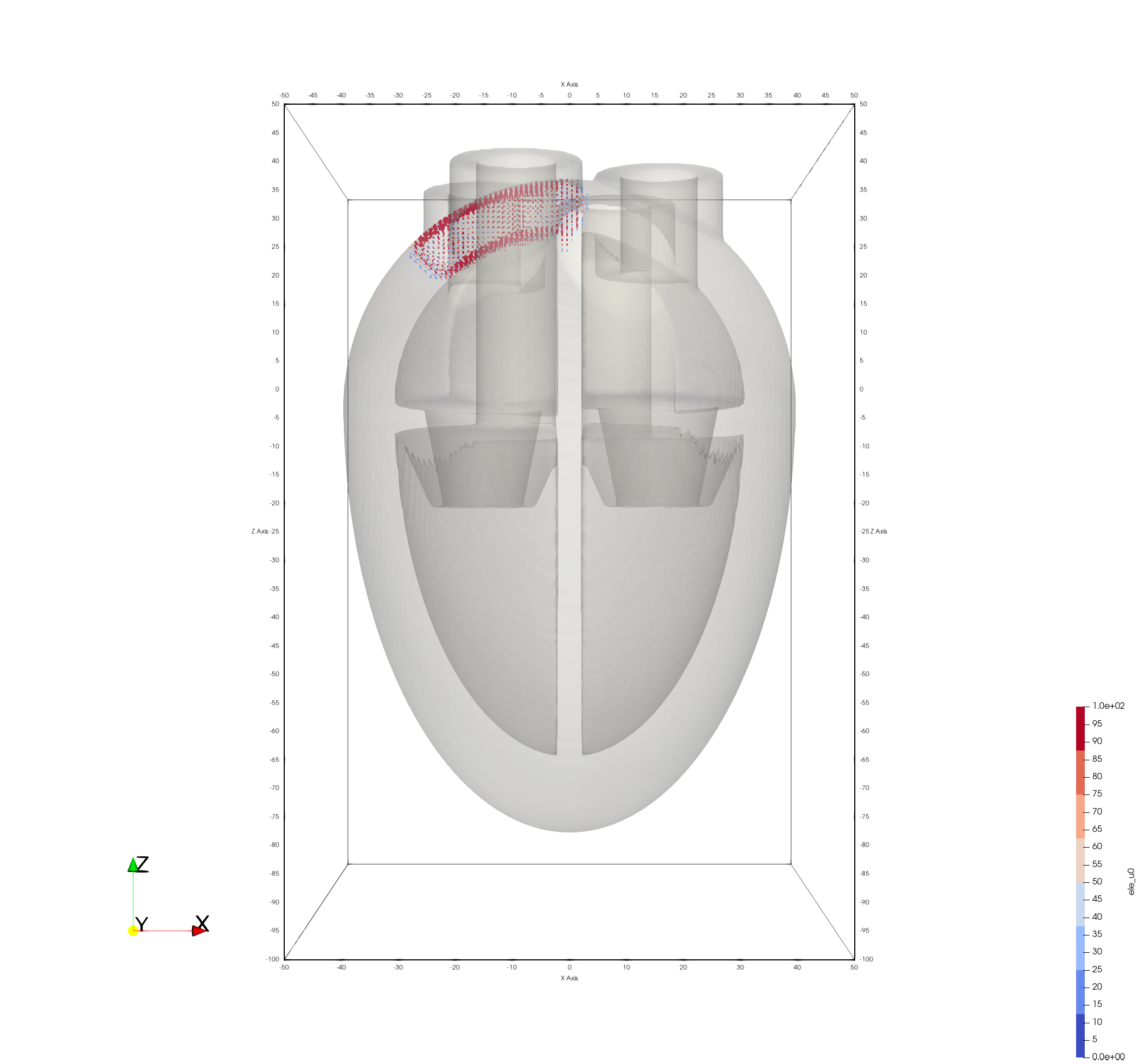}
\includegraphics[width=0.24\textwidth,trim={15cm 4cm 15cm 4cm},clip]{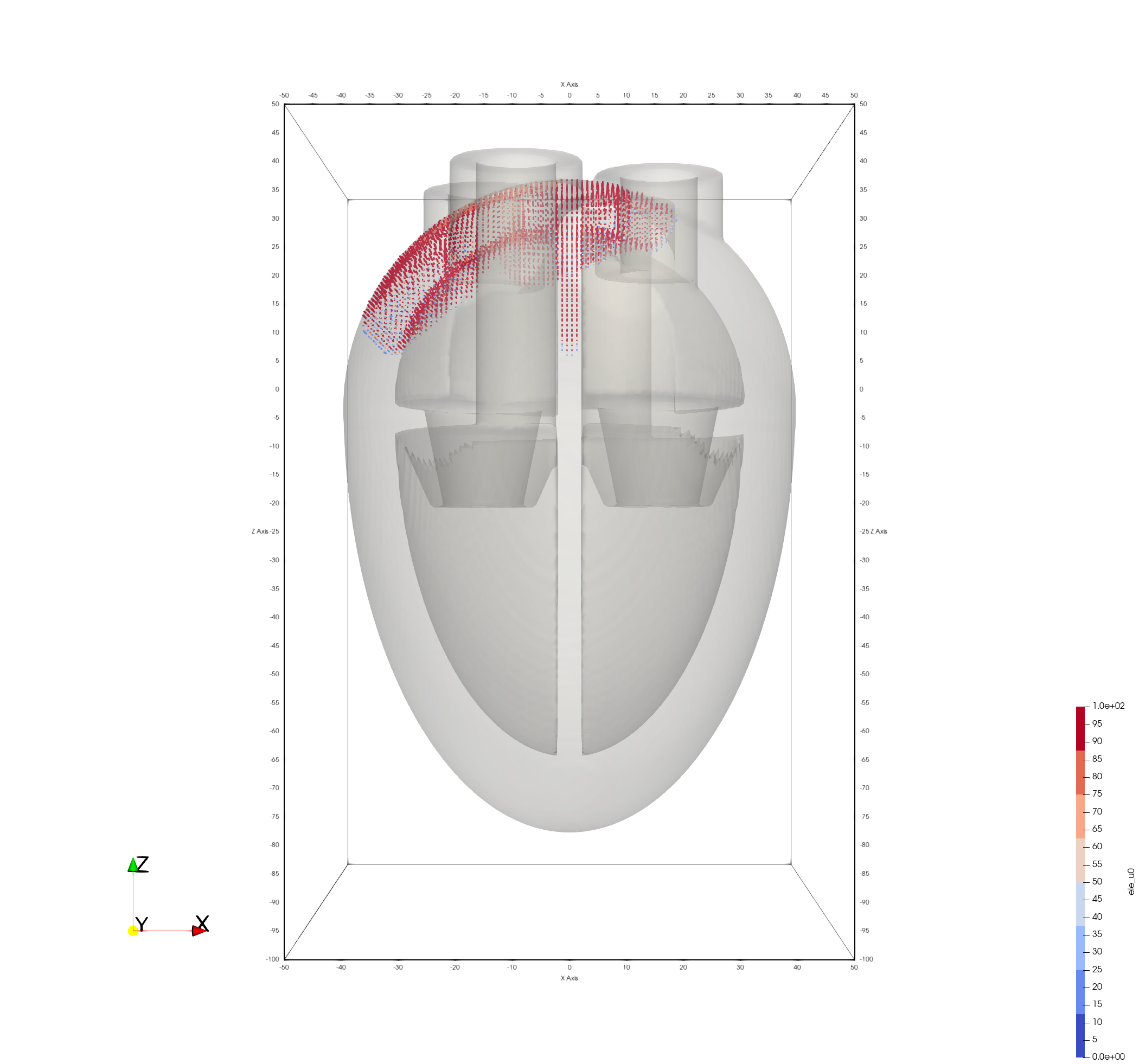}
\includegraphics[width=0.24\textwidth,trim={15cm 4cm 15cm 4cm},clip]{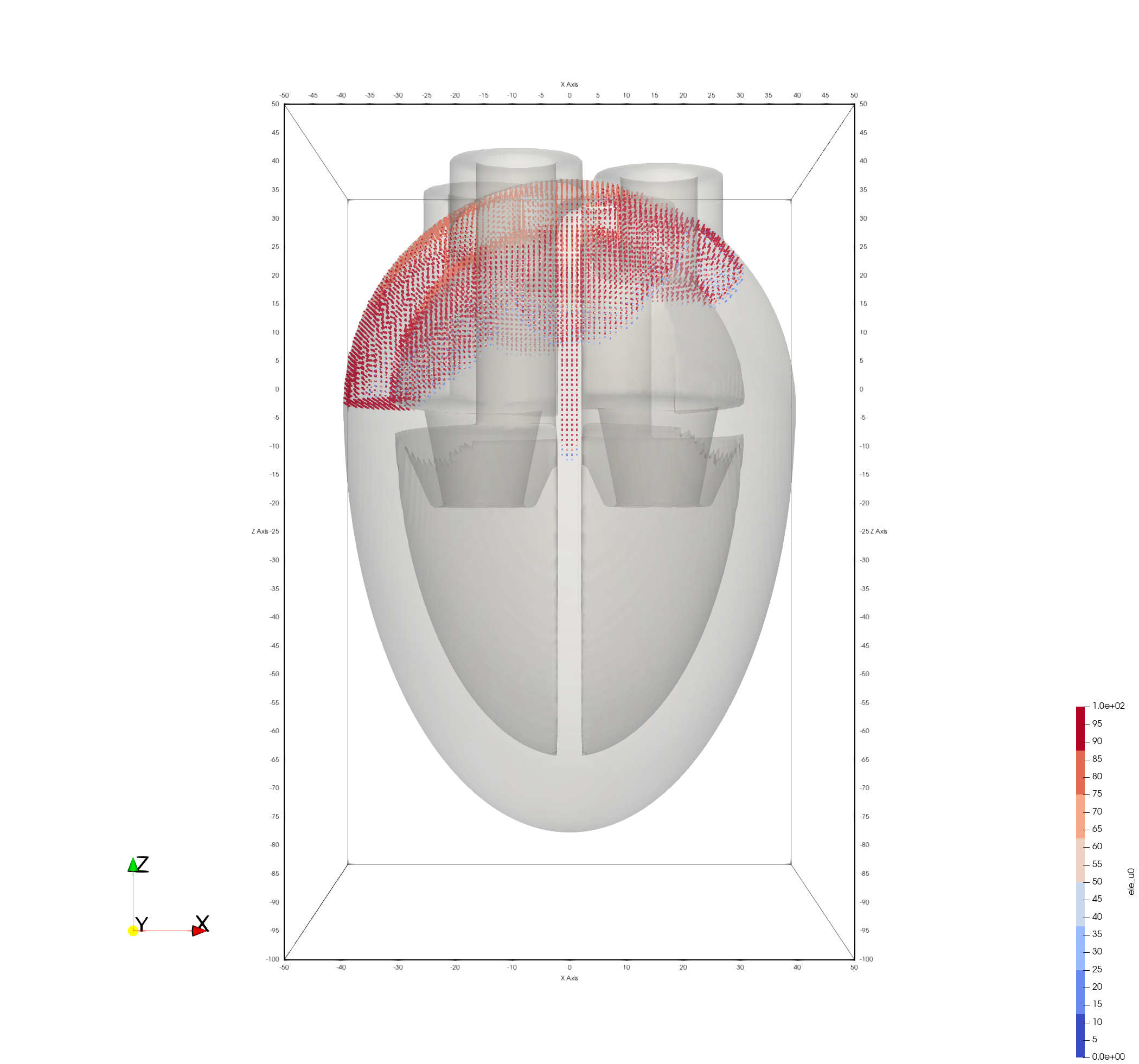}
\includegraphics[width=0.24\textwidth,trim={15cm 4cm 15cm 4cm},clip]{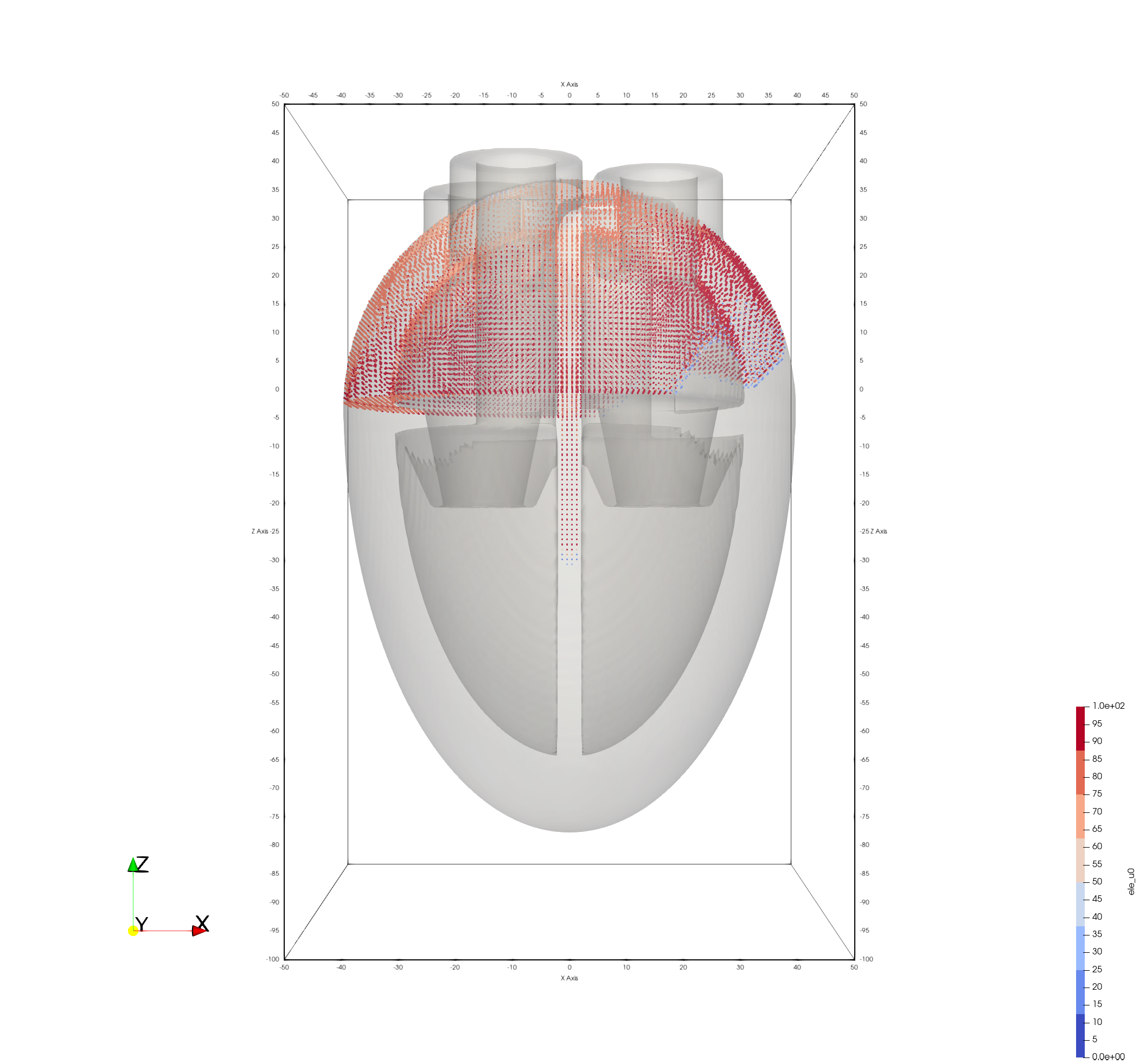}
\includegraphics[width=0.24\textwidth,trim={15cm 4cm 15cm 4cm},clip]{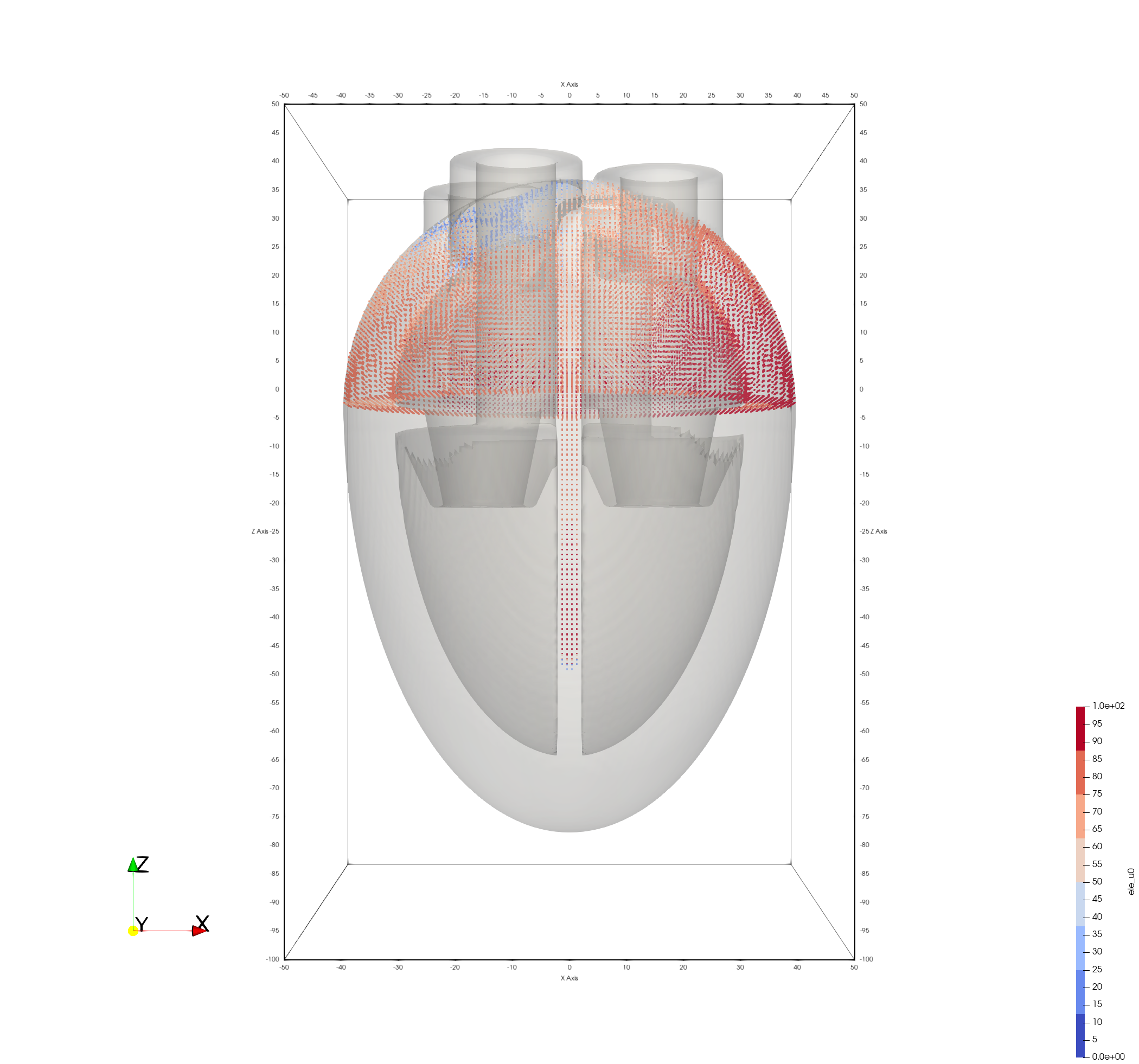}
\includegraphics[width=0.24\textwidth,trim={15cm 4cm 15cm 4cm},clip]{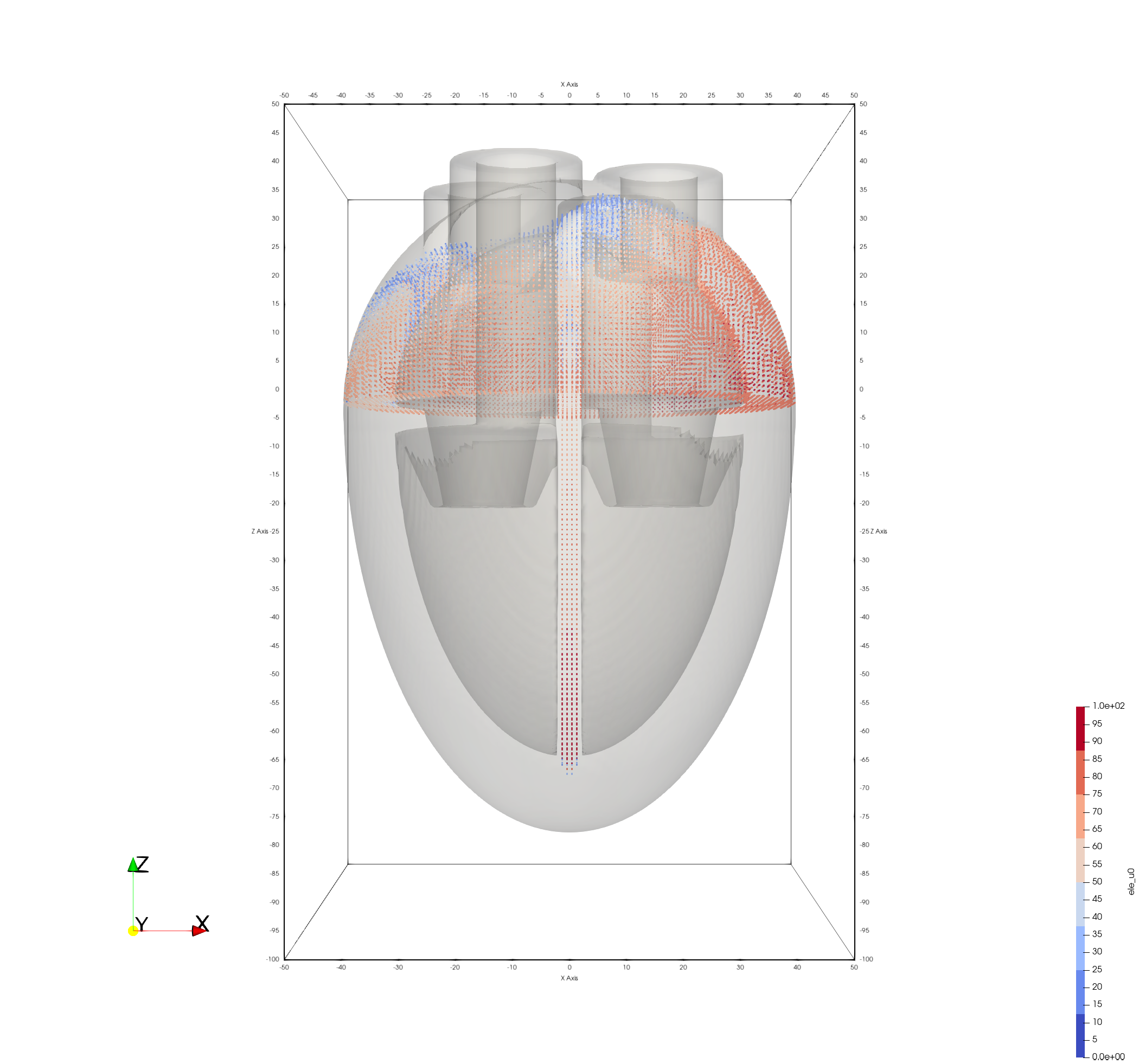}
\includegraphics[width=0.24\textwidth,trim={15cm 4cm 15cm 4cm},clip]{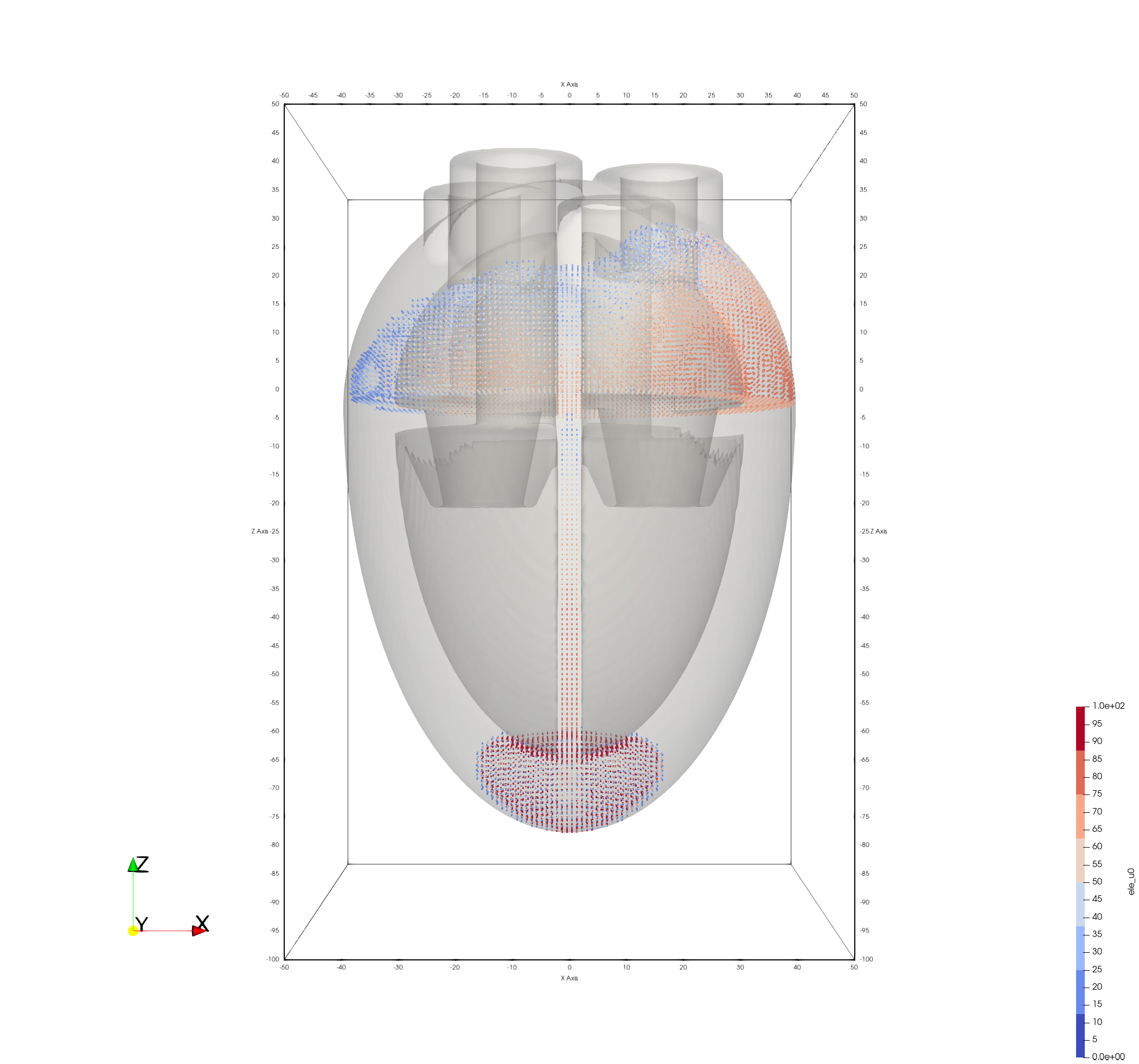}
\includegraphics[width=0.24\textwidth,trim={15cm 4cm 15cm 4cm},clip]{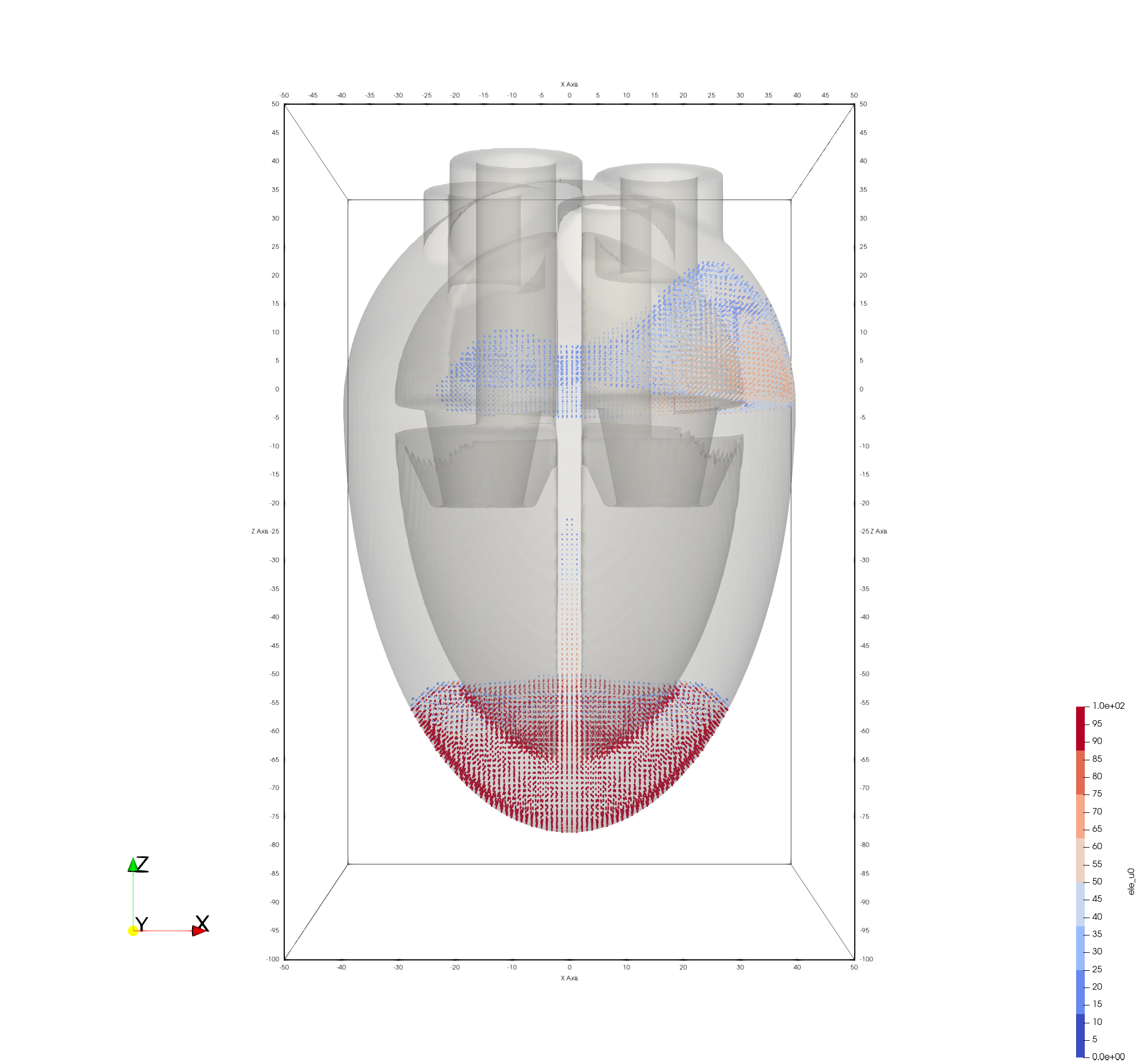}
\includegraphics[width=0.24\textwidth,trim={15cm 4cm 15cm 4cm},clip]{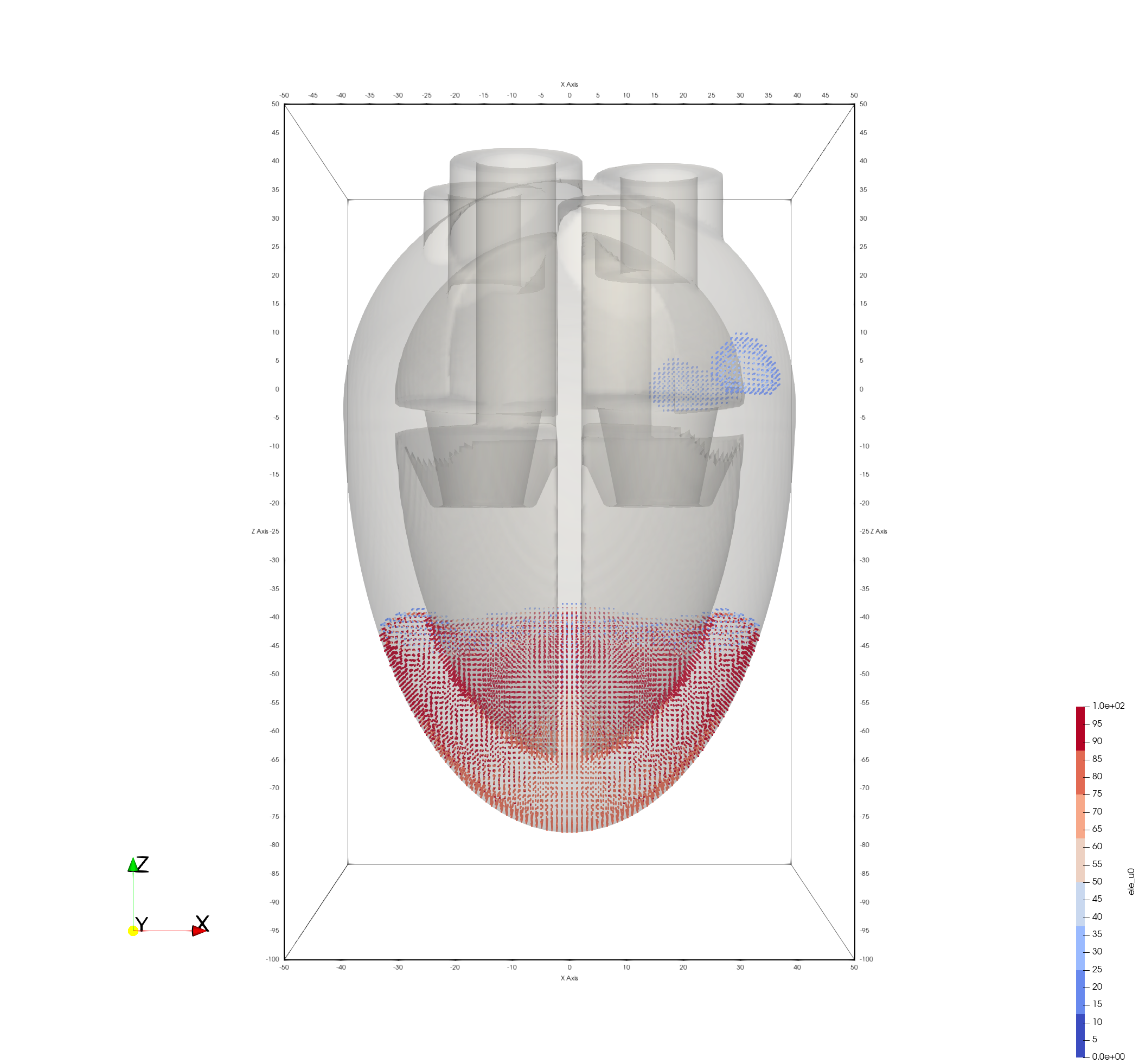}
\includegraphics[width=0.24\textwidth,trim={15cm 4cm 15cm 4cm},clip]{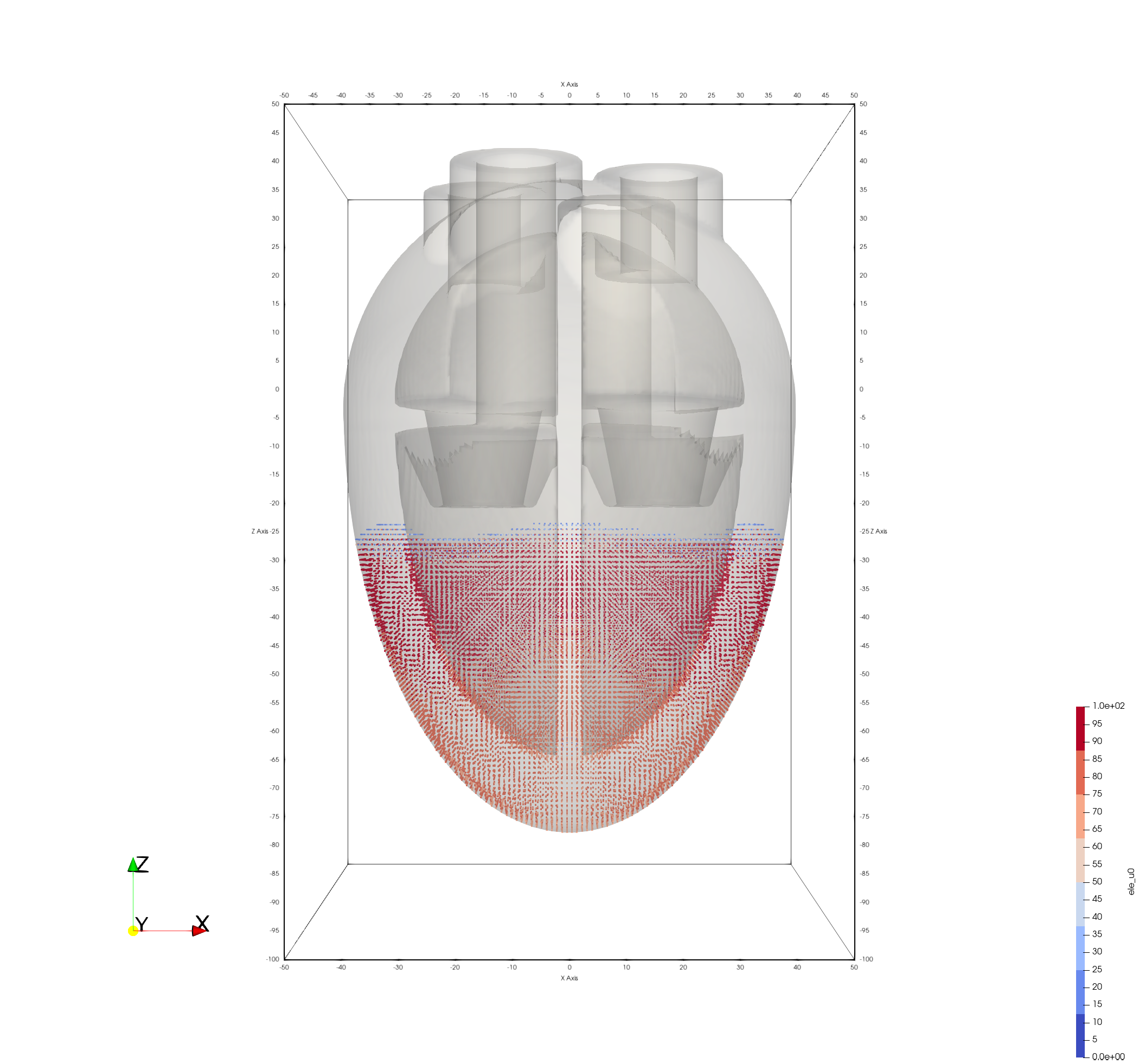}
\includegraphics[width=0.24\textwidth,trim={15cm 4cm 15cm 4cm},clip]{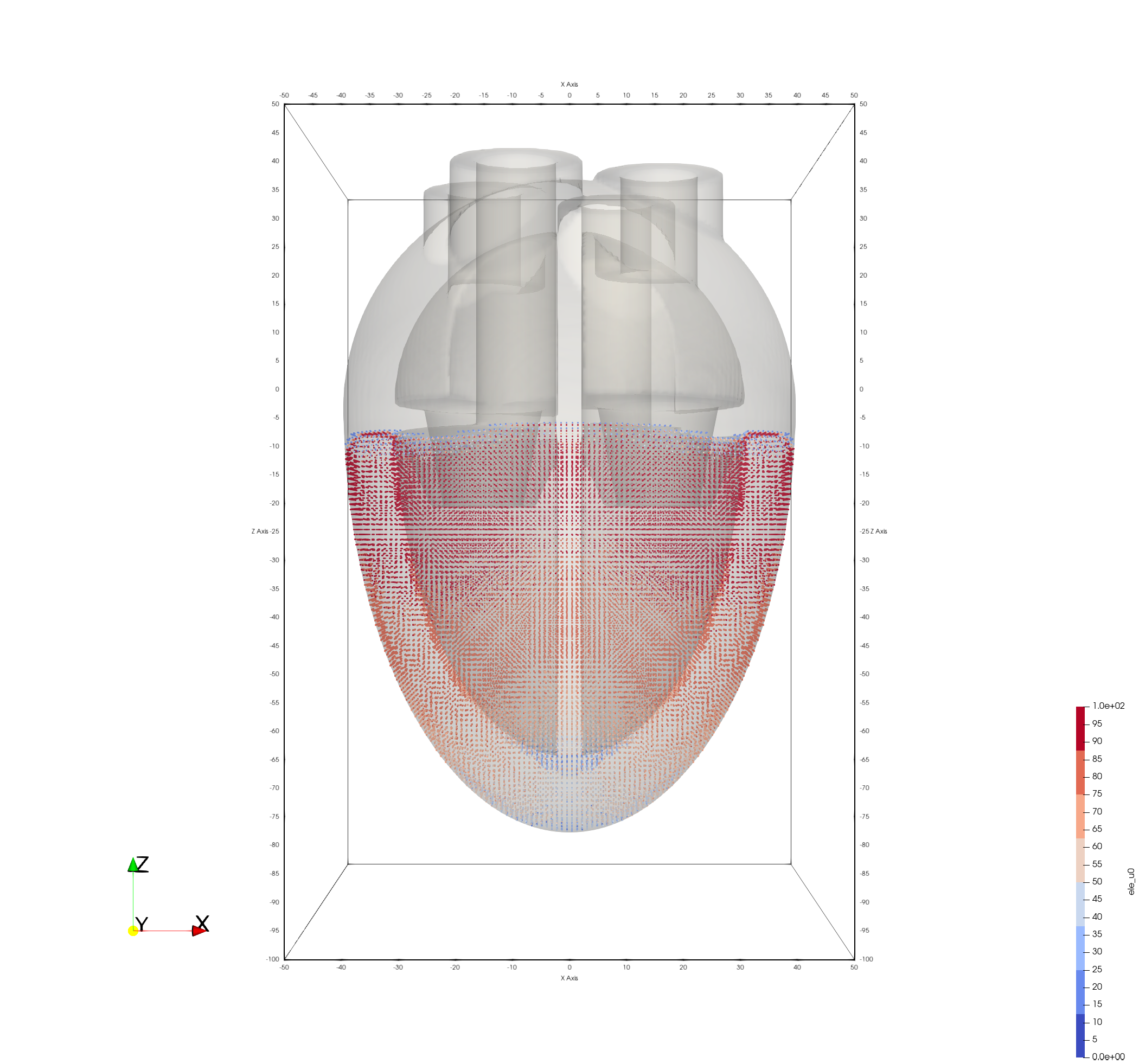}
\includegraphics[width=0.24\textwidth,trim={15cm 4cm 15cm 4cm},clip]{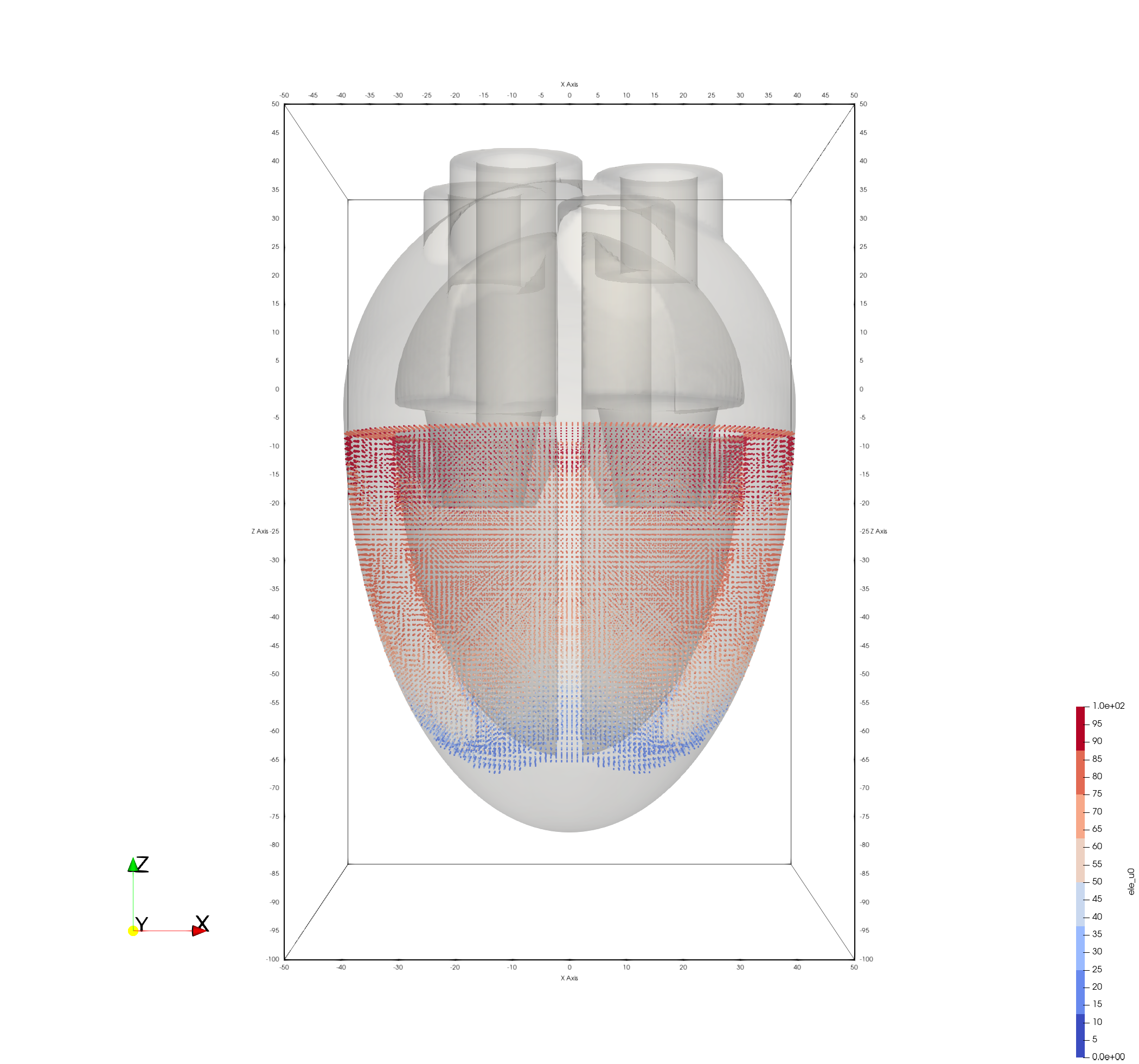}
\includegraphics[width=0.24\textwidth,trim={15cm 4cm 15cm 4cm},clip]{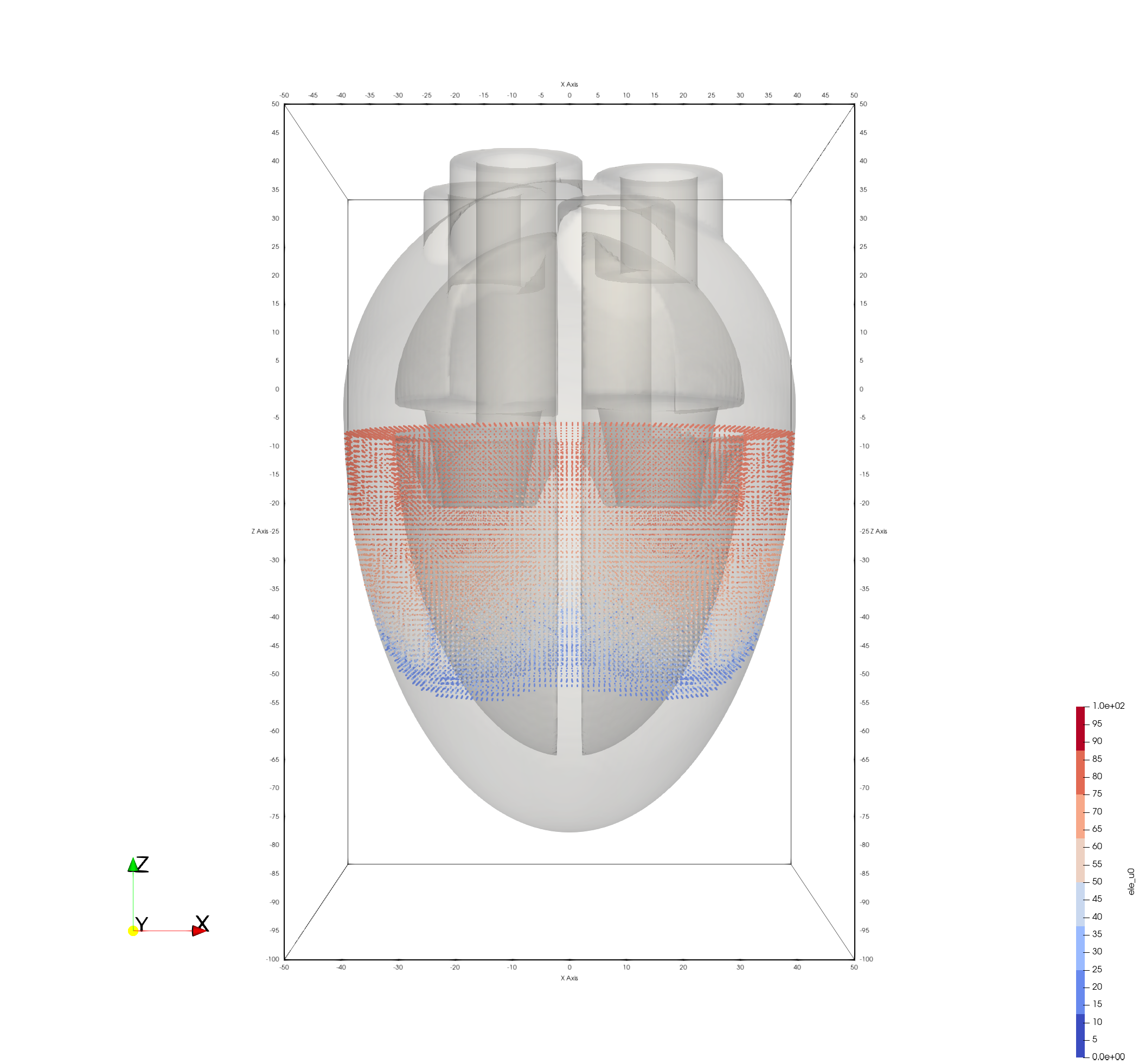}
\includegraphics[width=0.24\textwidth,trim={15cm 4cm 15cm 4cm},clip]{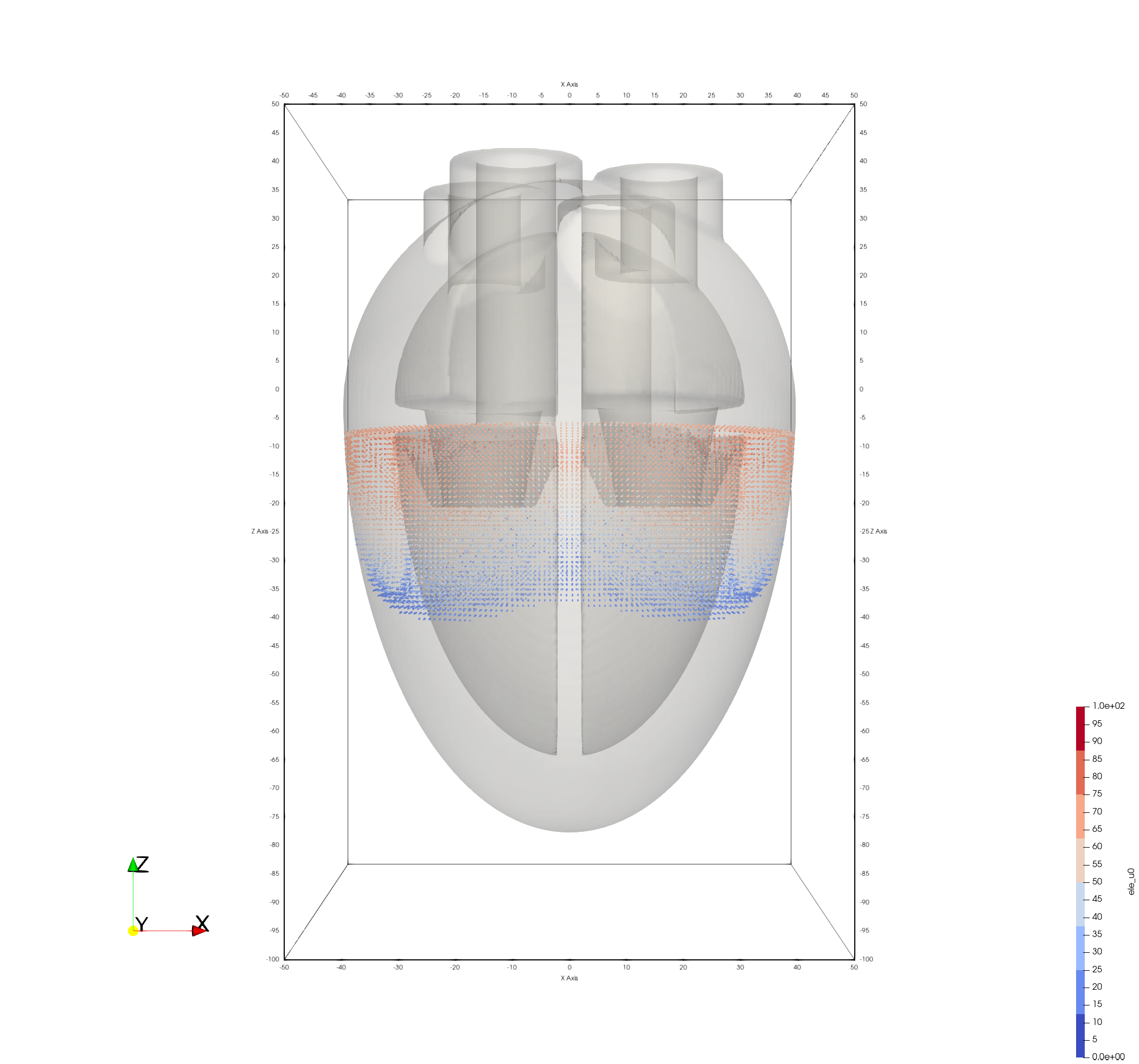}
\includegraphics[width=0.24\textwidth,trim={15cm 4cm 15cm 4cm},clip]{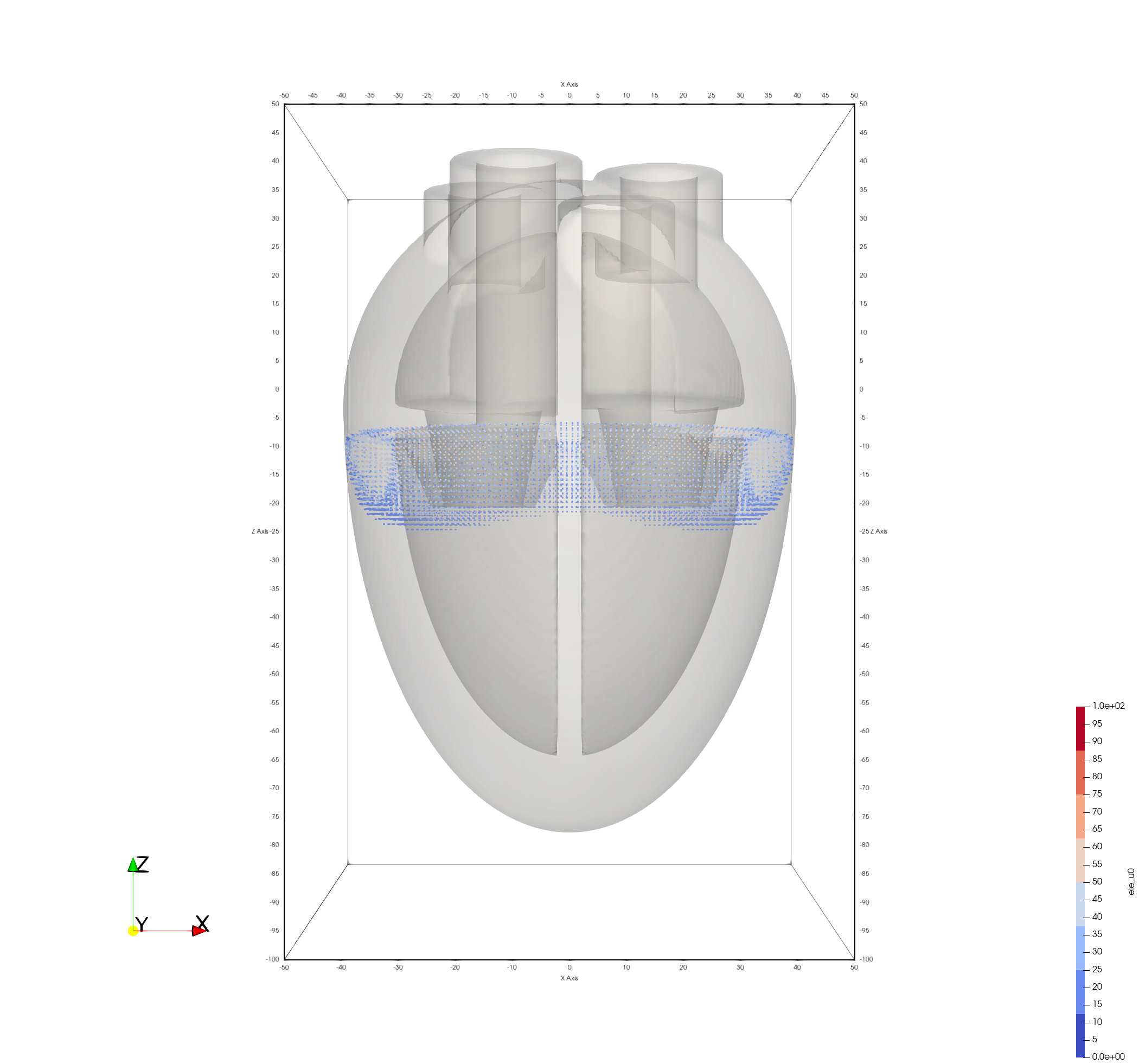}
\includegraphics[width=0.24\textwidth,trim={15cm 4cm 15cm 4cm},clip]{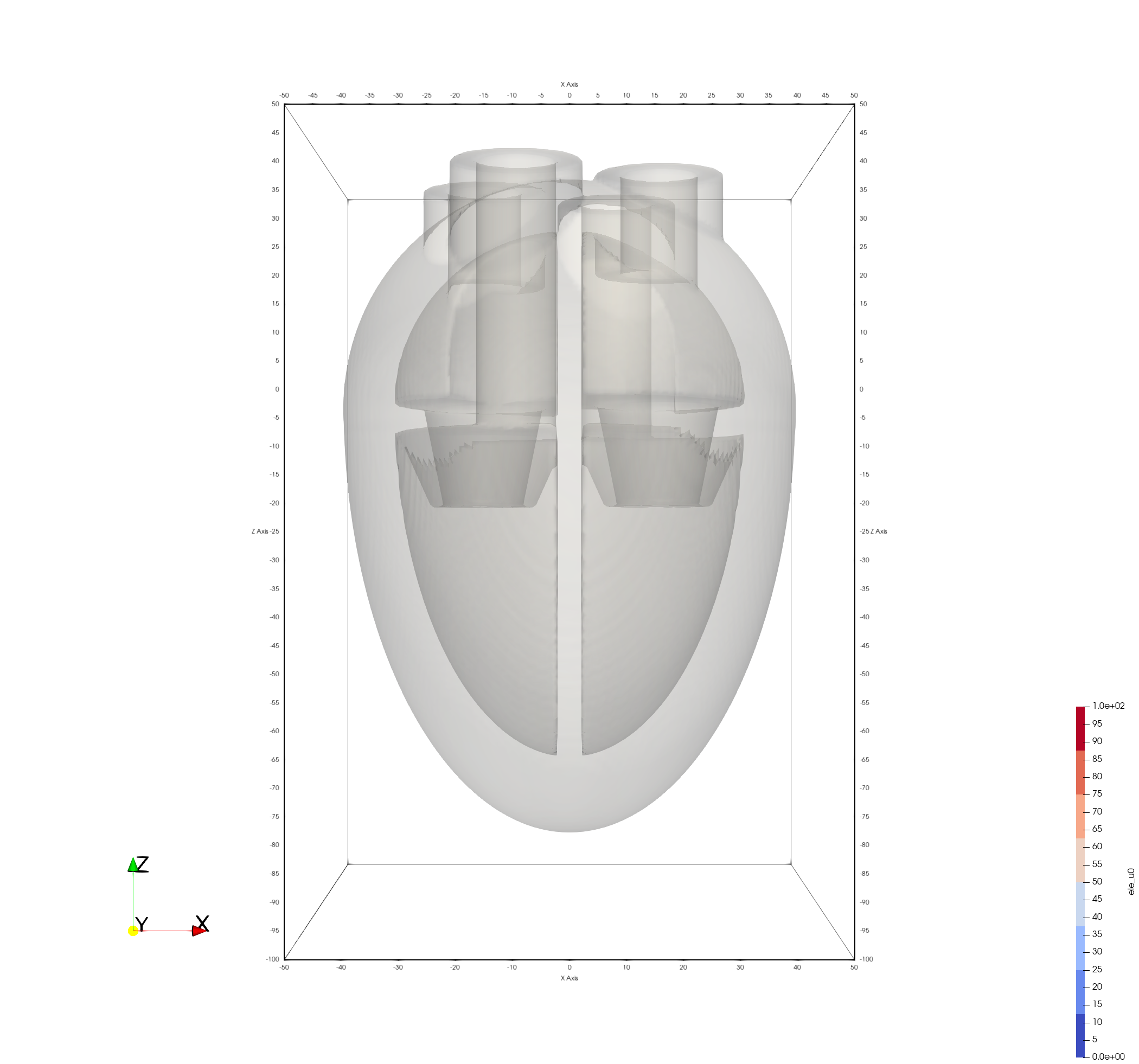}
\caption{The evolution of the cardiac action potential at intervals of 50ms ($\delta x = 1.0$mm).}
\label{fig:ap1}
\end{figure}

The simulation of the electrophysiology alone can be performed with a slightly longer time step of $\delta t = 0.05$ms. The results are shown in Figure \ref{fig:ap1} and reproduce an approximation on the cardiac action potential.  The transport of the depolarisation through the fast fibre on the central axis of the geometry can be seen.  The computational cost is quite low, and the computation can be performed quickly on a desktop computer.  In this case we used an iMac with its native Radeon Pro 580 8GB GPU and the simulation with 1 million degrees of freedom took 66 seconds. Many repeated simulations are therefore possible, and as such the model can be considered as the objective function of an optimisation problem. This allows for richer iterative experiments such as parameter estimation, neural network training and the testing of surgical or drug therapies. A more complex ionic model can also be substituted into the calculations.

\subsubsection{ECG}

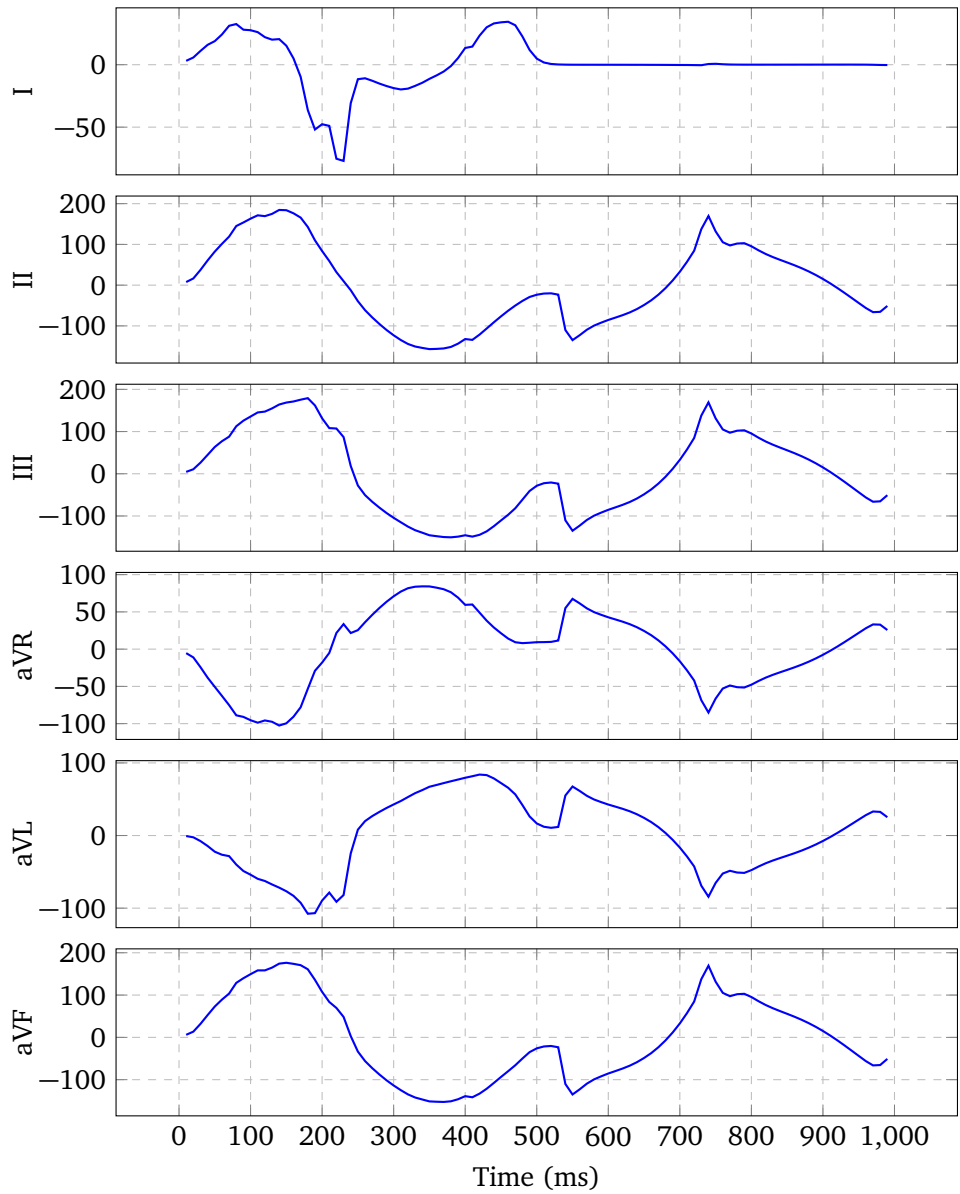
\begin{figure}[h!]
\centering
\begin{tikzpicture}
  \begin{groupplot}[
      group style={
        group name=my plots,
        group size=1 by 6,
        vertical sep=8pt
      },
	width=\textwidth,
	height=0.3\textwidth,
    xmajorgrids=true,
    ymajorgrids=true,
    grid style=dashed
    ]
    \nextgroupplot[ylabel={I},xticklabels=\empty]
    \addplot [color=blue  ,thick] table[x index=0, y index=4, col sep=comma] {ecg/ecg1.txt};
    \nextgroupplot[ylabel={II},xticklabels=\empty]
    \addplot [color=blue  ,thick] table[x index=0, y index=5, col sep=comma] {ecg/ecg1.txt};
    \nextgroupplot[ylabel={III},xticklabels=\empty]
    \addplot [color=blue  ,thick] table[x index=0, y index=6, col sep=comma] {ecg/ecg1.txt};
    \nextgroupplot[ylabel={aVR},xticklabels=\empty]
    \addplot [color=blue  ,thick] table[x index=0, y index=7, col sep=comma] {ecg/ecg1.txt};
    \nextgroupplot[ylabel={aVL},xticklabels=\empty]
    \addplot [color=blue  ,thick] table[x index=0, y index=8, col sep=comma] {ecg/ecg1.txt};
    \nextgroupplot[ylabel={aVF},xlabel={Time (ms)}]
    \addplot [color=blue  ,thick] table[x expr={\thisrowno{0}*10}, y index=9, col sep=comma] {ecg/ecg1.txt};
  \end{groupplot}
\end{tikzpicture}
\caption{The six standard outputs of a 3-lead Electrocardiogram.}
\label{fig:ecg1}
\end{figure}

In order to reconstruct the output of a synthetic three-lead Electrocardiogram (ECG), the transmembrane voltage was allowed to diffuse out of the heart and into the surrounding tissue.  The conductivity of the tissue was set to a high value and thus mimicked the movement of charge through the torso. 

The values at diametrically opposite corners of the domain were extracted from the data to represent the left arm (LA), right arm (RA) and left leg (LL) electrodes. The standard calculations were applied to generate the six outputs, as per the following formulae:
\begin{eqnarray*}
I   &=& LA - RA \\
II  &=& LL - RA \\
III &=& LL - LA \\
aVR &=& RA - \frac{1}{2}(LA + LL) \\
aVL &=& LA - \frac{1}{2}(RA + LL) \\
aVF &=& LL - \frac{1}{2}(RA + LA)
\end{eqnarray*}
The results are shown in Figure \ref{fig:ecg1}. They do not appear to contain much detail and are certainly open to interpretation. They do however contain peaks corresponding to the P and T waves, and the QRS complex is present but not pronounced. The output depends upon both the choice of conductivity through the torso and the position of the recording points and is greatly affected by the small domain on which the problem is solved.  These parameters can be chosen by experimenters, and since the ECG can be generated within around 3 minutes per heartbeat, there is much room for refinement.

\subsubsection{Simulation Times}

The following tables summarise electrophysiology simulation times for various resolutions and processors.  In each case the total simulation time was 1000ms (1 second) and the software output a set of files containing a complete description of the heart at 10ms intervals (100 sets of files in total).   The simulation is stable at the longer time step of 0.05ms, but the runs with shorter time step 0.01ms are shown for comparison with the whole heart timing data. File input/output (IO) and PCI bus operations dominate the calculation time for smaller simulations. When the ECG is required the iterative solution of the elliptic problem increases the total time depending on the resolution required.  For example on the iMac the solution time for the heart at a resolution of 100 increases to around 3.5 minutes.
\begin{table}[h]
\def\arraystretch{1.2}
\begin{center}
\begin{tabular}{crccc}
\multicolumn{5}{l}{\textbf{}} \\
\textbf{Dimensions} & \textbf{Elements} & \textbf{$\delta x$(mm)} & \textbf{$\delta t$(ms)} & \textbf{Time(mm:ss)}\\
\hline
50$\times$50$\times$75 		& 187,500		& 2.0 &	0.01 & 00:19\\
100$\times$100$\times$150 	& 1,500,000		& 1.0 &	0.01 & 01:58\\
200$\times$200$\times$300 	& 12,000,000	& 0.5 &	0.01 & 21:32\\
\hline
50$\times$50$\times$75 		& 187,500		& 2.0 &	0.05 & 00:07\\
100$\times$100$\times$150 	& 1,500,000		& 1.0 &	0.05 & 01:54\\
200$\times$200$\times$300 	& 12,000,000	& 0.5 &	0.05 & 12:00
\end{tabular}
\end{center}
\caption{AMD Radeon Pro 580 (2017 iMac).}
\end{table}

\begin{table}[h]
\def\arraystretch{1.2}
\begin{center}
\begin{tabular}{crccc}
\multicolumn{5}{l}{\textbf{}} \\
\textbf{Dimensions} & \textbf{Elements} & \textbf{$\delta x$(mm)} & \textbf{$\delta t$(ms)} & \textbf{Time(mm:ss)}\\
\hline
50$\times$50$\times$75 		& 187,500		& 2.0 &	0.01 & 00:10\\
100$\times$100$\times$150 	& 1,500,000		& 1.0 &	0.01 & 00:52\\
200$\times$200$\times$300 	& 12,000,000	& 0.5 &	0.01 & 05:15\\
\hline
50$\times$50$\times$75 		& 187,500		& 2.0 &	0.05 & 00:06\\
100$\times$100$\times$150 	& 1,500,000		& 1.0 &	0.05 & 00:30\\
200$\times$200$\times$300 	& 12,000,000	& 0.5 &	0.05 & 03:39
\end{tabular}
\end{center}
\caption{NVIDIA A100 PCIE 40GB (USI ICS Cluster).}
\end{table}

\FloatBarrier
When the output is ECG only there is no need to write files, only to copy the electrophysiology buffer to the CPU for post-processing. These timings are shown below: 
\begin{table}[h]
\def\arraystretch{1.2}
\begin{center}
\begin{tabular}{crccc}
\multicolumn{5}{l}{\textbf{}} \\
\textbf{Dimensions} & \textbf{Elements} & \textbf{$\delta x$(mm)} & \textbf{$\delta t$(ms)} & \textbf{Time(mm:ss)}\\
50$\times$50$\times$75 		& 187,500		& 2.0 &	0.01 & 00:16\\
100$\times$100$\times$150 	& 1,500,000		& 1.0 &	0.01 & 01:30\\
200$\times$200$\times$300 	& 12,000,000	& 0.5 &	0.01 & 16:22\\
\hline
50$\times$50$\times$75 		& 187,500		& 2.0 &	0.05 & 00:04\\
100$\times$100$\times$150 	& 1,500,000		& 1.0 &	0.05 & 00:20\\
200$\times$200$\times$300 	& 12,000,000	& 0.5 &	0.05 & 03:19
\end{tabular}
\end{center}
\caption{ECG only: AMD Radeon Pro 580 (2017 iMac).}
\end{table}

\begin{table}[h]
\def\arraystretch{1.2}
\begin{center}
\begin{tabular}{crccc}
\multicolumn{5}{l}{\textbf{}} \\
\textbf{Dimensions} & \textbf{Elements} & \textbf{$\delta x$(mm)} & \textbf{$\delta t$(ms)} & \textbf{Time(mm:ss)}\\
\hline
50$\times$50$\times$75 		& 187,500		& 2.0 &	0.01 & 00:04\\
100$\times$100$\times$150 	& 1,500,000		& 1.0 &	0.01 & 00:12\\
200$\times$200$\times$300 	& 12,000,000	& 0.5 &	0.01 & 01:33\\
\hline
50$\times$50$\times$75 		& 187,500		& 2.0 &	0.05 & <00:01\\
100$\times$100$\times$150 	& 1,500,000		& 1.0 &	0.05 & 00:04\\
200$\times$200$\times$300 	& 12,000,000	& 0.5 &	0.05 & 00:22
\end{tabular}
\end{center}
\caption{ECG only: NVIDIA A100 PCIE 40GB (USI ICS Cluster).}
\end{table}

\FloatBarrier
\subsection{Whole Heart}

\begin{figure}[h!]
\centering
\includegraphics[width=0.24\textwidth,trim={15cm 4cm 15cm 4cm},clip]{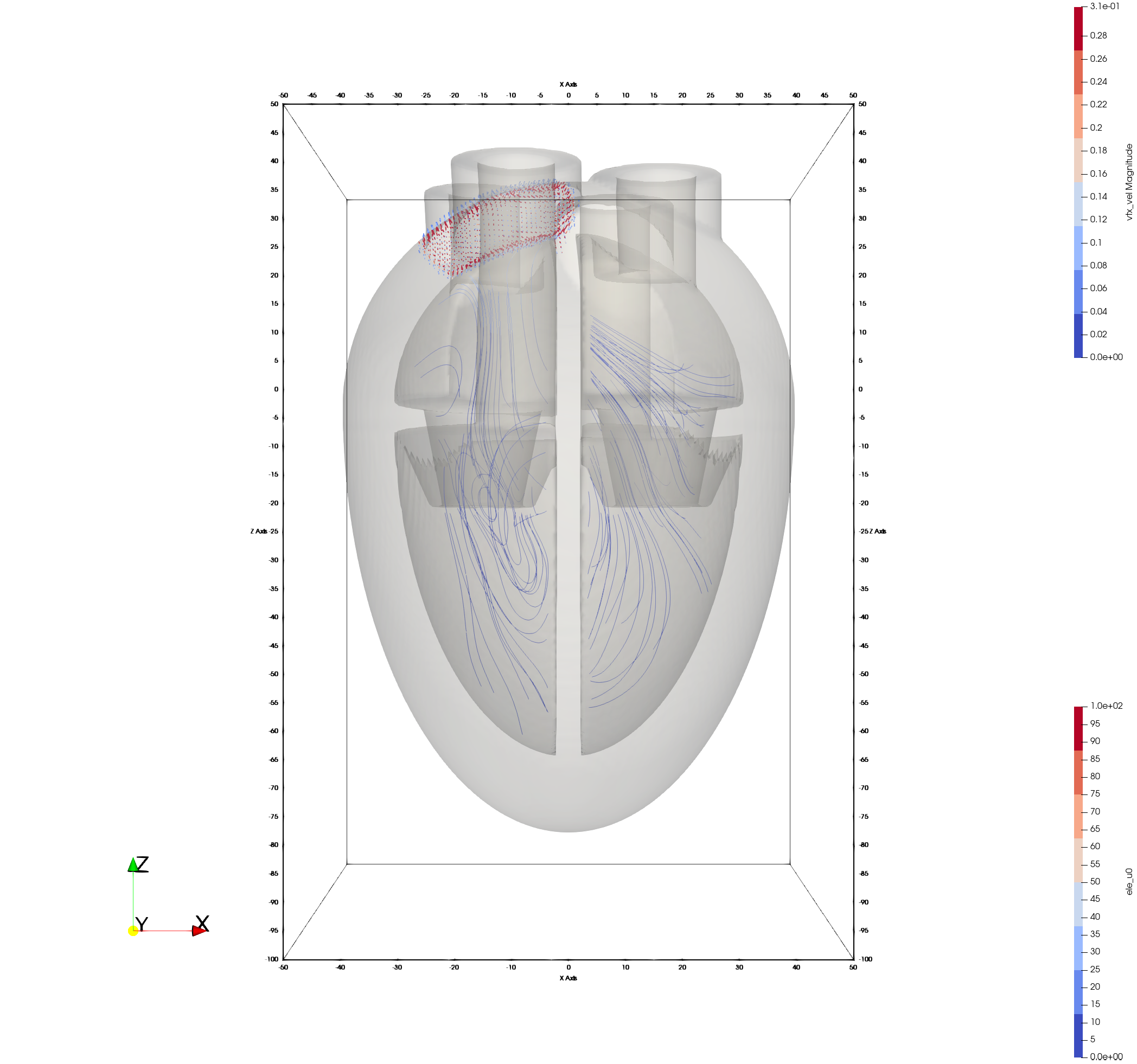}
\includegraphics[width=0.24\textwidth,trim={15cm 4cm 15cm 4cm},clip]{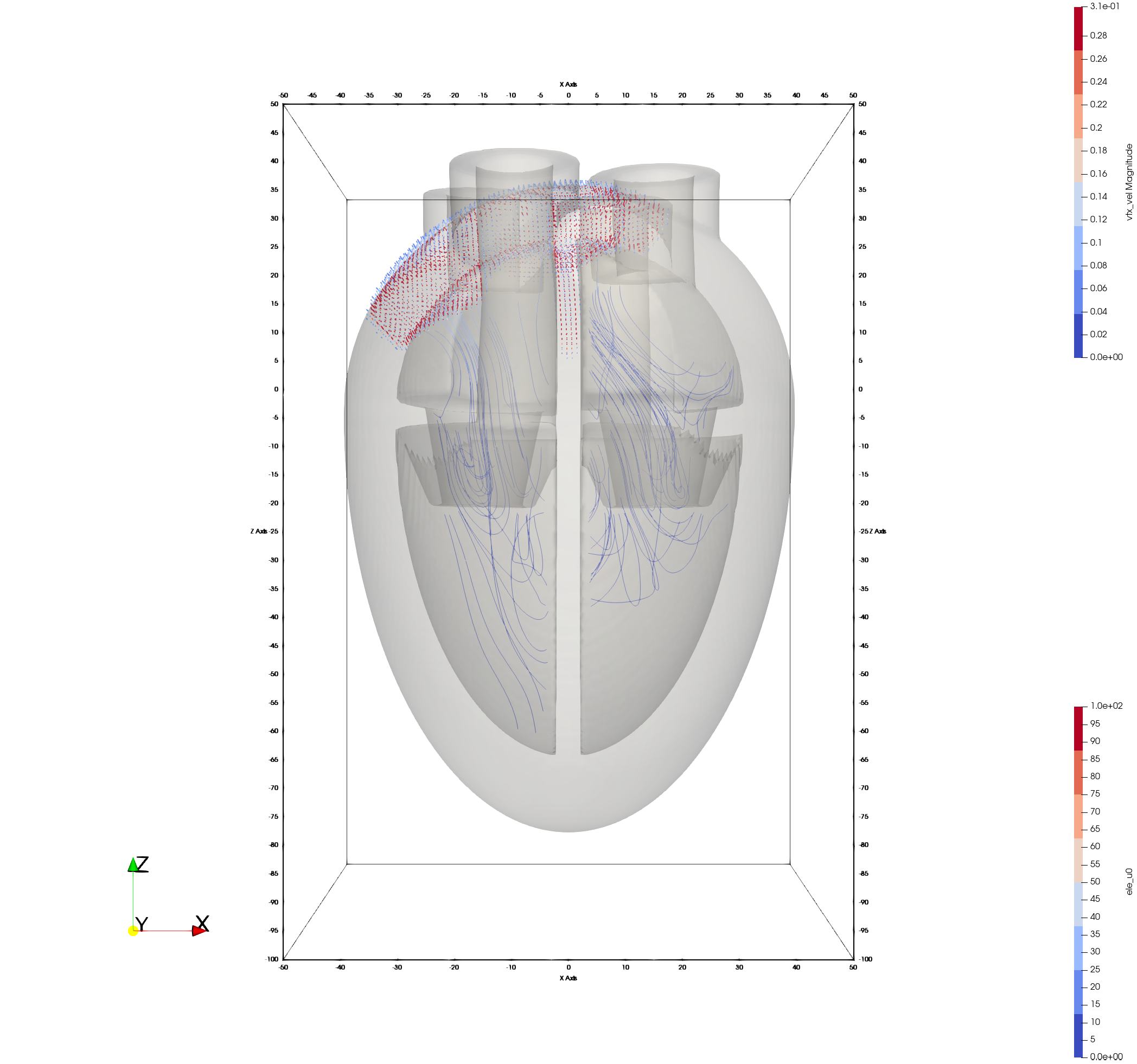}
\includegraphics[width=0.24\textwidth,trim={15cm 4cm 15cm 4cm},clip]{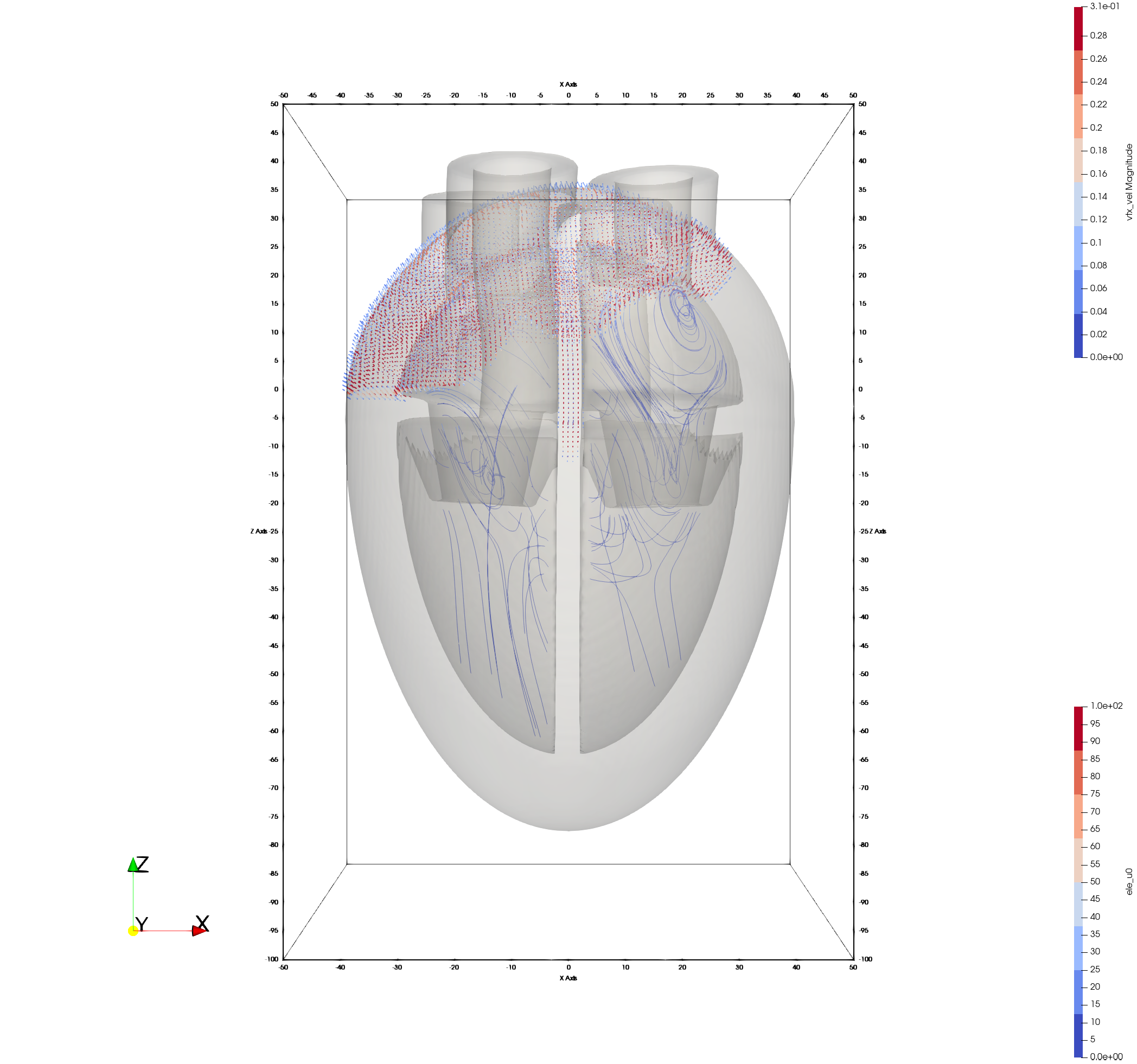}
\includegraphics[width=0.24\textwidth,trim={15cm 4cm 15cm 4cm},clip]{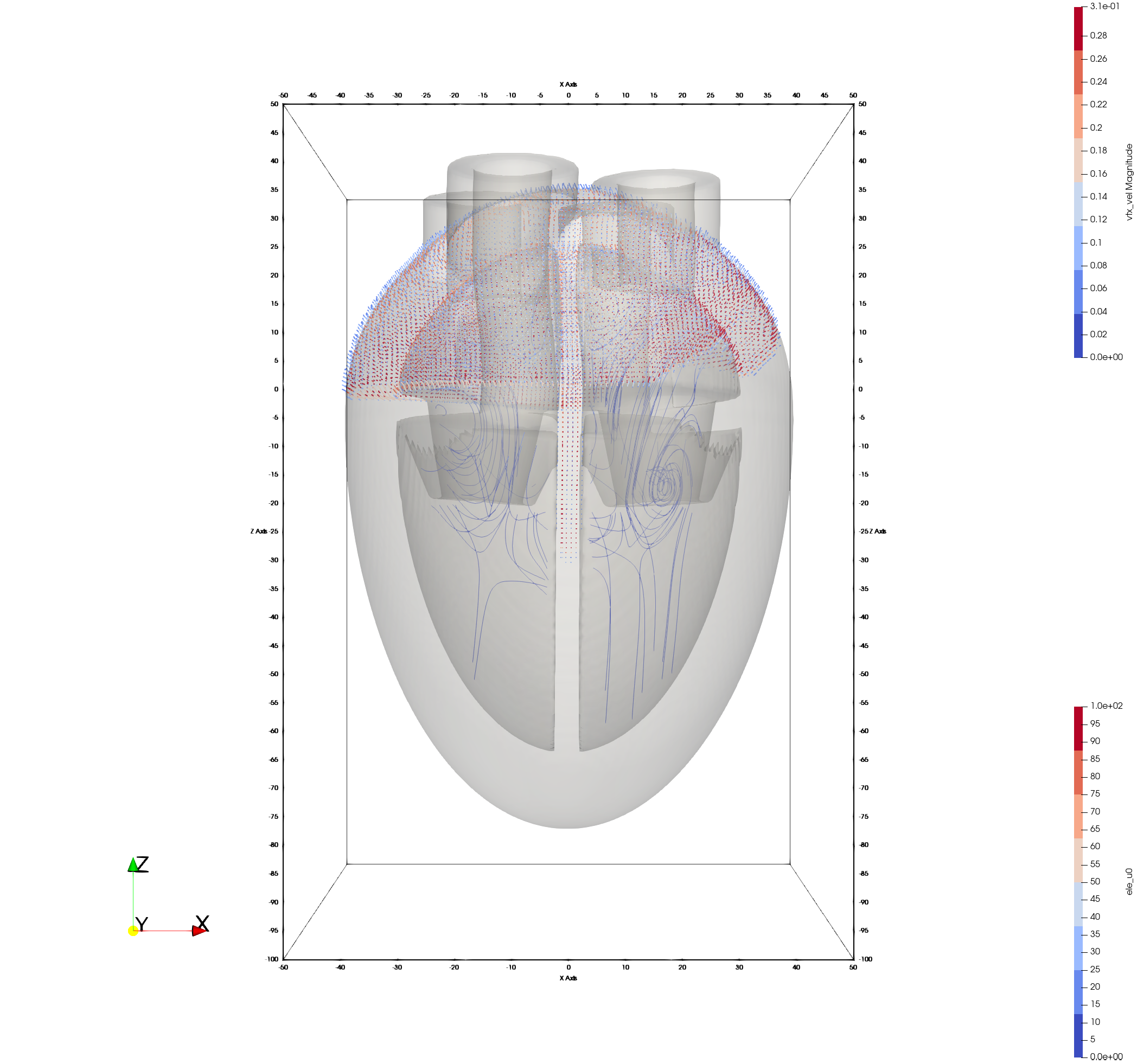}
\includegraphics[width=0.24\textwidth,trim={15cm 4cm 15cm 4cm},clip]{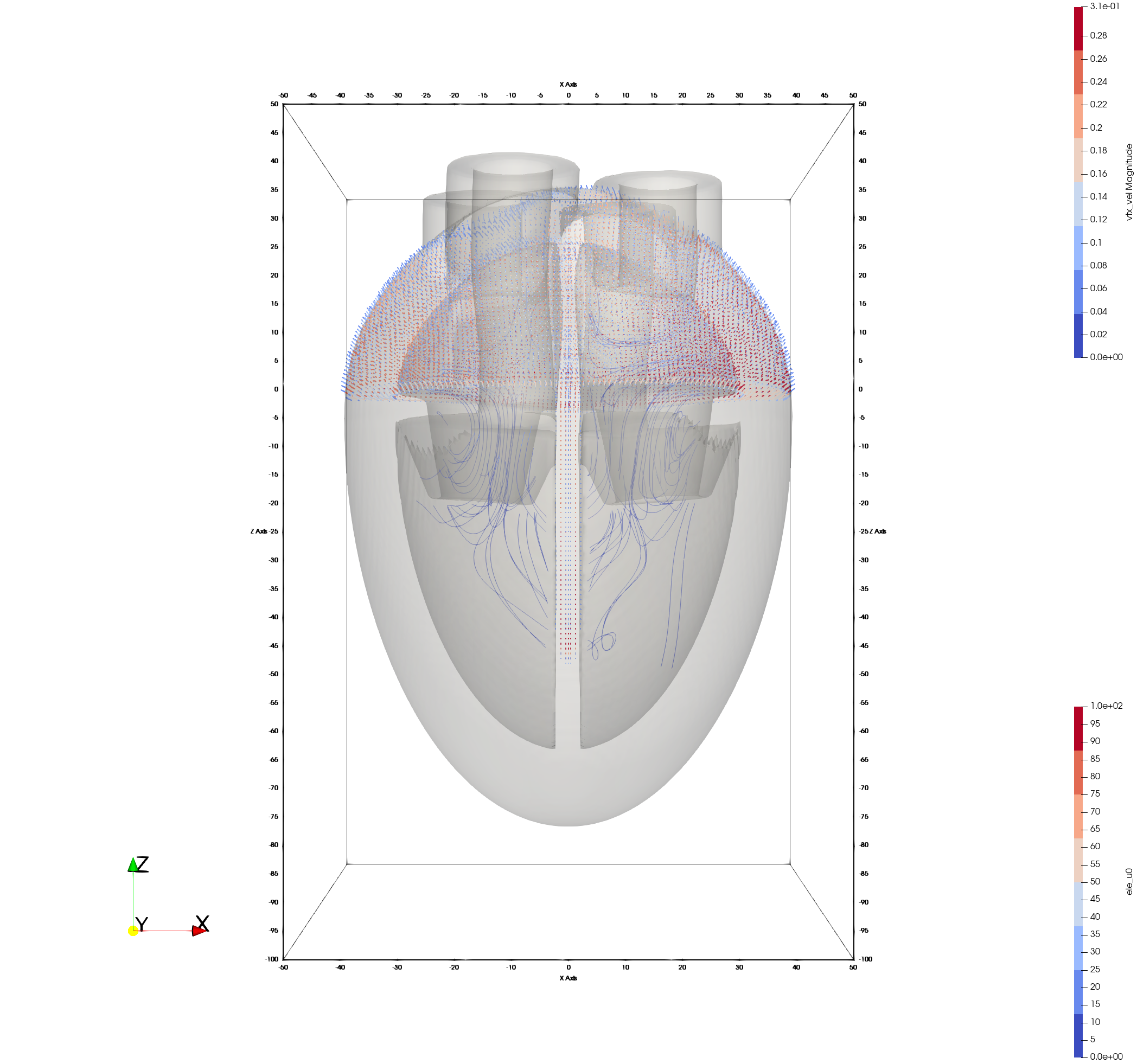}
\includegraphics[width=0.24\textwidth,trim={15cm 4cm 15cm 4cm},clip]{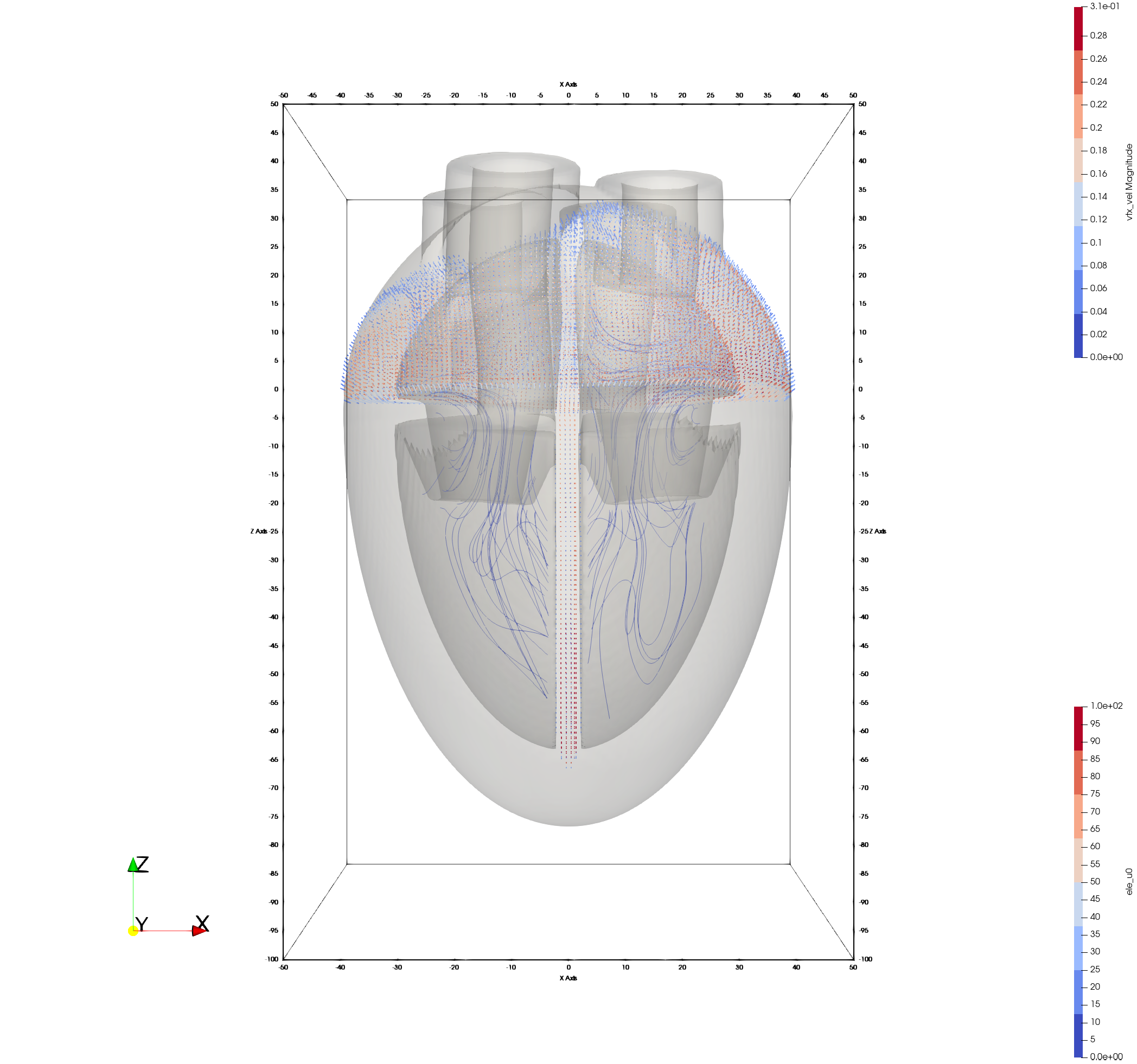}
\includegraphics[width=0.24\textwidth,trim={15cm 4cm 15cm 4cm},clip]{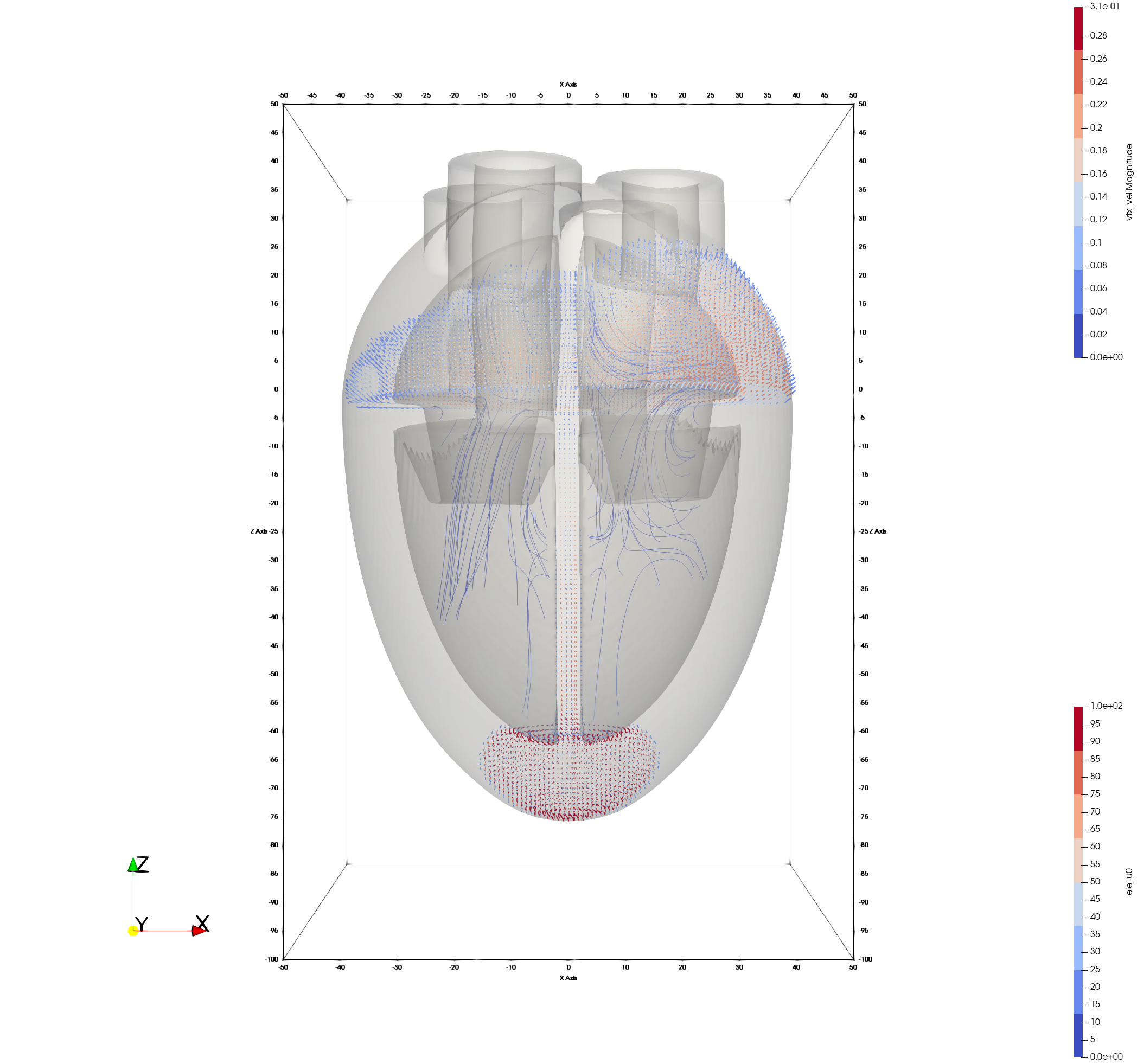}
\includegraphics[width=0.24\textwidth,trim={15cm 4cm 15cm 4cm},clip]{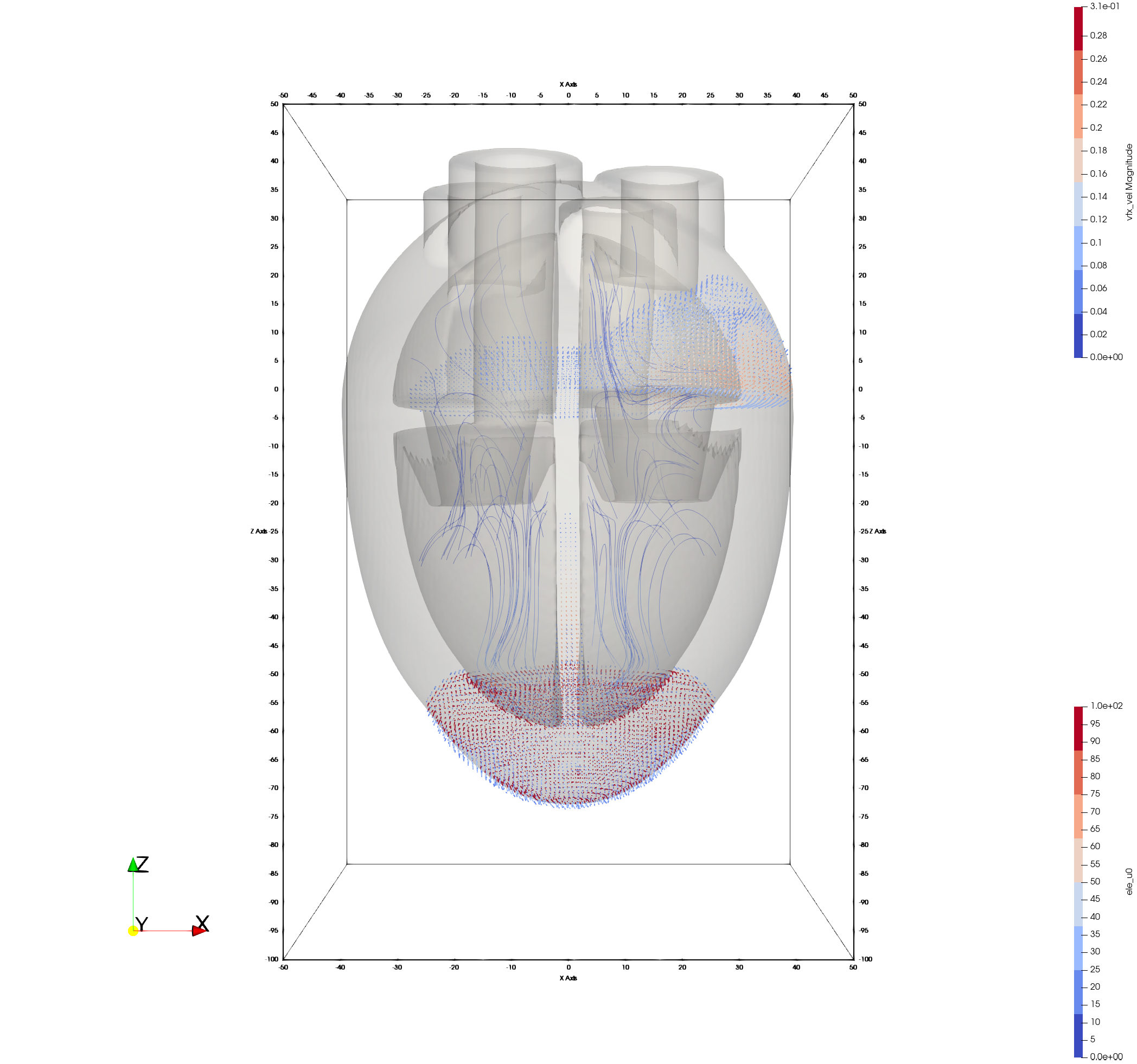}
\includegraphics[width=0.24\textwidth,trim={15cm 4cm 15cm 4cm},clip]{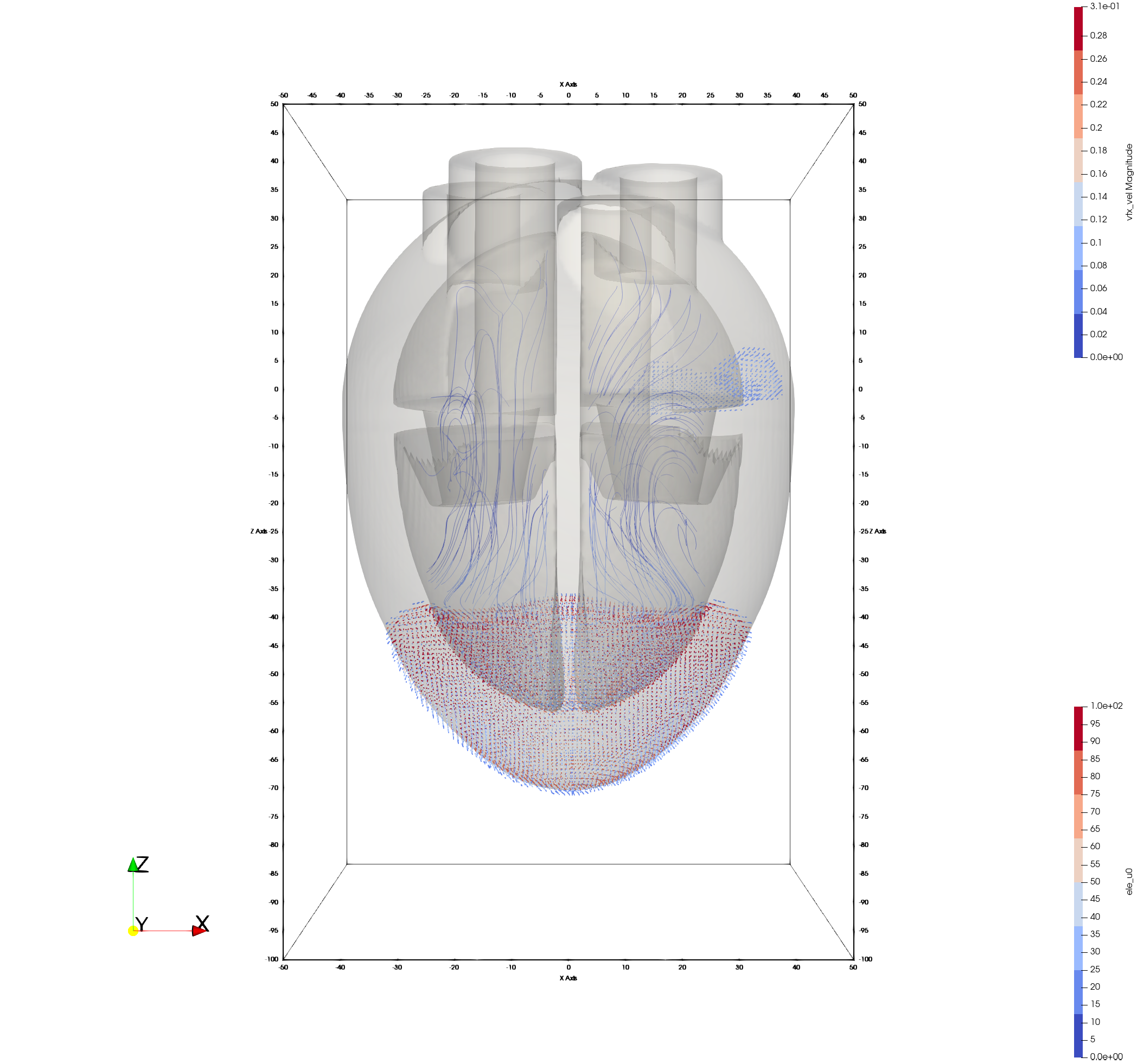}
\includegraphics[width=0.24\textwidth,trim={15cm 4cm 15cm 4cm},clip]{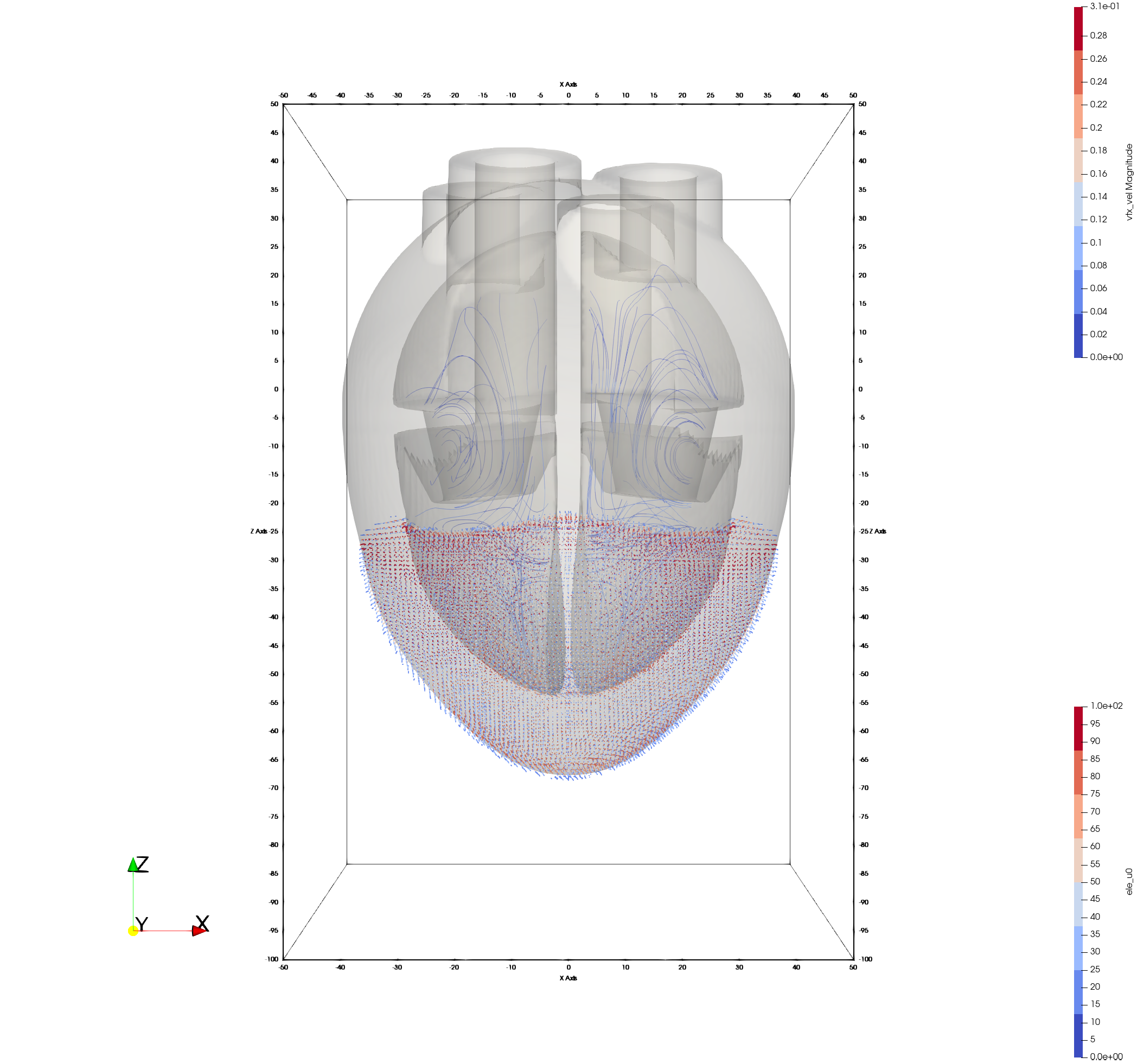}
\includegraphics[width=0.24\textwidth,trim={15cm 4cm 15cm 4cm},clip]{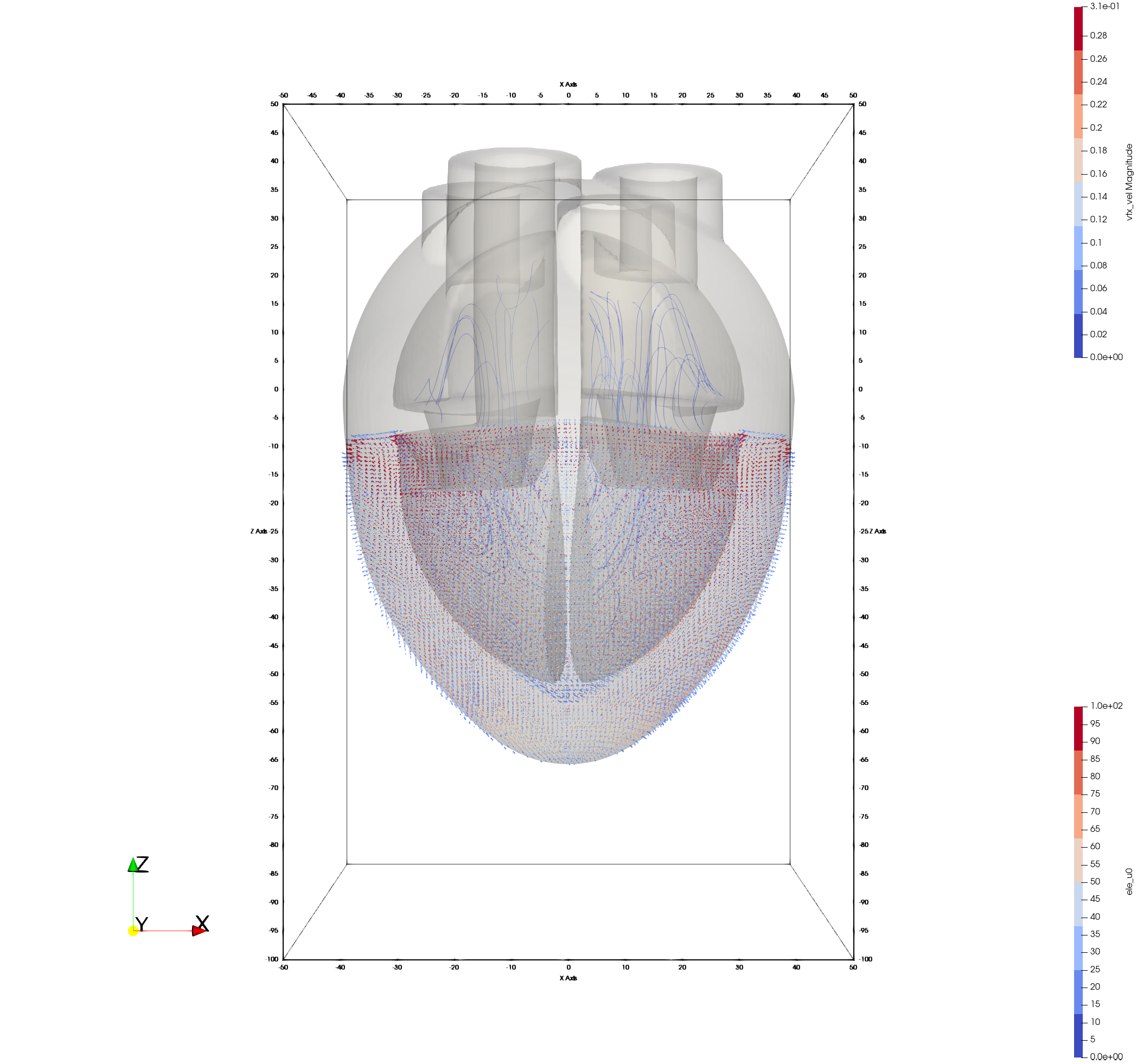}
\includegraphics[width=0.24\textwidth,trim={15cm 4cm 15cm 4cm},clip]{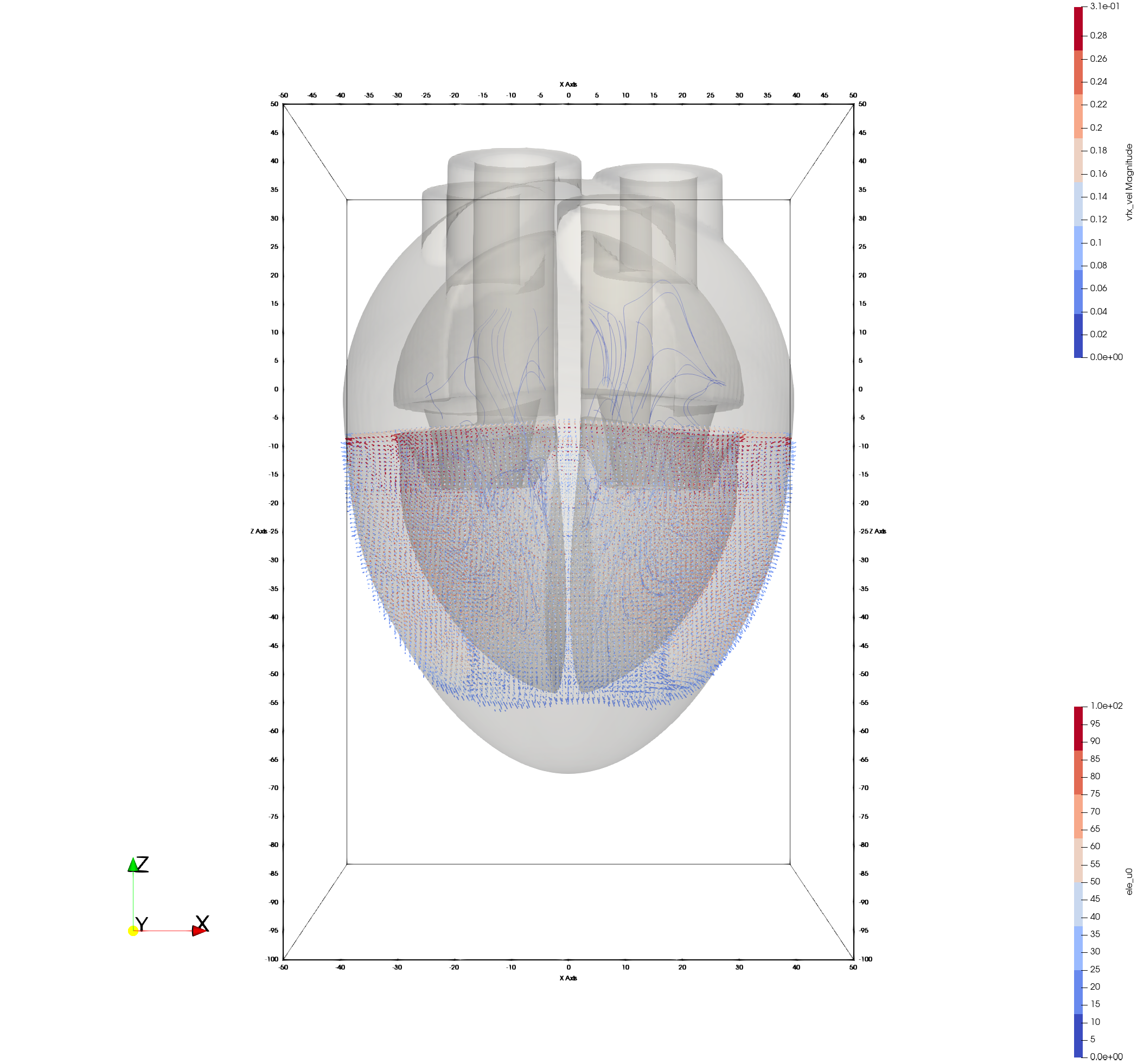}
\includegraphics[width=0.24\textwidth,trim={15cm 4cm 15cm 4cm},clip]{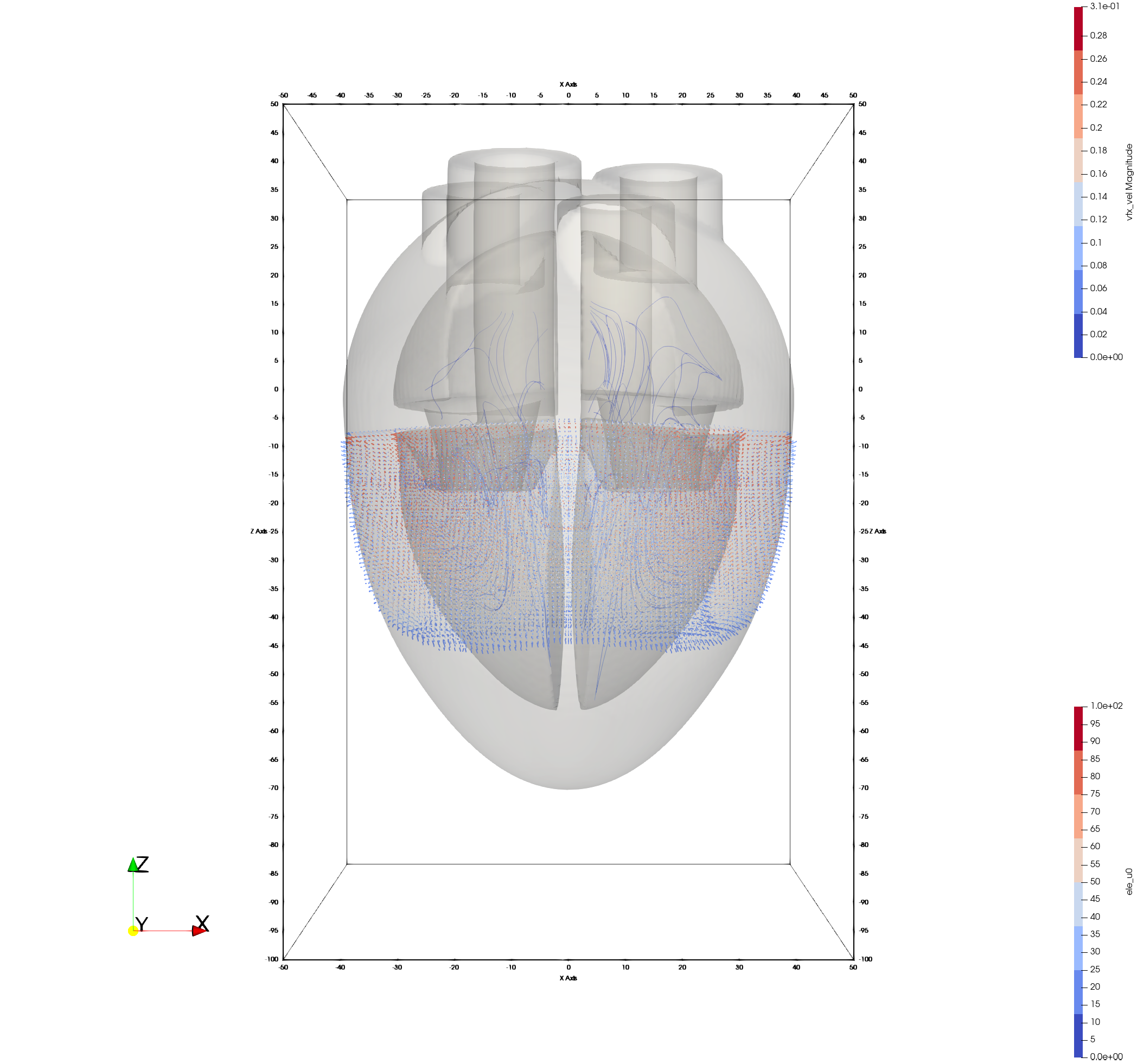}
\includegraphics[width=0.24\textwidth,trim={15cm 4cm 15cm 4cm},clip]{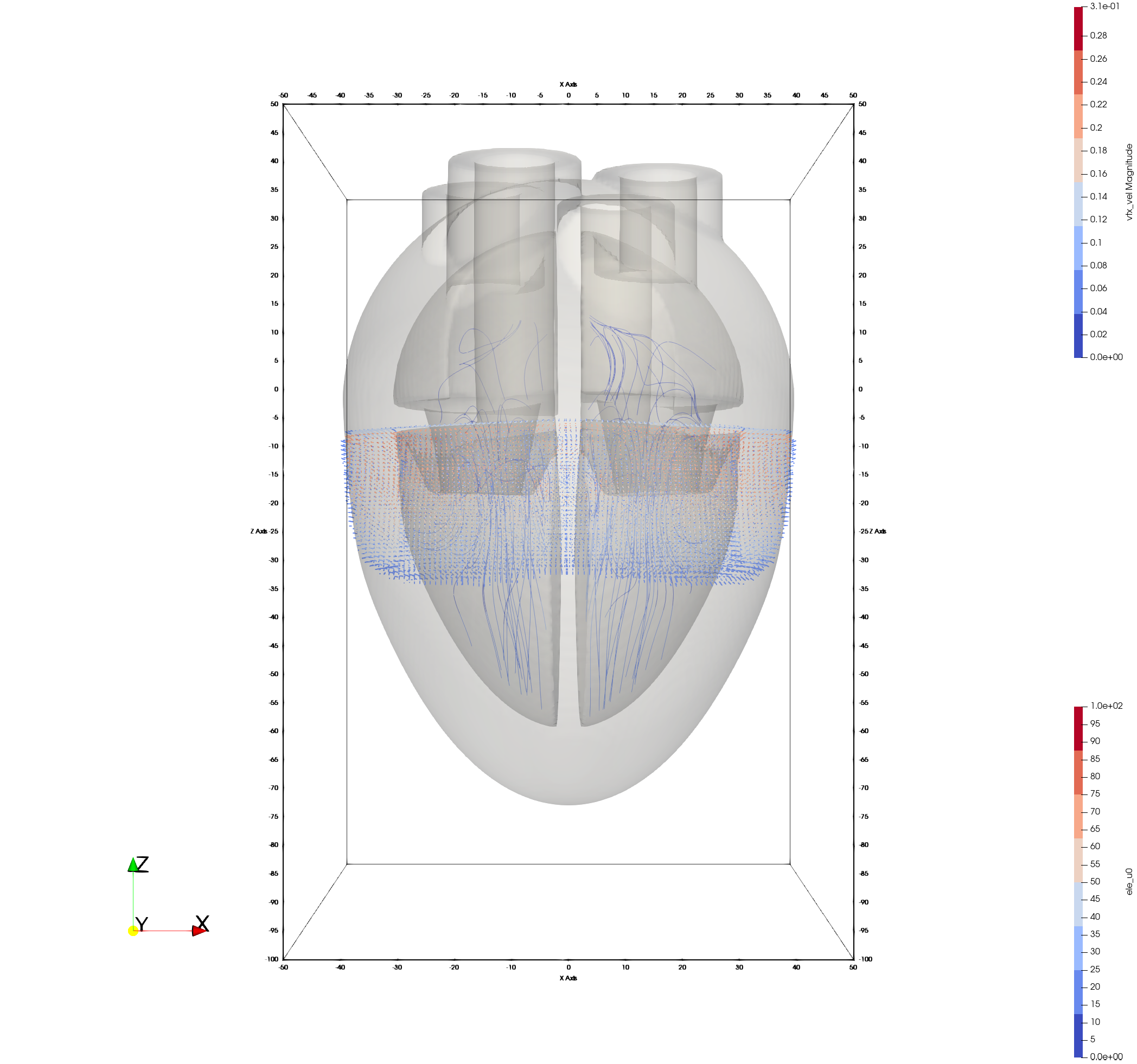}
\includegraphics[width=0.24\textwidth,trim={15cm 4cm 15cm 4cm},clip]{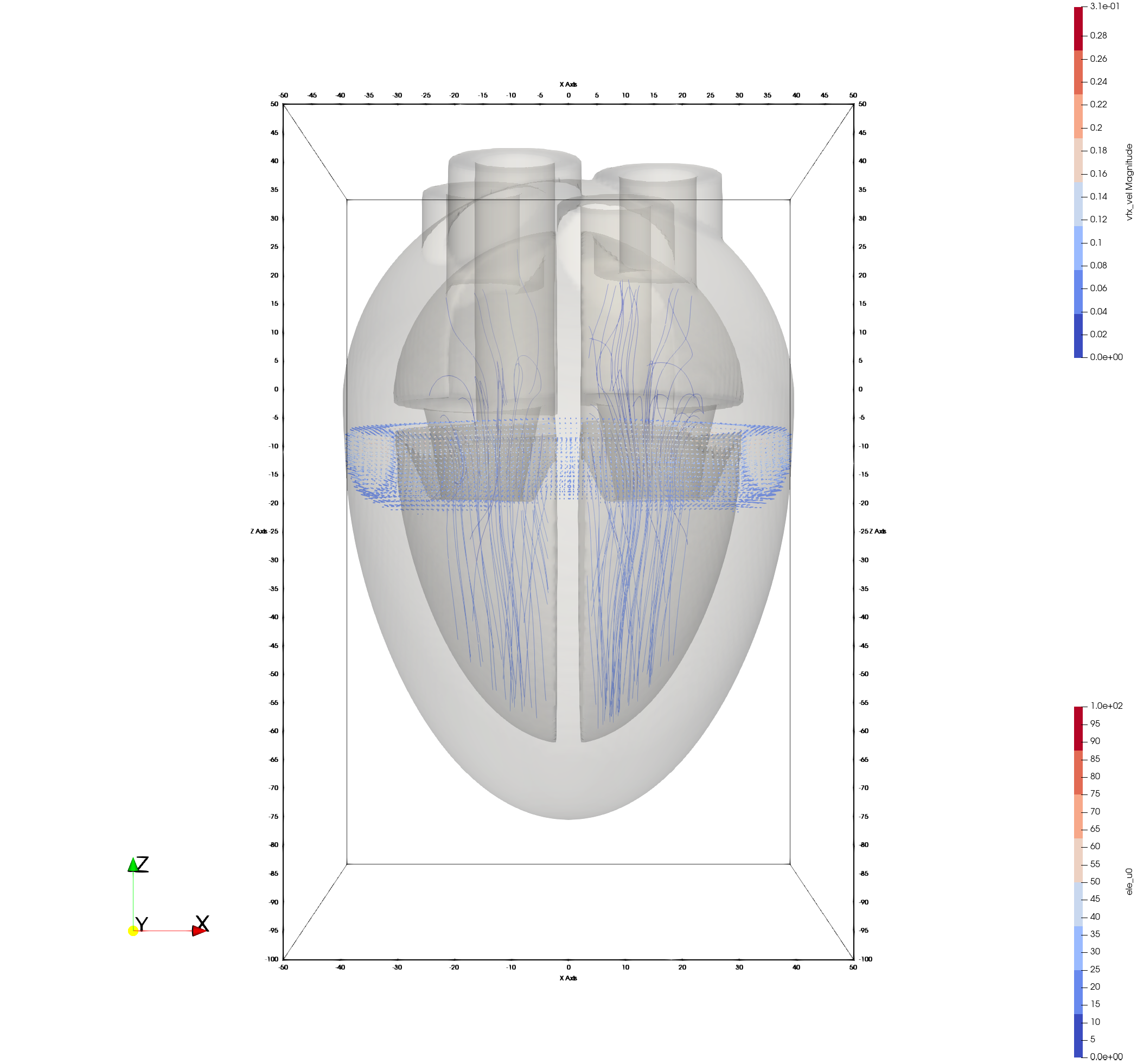}
\includegraphics[width=0.24\textwidth,trim={15cm 4cm 15cm 4cm},clip]{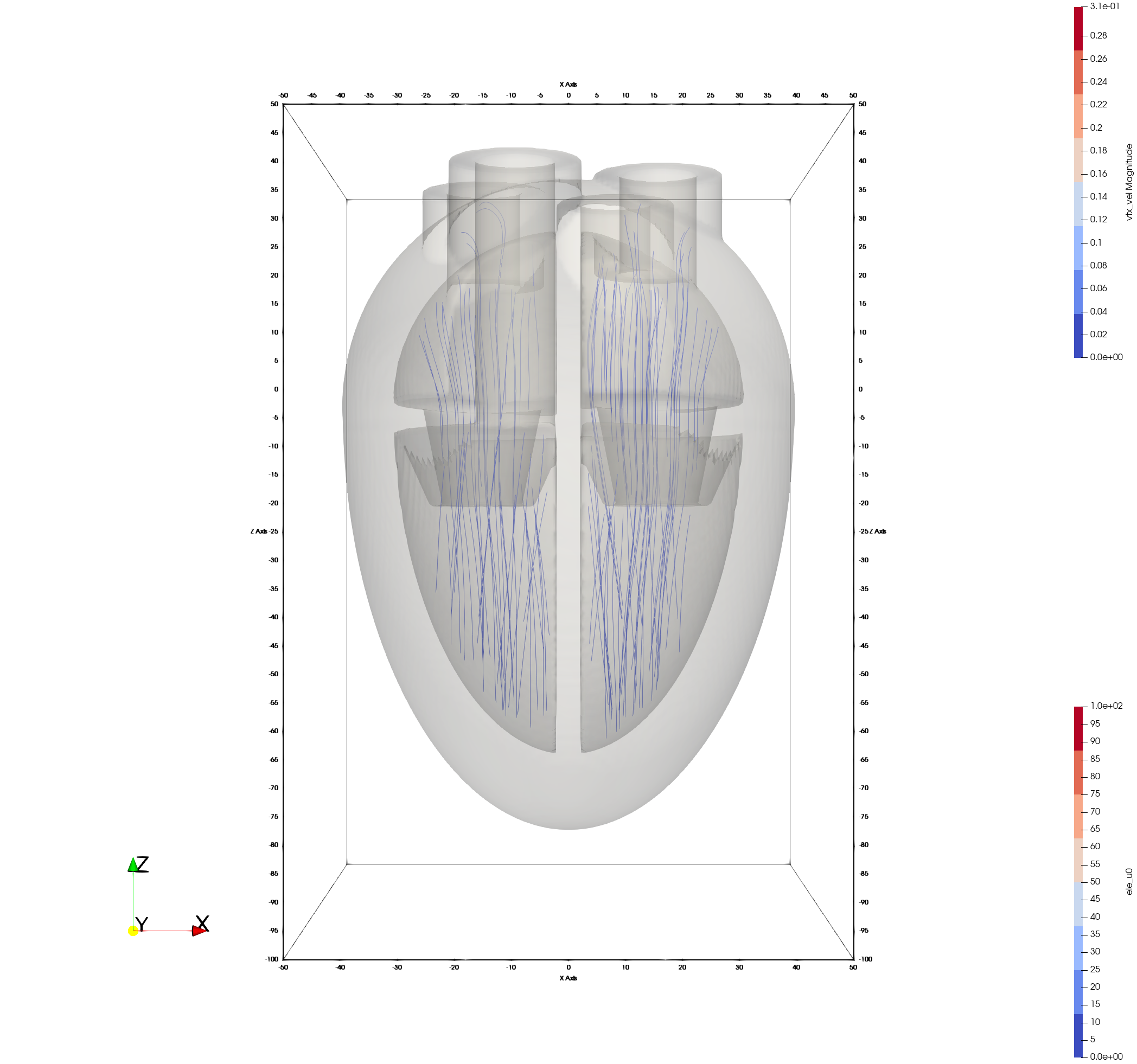}
\caption{The evolution of the whole heart simulation at intervals of 50ms ($\delta x = 1.0$mm).}
\label{fig:whole1}
\end{figure}

The whole heart simulation includes a fully-coupled representation of electrophysiology, solid and fluid dynamics. The algorithm was executed on a single NVIDIA A100 PCIE 40GB GPU at resolution $\delta x = 1.0$mm and $\delta t=0.001$ms. The simulation time was 6081 seconds or approximately 100 minutes. The results are shown in Figure \ref{fig:whole1}, with streamlines indicating the fluid velocity field.

The output files contain a complete description of position, velocity, membrane voltage and pressure throughout the rectangular domain. They capture the basic elements of the heartbeat, specifically a two-stage contraction of atria, followed by ventricles, and the resulting unidirectional flow of blood through the heart.  Due to the absence of valve functionality there is a considerable reverse flow of blood from ventricle to atrium during ventricular systole.

\subsubsection{PV Loop}

For the calculation of Pressure and Volume in the ventricle, a region of voxels were selected and tagged within the left ventricle. As the software generated data the voxels were counted and the pressure and volumes extracted.  The volume was then divided by the number of voxels to give average values.
 
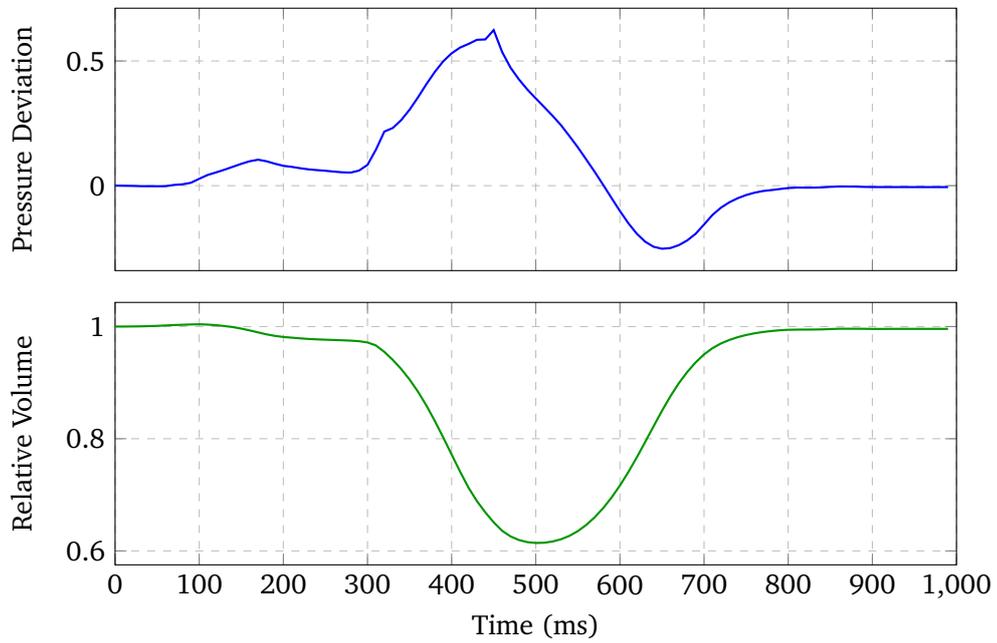
\begin{figure}[h!]
\centering
\begin{tikzpicture}
  \begin{groupplot}[
      group style={
        group name=my plots,
        group size=1 by 2,
        vertical sep=12pt
      },
	width=\textwidth,
	height=0.4\textwidth,
    xmajorgrids=true,
    ymajorgrids=true,
    grid style=dashed,
	xmin=0,
    xmax=1000,
    ]
    \nextgroupplot[ylabel={Pressure Deviation},xticklabels=\empty] 
    \addplot [color=blue  ,thick] table[x expr={\thisrowno{0}*10}, y index=3, col sep=comma] {pv/pv1.txt};
    \nextgroupplot[ylabel={Relative Volume},xlabel={Time (ms)}] 
	\addplot [color=green  ,thick] table[x expr={\thisrowno{0}*10}, y index=5, col sep=comma] {pv/pv1.txt};
  \end{groupplot}
\end{tikzpicture}
\caption{Deviation in Average Pressure and Relative Volume over a 1000ms simulation of the heartbeat.}
\label{fig:pv1}
\end{figure}

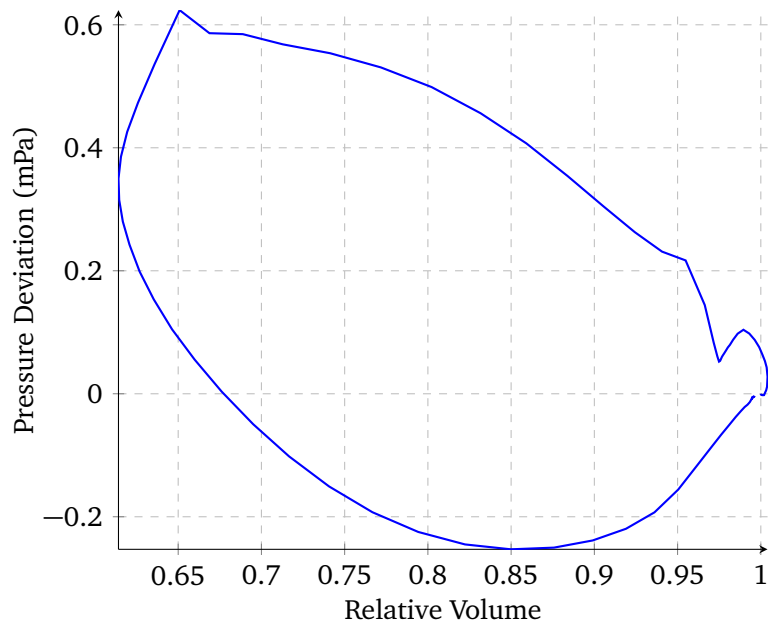
\begin{figure}[h]
\centering
\begin{tikzpicture}
\begin{axis}[
	width=0.8\textwidth,
    xlabel={Relative Volume},
    ylabel={Pressure Deviation (mPa)},
    legend pos=south east,
    legend style={font=\footnotesize},
    xmajorgrids=true,
    ymajorgrids=true,
    grid style=dashed,
    axis x line = bottom,
    axis y line = left]
\addplot [color=blue  ,thick] table[x index=5, y index=3, col sep=comma] {pv/pv1.txt};
\end{axis}
\end{tikzpicture}
\caption{Pressure-Volume Loop derived from the data in Figure \ref{fig:pv1}.}
\label{fig:pv2}
\end{figure}

The results are shown plotted against time in Figure \ref{fig:pv1} over 100 frames, each representing 10ms of simulation. They were then combined to plot a PV loop as shown in Figure  \ref{fig:pv2}.  From the plot it can be seen that the PV loop displays the basic characteristics of the behaviour of the ventricle. An increase in pressure followed by an increase in volume corresponds to the filling of the ventricle from the atrium. Then the decrease in volume coupled to an increase in pressure associated with ventricular contraction and expulsion.  The results are by no means physiologically accurate. This is partly due to the absence of valve function from the simulation, and also due to the damped and compressible fluid behaviour which results form the incomplete solution of the Poisson pressure equation.

\FloatBarrier

Again, the simulation was computed on a single NVIDIA A100 PCIE 40GB GPU at resolution $\delta x = 1.0$mm and $\delta t=0.001$ms. The simulation time was 6081 seconds or approximately 100 minutes.

\subsubsection{Simulation Times}

The following tables summarise whole heart simulation times for various resolutions and processors.  In each case the total simulation time was 1000ms (1 second) and the software output a set of files containing a complete description of the heart at 10ms intervals (100 sets of files in total). The column labelled $k$ is the number of poisson iterations per time step $\delta t$.

We also performed a simulation on a single node Piz Daint at the Swiss national supercomputing centre.  The machine has a faster file system but less powerful NVIDIA Tesla P100 16GB GPUs.  The timings were slightly slower than the NVIDIA A100 of the USI ICS cluster and so the results are not shown.

\begin{table}[h]
\def\arraystretch{1.2}
\begin{center}
\begin{tabular}{crcccc}
\multicolumn{5}{l}{\textbf{}} \\
\textbf{Dimensions} & \textbf{Elements} & \textbf{$\delta x$(mm)} & \textbf{$\delta t$(ms)} & \textbf{Poisson $k$} & \textbf{Time(mm:ss)}\\
\hline
50$\times$50$\times$75 		& 187,500	& 2.0 &	0.01 &	10 & 11:21\\
100$\times$100$\times$150 	& 1,500,000	& 1.0 &	0.01 &	10 & 68:02\\
\end{tabular}
\end{center}
\caption{AMD Radeon Pro 580 (2017 iMac).}
\end{table}

\begin{table}[h]
\def\arraystretch{1.2}
\begin{center}
\begin{tabular}{crcccc}
\multicolumn{5}{l}{\textbf{}} \\
\textbf{Dimensions} & \textbf{Elements} & \textbf{$\delta x$(mm)} & \textbf{$\delta t$(ms)} & \textbf{Poisson $k$} & \textbf{Time(mm:ss)}\\
\hline
50$\times$50$\times$75 		& 187,500	& 2.0 &	0.010 &	10 & 01:16\\
100$\times$100$\times$150 	& 1,500,000	& 1.0 &	0.010 &	10 & 09:57\\
\end{tabular}
\end{center}
\caption{NVIDIA A100 PCIE 40GB (USI ICS Cluster).}
\end{table}

%

\part{Conclusion}

\chapter{Review \& Outlook}\label{chp:rev}

\section{Review}

We have given details of a mathematical model and computational algorithm that can provide a complete description of a single heartbeat, along with physiological metrics such as and ECG and PV loop.  The computational results are available within seconds for a small model of electrophysiology up to around one hour for a complete model, using only a single GPU processor.

The design of the mathematical model and its discretisation is made with computational constraints in mind. It is based upon classically valid mathematical approaches and principles,  present in textbooks rather than journal papers. There is a deliberate avoidance of complicated mathematics which gives a benefit in simplicity, both to the user and the processor. The algorithm is both mesh and matrix-free, relying on assembly per vertex or element as required. As a result the design is readily configurable and extensible to include more complex geometry or mathematical models in future.

The code is extremely short, and has no software or hardware dependence. There is no make file, and the compile line:
\begin{equation}
 \texttt{gcc mesh.c ocl.c main.c -lOpenCL -lm}
\end{equation}
references only the C source code, OpenCL and the native C mathematics library. There are functions for data retrieval, calculus and linear algebra, as well as the signed distance functions of the geometry.  For each of the operators there is a single computational kernel, such as for the update of membrane potential, diffusion of electrical charge, the calculation of divergence of stress etc.

The simplicity of the code contributes to speed of execution and low consumption of computational and human resources, and ultimately time and money. As a result it can be used iteratively by a single competent user, for repeated experiments, parameter fitting or the training of Artificial Neural Networks (ANN). It thus has application to both academic research and clinical use. The configurable and extensible nature allows it to be adapted to different pathologies, interventions and ultimately individual subjects.

The result of this work is therefore a proof of concept of a novel computational approach to cardiac simulation which is practical, not theoretical.

\section{Assesment}

The work is at a very early stage.  The aim was to arrive at a complete working model, accepting compromises along the way, without losing sight the ultimate objective. As a result there are many sources of error, but the process of completing the model allows us to consider these errors and to make judgements about the relative costs and benefits associated with their correction. 

Again it is important to note that we are not presenting the cardiac simulation itself as state-of-the-art, but showing that it is possible to make a cardiac simulation with this novel computational and algorithmic approach.

The following sections give a brief discussions of the strengths and weaknesses of the work, grouping them together according to the structure of the algorithm.

\subsection{Electrophysiology}

The electrophysiology problem is a \textit{reaction-diffusion} equation handled via operator splitting. The Mitchell-Schaeffer ionic membrane model is highly non-linear and as a result is updated using an Explicit Euler time integration. The diffusive mono- and bi-domain models are integrated with Implicit Euler time integration.

The stiffness of the explicit problem demands a relatively short time step, and as a result the linear system for the implicit integration is dominated by the mass matrix.  Thus it is extremely well-conditioned and thus suitable for an iterative solver such as a Jacobi or Richardson iteration.  The properties of these solvers are well-known, removing high frequencies quickly and efficiently, but effectively failing for low frequencies. 

At this point it is important to reconsider the reactive part of the equation.  Charge diffuses to neighbouring tissue and will fire an action potential if it is over the stimulation threshold of the model.  Thereafter the problem is completely changed and dominated by the newly depolarised region.  As a result the diffusion of charge at a distance, corresponding to the low-frequency solution is no longer relevant.  The short time step and local nature of the problem lead to a well-conditioned system, requiring few iterations to reach a solution accurate in high frequencies.

This is an example of how consideration of the dynamic nature of the problem guides the design and efficiency of the algorithm.

\subsection{Mechanics}

The solid mechanics problem is again handled via operator-splitting. A Cauchy stress tensor field dependent upon both solid deformation via the nonlinear Finite Strain Theory, as well as fluid pressure is assembled.  The divergence of the field is then calculated to give a resultant force which is used to update both solid and fluid velocity via the first step of a Leapfrog time integration scheme. This is a semi-implicit step, because it contains the implicit pressure correction provided by the fluid dynamics model. The Leapfrog is symplectic in the phase plane and as such conserves angular momentum, but the time step is limited by stability considerations.

The solid damping and fluid viscosity operators are then applied via an Implicit Euler scheme as for the diffusive part of the electrophysiology problem.  Again, they benefit in exactly the same way.  The sort time step and local behaviour of the dynamic problem allow for relatively few iterations of a Jacobi or Richardson solver.

\subsection{Fluid-Stucture Interaction}

There is a theme emerging in the design of the solver.  If we accept that the nonlinear parts of the mathematical model are to be solved in an explicit manner, then we must in turn accept a short time step.  The matrix- and mesh-free nature of the discretisation, combined with the suitability of the GPU for parallel iteration make this feasible. This in turn benefits the solution of the linear parts of the model.

The fluid-structure interaction problem takes this to its extreme.  In order to know the movement of the surface it is necessary to solve the entire fluid dynamics problem to gain a reaction force.  This in turn requires a knowledge of the solid velocity.  As such there is a huge, highly nonlinear problem right at the centre of the entire cardiac simulation. Its solution dominates all other computational effort. The decision in design was to handle this problem in an explicit manner.  This in turn limits the maximum time step for the entire model.

There is a benefit which comes from the the choice of Chorin projection in combination with Finite Volume Method.  Since we update the solution with an explicit step then solve the Poisson Pressure equation, we are applying a \textit{predictor-corrector} method. Firstly, the short time step provides and excellent initial guess for the pressure field from the previous iteration, and reduces the number of iterations to reach a reasonable solution. Secondly, if we do not make a complete solve of the pressure equation the error in the corrected solution is not lost, but remains due to the conservative nature of FVM into the next time step.  Rather than a uniform pressure field we see a pressure wave travelling through the fluid as the tissue around it contracts.  This is the speed of sound, a synthetic compressibility which enters the fluid model. Other methods, including the half stepping of the semi-implicit formulation of \cite{Formaggia2010} have the same effect. Since the fluid does not correctly store the energy of compression, this is the largest source of error in our work. We hope to address this problem through the application of multi-resolution methods which are discussed in Section \ref{sec:fut}. 

The advection kernel uses an interpolated gradient, and as such is subject to the well-known problem of 'false diffusion', where the solution of the advection problem also contains some mesh-dependent diffusion.  In this case the diffusion is some artificial viscosity, which can be dealt with by underestimating the true viscosity.  At higher mesh resolutions the problem is less significant.

\subsection{GPU Processing}

The use of the GPU itself brings with it some sources of compromise. In general, single precision arithmetic is the standard for graphics processing.  This brings with it the obvious reduction in numerical precision, but also require care in the design calculation to avoid overflow.  This can be seen in the convergence studies, where a finer mesh leads to division by a machine-zero volume. This problem can be overcome but is left in the work to demonstrate that whilst invisible it is always present.

Conversely there is a benefit in speed of processing and memory access associated with the smaller sized single precision variables. The design process aims to provide a tool of practical use, and thus the benefit in performance is seen to outweigh absolute numerical precision, especially since part of its objective is to determine the largely unknown parameters of the model.

The memory capacity of a single GPU processor limits the total size and resolution of the simulation. Since there is a minimum spatial mesh width $\delta x$ in space, there follows an absolute minimum time step $\delta t$. 

The cost of communication across the PCI bus from GPU to CPU is too high to allow multi-processing per time step via MPI. There have been advances in shared memory GPU multiprocessors, such as NVIDIA's NVLINK, but they are also limited in bandwidth. It is better to consider that it would be possible to run thousands of cardiac simulations in parallel on a supercomputer.  As mentioned earlier, this makes the entire simulation an objective function for a set of optimisation problems, most notably parameter estimation. 

\section{Future Work}\label{sec:fut}

The title of the work states its aim, to provide a set of tools for cardiac simulation. The extensibility and simplicity of the code should allow for researchers in this field and others to adapt these techniques to their own work.

The development of the software has given insight into the behaviour of the tissue surrounding the heart and its potential contribution to AF and tachycardia. This will be investigated.

The entire design of the discretisation and solver make them ideal for multi-level solution.  The current problem can be decomposed into many resolutions on both space and time.  The representation of geometry via signed distance functions and the mesh-free assembly allow the entire problem to be reconstructed fully at any mesh size.

In order to generate a more physiologically accurate PV loop the design of valves must be improved.  This can be done either physically or through the introduction of diodes into the fluid dynamics calculation.

The software allows for extension and refinement to include richer and more precise mathematical models.  It should be possible in future to validate the simulation itself against physiological indicators.  At this point the tools may be able to make a valid contribution to research and clinical practice.


\bibliographystyle{apalike}
\bibliography{bib1}

\end{document}